\documentclass{article}

\usepackage{arxiv}

\usepackage[utf8]{inputenc} 
\usepackage[T1]{fontenc}    
\usepackage{hyperref}       
\usepackage{url}            
\usepackage{booktabs}       
\usepackage{amsfonts}       
\usepackage{nicefrac}       
\usepackage{microtype}      
\usepackage{lipsum}		
\usepackage{graphicx}
\usepackage[square,sort,numbers]{natbib}
\usepackage{doi}
\usepackage{graphicx}
\usepackage{hyperref}
\usepackage{amsmath}
\usepackage{amsthm}
\usepackage{amssymb}
\usepackage{amsfonts}
\usepackage{bm}
\usepackage{epsdice}
\usepackage{algorithm}
\usepackage[]{algorithmic}
\usepackage{subfig}
\usepackage{bbding}
\usepackage{makecell}

\newtheorem{definition}{Definition}

\usepackage{tabularx}
\usepackage{multirow}
\usepackage{float} 
\usepackage[table,dvipsnames]{xcolor}

\usepackage{colortbl}
\usepackage{soul}

\newcolumntype{C}[1]{>{\centering\arraybackslash}p{#1}}
\newcolumntype{L}[1]{>{\raggedright\arraybackslash}p{#1}}

\newcommand{\thickhline}{\Xhline{3\arrayrulewidth}}
\newcommand{\nashhighlight}{\cellcolor{SpringGreen}}
\newcommand{\cnashhighlight}{\cellcolor{cyan}}
\newcommand{\lehighlight}{\cellcolor{pink}}

\title{Preference Communication in Multi-Objective Normal-Form Games\thanks{Part of this work was carried out by the first author for his master thesis \cite{ropke2021thesis} under the supervision of the other authors. Some preliminary results in this article were presented in the Adaptive and Learning Agents Workshop 2021 \cite{ropke2021communication}.}}

\author{ \href{https://orcid.org/0000-0001-5045-6127}{\includegraphics[scale=0.06]{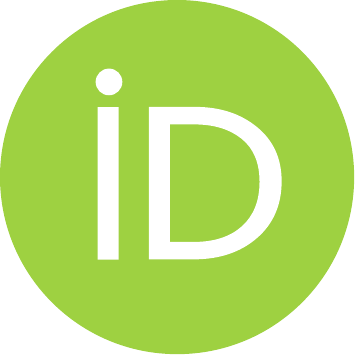}\hspace{1mm}Willem R\"{o}pke \Envelope}\\
	Artificial Intelligence Lab \\
	Vrije Universiteit Brussel, Belgium \\
	\texttt{willem.ropke@vub.be} \\
	\And
	Diederik M.\ Roijers\\
	Artificial Intelligence Lab \\
	Vrije Universiteit Brussel, Belgium \&\\
	Microsystems Technology \\
	HU~University~of~Applied~Sciences~Utrecht The~Netherlands\\
	\texttt{diederik.yamamoto-roijers@hu.nl} \\
	\And
	Ann Now\'{e} \\
	Artificial Intelligence Lab \\
	Vrije Universiteit Brussel, Belgium \\
	\texttt{ann.nowe@vub.be} \\
	\And
	Roxana R\u{a}dulescu \\
	Artificial Intelligence Lab \\
	Vrije Universiteit Brussel, Belgium \\
	\texttt{roxana.radulescu@vub.be} \\
}

\renewcommand{\shorttitle}{\textit{arXiv} Template}

\hypersetup{
pdftitle={Preference Communication in Multi-Objective Normal-Form Games},
pdfsubject={cs.MA, cs.LG},
pdfauthor={Willem R\"{o}pke, Diederik M.\ Roijers, Ann Now\'{e}, Roxana R\u{a}dulescu},
pdfkeywords={Multi-Agent Reinforcement Learning, Multi-Objective Optimization, Nash Equilibrium},
}

\begin{document}
\maketitle

\begin{abstract}
We consider preference communication in two-player multi-objective normal-form games. In such games, the payoffs resulting from joint actions are vector-valued. Taking a utility-based approach, we assume there exists a utility function for each player which maps vectors to scalar utilities and consider agents that aim to maximise the utility of expected payoff vectors. As agents typically do not know their opponent's utility function or strategy, they must learn policies to interact with each other. Inspired by Stackelberg games, we introduce four novel preference communication protocols to aid agents in arriving at adequate solutions. Each protocol describes a specific approach for one agent to communicate preferences over their actions and how another agent responds. Additionally, to study when communication emerges, we introduce a communication protocol where agents must learn when to communicate. These protocols are subsequently evaluated on a set of five benchmark games against baseline agents that do not communicate. We find that preference communication can alter the learning process and lead to the emergence of cyclic policies which had not been previously observed in this setting. We further observe that the resulting policies can heavily depend on the characteristics of the game that is played. Lastly, we find that communication naturally emerges in both cooperative and self-interested settings.
\end{abstract}

\keywords{Multi-Agent Reinforcement Learning \and Multi-Objective Optimization \and Nash Equilibrium}

\section{Introduction}
\label{sec:introduction}
The field of Multi-Agent Reinforcement Learning (MARL) enables learning agents to optimise their strategies by interacting with the environment as well as other agents \cite{nowe2012game}. In recent years, MARL achieved impressive results in domains such as adaptive traffic signal control \cite{mannion2016experimental} and managed to beat top-level humans in complex multi-player games such as StarCraft \cite{vinyals2019grandmaster}. 

A key complicating factor is that it is often needed to optimise for multiple objectives at once. An example can be found in drug design where we aim to find new molecules that are similar to an existing one but with better performance \cite{zhou2019optimization}.

Many real-world problems consist of both multiple agents and multiple objectives. By developing techniques focusing on only one of these factors, we risk oversimplifying these complexities. As an illustrative example, consider a logistical service provider with multiple trucks. These trucks must coordinate how to divide the packages, as this influences their driving routes. In this scenario, each truck has the (possibly conflicting) objectives to deliver all goods as fast as possible to their destination, with the lowest associated cost, while also minimising carbon emissions to minimise its strain on the environment. This results in a combinatorial optimisation problem that can quickly become intractable to plan in a centralised manner. Taking a learning approach, and specifically multi-objective multi-agent, allows trucks to \emph{learn} how to optimise for the different objectives in the presence of the other trucks \cite{mazyavkina2021reinforcement}.

In this article, we study two-player Multi-Objective Normal-Form Games (MONFGs), which are stateless games with multiple objectives. Following a utility-based perspective, we assume that every agent has a possibly non-linear utility function that maps vectors to scalar utilities \cite{roijers2013survey}. Previous work in this area has shown that in settings where agents aim to maximise the utility from expected vectorial payoffs, no Nash equilibria need exist under non-linear utility functions \cite{radulescu2020utility}. We study this setting further as it remains to be determined how agents can efficiently learn when Nash equilibria are present as well as what behaviour is learned when these equilibria do not exist.

We design a set of preference communication protocols, inspired by Stackelberg games, to aid convergence and promote collaboration. Agents learn policies through repeatedly playing a base MONFG and following the protocol which alternates them between being a leader that communicates a certain preference and a follower that responds to this communication. We contribute protocols for cooperative communication, self-interested communication and also examine a hierarchical setting where agents must learn when to communicate. We empirically evaluate our protocols on a set of five two-player MONFGs. These benchmarks include both games with and without Nash equilibria to study the effect of communication under different dynamics. Furthermore, we provide baseline results for independent learners without the ability to communicate to provide the necessary context and comparisons. All code used for this article can be found at \url{https://github.com/wilrop/communication_monfg}. Concretely, we contribute the following:

\begin{enumerate}
    \item We design four novel communication protocols, inspired by Stackelberg games, for two-player MONFGs. Each protocol alternates players between being a leader that communicates a certain preference and being a follower that conditions their strategy on this communication.
    \item We empirically evaluate our algorithms on a set of five benchmarks. We also provide a baseline implementation and evaluation of agents that are unable to communicate.
    \item We find that agents that are looking to cooperate receive a boost in convergence rate when employing preference communication.
    \item We show that self-interested agents can end up cycling through an optimal set of stationary policies. This leads us to consider cyclic Nash equilibria as a relevant solution concept. 
    \item We show that in a setting where agents learn when to communicate, cyclic policies can arise for both cooperative and self-interested agents. 
    \item We find that communication emerges both in cooperative as well as self-interested settings and that the final communication policy is determined by the learning rate and the characteristics of the underlying game.
\end{enumerate}

The remainder of this article is structured as follows. In Section \ref{sec:background}, we present the necessary background. Section \ref{sec:preference-communication} introduces the novel preference communication protocols and provides an analysis of each protocol in terms of target behaviour. Next, Section \ref{sec:experiments} provides an empirical evaluation of each proposed protocol on a set of benchmark games and discusses the results. In Section \ref{sec:related-work} we discuss related work in adjacent fields and highlight the relevant state of the art. Lastly, Section \ref{sec:conclusion} provides a conclusion and discusses possible directions for future work.

\section{Background}
\label{sec:background}

\subsection{Multi-Objective Normal-Form Games}
A common model of interactions between rational actors is the (single-objective) Normal-Form Game (NFG). In these games, each player receives a scalar payoff based on the joint strategy of all players. Intuitively, a Multi-Objective Normal-Form Game (MONFG) is the multi-objective counterpart to NFGs where agents now receive a vectorial payoff consisting of a payoff for each objective. We formally define an MONFG as follows:  

\begin{definition}[Multi-objective normal-form game]
\label{def:MONFG}
A (finite, n-player) multi-objective normal-form game is a tuple $(N, \mathcal{A}, \bm{p})$, with $d \geq 2$ objectives, where: 
\begin{itemize}
    \item $N$ is a finite set of $n$ players, indexed by $i$;
    \item $\mathcal{A} = A_1 \times \dots \times A_n$, where $A_i$ is a finite set of actions available to player $i$. Each vector $a = (a_1, \dots , a_n) \in \mathcal{A}$ is called an action profile;
    \item $\bm{p} = (\bm{p}_1, \dots , \bm{p}_n)$ where $\bm{p}_i : \mathcal{A} \to \mathbb{R}^d$ is the vectorial payoff function for player $i$, given an action profile.
\end{itemize}
\end{definition}

In the case of two player MONFGs, we can represent such games in the form of a matrix as shown in Table \ref{tab:monfg}. In this example, if the row player opts to play action A and the column player opts for action B, the row player receives a payoff vector of $(0, 1)$ while the column player obtains $(1, 0)$.

\begin{table}[h]
\centering
\begin{tabular}{lll}
                       & \multicolumn{1}{c}{A}       & \multicolumn{1}{c}{B}       \\ \cline{2-3} 
\multicolumn{1}{l|}{A} & \multicolumn{1}{l|}{(1, 1); (0, 0)} & \multicolumn{1}{l|}{(0, 1); (1, 0)} \\ \cline{2-3} 
\multicolumn{1}{l|}{B} & \multicolumn{1}{l|}{(1, 0); (0, 1)} & \multicolumn{1}{l|}{(0, 0); (1, 1)} \\ \cline{2-3} 
\end{tabular}
\caption{A matrix representation of a multi-objective normal-form game. Each cell holds the vectorial payoff for both agents under the corresponding action profile.}
\label{tab:monfg}
\end{table}

In general, players are not restricted to deterministically play a single action, known as playing a \emph{pure strategy}. Rather, players may select actions according to a probability distribution over their actions, known as playing a \emph{mixed strategy}.
\begin{definition}[Mixed strategy]
\label{def:mixed-strategy}
Let $(N, \mathcal{A}, \bm{p})$ be an MONFG, and for any set $X$ let $\Pi(X)$ be the set of all probability distributions over $X$. Then the set of mixed strategies for player $i$ is $S_i = \Pi(A_i )$.
\end{definition}
We note that in the context of MARL, strategies are often referred to as policies and denoted by $\pi$. We focus on the learning aspect of agents in multi-objective games and use the terms interchangeably for the remainder of this article.

\subsection{Utility-Based Approach}
In this work, we assume a utility-based approach to deal with the multi-objective nature of our games \cite{roijers2013survey,roijers2017multi,radulescu2020multi}. Concretely, we assume that each agent $i$ can derive a scalar utility from a reward vector by applying their utility function $u_i: \mathbb{R}^d \rightarrow \mathbb{R}$. This implies a total order over vectors. We believe this is a reasonable assumption, as an agent will ultimately have to be able to select a payoff vector out of (a subset of) the available alternatives. 

An intuitive example of a utility function is the linear utility function which assigns a weight $w \in [0, 1]$ to each objective $o$ from the set of objectives $O$ and subsequently sums over these weighted returns as shown below.
\begin{equation}
    u_i\left(\bm{p}_i(\pi)\right) = \sum_{o \in O}w_{i,o}p_{i,o}(\pi)
\end{equation}
with $\bm{p}_i(\pi)$ the result from applying the joint policy $\pi$ to player $i$'s payoff function $\bm{p}_i$ and $p_{i,o}(\pi)$ the specific payoff for objective $o$ from the same payoff function and joint policy.

In general, utility functions can be highly non-linear and are only assumed to be monotonically increasing. This assumption translates to the fact that each agent should always prefer more of an objective over less, given equal rewards for all other objectives. Formally: 
\begin{equation}
(\forall o, p_{i,o}(\pi) \geq p_{i,o}(\pi')) \implies u_i\left(\bm{p}_i\left(\pi\right)\right) \geq u_i\left(\bm{p}_i\left(\pi'\right)\right)
\end{equation}

We assume that all agents are rational such that they aim to maximise their utility as defined by their utility functions. There is however not a unique choice for how to apply the utility function to the payoffs. The literature identifies two alternatives \cite{roijers2013survey}. As an illustration, assume the logistical example that was provided before. One option is that we aim to optimise the average utility that we can derive from a package delivery. In that case, we are optimising for the utility of each individual policy execution, resulting in what is called the Expected Scalarised Returns (ESR) criterion:
\begin{eqnarray}
\label{eq:ESR}
   p_{i} = \mathbb{E}\left[u_i\left(\bm{p}_i\left(\pi\right)\right)\right] = \sum_{a \in \mathcal{A}}u_i\left(\bm{p}_i(a)\right)\prod_{j = 1}^n \pi_j(a_j)
\end{eqnarray}
with $p_{i}$ the expected utility for agent $i$ with utility function $u_i$ and $\bm{p}_i\left(\pi\right)$ the result from applying the joint policy $\pi$ to player $i$'s payoff function $\bm{p}_i$.

Alternatively, it is possible that we aim to optimise the utility we can derive from the average package delivery, in which case we first calculate the expectation over the vectorial returns before scalarising. This is called the Scalarised Expected Returns (SER) criterion:
\begin{eqnarray}
\label{eq:SER}
    p_{i} = u_i\left(\mathbb{E}\left[\bm{p}_i(\pi)\right]\right) = u_i\left(\sum_{a \in \mathcal{A}}\bm{p}_i(a)\prod_{j = 1}^n \pi_j(a_j)\right)
\end{eqnarray}
where $p_{i}$ is now the utility of the expected returns. 

It has been shown that these optimisation criteria can lead to different outcomes in both single-agent as well as multi-agent settings \cite{vamplew2021impact,radulescu2020utility}. Furthermore, recent work has found that an MONFG under ESR with known utility functions can be reduced to a single-objective \emph{trade-off} game \cite{radulescu2020utility}. This implies that regular MARL techniques can be used to solve such problems. The SER criterion on the other hand cannot easily be solved by such techniques and has been understudied thus far \cite{radulescu2020multi}. For these reasons, we concern ourselves with the SER optimisation criterion in this work.


\subsection{Solution Concepts}
\label{sec:solution-concepts}
We consider a number of different solution concepts. Below, we introduce Nash equilibria and cyclic Nash equilibria in the context of MONFGs under SER. 

\subsubsection{Nash Equilibria}
A Nash equilibrium (NE) is a joint policy from which no agent can unilaterally deviate and increase their payoff. The original definition of this solution concept in the context of scalar payoffs can be reformulated for vectorial payoffs \cite{radulescu2020utility}. For the purpose of notation, we define $\pi_{-i} = (\pi_1, \cdots, \pi_{i-1}, \pi_{i+1}, \cdots, \pi_n)$ as the joint policy $\pi$ without the policy of agent $i$, so that we may write $\pi = (\pi_i, \pi_{-i})$. A Nash equilibrium is then defined as follows: 

\begin{definition}[Nash equilibrium for scalarised expected returns]
\label{def:MOMA-NE-SER}
A joint policy $\pi^{NE}$ is a Nash equilibrium in an MONFG under the scalarised expected returns criterion if for all players $i \in \{1, \cdots , n\}$ and all alternative policies $\pi_i$:
\[
u_i\left(\mathbb{E} \bm{p}_i \left(\pi^{NE}_i, \pi^{NE}_{-i} \right)\right) \geq u_i\left(\mathbb{E} \bm{p}_i \left(\pi_i, \pi^{NE}_{-i} \right)\right) 
\]
\end{definition}

Whenever the NE consists solely of pure strategies, we call this a pure strategy Nash equilibrium. As an example, consider the MONFG in Table \ref{tab:nash} and the utility function $u(x, y) = x + y$ for both players. We show the (pure strategy) NE in the highlighted cell. In this game, playing A is the best response for player 1 (row player) to either action from player 2 (column player), as the reward vector dominates all other rewards. Player 2 on the other hand has a best response strategy of playing B for every action from player 1. The joint strategy (A, B) thus presents an NE in this game. 

\begin{table}[h]
    \centering
    \begin{tabular}{lll}
                       & \multicolumn{1}{c}{A}       & \multicolumn{1}{c}{B}       \\ \cline{2-3} 
\multicolumn{1}{l|}{A} & \multicolumn{1}{l|}{(1, 1); (0, 0)} & \multicolumn{1}{l|}{\nashhighlight{(1, 1); (1, 1)}} \\ \cline{2-3} 
\multicolumn{1}{l|}{B} & \multicolumn{1}{l|}{(0, 0); (0, 0)} & \multicolumn{1}{l|}{(0, 0); (1, 1)} \\ \cline{2-3} 
\end{tabular}
\caption{A two-player MONFG containing a pure strategy Nash equilibrium which is highlighted in green.}
\label{tab:nash}
\end{table}

It is well known that in single-objective NFGs, each game must have at least one mixed strategy NE. In MONFGs however, it has been shown that when optimising for the SER objective, and specifically when using non-linear utility functions, NE need not exist \cite{radulescu2020utility}.

\subsubsection{Cyclic Nash Equilibria}
A second solution concept that is worth discussing is the cyclic Nash equilibrium (CNE). A cyclic policy can be defined as a sequence of stationary policies $\pi = \{\pi_1, \cdots, \pi_k\}$ that is continuously cycled through \cite{zinkevich2005cyclic}. Whenever agents have no incentive to deviate from this cycle, we denote the joint policy as a CNE. Cyclic Nash equilibria were first described by Zinkevich et al. \cite{zinkevich2005cyclic} in the context of Markov games. We show in Section \ref{sec:experiments} that the protocols we contribute can naturally lead to cycling amongst policies over repeated plays of the game. Moreover, this cycling may be optimal and thus leads us to consider cyclic Nash equilibria. As far as the authors are aware, this has not previously been noted in MONFGs. We contribute a definition of a CNE under SER below.

\begin{definition}[Cyclic Nash equilibrium for scalarised expected returns]
\label{def:MOMA-CNE-SER}
A joint cyclic policy $\pi^{NE}$, with $\pi_i^{NE} = \{\pi^{NE}_{i,1}, \cdots, \pi^{NE}_{i,k}\}$, is a cyclic Nash equilibrium in an MONFG under the scalarised expected returns criterion if for all players $i \in \{1, \cdots , n\}$ and all alternative cyclic policies $\pi_i$:
\[
u_i\left(\mathbb{E} \bm{p}_i \left(\pi^{NE}_{i}, \pi^{NE}_{-i} \right)\right) \geq u_i\left(\mathbb{E} \bm{p}_i \left(\pi_{i}, \pi^{NE}_{-i} \right)\right)
\]
\end{definition}

As an example, consider the game in Table \ref{tab:cyclic-ne} with both players employing the utility function $u(p_1, p_2) = p_1 \cdot p_2$. If both players assume the cyclic strategy $\pi_1 = \pi_2 = \{A, B\}$, this leads to an expected return of (6, 6) and a utility of 36. Observe that deviating to any other strategy only decreases their utility as players will sometimes obtain a payoff vector of (0, 0). Therefore, the joint cyclic strategy is a cyclic Nash equilibrium.

\begin{table}[h]
\centering
\begin{tabular}{crr}
\multicolumn{1}{l}{}   & \multicolumn{1}{c}{A}          & \multicolumn{1}{c}{B}          \\ \cline{2-3} 
\multicolumn{1}{c|}{A} & \multicolumn{1}{r|}{\cnashhighlight{$(10, 2); (10, 2)$}} & \multicolumn{1}{r|}{$(0, 0); (0, 0)$}  \\ \cline{2-3} 
\multicolumn{1}{c|}{B} & \multicolumn{1}{r|}{$(0, 0); (0, 0)$}  & \multicolumn{1}{r|}{\cnashhighlight{$(2, 10); (2, 10)$}} \\ \cline{2-3} 
\end{tabular}
\caption{A two-player MONFG where committing to a cyclic strategy is optimal. The cyclic Nash equilibrium is highlighted in blue.}
\label{tab:cyclic-ne}
\end{table}

\subsection{Learning in Multi-Objective Games}
\label{sec:learning-monfg}
We first introduce the single-agent algorithm that we use for learning in multi-objective games. Next, we describe how this algorithm is used in multi-agent settings and the challenges that come with this.

\subsubsection{Actor-Critic}
For our agents to learn in MONFGs, we employ a well-known algorithm from single-agent reinforcement learning called actor-critic. The actor-critic method itself is built up from two key ideas, namely policy gradient and value based methods.

Policy gradient attempts to learn parameters $\bm{\theta}$ to a policy $\pi_\theta$ that maximise expected returns. It does this by performing gradient ascent on an objective function $J(\bm{\theta})$. We show this update in Equation \ref{eq:policy-gradient-update}.
\begin{eqnarray}
\label{eq:policy-gradient-update}
    \bm{\theta}_{t+1} = \bm{\theta}_t + \alpha_\theta \nabla J(\bm{\theta}_t)
\end{eqnarray}
where $\alpha_\theta$ is the learning rate for the parameters $\bm{\theta}$. 

Value-based methods on the other hand attempt to learn a value for states or state-action pairs. These values are later used to select actions which are expected to maximise returns. In this work, we consider Q-learning specifically which estimates the value of state-action pairs by using the update rule in Equation \ref{eq:q-update}. Please note that we explicitly show the multi-objective update rule for Q-learning in stateless settings \cite{radulescu2020utility}.
\begin{eqnarray}
\label{eq:q-update}
    \bm{Q}(a_t) \gets \bm{Q}(a_t) + \alpha_Q \left[\bm{p}_t - \bm{Q}(a_t) \right]
\end{eqnarray}
with $\alpha_Q$ the learning rate for the Q-values, $a_t$ the action taken at timestep $t$ and $\bm{p}_t$ the obtained vectorial reward.

The actor-critic method combines both approaches in one algorithm. Specifically, learned Q-values (the critic) are used as a baseline in the objective function for updating the policy learned through policy gradient (the actor). In multi-objective games, and specifically under SER, we can formulate the objective function in terms of maximising utility from expected returns which are captured in vectorial Q-values \cite{zhang2020opponent}: 
\begin{eqnarray}
\label{eq:objective-func}
    J(\bm{\theta}) = u_i \left(\sum_{a \in A_i} \pi_i(a | \bm{\theta})\bm{Q}(a)\right)
\end{eqnarray}
In Equation \ref{eq:objective-func}, $a \in A_i$ are the actions available to agent $i$ and $\bm{Q}(a)$ is the vectorial Q-value associated with an action. This makes it so that $\sum_{a \in A_i} \pi_i(a | \bm{\theta})\bm{Q}(a)$ represents the expected multi-objective return of the current policy. By applying the utility function $u_i$ to this expected multi-objective return, we are effectively calculating the SER.

\subsubsection{Independent Actor-Critic}
To extend reinforcement learning to multi-agent settings, an attractive solution is to use independent learners which use algorithms from single-agent reinforcement learning independently \cite{leslie2005individual,tampuu2017multi}. It is often (implicitly) assumed that agents attempt to achieve a Nash equilibrium \cite{nowe2012game}. This is not guaranteed to succeed with independent learning as many of the convergence guarantees from single-agent reinforcement learning rely on the Markov property of their environment, which is violated for independent learners \cite{laurent2011world}. However, great results have been obtained in practice \cite{bowling2002multi}.

The communication protocols we propose are based on independent learners that use the multi-objective actor-critic algorithm shown in Algorithm \ref{alg:indep-actor-critic} \cite{zhang2020opponent}. Note that each agent's policy is defined as a softmax over their policy parameters:
\begin{equation*}
    \pi(a=a_i|\bm{\theta}) = \frac{e^{\theta_i}}{\sum^{|A|}_{j=1} e^{\theta_j}}
\end{equation*}
The result of applying this function is a probability distribution over actions. In the learning loop, agents sample an action from this distribution and observe their reward vector. They subsequently update the Q-value of their action and calculate the objective function. Finally, the policy parameters $\bm{\theta}$ are updated by ascending on the gradient of the objective function.

\begin{algorithm}[h]
\caption{Independent multi-objective actor-critic under SER}
\label{alg:indep-actor-critic}
\begin{algorithmic}[1]
\STATE \textbf{Input:} learning rates $\alpha_Q$ and $\alpha_\theta$
\FOR {each player}
\STATE For each action $a \in A$ and with $d \geq 2$ objectives, initialise $\bm{Q}(a) \gets \bm{0}$
\STATE Initialise $\bm{\theta} = \bm{0}$ and $\pi(a=a_i|\bm{\theta}) = \frac{e^{\theta_i}}{\sum^{|A|}_{j=1} e^{\theta_j}}$
\ENDFOR

\FOR{each episode}
\FOR{each player}
\STATE Play action $a \sim \pi(a|\bm{\theta})$
\ENDFOR
\FOR{each player}
\STATE Observe payoff vector $\bm{p} \in \mathbb{R}^d$
\STATE $\bm{Q}(a) \gets \bm{Q}(a) + \alpha_Q \left[\bm{p} - \bm{Q}(a)\right]$
\STATE calculate objective function: $J(\bm{\theta}) = u\left(\sum_{a \in A} \pi(a|\bm{\theta}) \bm{Q}(a)\right)$
\STATE Update policy parameters: $\bm{\theta} \gets \bm{\theta} + \alpha_{\theta} \nabla J(\bm{\theta})$
\ENDFOR
\ENDFOR
\end{algorithmic}
\end{algorithm}

\subsection{Stackelberg Games}
We design several communication protocols that are inspired by Stackelberg games. A Stackelberg game is an $n$-player game where one or more players are designated as leaders that publicly commit to playing a certain strategy. Other players are assumed to be followers who condition their own strategy on the received commitment. Note that we refer to the game that is being played without commitment as the simultaneous move game.

When considering commitment, it becomes useful to consider a third type of solution concept called a Leadership equilibrium (LE) or Stackelberg equilibrium. We restrict ourselves to two-player games, for which we define an LE as follows:
\begin{definition}[Leadership equilibrium]
\label{def:MOMA-LE-SER}
A joint strategy $s$ is a leadership equilibrium in a two-player MONFG if the leader’s mixed strategy $s_1 \in S_1$ maximises their utility $u_1$, given that for each $s'_1 \in S_1$ the follower plays a strategy $s_2 \in S_2$ maximising $u_2$ given~$s'_1$.
\end{definition}
While this definition does not explicitly specify an optimality criterion for either player, the utility obtained by them does implicitly depend on their optimality criterion.

As an example of an LE, consider the game in Table \ref{tab:leadership-equilibrium} where the row player is the leader and the column player the follower. Assume both players use the utility function $u(p_1, p_2) = p_1^2 + p_2^2$. The leadership equilibrium in this case is to commit to the pure strategy A, which induces the follower to respond with A as well. This joint strategy is not a Nash equilibrium in the simultaneous move game, as the row player would have an incentive to deviate to B.

\begin{table}[h]
\centering
\begin{tabular}{crr}
\multicolumn{1}{l}{\textbf{}} & \multicolumn{1}{c}{A}                 & \multicolumn{1}{c}{B}                 \\ \cline{2-3} 
\multicolumn{1}{c|}{A}        & \multicolumn{1}{r|}{\lehighlight{$(2, 0); (2, 0)$}} & \multicolumn{1}{r|}{$(0, 2); (0, 1)$} \\ \cline{2-3} 
\multicolumn{1}{c|}{B}        & \multicolumn{1}{r|}{$(3, 0); (1, 0)$} & \multicolumn{1}{r|}{$(0, 1); (0, 2)$} \\ \cline{2-3} 
\end{tabular}
\caption{A two-player Stackelberg game. The leadership equilibrium is highlighted in pink for the row player as leader.}
\label{tab:leadership-equilibrium}
\end{table}

In two-player single-objective games, leadership equilibria are guaranteed a payoff at least as high as a Nash equilibrium in the simultaneous move game and may even be strictly better than all NE \cite{stengel2010leadership,letchford2014value}. This motivates exploring the value of commitment for learning agents in multi-objective settings as well.

\section{Preference Communication}
\label{sec:preference-communication}
In this section, we introduce the novel preference communication protocols. The goal of these protocols is to foster cooperation when possible and allow for improvements in learning speed through increased coordination. In total, we contribute four novel approaches that can be used in a range of different settings. 

\subsection{Communication Protocols}
All communication protocols revolve around a central idea inspired by Stackelberg games. In every play of the game, one agent is selected as the \emph{leader} while the other is determined a \emph{follower}. Agents switch every round between being the leader and being the follower. The leader is required to commit to a pure or mixed strategy depending on the specified protocol. As we study \emph{preference communication}, the commitment made by the leader always represents some sort of preference over the future (pure) strategy they intend to play. The follower in turn conditions their own strategy on this commitment. We hypothesise that having players alternate commitment helps coordinate joint-policies and increases convergence rates. 

For agents to converge to a Nash equilibrium or otherwise stable utility, we gradually move a player's policy in the direction of their best response. Best-response iteration is often used in single-objective games where agents calculate a best response to the current (empirically recorded) joint policy \cite{anthony2020learning}. Rather than explicitly calculating a best response, we perform gradient ascent using the actor-critic algorithm shown in Section \ref{sec:learning-monfg}. Gradient ascent has also been applied with success for learning in single-objective games \cite{bowling2002multi}, further motivating this approach in our setting as well.

While we focus on the empirical performance of the protocols, it is interesting to briefly consider theoretical properties as well. In single-objective games, gradient ascent is not guaranteed to converge \cite{singh2000nash}. In the case of multi-objective games, and specifically under SER, an additional issue is that Nash equilibria need not exist. Therefore, convergence to a Nash equilibrium in general already becomes impossible. Empirical evidence from single-objective games suggests that adaptations to naive gradient ascent and best-response iteration often converge to NE or reach quantitatively high performance \cite{bowling2002multi,anthony2020learning}. In Section \ref{sec:experiments}, we present the same observation that gradient ascent works well in practice for multi-objective games. The precise classes of games for which convergence can be proven are left for future work.

\subsection{Cooperative Action Communication}
\label{sec:coop-pref-com}
Cooperative action communication enables agents to cooperate by optimising for a single joint policy. The protocol contains two main components. First, the leader samples an action from their current policy and publicly commits to playing this in the round. Next, the follower updates their own policy based on the commitment. This update can be considered as anticipation to reach an adequate joint policy faster. We hypothesise that action communication can coordinate avoiding joint-actions that are bad for both players.

We stress that limiting agents to a single stationary policy under both roles is necessarily a measure to induce cooperation. By enforcing agents to lead and follow with the same policy, we limit the level of exploitation. This is because any policy update towards an unfavourable joint policy for the opponent will have to be used when committing. This enables said opponent to counter this policy, thus bringing the joint policy back towards a suitable middle ground. This will be specifically useful in games without Nash equilibria, where stable solutions need not exist. We hypothesise that alternating commitment may help them converge to a better joint policy.

\begin{algorithm}[h]
\caption{Cooperative action communication}
\label{alg:coop-action-alg}
\begin{algorithmic}[1]
\STATE \textbf{Input:} learning rates $\alpha_Q$ and $\alpha_\theta$
\FOR {each player}
\STATE For each action $a \in A$, opponent action $a' \in A'$ and with $d \geq 2$ objectives, initialise $\bm{Q}(a, a') \gets \bm{0}$
\STATE Initialise $\bm{\theta} = \bm{0}$ and $\pi(a=a_i|\bm{\theta}) = \frac{e^{\theta_i}}{\sum^{|A|}_{j=1} e^{\theta_j}}$
\ENDFOR

\FOR{each episode}
\FOR{each player}
\IF{player is the leader}
\STATE Generate a new message $m$ by sampling from the policy $m = a \sim \pi(a|\bm{\theta})$
\ELSE
\STATE Observe message $m$
\STATE calculate objective function: $J(\bm{\theta}) = u\left(\sum_{a \in A} \pi(a|\bm{\theta}) \bm{Q}(a, m)\right)$
\STATE Update policy parameters: $\bm{\theta} \gets \bm{\theta} + \alpha_{\theta} \nabla J(\bm{\theta})$
\ENDIF
\ENDFOR
\FOR{each player}
\IF{player is the leader}
\STATE Play action $a = m$
\ELSE
\STATE Play action $a \sim \pi(a|\bm{\theta})$
\ENDIF
\ENDFOR
\FOR{each player}
\STATE Observe payoff vector $\bm{p} \in \mathbb{R}^d$ and opponent action $a'$
\STATE $\bm{Q}(a, a') \gets \bm{Q}(a, a') + \alpha_Q \left[\bm{p} - \bm{Q}(a, a')\right]$
\STATE calculate objective function: $J(\bm{\theta}) = u\left(\sum_{a \in A} \pi(a|\bm{\theta}) \bm{Q}(a, a')\right)$
\STATE Update policy parameters: $\bm{\theta} \gets \bm{\theta} + \alpha_{\theta} \nabla J(\bm{\theta})$
\ENDFOR
\ENDFOR
\end{algorithmic}
\end{algorithm}

We introduce the complete communication protocol in Algorithm \ref{alg:coop-action-alg}. Agents first perform the initialisation step as in the original actor-critic setting shown in Algorithm \ref{alg:indep-actor-critic}, with the difference that now a joint action Q-table is defined in line 3. This makes agents following this protocol related to work on joint-action learners \cite{claus1998dynamics}. 

In line 9, the leader samples their next action and communicates it to the follower. From line 11 to 13, the follower updates their own policy by selecting the applicable Q-values from their joint-action Q-table. After this, both agents play their selected actions and perform a regular actor-critic update to account for the new payoff observations. Note that because we assume agents are looking to cooperate, we allow actions to be completely observable. This ensures that agents can always update the correct joint-action in their Q-table.

\begin{figure}[h]
    \centering
    \includegraphics[width=.61\linewidth]{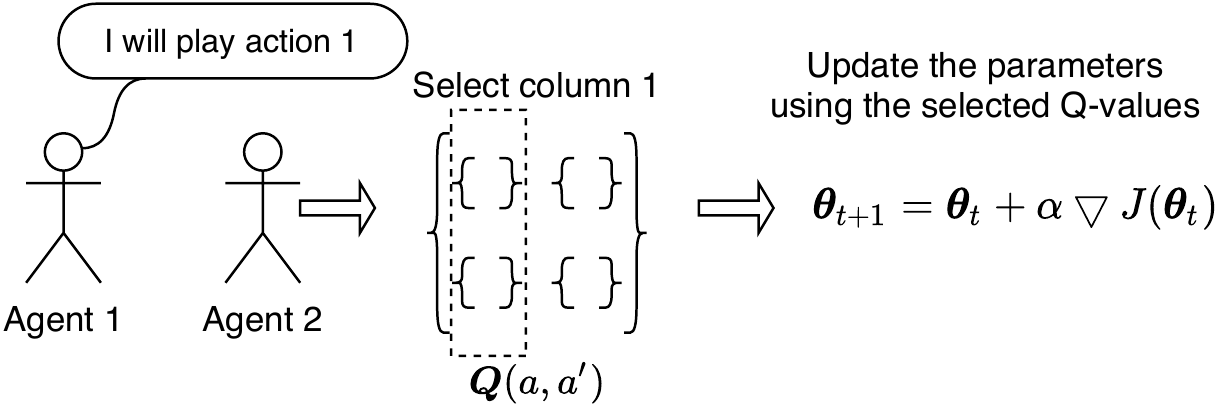}
    \caption{The cooperative action communication protocol. Here, the leader commits to playing an action, after which the follower updates their own policy.}
    \label{fig:coop-alg}
\end{figure}

We show an example scenario in Figure \ref{fig:coop-alg}. Assume agent 1 is the leader and commits to playing action 1. Agent 2 observes this and selects the correct column of Q-values. These values are then used to update their policy after which the agent can sample an action from their updated policy.

\subsection{Self-interested Action Communication}
The self-interested action communication protocol allows agents to learn a non-stationary policy conditioned on their current role, which switches with each play of the game, and perceived communication. This means that when leading, agents use a communication policy and select their preferred action to commit to. While following, agents learn a best response policy to each message that can be received. This implies that both players will learn a, possibly distinct, leadership equilibrium. Furthermore, as commitment is alternated, we expect agents to cycle through their LE. We note that in single-objective games, it would suffice to learn the optimal counter action to any committed action from the leader. However, in multi-objective games it is well known that a stochastic policy can outperform deterministic policies \cite{vamplew2009constructing}.

\begin{algorithm}[h!]
\caption{Self-interested action communication}
\label{alg:self-interested-alg}
\begin{algorithmic}[1]
\STATE \textbf{Input:} learning rates $\alpha_Q$ and $\alpha_\theta$
\FOR {each player}
\STATE For each observation $o \in O = A \bigcup \{None\}$, action $a \in A$ and with $d \geq 2$ objectives, initialise $\bm{Q}(o, a) \gets \bm{0}$
\STATE Initialise $\bm{\theta} = 0$ and $\pi(a=a_i|r, o, \bm{\theta}) = \frac{e^{\theta_{r, o, i}}} {\sum^{|O|}_{j=1} e^{\theta_{r, o, j}}}$ with role $r$ and observation $o$
\ENDFOR

\FOR{each episode}
\FOR{each player}
\IF{player is the leader}
\STATE Observation $o = None$
\STATE Generate a new message $m = a \sim \pi(a|leader, o, \bm{\theta})$
\ELSE
\STATE Observe the message $o = m$
\ENDIF
\ENDFOR
\FOR{each player}
\IF{player is the leader}
\STATE Play action $a = m$
\ELSE
\STATE Play action $a \sim \pi(a|follower, o, \bm{\theta})$
\ENDIF
\ENDFOR
\FOR{each player}
\STATE Observe payoff vector $\bm{p} \in \mathbb{R}^d$
\STATE $\bm{Q}(o, a) \gets \bm{Q}(o, a) + \alpha_Q \left[\bm{p} - \bm{Q}(o, a)\right]$
\STATE calculate objective function: $J(\bm{\theta}) = u\left(\sum_{a \in A} \pi(a|r, o, \bm{\theta}) \bm{Q}(o, a)\right)$
\STATE Update policy parameters: $\bm{\theta}_{r,o} \gets \bm{\theta}_{r,o} + \alpha_{\theta} \nabla_{r,o} J(\bm{\theta})$
\ENDFOR
\ENDFOR
\end{algorithmic}
\end{algorithm}

We show the protocol in Algorithm \ref{alg:self-interested-alg}. Note that we adopt the notation of a single non-stationary policy that is conditioned on the current role and observation. The practical implementation separates this into a leader policy and a collection of follower policies, which is conceptually equivalent.

The communication section of the algorithm is contained between lines 7 and 14. As the leader receives no communication, they set their observation to $None$ in line 9 and generate a new message from their current policy in line 10. The follower subsequently receives this message and stores it as an observation in line 12. The episode continues with the leader playing their committed action and the follower using the message to sample a best-response in line 19. The episode concludes by updating the Q-table and policy parameters. Note that because agents are self-interested, we assume that they do not share what action they selected in line 23. This only makes a difference for the leader, as the follower already observed the commitment from the leader.

\begin{figure}[ht]
    \centering
    \includegraphics[width=.6\linewidth]{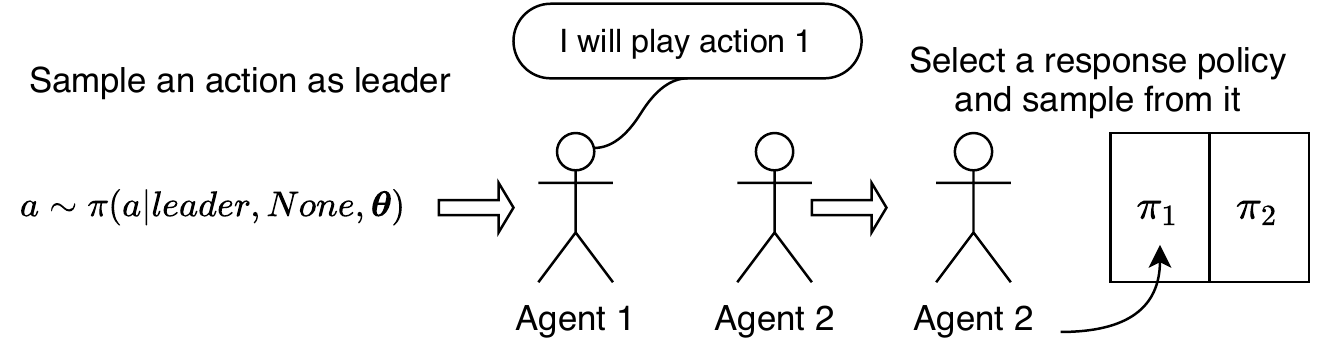}
    \caption{The self-interested action communication protocol. The leader commits to an action by sampling from their non-stationary policy. The follower then selects the best-response policy to action 1.}
    \label{fig:comp-alg}
\end{figure}

Consider the example scenario in Figure \ref{fig:comp-alg}. The leader commits to action 1 by sampling from their (non-stationary) policy. The follower observes this and selects the best response policy to action 1. After this, the follower would sample from their policy, i.e. $a \sim \pi(a|follower, 1, \bm{\theta})$, analogous to how the leader selected their action. The episode continues by each agent playing their selected action and updating their policy.

\subsection{Cooperative Policy Communication}
Cooperative policy communication is a cooperative protocol similar to the protocol described in Section \ref{sec:coop-pref-com}. While agents are still optimising a single joint policy, the leader now commits to their entire policy rather than to their preferred action. The advantage of communicating policies, rather than the next action, is that followers can better gauge the current preferences of the leader. The advantage for the leader is that they need not share the exact action that will be played which might be exploited by the follower. We show the protocol in Algorithm \ref{alg:policy-com-alg}. 

\begin{algorithm}[h]
\caption{Policy communication actor-critic}
\label{alg:policy-com-alg}
\begin{algorithmic}[1]
\STATE \textbf{Input:} learning rates $\alpha_Q$ and $\alpha_\theta$
\FOR {each player}
\STATE For each action $a \in A$, opponent action $a' \in A'$ and with $d \geq 2$ objectives, initialise $\bm{Q}(a, a') \gets \bm{0}$
\STATE Initialise $\bm{\theta} = \bm{0}$ and $\pi(a=a_i|\bm{\theta}) = \frac{e^{\theta_i}}{\sum^{|A|}_{j=1} e^{\theta_j}}$
\STATE Initialise an opponent policy: $\pi'(a') = \frac{1}{|A'|}$
\ENDFOR

\FOR{each episode}
\FOR{each player}
\IF{player is the leader}
\STATE Select the current policy $\pi(a=a_i|\bm{\theta})$ as the message $m$ 
\ELSE
\STATE Observe message $m$
\STATE Update latest opponent policy $\pi' = m$
\STATE calculate objective function: $J(\bm{\theta}) = u\left(\sum_{a \in A} \pi(a|\bm{\theta}) \sum_{a' \in A'} \pi'(a')\bm{Q}(a, a')\right)$
\STATE Update policy parameters: $\bm{\theta} \gets \bm{\theta} + \alpha_{\theta} \nabla J(\bm{\theta})$
\ENDIF
\ENDFOR
\FOR{each player}
\STATE Play action $a \sim \pi(a|\bm{\theta})$
\ENDFOR
\FOR{each player}
\STATE Observe payoff vector $\bm{p} \in \mathbb{R}^d$ and opponent action $a'$
\STATE $\bm{Q}(a, a') \gets \bm{Q}(a, a') + \alpha_Q \left[\bm{p} - \bm{Q}(a, a')\right]$
\STATE calculate objective function: $J(\bm{\theta}) = u\left(\sum_{a \in A} \pi(a|\bm{\theta}) \sum_{a' \in A'} \pi'(a')\bm{Q}(a, a')\right)$
\STATE Update policy parameters: $\bm{\theta} \gets \bm{\theta} + \alpha_{\theta} \nabla J(\bm{\theta})$
\ENDFOR
\ENDFOR
\end{algorithmic}
\end{algorithm}

The leader communicates their current policy in line 10. The follower responds analogous to Algorithm \ref{alg:coop-action-alg} by marginalising over the joint-action Q-table table to derive the expected Q-values. These Q-values are used to calculate the objective function in line 14. After updating the policy, the episode continues regularly. Note that in order for the leader to also update their policy using joint-action Q-values, both agents record the latest opponent policy. This is then used to marginalise over Q-values when leading (see line 5 and 24).

\begin{figure}[h]
    \centering
    \includegraphics[scale=0.6]{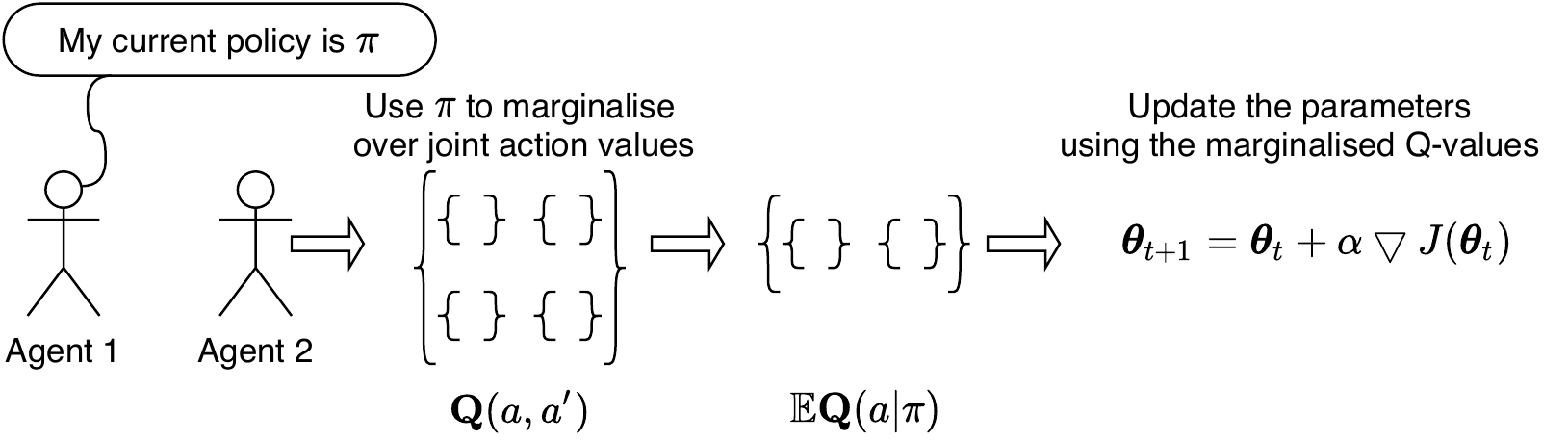}
    \caption{An illustration of the cooperative policy communication protocol.}
    \label{fig:policy-com}
\end{figure}

Consider again an example scenario in Figure \ref{fig:policy-com}. Assume the leader communicates their current policy $\pi$. The follower uses this policy to derive expected Q-values for their actions. Given these expected Q-values, they perform an update of their parameters to obtain a better policy.

We note that, in contrast to the setting with cooperative action communication, the policy communication approach cannot easily be adjusted for self-interested dynamics. The reason is that self-interested dynamics force followers to learn a best-response policy to every distinct message. When communicating policies this is impossible as policies are by definition a collection of continuous preferences over actions. We discuss solutions to this problem as a direction for future work in Section \ref{sec:conclusion}.

\subsection{A Hierarchical Approach to Communication}
\label{sec:hierarchical-com}
The final communication protocol proposes a second layer in the decision making process. Rather than forcing the leading agent to commit to certain preferences, agents learn \emph{when} to communicate as well as \emph{what} to communicate. We accomplish this by introducing a hierarchy in the policies of the agents. Agents learn an upper-level policy that determines whether they will communicate when leading. The lower layer has two protocols, a no communication protocol and a communication protocol. Naturally, the no communication protocol is followed when the upper-layer decides not to communicate and the communication protocol is used otherwise. The goal of this approach is to study what dynamics lead agents to benefit from communication.

\begin{algorithm}[h]
\caption{Hierarchical communication actor-critic}
\label{alg:hierarchical-com-alg}
\begin{algorithmic}[1]
\STATE \textbf{Input:} learning rates $\alpha_Q$ and $\alpha_\theta$
\FOR {each player}
\STATE Initialise low-level protocols $P$: A no-communication and a communication protocol
\STATE For each protocol $p \in P$ and with $d \geq 2$ objectives, initialise $\bm{Q}(p) \gets \bm{0}$
\STATE Initialise $\bm{\theta} = \bm{0}$ and a top-level communication policy $\pi(p=p_i|\bm{\theta}) = \frac{e^{\theta_i}}{\sum^{2}_{j=1} e^{\theta_j}}$
\ENDFOR

\FOR{each episode}
\FOR{each player}
\IF{player is the leader}
\STATE Select the low-level protocol using the top-level policy: $p \sim \pi(p|\bm{\theta})$
\STATE Select a message $m$ from $p$
\ELSE
\STATE Observe message $m$
\ENDIF
\ENDFOR
\FOR{each player}
\STATE Play action $a$ from same protocol $p$ given $m$
\ENDFOR
\FOR{each player}
\STATE Observe payoff vector $\bm{p} \in \mathbb{R}^d$ and opponent action $a'$
\STATE $\bm{Q}(p) \gets \bm{Q}(p) + \alpha_Q \left[\bm{p} - \bm{Q}(p)\right]$
\STATE calculate objective function: $J(\bm{\theta}) = u\left(\sum_{p}^2 \pi(p|\bm{\theta}) \bm{Q}(p)\right)$
\STATE Update policy parameters: $\bm{\theta} \gets \bm{\theta} + \alpha_{\theta} \nabla J(\bm{\theta})$
\STATE Update protocol $p$ used in this episode
\ENDFOR
\ENDFOR
\end{algorithmic}
\end{algorithm}

We contribute a pseudo-code implementation in Algorithm \ref{alg:hierarchical-com-alg}. Agents first initialise the lower level protocols in line 3 and define the top level policy in line 5. The leader selects which protocol to use in line 10, after which a new message is generated. We note that if the no communication policy is chosen, the generated message will always be $None$. In line 17, both players play an action according to the protocol that was selected by the leader. The episode concludes by observing the payoff and updating the Q-values and top-level policy, as well as cascading an update to the lower level protocol in line 24. We design this setting so that each communication protocol described in previous sections can be used in the second layer. Note that selecting a specific protocol thus also implies whether agents will aim for cooperation or behave self-interested.

\begin{figure}[h!]
    \centering
    \includegraphics[scale=0.62]{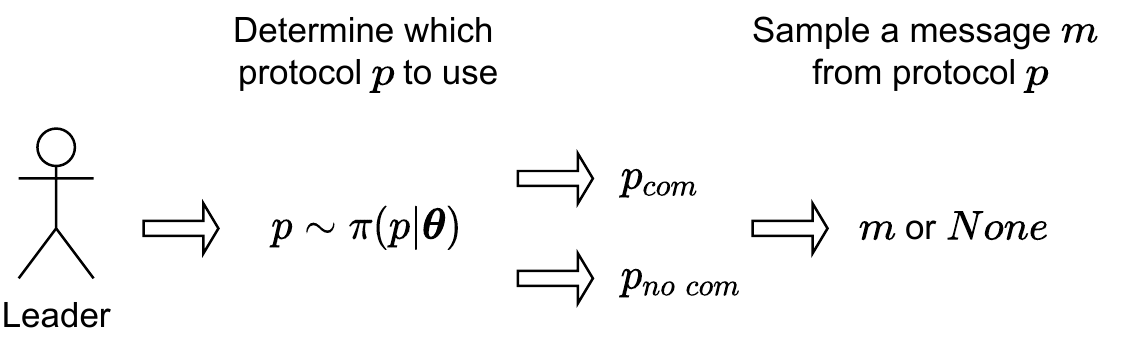}
    \caption{A hierarchical approach to communication in which the leader decides to communicate or not.}
    \label{fig:hierarchical-com}
\end{figure}

We add a final example scenario for the decision process of the leader in Figure \ref{fig:hierarchical-com}. The leader first uses their top-level policy to determine whether to communicate or not. Next, they select the correct protocol from the lower layer and either prepare a message or refuse to communicate by sending $None$.

\section{Experimental Setup and Results}
\label{sec:experiments}
To evaluate the proposed communication protocols, we perform experiments on a set of five benchmark games from the MONFG literature \cite{radulescu2020utility,radulescu2021opponent,zhang2020opponent}. We include two types of games: one where no NE under SER with the given utility functions exists and one where at least one pure strategy NE exists. We show in the discussion of our results that the presence of NE is a relevant factor for the learned policies of agents. 

In each game, the agents get the same payoff vectors $\bm{p} = (p_1, p_2)$, but value these payoffs differently, due to having different utility functions. This is known as the team reward, individual utility setting \cite{radulescu2020multi}. The utility function for agent 1 (the row player) is:
\begin{eqnarray}
u_1(p_1, p_2) = p_1^2 + p_2^2
\end{eqnarray}
The intuition behind this utility function is an agent that values extremes on either objective higher than an average return on both. Agent 2 (the column player) values moderate payoffs in both objectives over high payoffs in only one objective and uses the following utility function:
\begin{eqnarray}
u_2(p_1, p_2) = p_1 \cdot p_2
\end{eqnarray}

We show the two games without NE in Tables \ref{tab:g1} and \ref{tab:g2}. The games with NE, given the above utility functions, are shown in Tables \ref{tab:g3}, \ref{tab:g4} and \ref{tab:g5}. The highlighted cells represent the pure strategy Nash equilibria. Note that Game 4 and 5 are slightly different from previous work to allow for pure strategy Nash equilibria rather than $\varepsilon$-Nash equilibria.

\begin{table}[ht]
\centering
\begin{tabular}{llll}
                       & \multicolumn{1}{c}{L}       & \multicolumn{1}{c}{M}       & \multicolumn{1}{c}{R}       \\ \cline{2-4} 
\multicolumn{1}{l|}{L} & \multicolumn{1}{l|}{$(4, 0); (4, 0)$} & \multicolumn{1}{l|}{$(3, 1); (3, 1)$} & \multicolumn{1}{l|}{$(2, 2); (2, 2)$} \\ \cline{2-4} 
\multicolumn{1}{l|}{M} & \multicolumn{1}{l|}{$(3, 1); (3, 1)$} & \multicolumn{1}{l|}{$(2, 2); (2, 2)$} & \multicolumn{1}{l|}{$(1, 3); (1, 3)$} \\ \cline{2-4} 
\multicolumn{1}{l|}{R} & \multicolumn{1}{l|}{$(2, 2); (2, 2)$} & \multicolumn{1}{l|}{$(1, 3); (1, 3)$} & \multicolumn{1}{l|}{$(0, 4); (0, 4)$} \\ \cline{2-4} 
\end{tabular}
\caption{Game 1 - The (im)balancing act game. This game has no NE under SER.}
\label{tab:g1}

\begin{tabular}{lll}
                       & \multicolumn{1}{c}{L}       & \multicolumn{1}{c}{R}       \\ \cline{2-3} 
\multicolumn{1}{l|}{L} & \multicolumn{1}{l|}{$(4, 0); (4, 0)$} & \multicolumn{1}{l|}{$(2, 2); (2, 2)$} \\ \cline{2-3} 
\multicolumn{1}{l|}{R} & \multicolumn{1}{l|}{$(2, 2); (2, 2)$} & \multicolumn{1}{l|}{$(0, 4); (0, 4)$} \\ \cline{2-3} 
\end{tabular}
\caption{Game 2 - The (im)balancing act game without M. This game has no NE under SER.}
\label{tab:g2}

\begin{tabular}{lll}
                       & \multicolumn{1}{c}{L}       & \multicolumn{1}{c}{M}       \\ \cline{2-3} 
\multicolumn{1}{l|}{L} & \multicolumn{1}{l|}{$(4, 0); (4, 0)$} & \multicolumn{1}{l|}{\nashhighlight$(3, 1); (3, 1)$} \\ \cline{2-3} 
\multicolumn{1}{l|}{M} & \multicolumn{1}{l|}{$(3, 1); (3, 1)$} & \multicolumn{1}{l|}{$(2, 2); (2, 2)$} \\ \cline{2-3} 
\end{tabular}
\caption{Game 3 - The (im)balancing act game without R. This game has one pure NE under SER, (L, M), with a utility of 10 for agent 1 and 3 for agent 2.}
\label{tab:g3}

\begin{tabular}{lll}
                       & \multicolumn{1}{c}{L}       & \multicolumn{1}{c}{M}       \\ \cline{2-3} 
\multicolumn{1}{l|}{L} & \multicolumn{1}{l|}{\nashhighlight$(4, 1); (4, 1)$} & \multicolumn{1}{l|}{$(1, 1.5); (1, 1.5)$} \\ \cline{2-3} 
\multicolumn{1}{l|}{M} & \multicolumn{1}{l|}{$(3, 1); (3, 1)$} & \multicolumn{1}{l|}{\nashhighlight$(3, 2), (3, 2)$} \\ \cline{2-3} 
\end{tabular}
\caption{Game 4 - A 2-action game with pure NE under SER. (L, L) has a utility of 17 and 4, while (M, M) has a utility of 13 and 6.}
\label{tab:g4}

\begin{tabular}{llll}
                       & \multicolumn{1}{c}{L}       & \multicolumn{1}{c}{M}       & \multicolumn{1}{c}{R}       \\ \cline{2-4} 
\multicolumn{1}{l|}{L} & \multicolumn{1}{l|}{\nashhighlight$(4, 1); (4, 1)$} & \multicolumn{1}{l|}{$(1, 1.5); (1, 1.5)$} & \multicolumn{1}{l|}{$(2, 1); (2, 1)$} \\ \cline{2-4} 
\multicolumn{1}{l|}{M} & \multicolumn{1}{l|}{$(3, 1); (3, 1)$} & \multicolumn{1}{l|}{\nashhighlight$(3, 2); (3, 2)$} & \multicolumn{1}{l|}{$(1, 2); (1, 2)$} \\ \cline{2-4} 
\multicolumn{1}{l|}{R} & \multicolumn{1}{l|}{$(1, 2); (1, 2)$} & \multicolumn{1}{l|}{$(2, 1.5); (2, 1.5)$} & \multicolumn{1}{l|}{\nashhighlight$(1.5, 3); (1.5, 3)$} \\ \cline{2-4} 
\end{tabular}
\caption{Game 5 - A 3-action game with three pure NE under SER. (L, L) with utilities of 17 and 4, (M, M) with utilities of 13 and 6 and (R, R) with utilities of $11.25$ and $4.5$. Note that the utility for (R, R) is lower than that of (M, M) for both agents.}
\label{tab:g5}
\end{table}

During experiments, we log the SER and action probabilities of both agents. In addition, we record the joint-action probabilities from the last 10\% of episodes, to examine the joint policies agents converge to. In experiments where agents are free to learn when to communicate, we also include their communication probabilities to analyse communication policies over time.

Each experiment is executed for 5000 plays of the game and averaged over 100 trials. We include a period of 100 rollouts of the same play using Monte-Carlo simulation to measure the SER, action selection and communication probabilities at each timestep. For all figures in the forced communication experiments, we use learning rates of 0.01. Note that in the figures for the SER, we only show the first 2500 episodes for games without NE as at this point the SER had already converged.


\subsection{No communication}
\label{sec:no-com-exp}
We first show the results for agents without the use of communication. The goal of these experiments is to define a baseline performance to later compare other experiments against. 

\begin{figure}[h!tb]%
    \centering
    \subfloat[Game 1]{{\includegraphics[width=.28\linewidth]{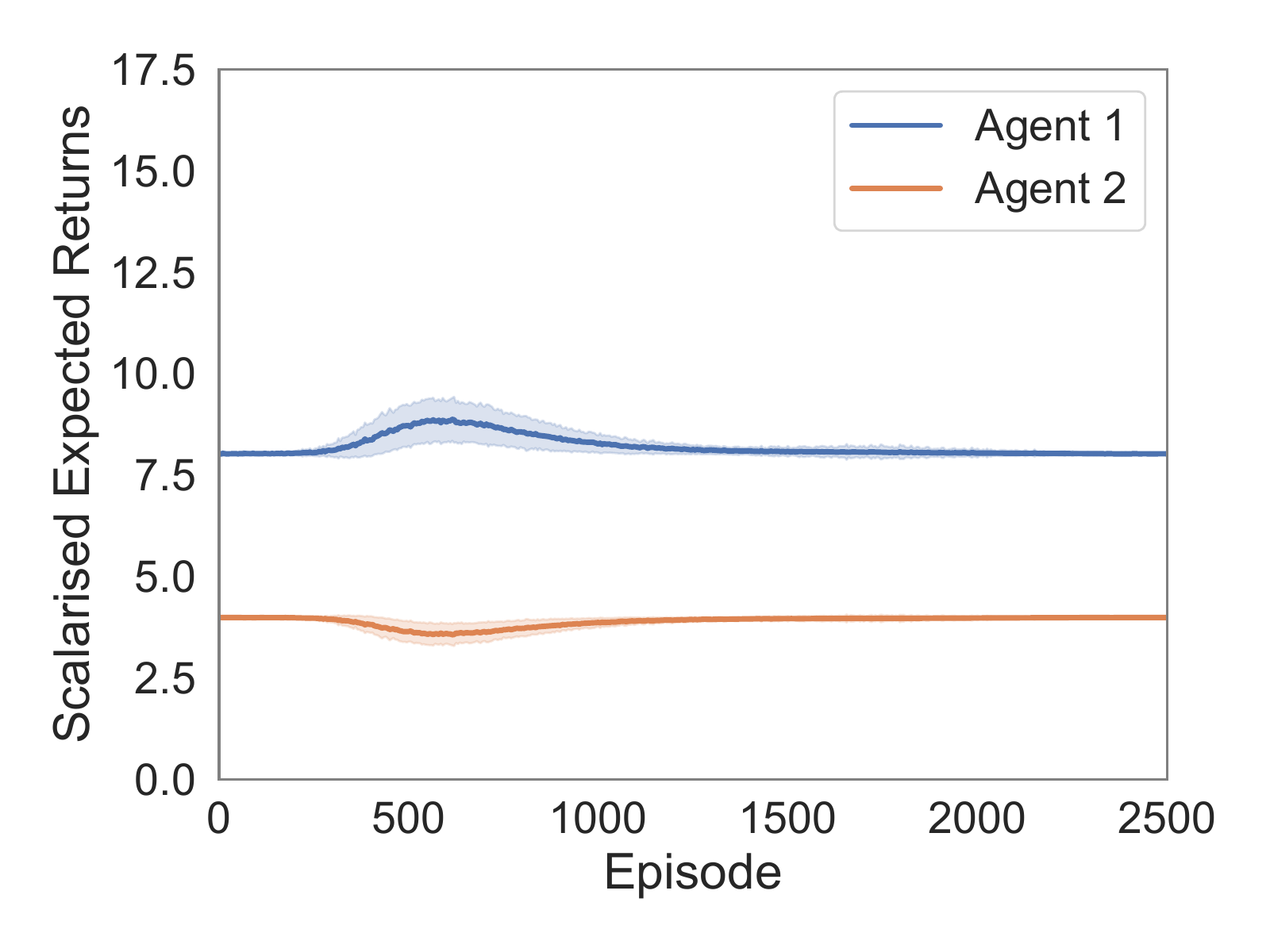}\label{fig:g1-ser-no-com} }}%
    \quad
    \subfloat[Game 2]{{\includegraphics[width=.28\linewidth]{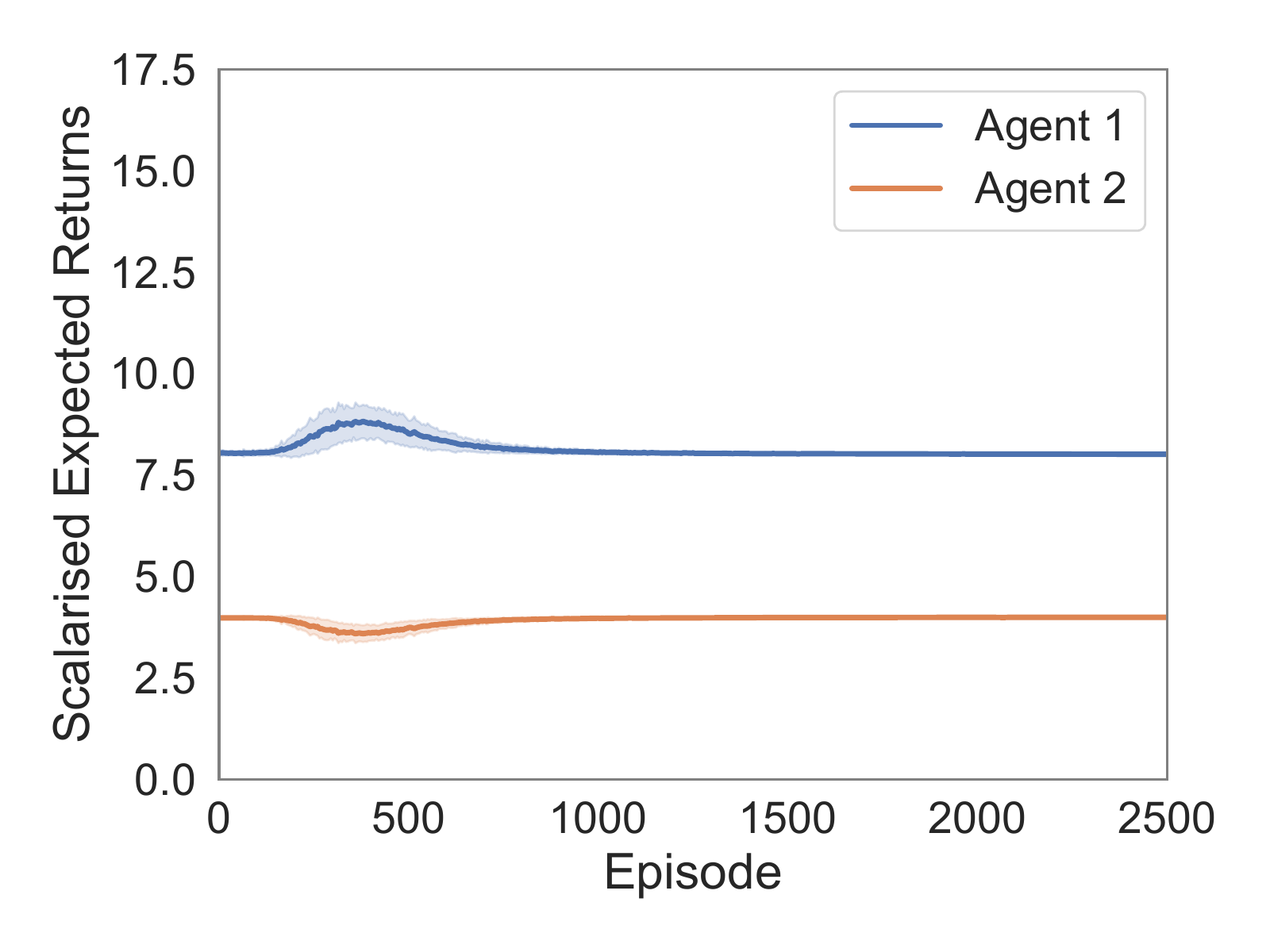}\label{fig:g2-ser-no-com} }}%
    \quad
    \subfloat[Game 3]{{\includegraphics[width=.28\linewidth]{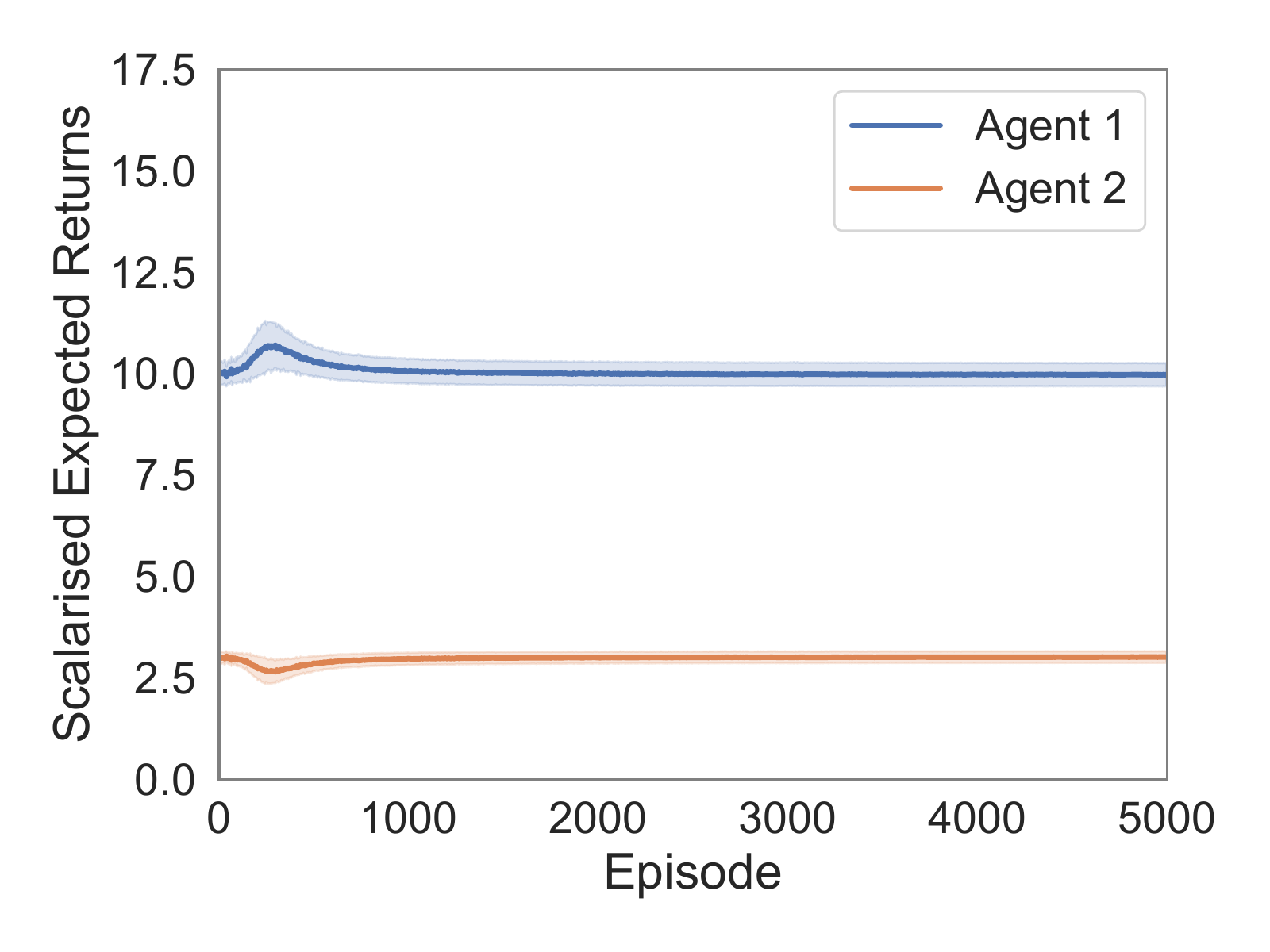}\label{fig:g3-ser-no-com} }}%
    \quad
    \subfloat[Game 4]{{\includegraphics[width=.28\linewidth]{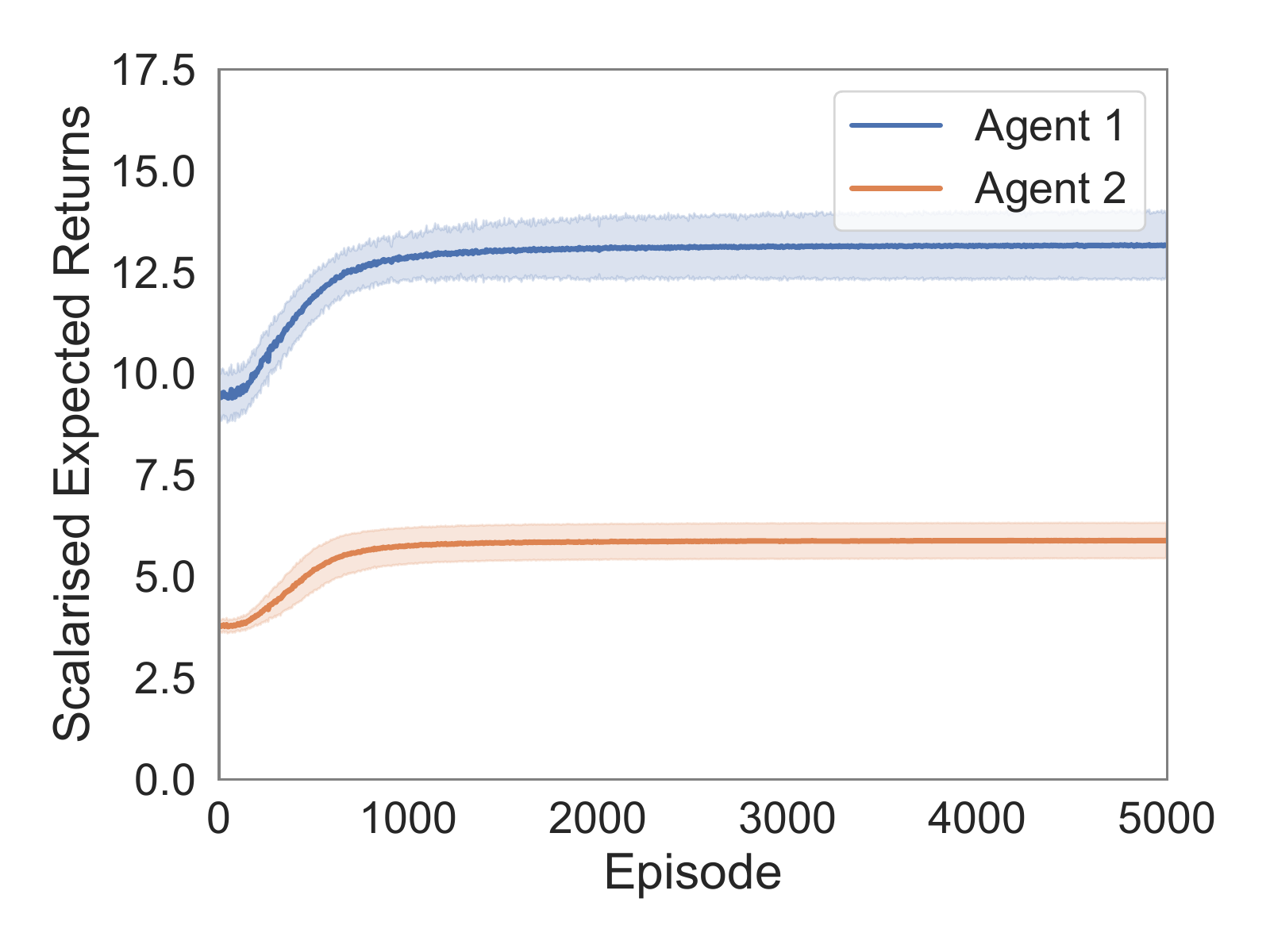}\label{fig:g4-ser-no-com} }}%
    \quad
    \subfloat[Game 5]{{\includegraphics[width=.28\linewidth]{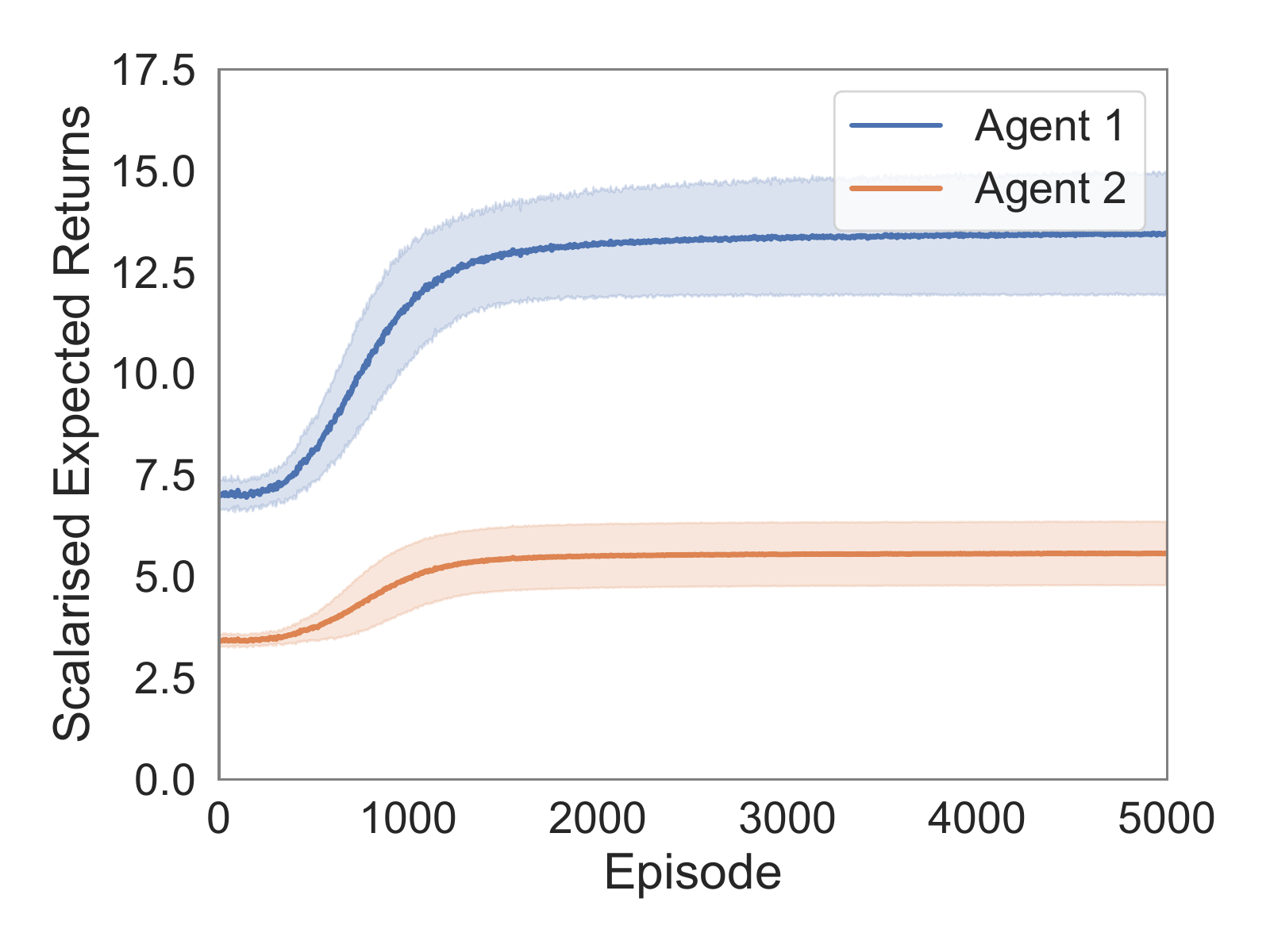}\label{fig:g5-ser-no-com} }}%
    \caption{The scalarised expected returns for both agents when learning without the use of communication.}%
    \label{fig:no-com-ser}%
\end{figure}

\begin{figure}[h!tb]%
    \centering
    \subfloat[Game 1]{{\includegraphics[width=.17\linewidth]{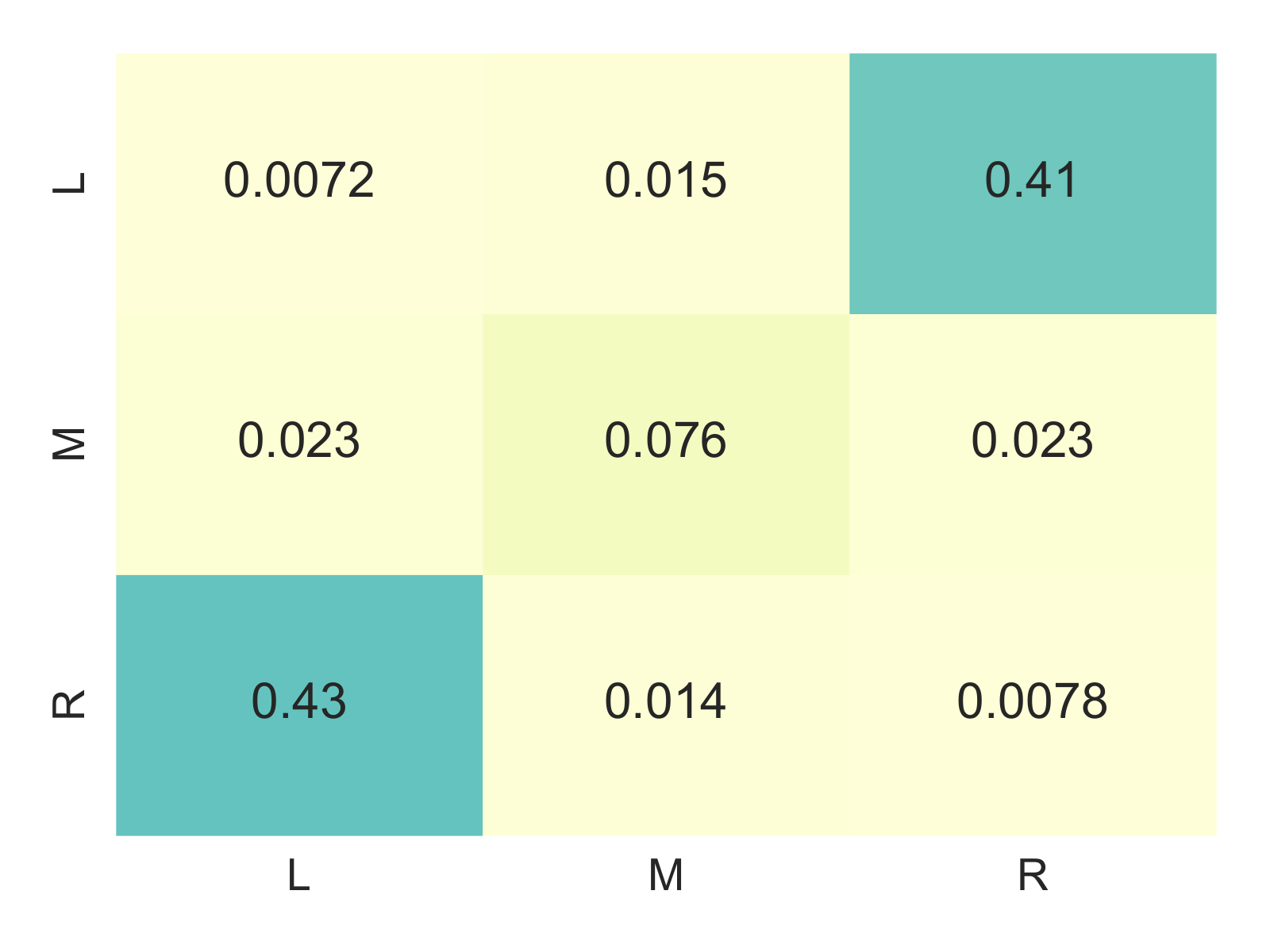}\label{fig:g1-states-no-com} }}%
    \quad
    \subfloat[Game 2]{{\includegraphics[width=.17\linewidth]{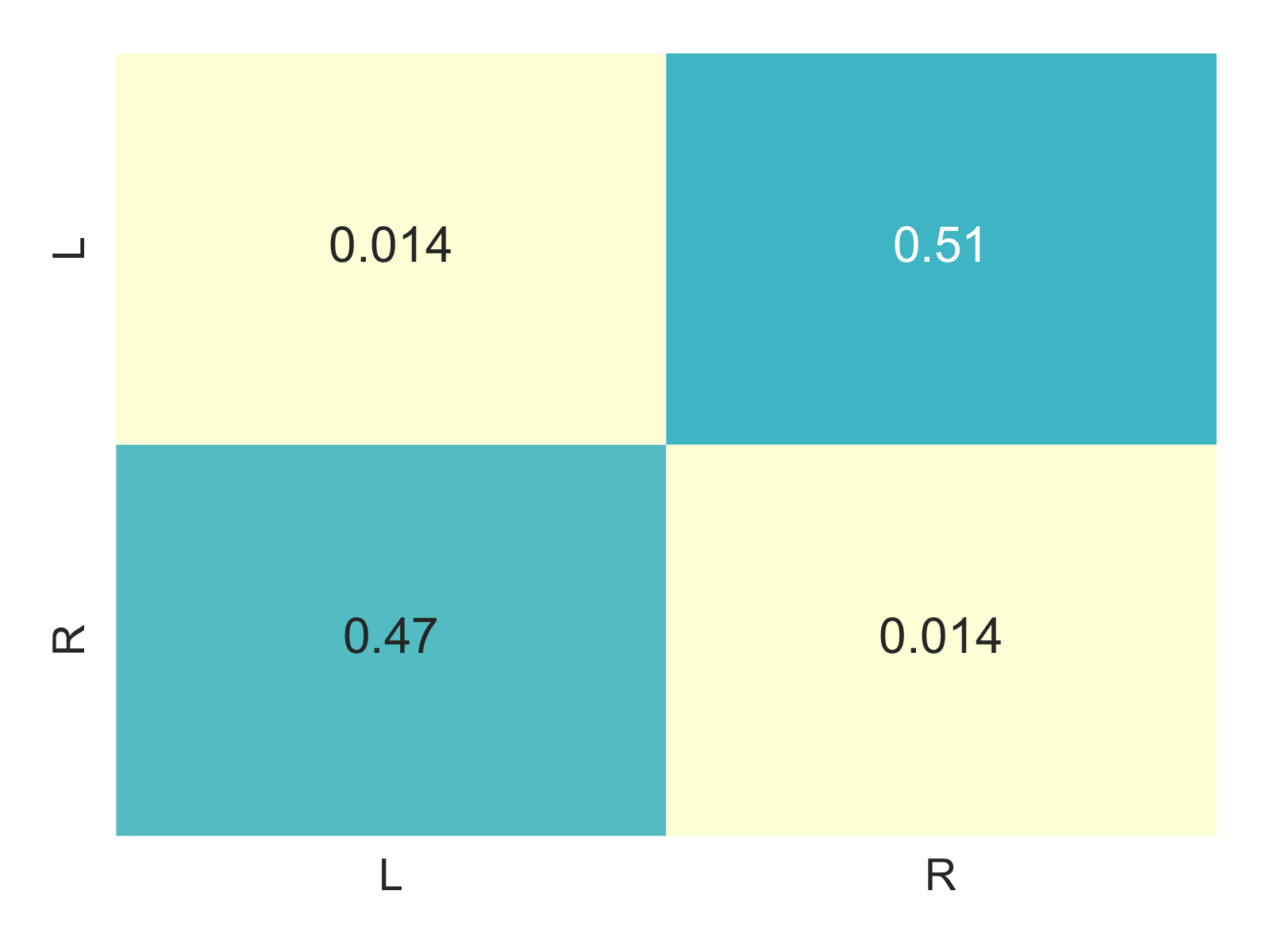}\label{fig:g2-states-no-com} }}%
    \quad
    \subfloat[Game 3]{{\includegraphics[width=.17\linewidth]{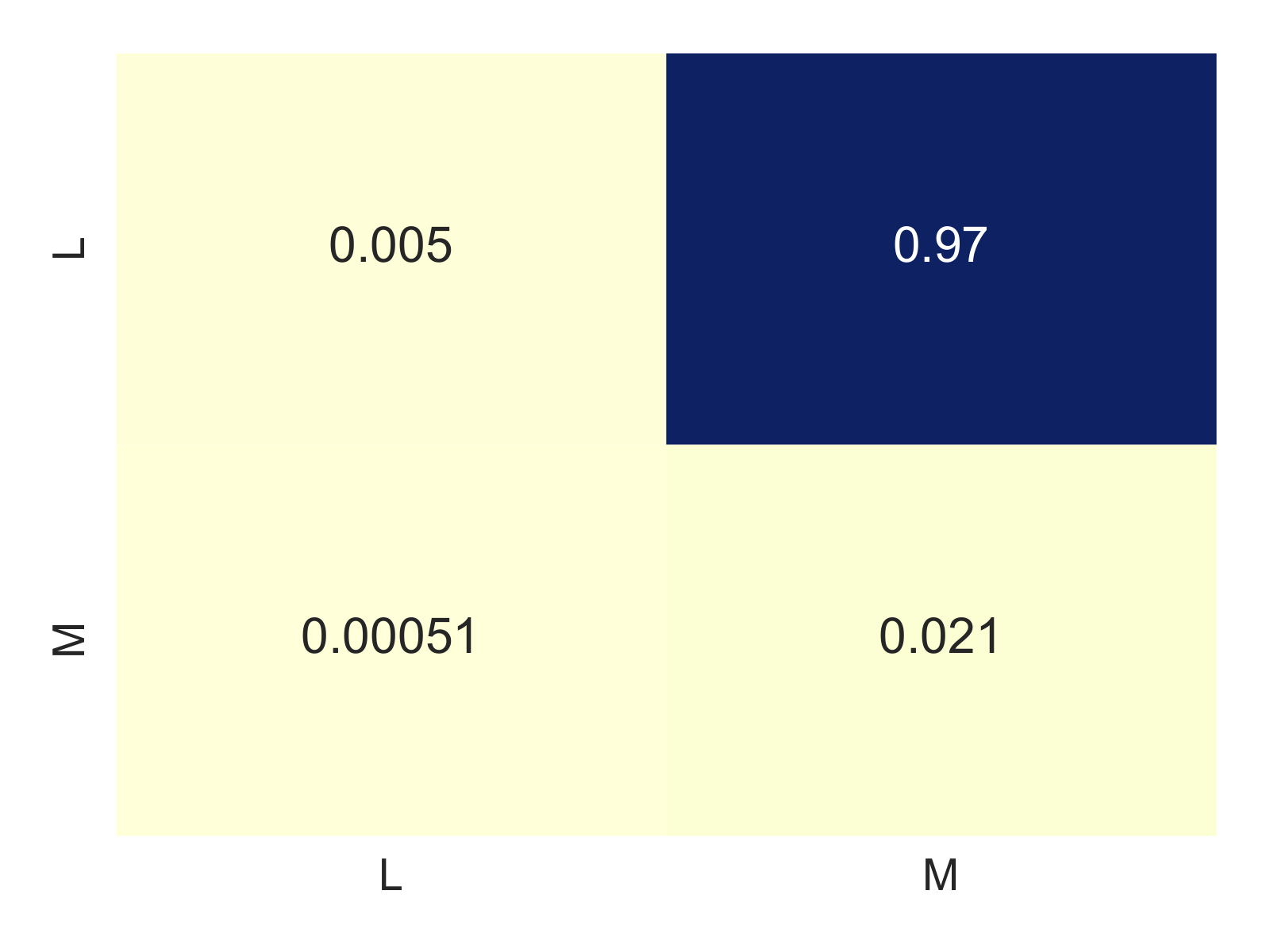}\label{fig:g3-states-no-com} }}%
    \quad
    \subfloat[Game 4]{{\includegraphics[width=.17\linewidth]{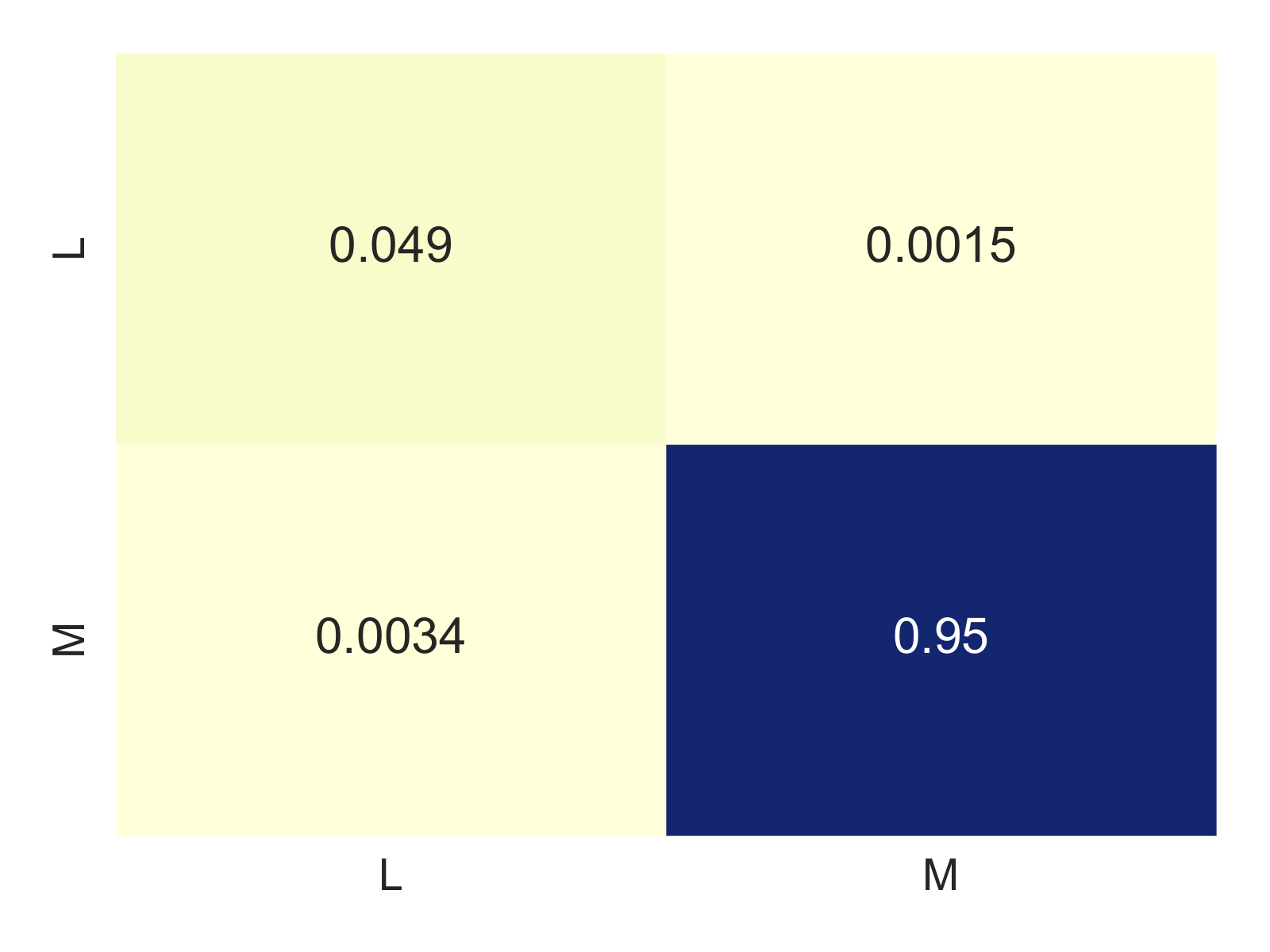}\label{fig:g4-states-no-com} }}%
    \quad
    \subfloat[Game 5]{{\includegraphics[width=.17\linewidth]{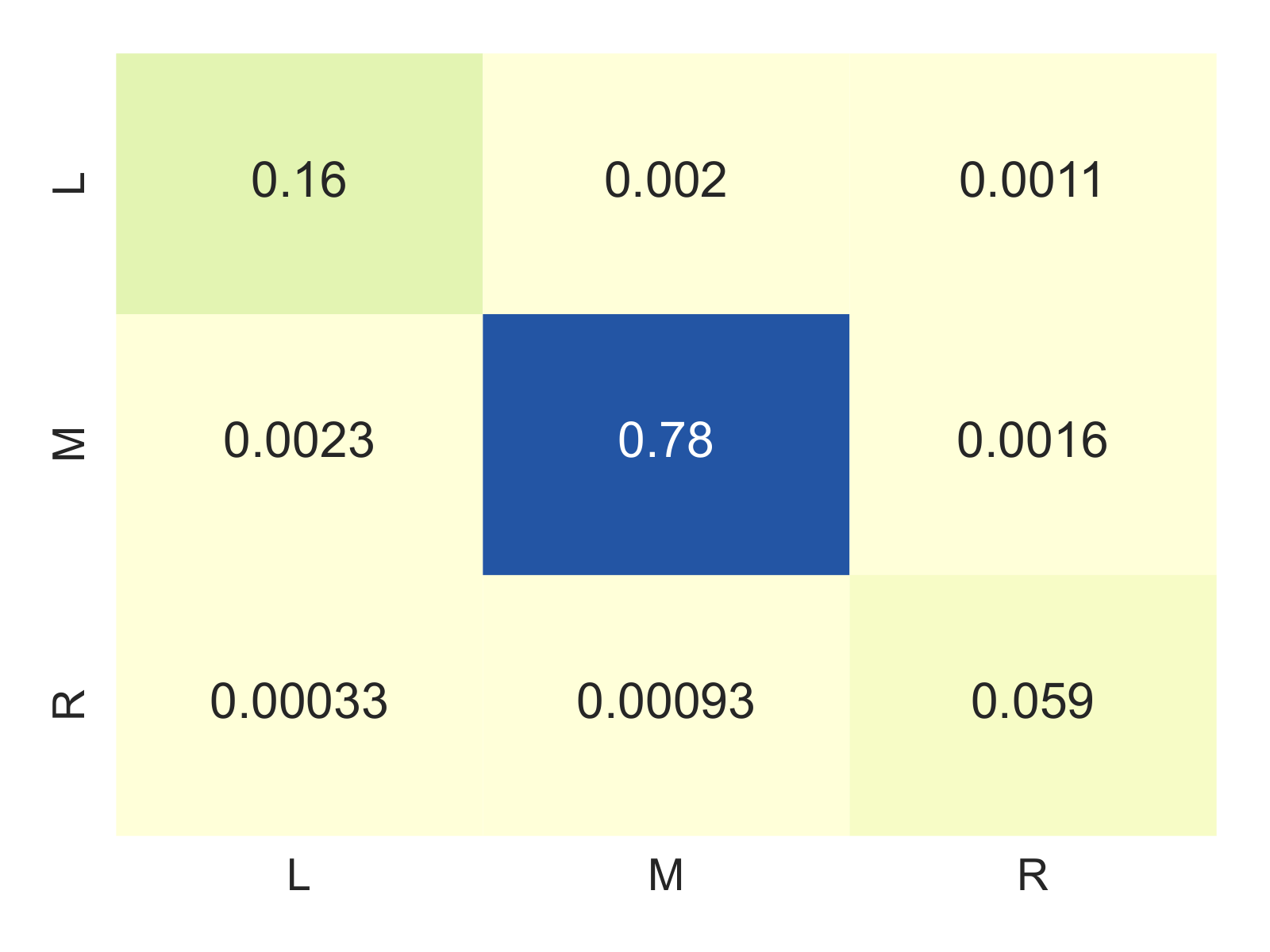}\label{fig:g5-states-no-com} }}%
    \caption{The empirical joint-action distributions in the last 10\% of episodes when learning without the use of communication.}%
    \label{fig:no-com-states}%
\end{figure}

Figure \ref{fig:no-com-ser} shows that agents reach a stable utility with low standard deviation in each game. These results are confirmed through previous work by R\u{a}dulescu et al. \cite{radulescu2020utility} and persist when repeating the experiments with a higher learning rate. In games with NE, this implies that independent learners learn the NE. Observe from Figure \ref{fig:no-com-states} that this is indeed the case as the NE are played with a considerably higher probability than other joint-actions. 

In games without NE, it is impossible for agents to converge on a stable strategy as there always is an incentive to deviate. We observe, however, that agents still reach a stable utility. We find that agents show cyclic behaviour, which averages out over different runs. Specifically, at each timestep at least one agent has an incentive to deviate from their current strategy. However, when an agent deviates from the joint strategy, this provides an incentive for their opponent to also deviate to ``counter'' this shift. This pattern maintains for several episodes until agents cycle back to their original policies. Similar results have also been found in single-objective games. Best-response iteration algorithms are known to show cyclic behaviour \cite{anthony2020learning}, while gradient ascent is not guaranteed to converge to an NE but average rewards do \cite{singh2000nash}.


\begin{figure}[h!tb]%
    \centering
    \subfloat[Game 1]{{\includegraphics[width=.28\linewidth]{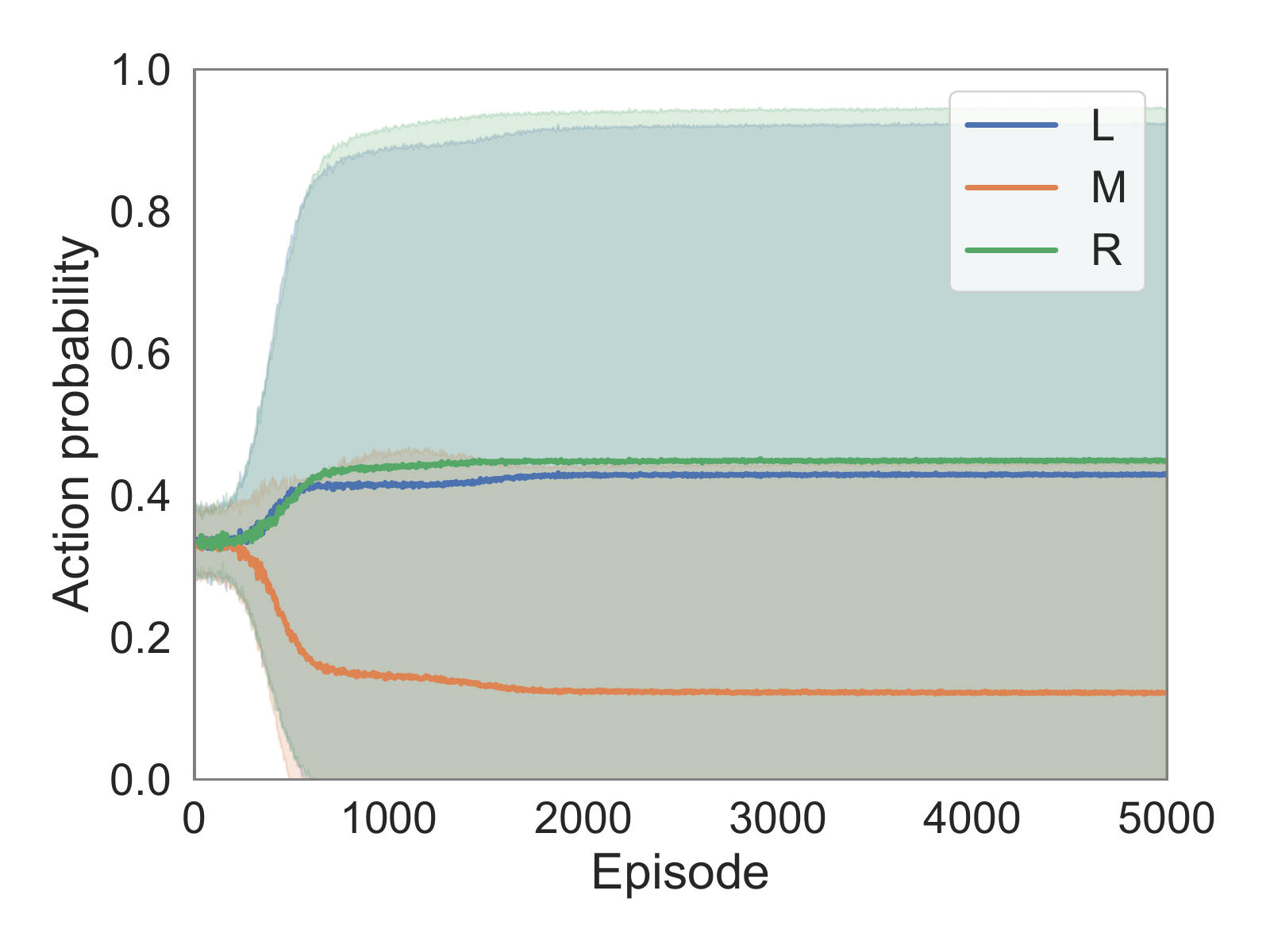}\label{fig:g1-A1-probs-no-com} }}%
    \quad
    \subfloat[Game 2]{{\includegraphics[width=.28\linewidth]{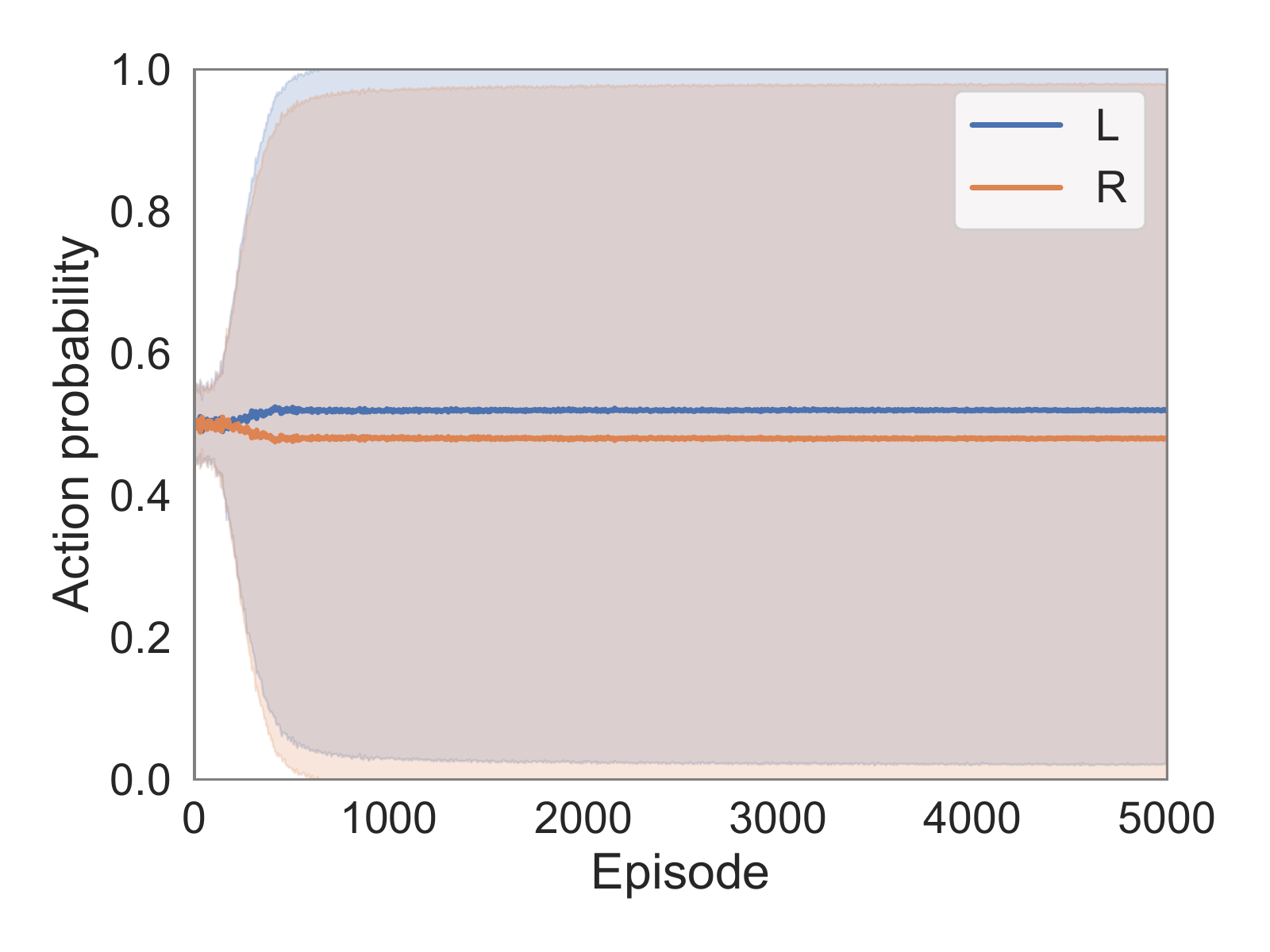}\label{fig:g2-A1-probs-no-com} }}%
    \quad
    \subfloat[Game 3]{{\includegraphics[width=.28\linewidth]{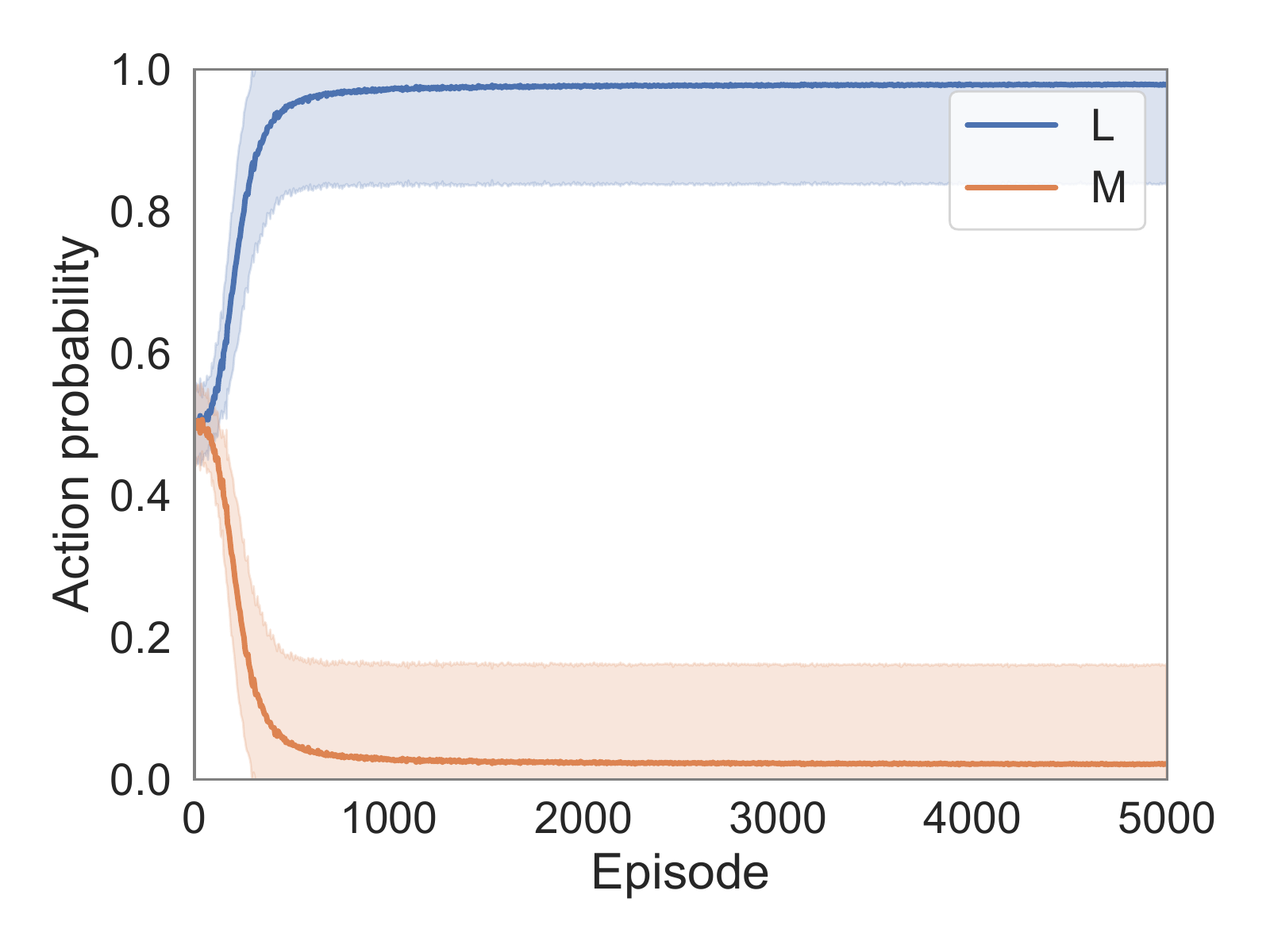}\label{fig:g3-A1-probs-no-com} }}%
    \quad
    \subfloat[Game 4]{{\includegraphics[width=.28\linewidth]{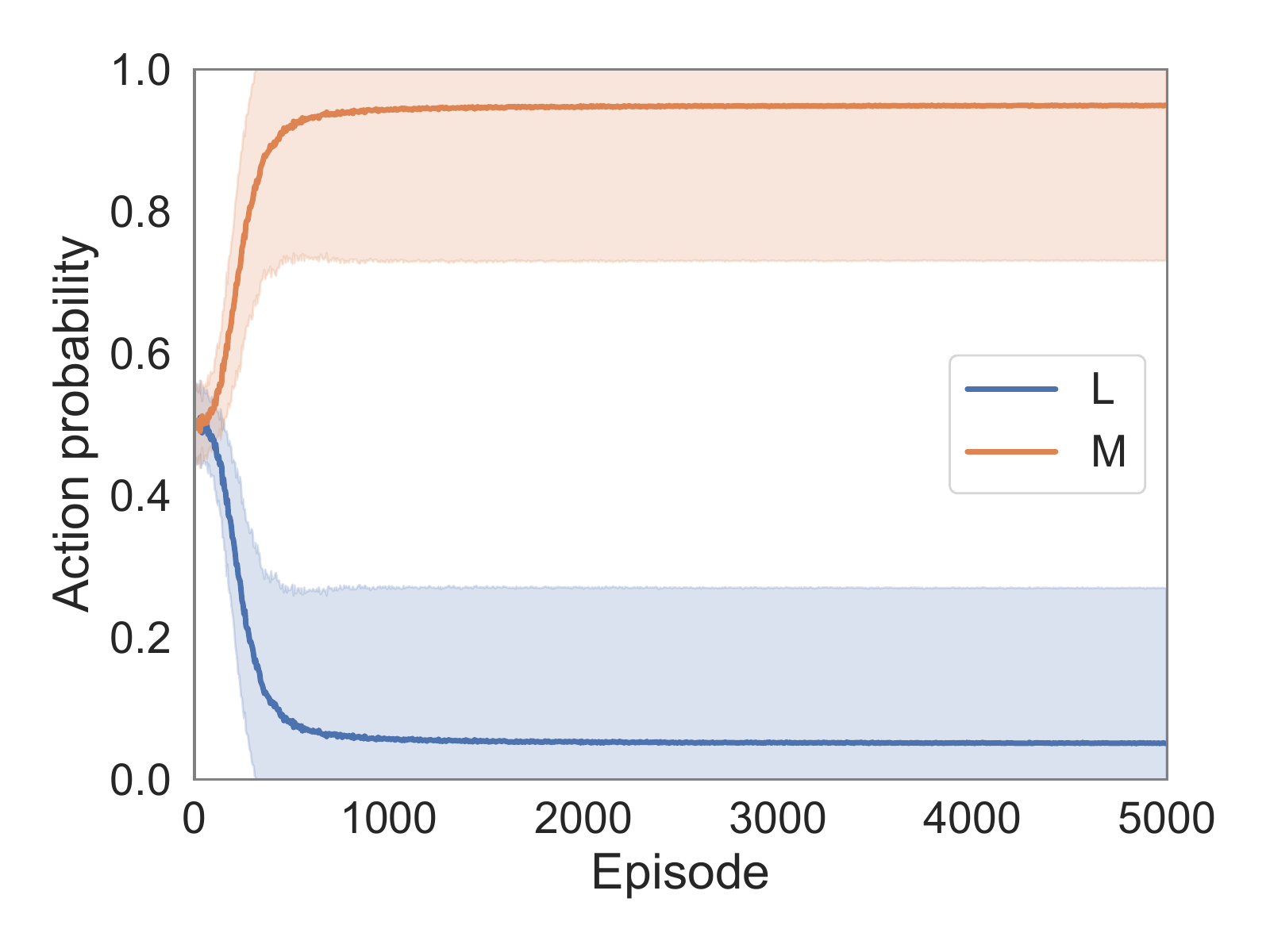}\label{fig:g4-A1-probs-no-com} }}%
    \quad
    \subfloat[Game 5]{{\includegraphics[width=.28\linewidth]{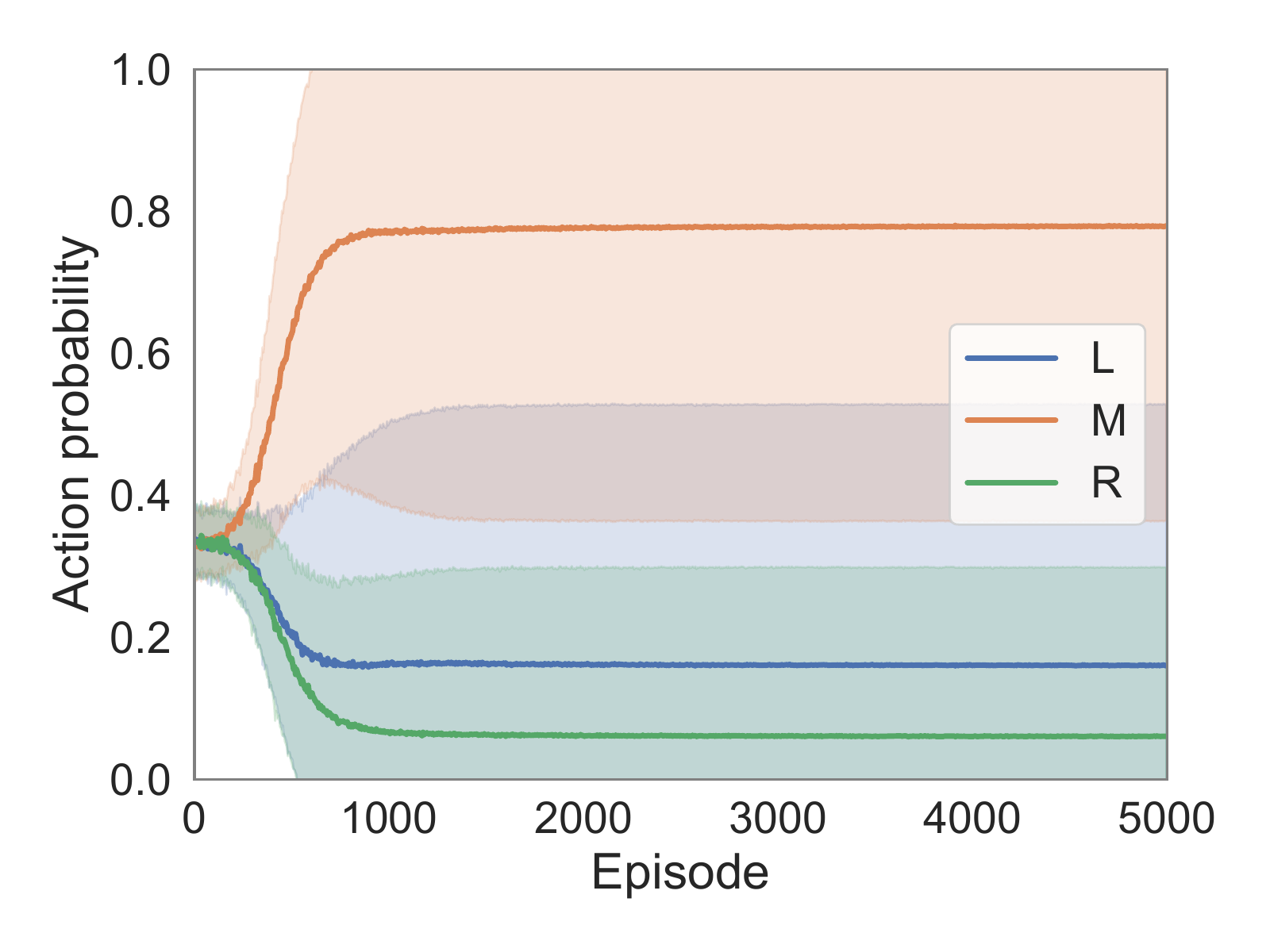}\label{fig:g5-A1-probs-no-com} }}%
    \caption{The action probabilities for agent 1 when learning without the use of communication.}%
    \label{fig:no-com-A1-probs}%
\end{figure}

\begin{figure}[h!tb]%
    \centering
    \subfloat[Game 1]{{\includegraphics[width=.28\linewidth]{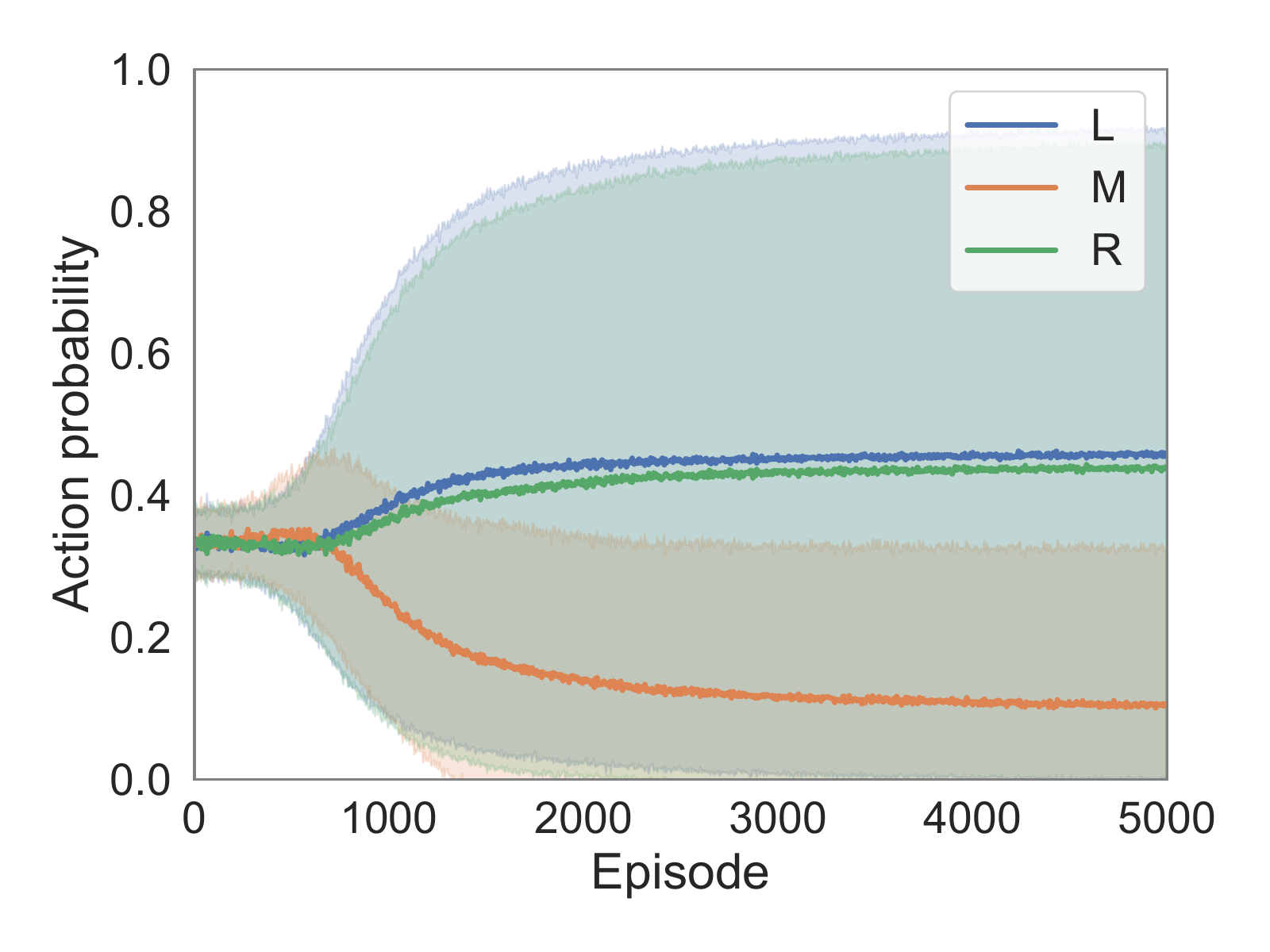}\label{fig:g1-A2-probs-no-com} }}%
    \quad
    \subfloat[Game 2]{{\includegraphics[width=.28\linewidth]{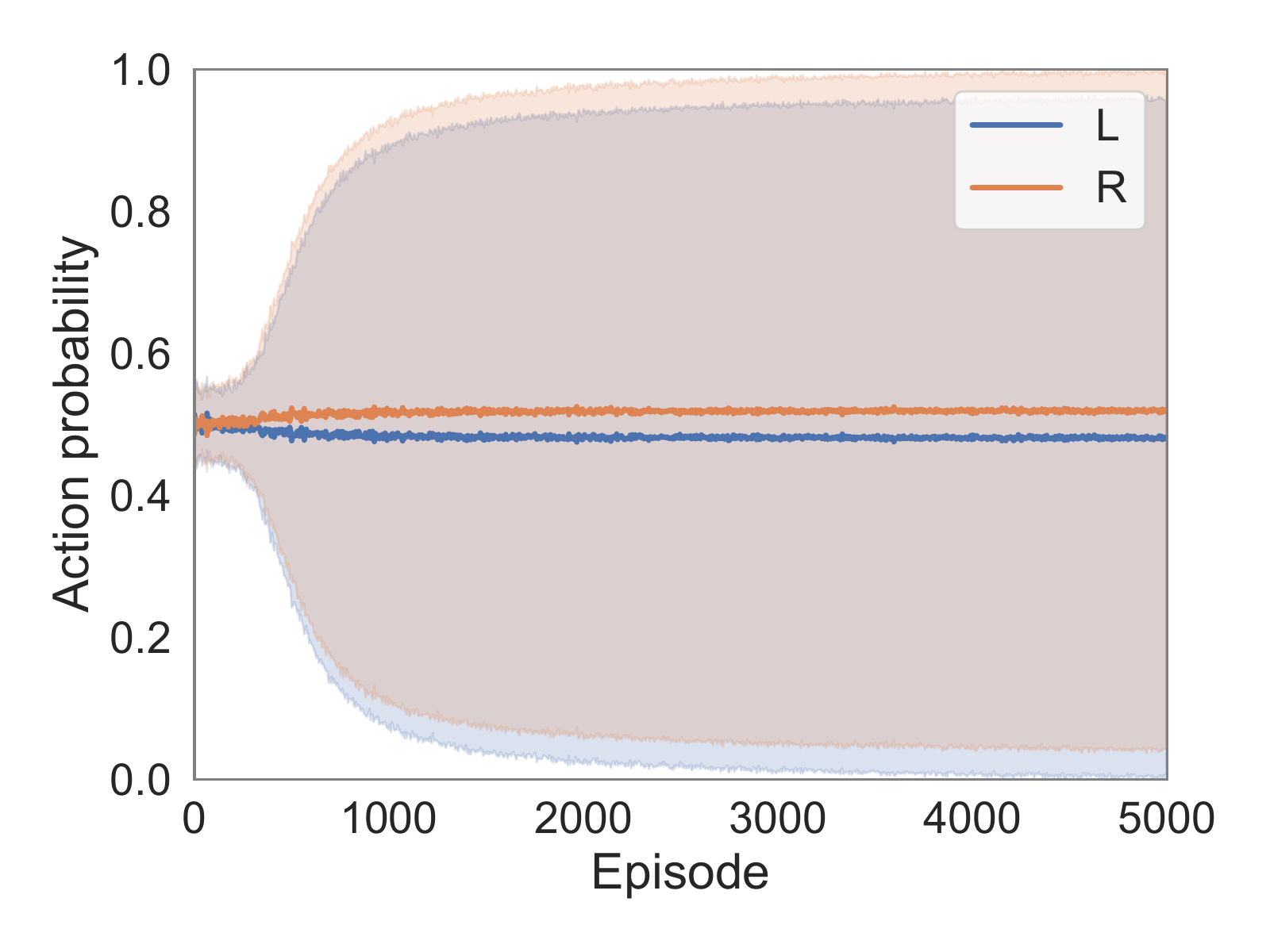}\label{fig:g2-A2-probs-no-com} }}%
    \quad
    \subfloat[Game 3]{{\includegraphics[width=.28\linewidth]{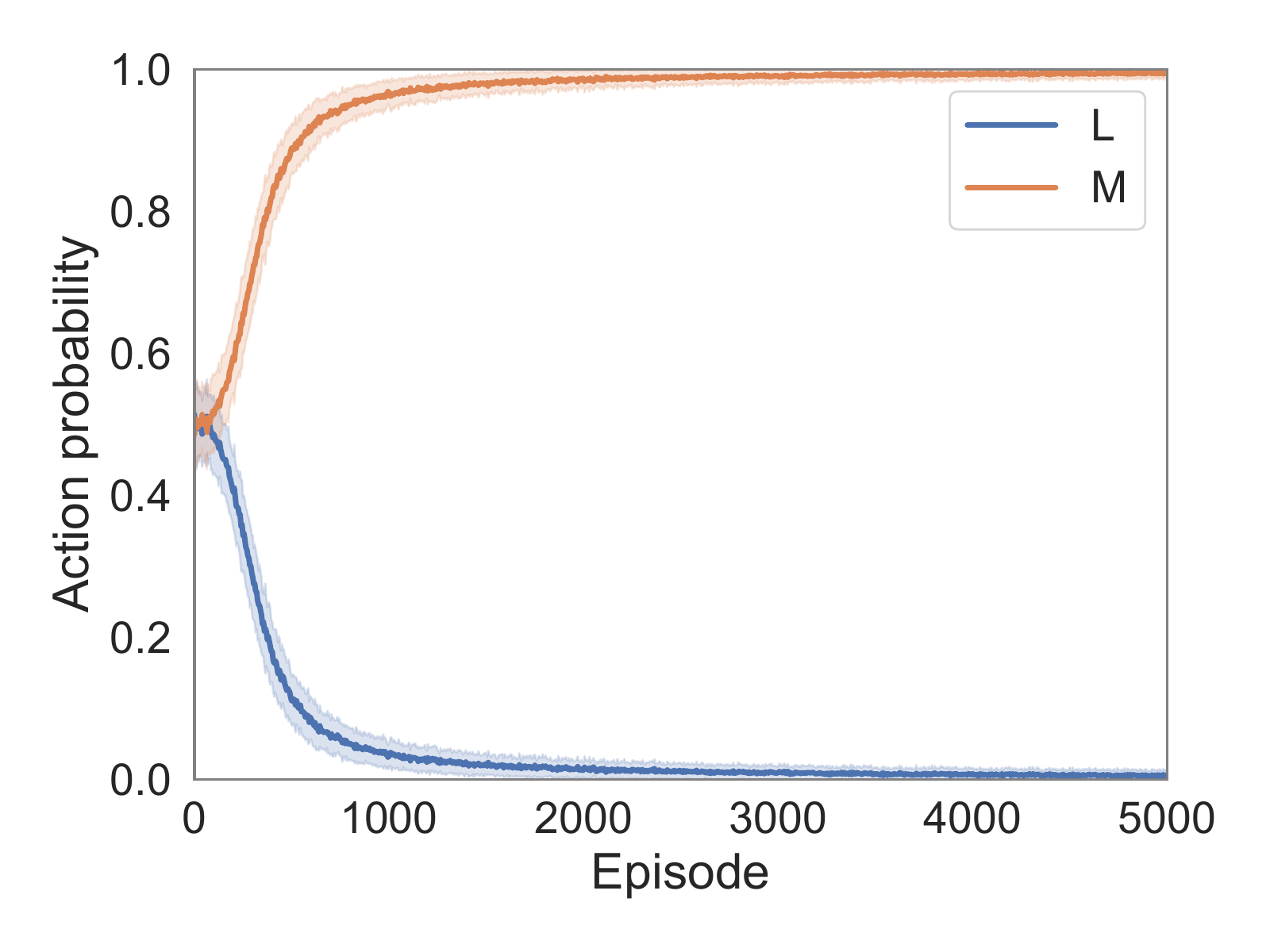}\label{fig:g3-A2-probs-no-com} }}%
    \quad
    \subfloat[Game 4]{{\includegraphics[width=.28\linewidth]{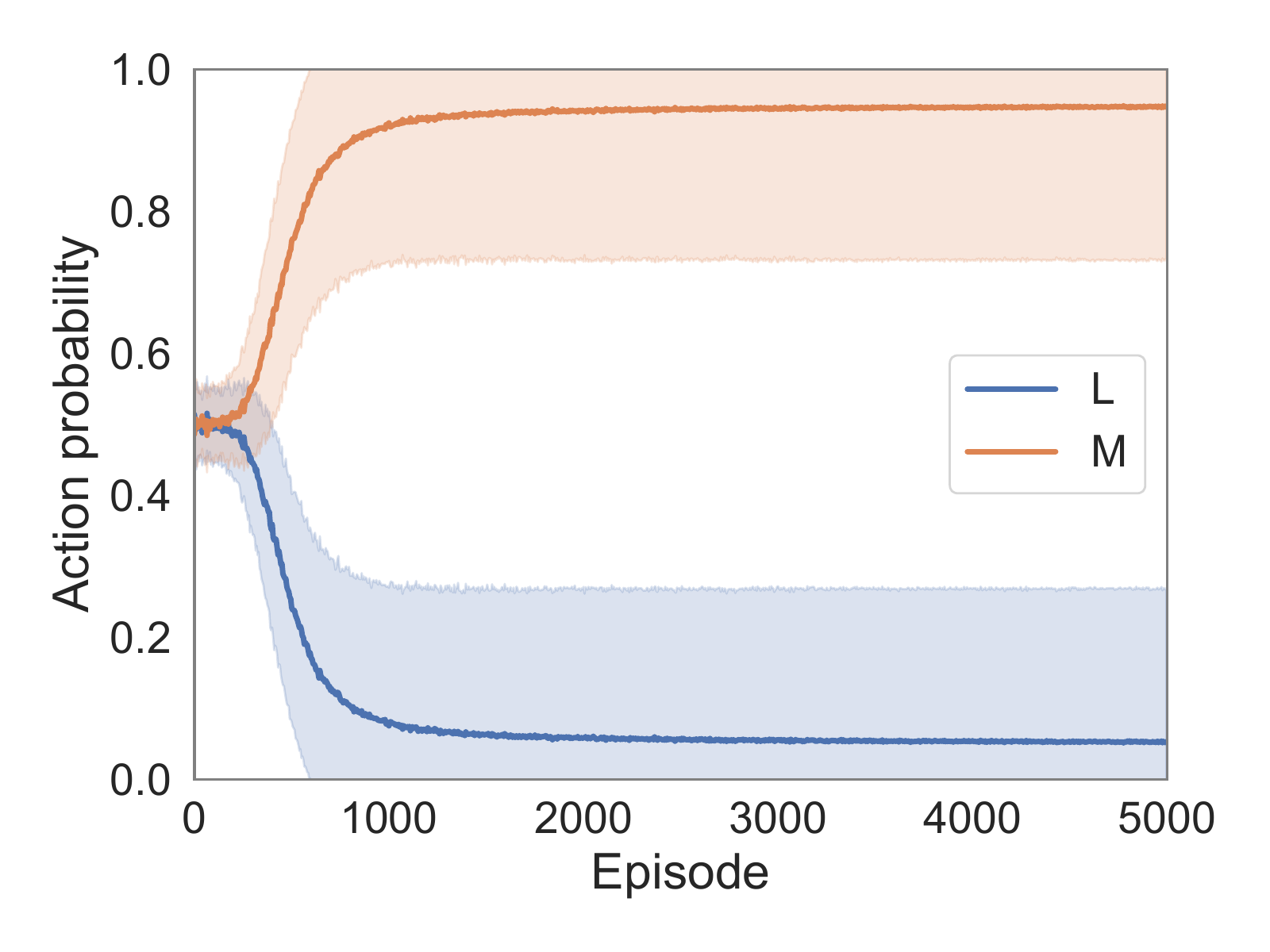}\label{fig:g4-A2-probs-no-com} }}%
    \quad
    \subfloat[Game 5]{{\includegraphics[width=.28\linewidth]{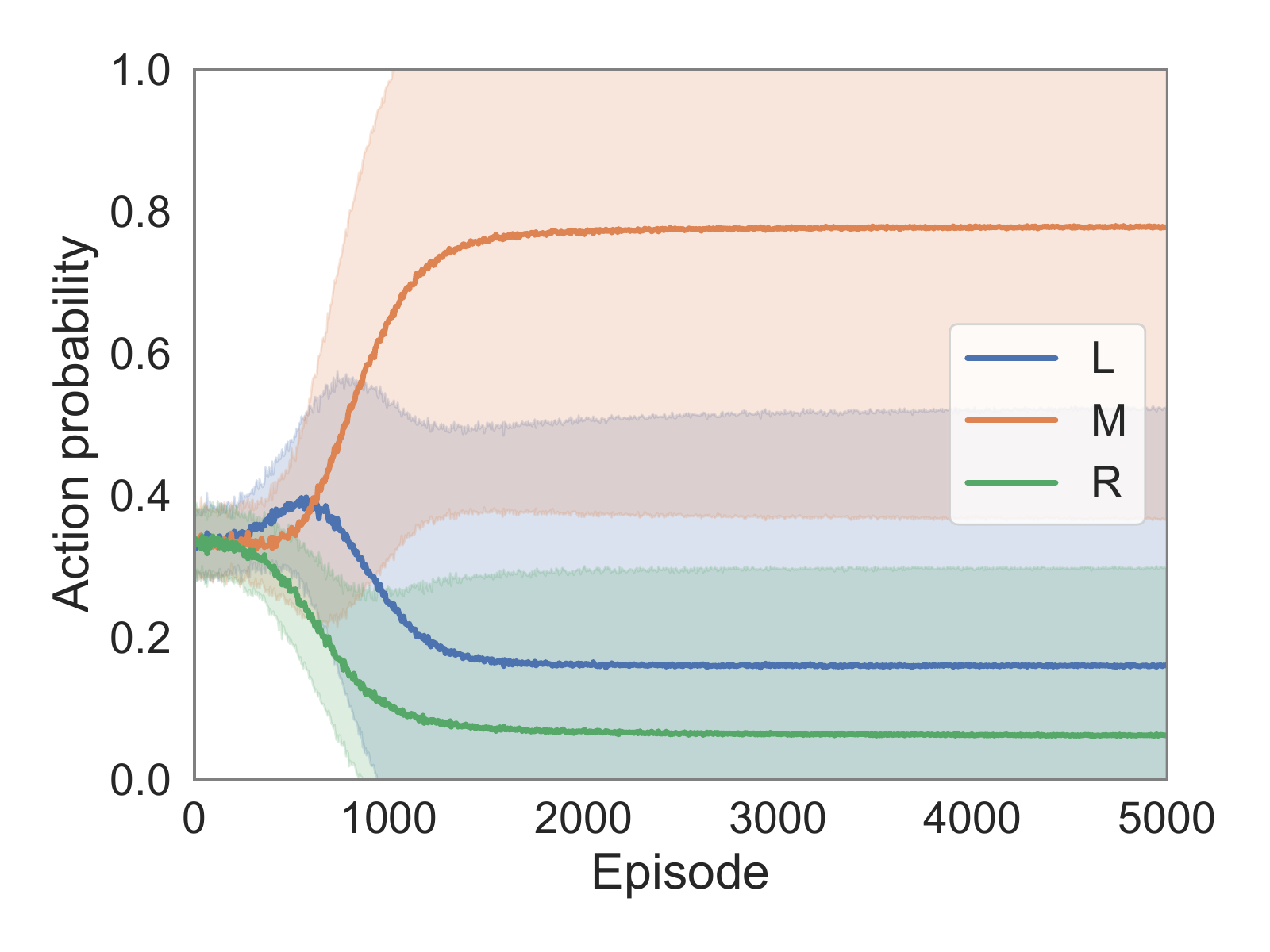}\label{fig:g5-A2-probs-no-com} }}%
    \caption{The action probabilities for agent 2 when learning without the use of communication.}%
    \label{fig:no-com-A2-probs}%
\end{figure}

These conclusions are also supported when considering the action probabilities of both players in Figures \ref{fig:no-com-A1-probs} and \ref{fig:no-com-A2-probs}. First, in games without NE we see that standard deviations around action probabilities are high. This is because no stable joint strategy exists, so that learning trajectories for strategies can vary across runs. In Game 3, which has a unique NE, agents converge to this. The standard deviation around agent 1 can be attributed to the limitations of independent learning. Specifically, in several runs agent 1 did not learn the full Q-value of L and got stuck in a suboptimal policy of playing R. In Game 4 and 5, multiple NE exist. As we repeat the experiment for multiple runs, agents converge on different NE which explains the standard deviation.

Lastly, Figure \ref{fig:no-com-states} shows what joint strategies are played after agents have converged. In the games without NE, we see that (R, L) and (L, R) are played most often. As mentioned before, agents go through cycles of policy updates that lead back to these policies. Because the average utility of these joint-strategies converges, we consider them compromise joint-policies given the independent learners. We expect such compromise joint-policies to be impossible with self-interested action communication, as any communication can be fully exploited.

In general, we observe that in games with NE agents converge to them. In Game 3, agents converge on the unique NE. We observe, however, the same conclusion as from Figure \ref{fig:no-com-A1-probs}, namely that agent 1 can in some cases get stuck playing R. 

In Game 4, agents mostly converge to the preferred NE of player 2. We attribute this to two factors. First, it appears harder to learn an NE with unbalanced payoffs. We hypothesise that in earlier episodes balanced payoffs seem more attractive as both objectives contribute to the utility. It is only once agents learn the unbalanced payoff to an adequate degree that its utility outperforms that of a balanced payoff. This is an important nuance in multi-objective games which does not exist for single-objective games. The second factor we identify is due to the utility functions and starting policies for the agents. With independent learning, agents learn the (vectorial) Q-values for their actions. In earlier episodes, both players employ an (approximately) uniform distribution over the actions. The result for player 1 is that the Q-value for M dominates L, making M their optimal policy. When player 1 plays a uniform distribution, the best-response for player 2 is to keep playing a uniform distribution. As such, the only way for agent 1 to converge to an equilibrium is to play M which induces (M, M). We expect the equilibrium of (L, L) to be played more frequently when introducing cooperative communication which allows for additional coordination or self-interested action communication where players have the opportunity to learn which equilibrium they prefer.

The same results are present for Game 5. Note however that (L, L) is played more frequently than in Game 4. We can attribute the higher convergence to (L, L) compared to Game 4 by considering the two previous factors. Observe that while the first factor still holds, i.e. imbalanced NE are harder to learn, the second does not. Specifically, player 2 now has an incentive to play L whenever player 1 assumes a uniform strategy. This explains why convergence to (L, L) is higher compared to Game 4, but not at the same level as (M, M). We note that while (R, R) is generally avoided here, in experiments with a higher learning rate agents increasingly converge on (R, R) due to fewer opportunities for exploration.

\subsection{Cooperative Action Communication}
\label{sec:coop-com-exp}
Cooperative action communication drives agents to cooperate by optimising for a single joint policy. This is done by letting the leader commit to their next action. The follower uses this commitment to update their policy in advance. Note that given the individual utility functions, the game itself is non-cooperative. The setting is made cooperative by enforcing the proposed protocol which ensures the optimisation of a single joint policy.

\begin{figure}[h!tb]%
    \centering
    \subfloat[Game 1]{{\includegraphics[width=.28\linewidth]{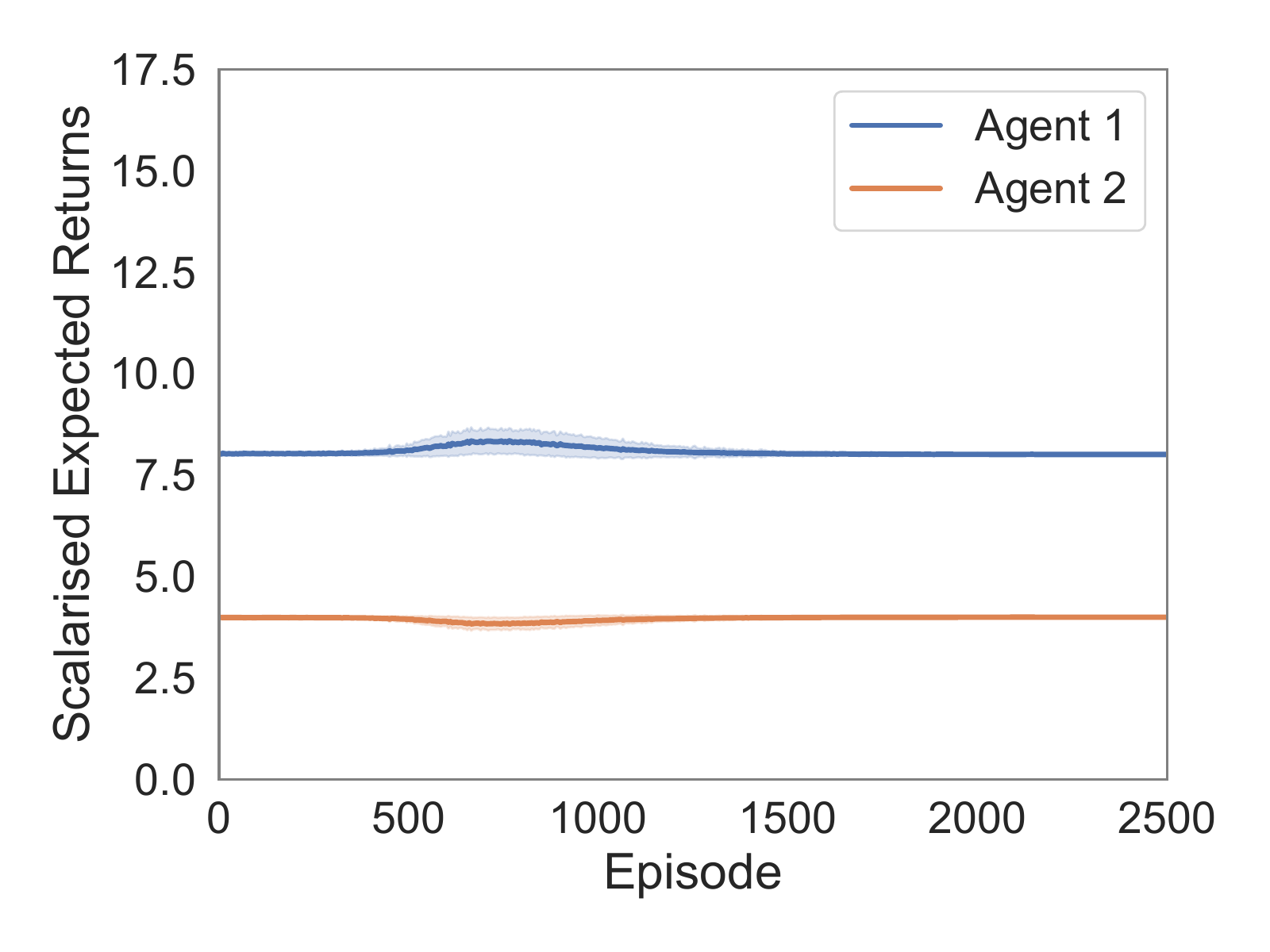}\label{fig:g1-ser-coop-action} }}%
    \quad
    \subfloat[Game 2]{{\includegraphics[width=.28\linewidth]{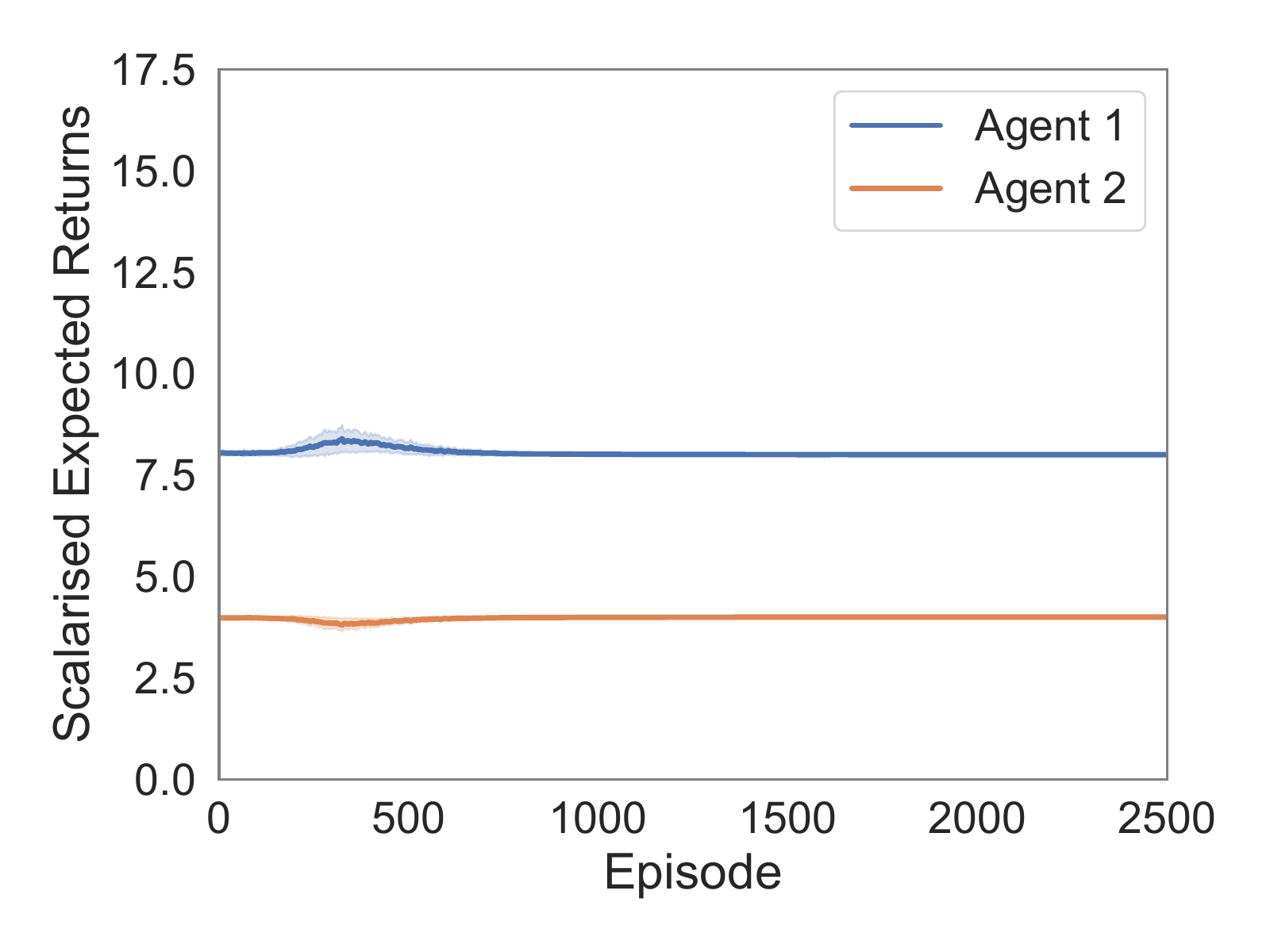}\label{fig:g2-ser-coop-action} }}%
    \quad
    \subfloat[Game 3]{{\includegraphics[width=.28\linewidth]{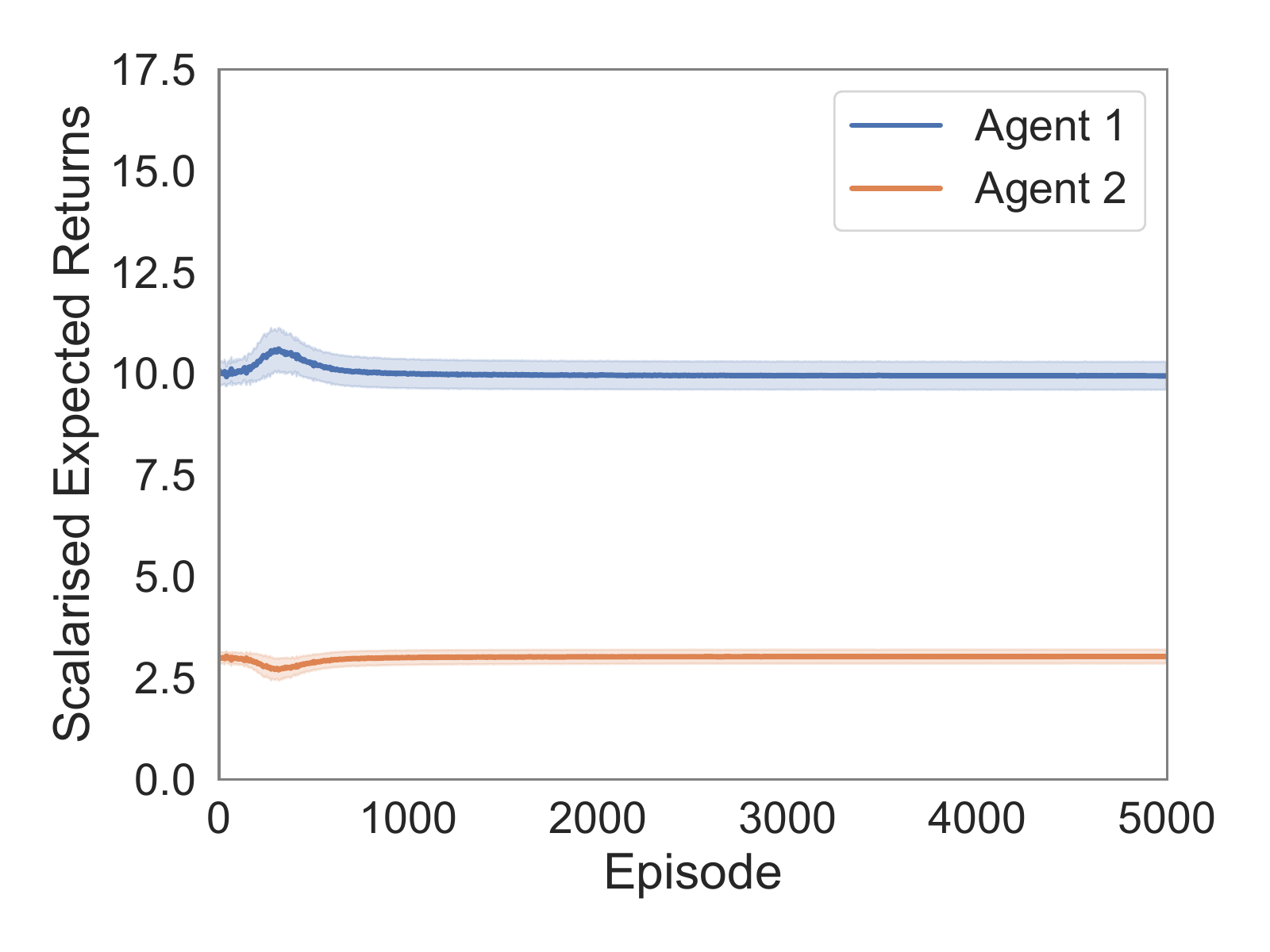}\label{fig:g3-ser-coop-action} }}%
    \quad
    \subfloat[Game 4]{{\includegraphics[width=.28\linewidth]{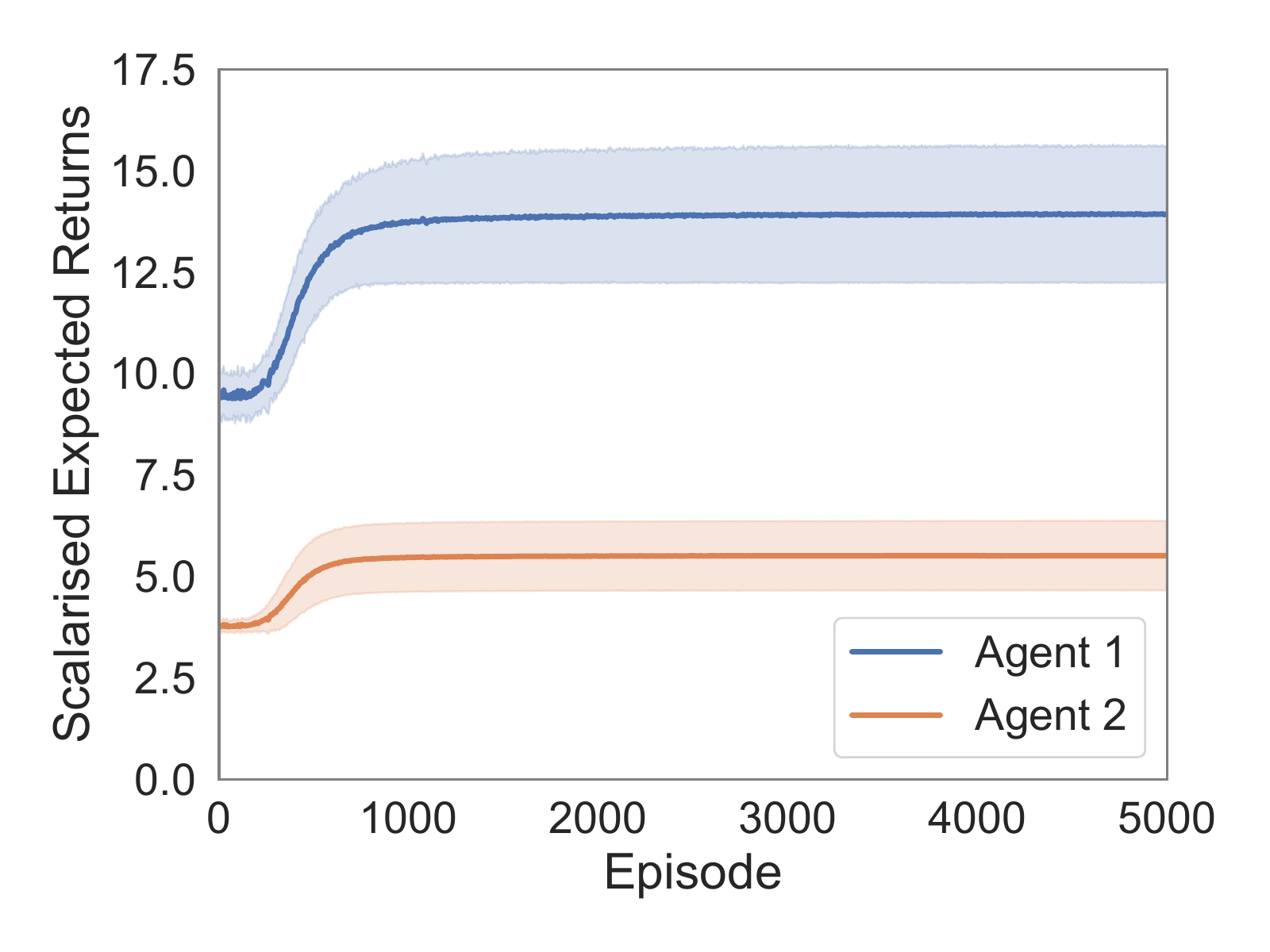}\label{fig:g4-ser-coop-action} }}%
    \quad
    \subfloat[Game 5]{{\includegraphics[width=.28\linewidth]{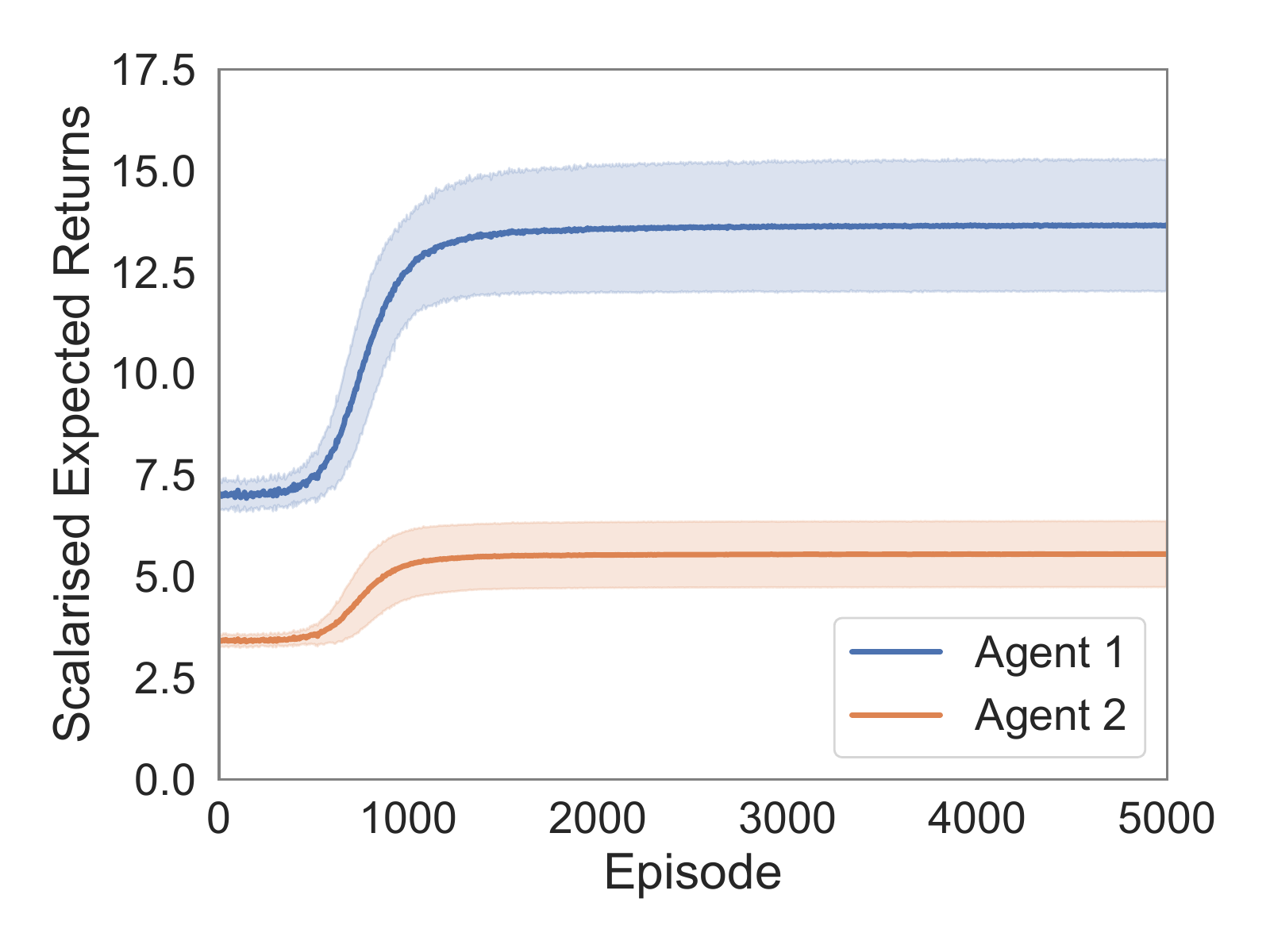}\label{fig:g5-ser-coop-action} }}%
    \caption{The SER for both agents when learning with cooperative action communication.}%
    \label{fig:coop-action-ser}%
\end{figure}

We first discuss the results for the SER in Figure \ref{fig:coop-action-ser}. In the games without NE, we see that there is less divergence from the final stable utility, compared to experiments without communication. Specifically, the ``bump" where agents are exploring different strategies is flatter. When no NE exist, committing to an action can always be exploited by the follower. This quickly teaches both agents that deviating from the middle ground strategies (R, L) or (L, R) can never work. 

In the games with NE, we observe that it provides a moderate increase in learning speed. Concretely, the steeper learning curve shows that agents converge to the final strategies faster. Since the follower updates their policy with regards to the commitment from the leader, this can move them to discover and converge to the NE faster. We highlight that this increase in learning speed is distinct from simply raising the learning rate for independent learners as this can result in more frequent convergence to dominated Nash equilibria.

\begin{figure}[h!tb]%
    \centering
    \subfloat[Game 1]{{\includegraphics[width=.17\linewidth]{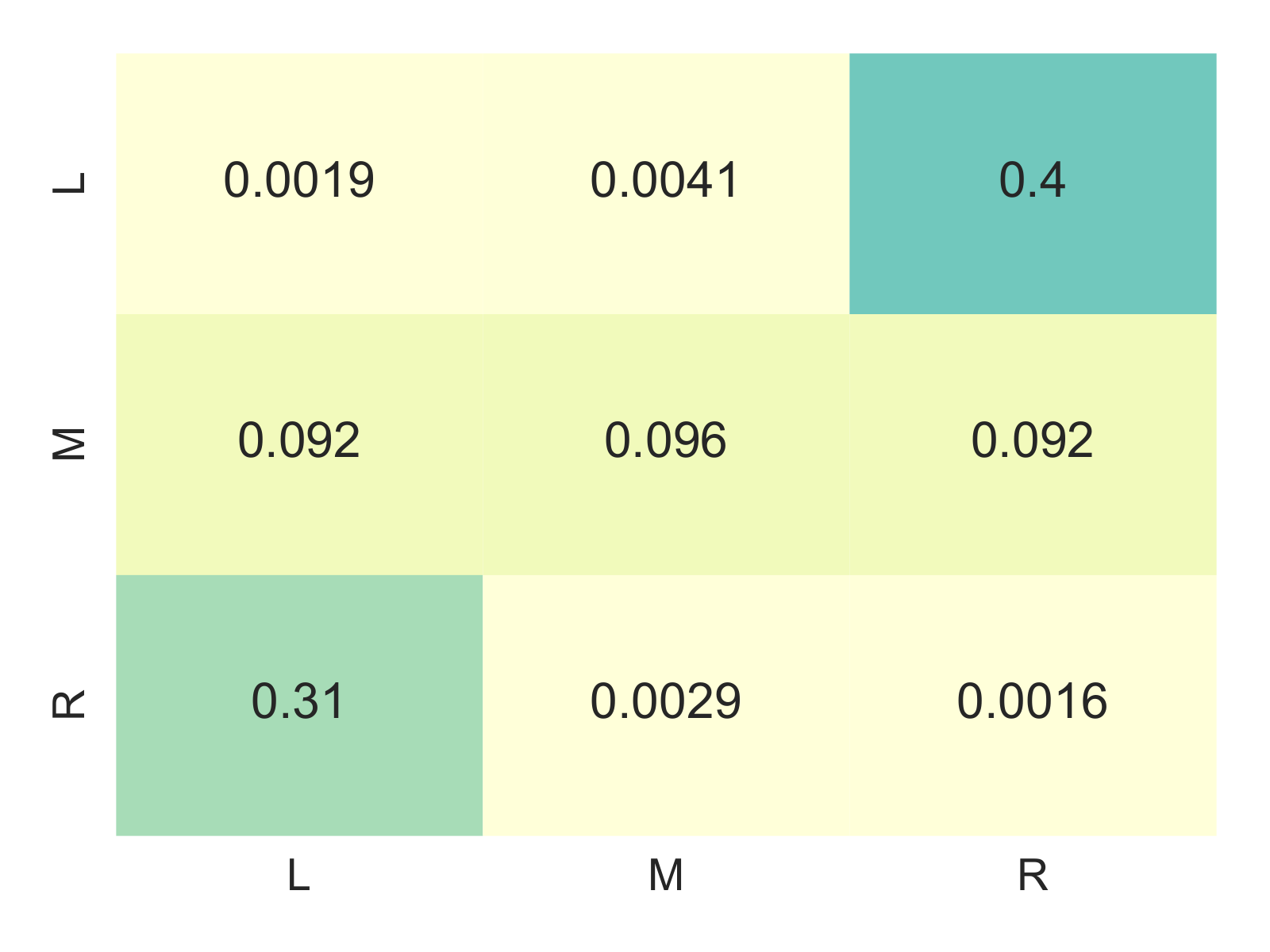}\label{fig:g1-states-coop-action} }}%
    \quad
    \subfloat[Game 2]{{\includegraphics[width=.17\linewidth]{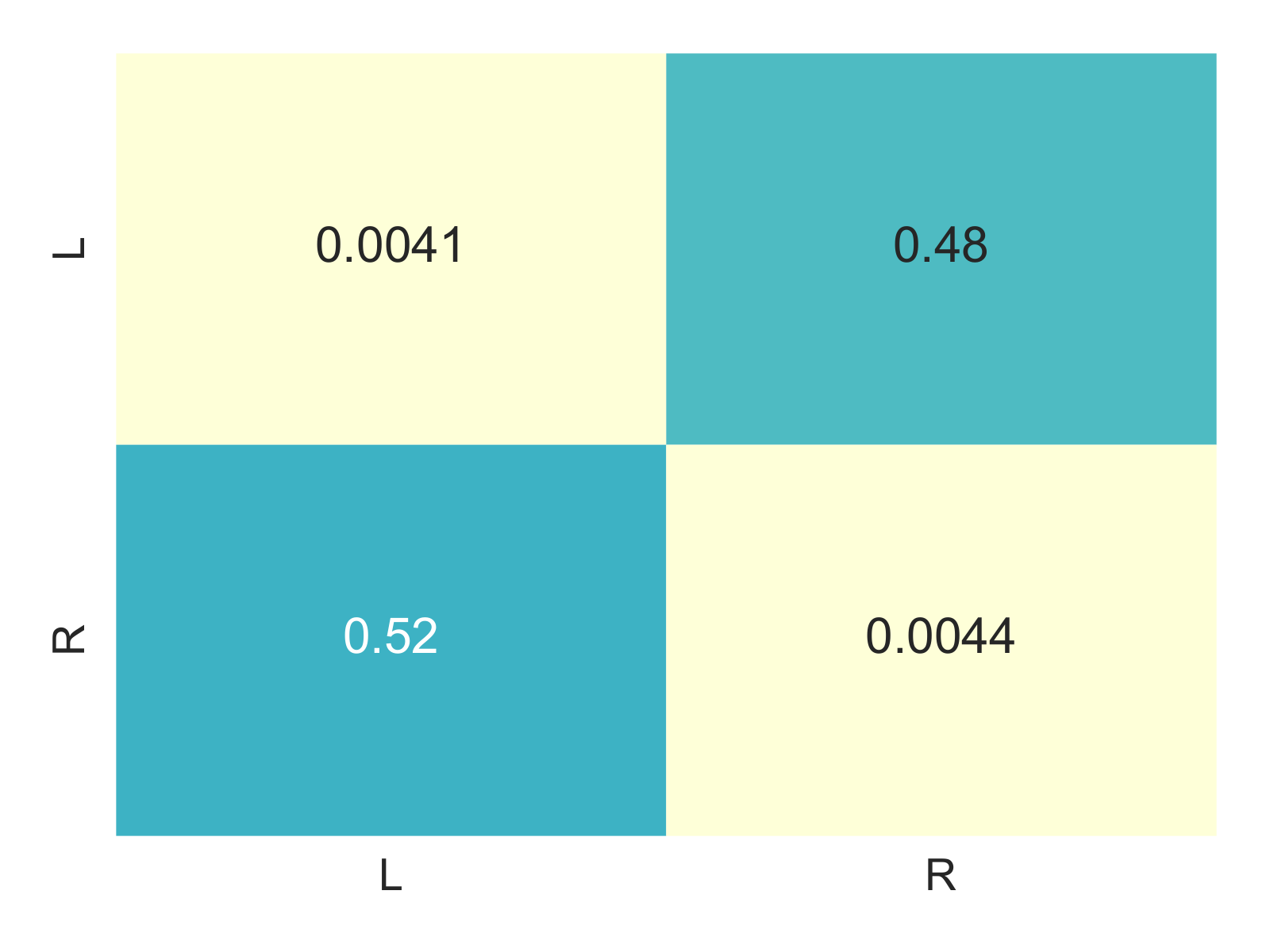}\label{fig:g2-states-coop-action} }}%
    \quad
    \subfloat[Game 3]{{\includegraphics[width=.17\linewidth]{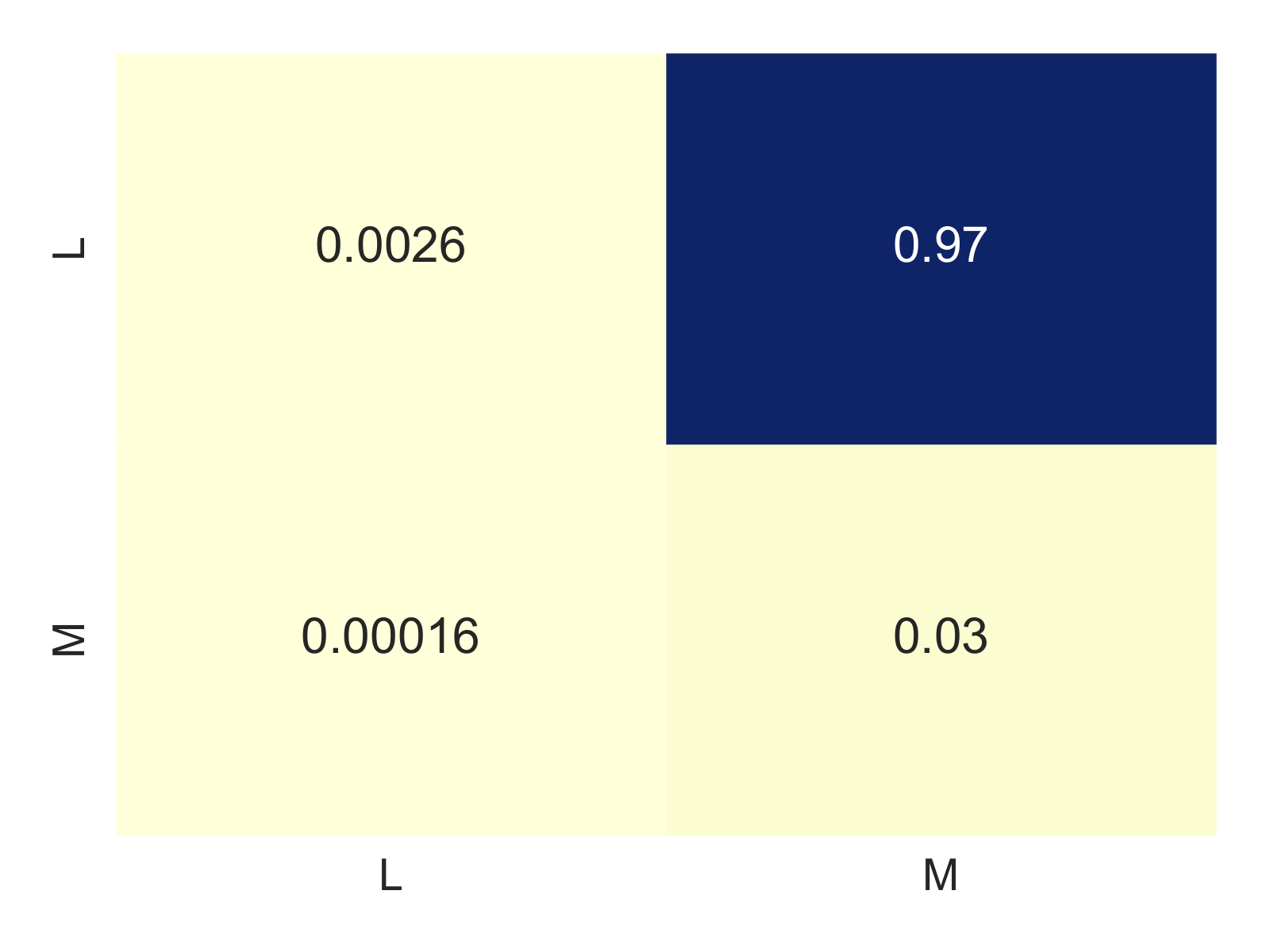}\label{fig:g3-states-coop-action} }}%
    \quad
    \subfloat[Game 4]{{\includegraphics[width=.17\linewidth]{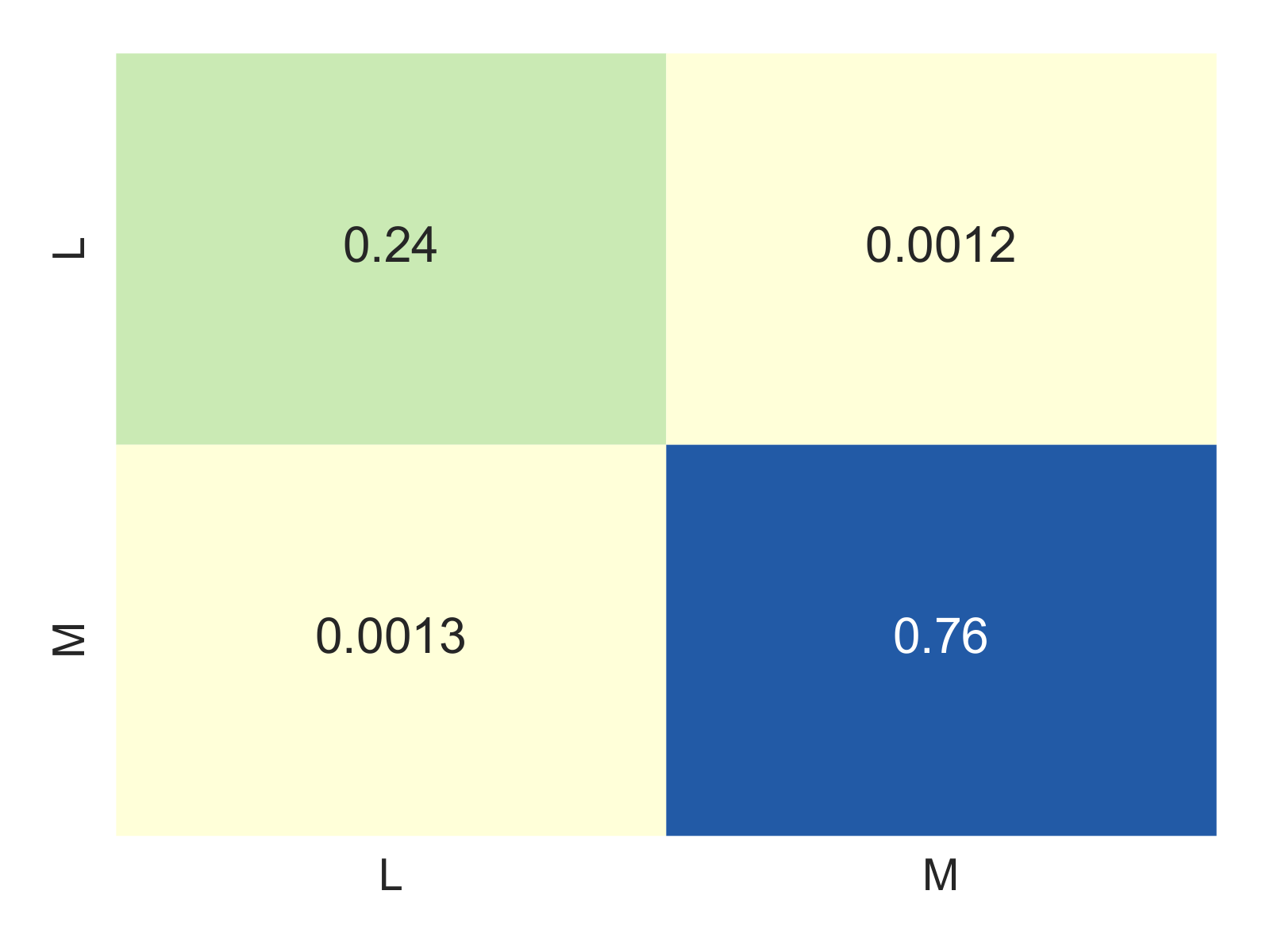}\label{fig:g4-states-coop-action} }}%
    \quad
    \subfloat[Game 5]{{\includegraphics[width=.17\linewidth]{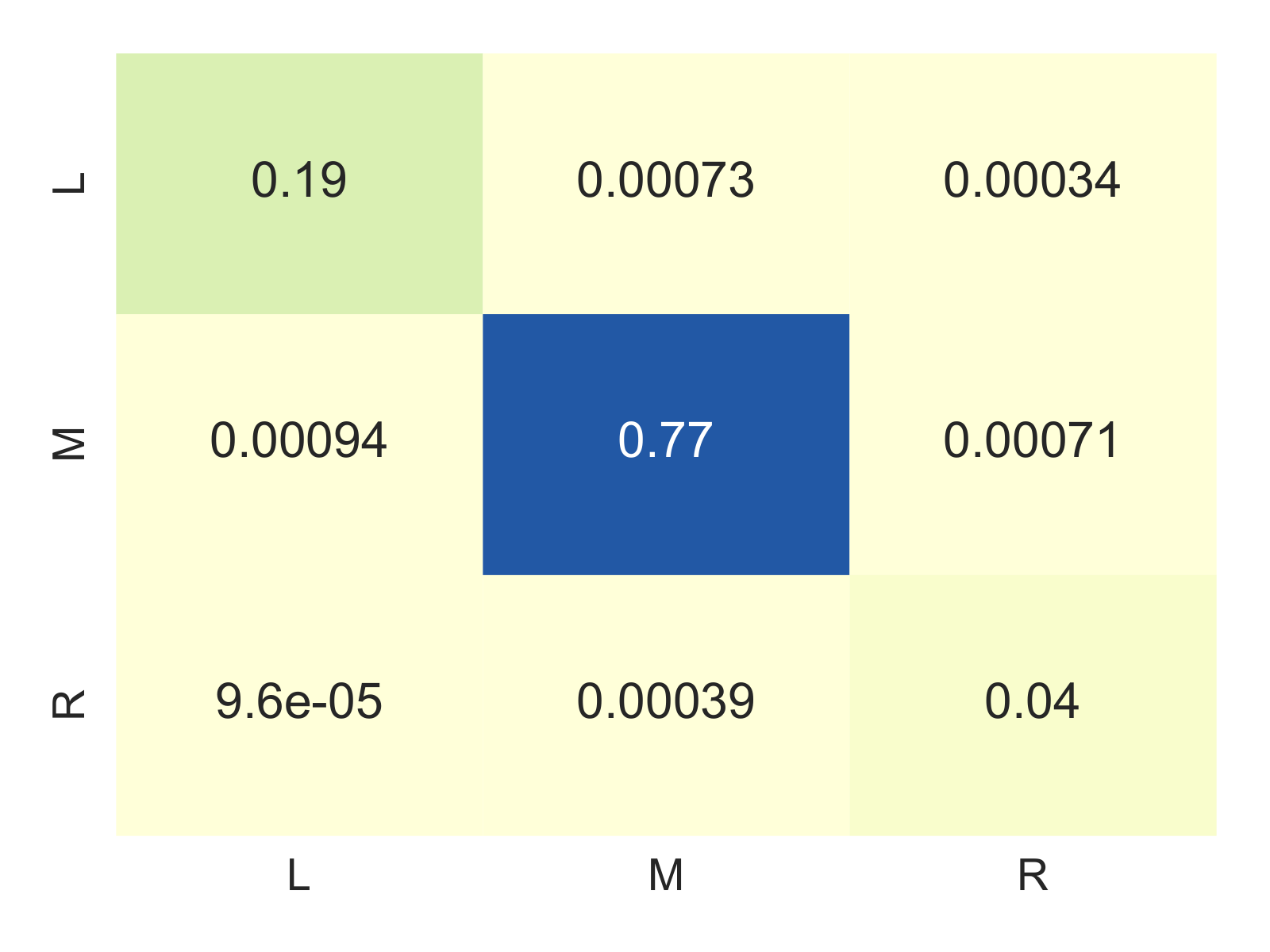}\label{fig:g5-states-coop-action} }}%

    \caption{The empirical joint-action distributions in the last 10\% of episodes when learning with cooperative action communication.}%
    \label{fig:coop-action-states}%
\end{figure}

Lastly, we consider the joint-action distribution in Figure \ref{fig:coop-action-states}. In games without NE, we observe little difference from strategies learned without communication. The joint strategies (R, L) and (L, R) are still preferred. 

In games with NE, we observe similar results as in Section \ref{sec:no-com-exp}. An interesting difference however is that in Game 4, players converge on (L, L) more often. We hypothesise this is due to players coordinating their policies better. Player 2 specifically can use a commitment of L by player 1 to move their policy closer to L as well, thus increasing the likelihood of converging to (L, L). We note however that the imbalanced NE is still less likely to be played than the balanced NE.

We omit the figures for the action selection probabilities from this discussion as they are highly similar to Figure \ref{fig:no-com-ser} from the experiments without communication.

\subsection{Self-Interested Action Communication}
\label{sec:comp-com-exp}
\begin{figure}[h!tb]%
    \centering
    \subfloat[Game 1]{{\includegraphics[width=.28\linewidth]{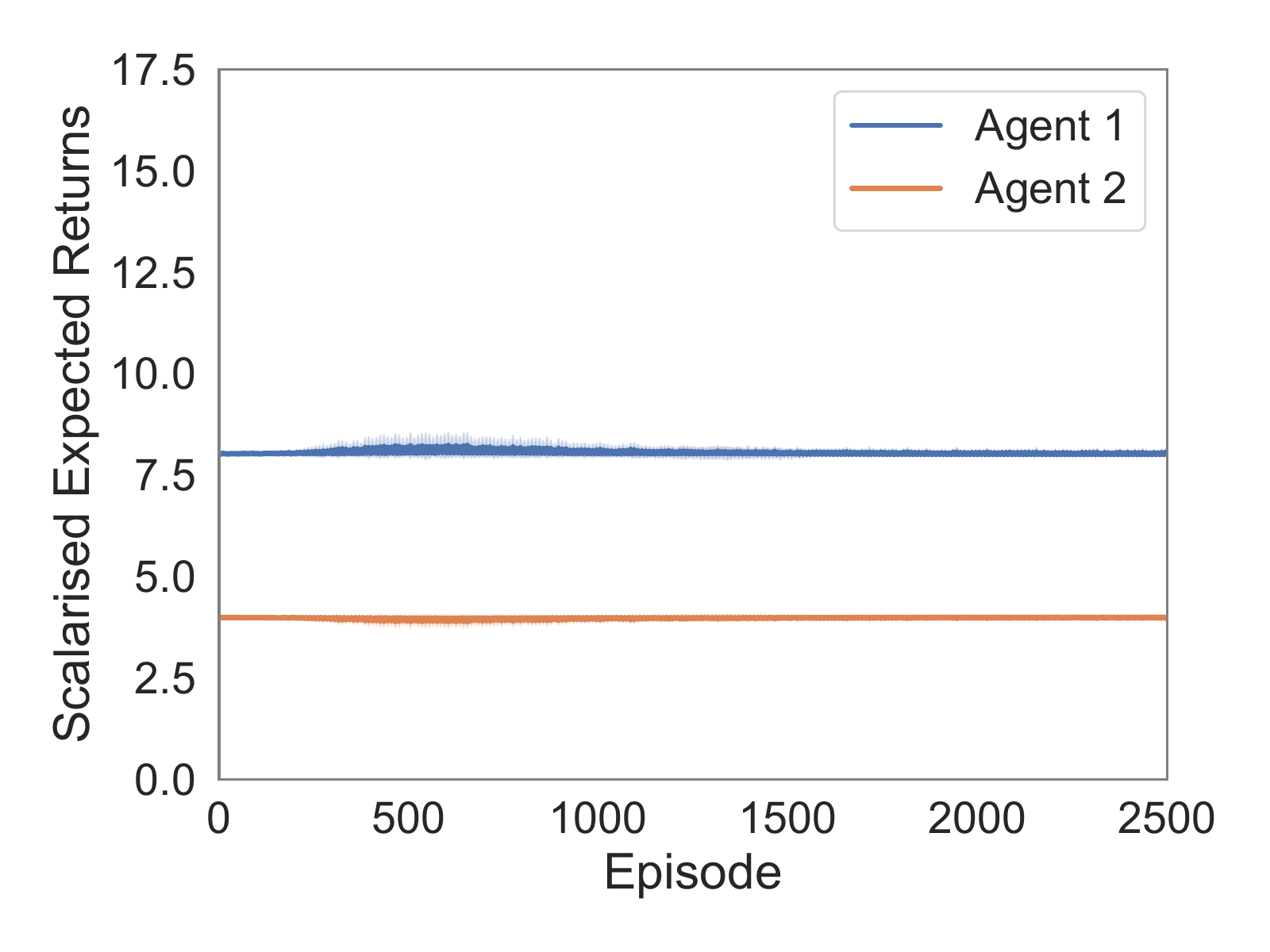}\label{fig:g1-ser-comp-action} }}%
    \quad
    \subfloat[Game 2]{{\includegraphics[width=.28\linewidth]{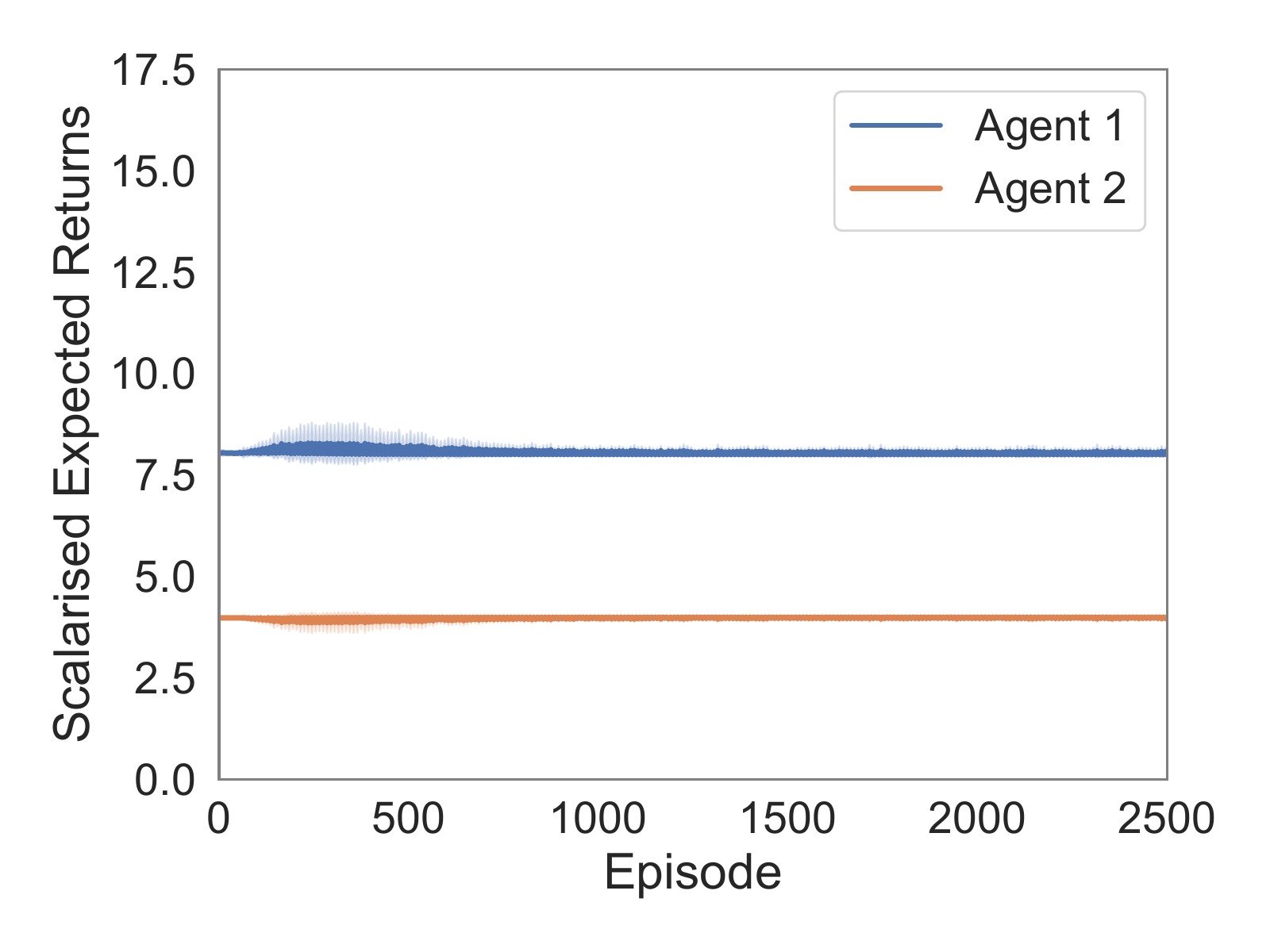}\label{fig:g2-ser-comp-action} }}%
    \quad
    \subfloat[Game 3]{{\includegraphics[width=.28\linewidth]{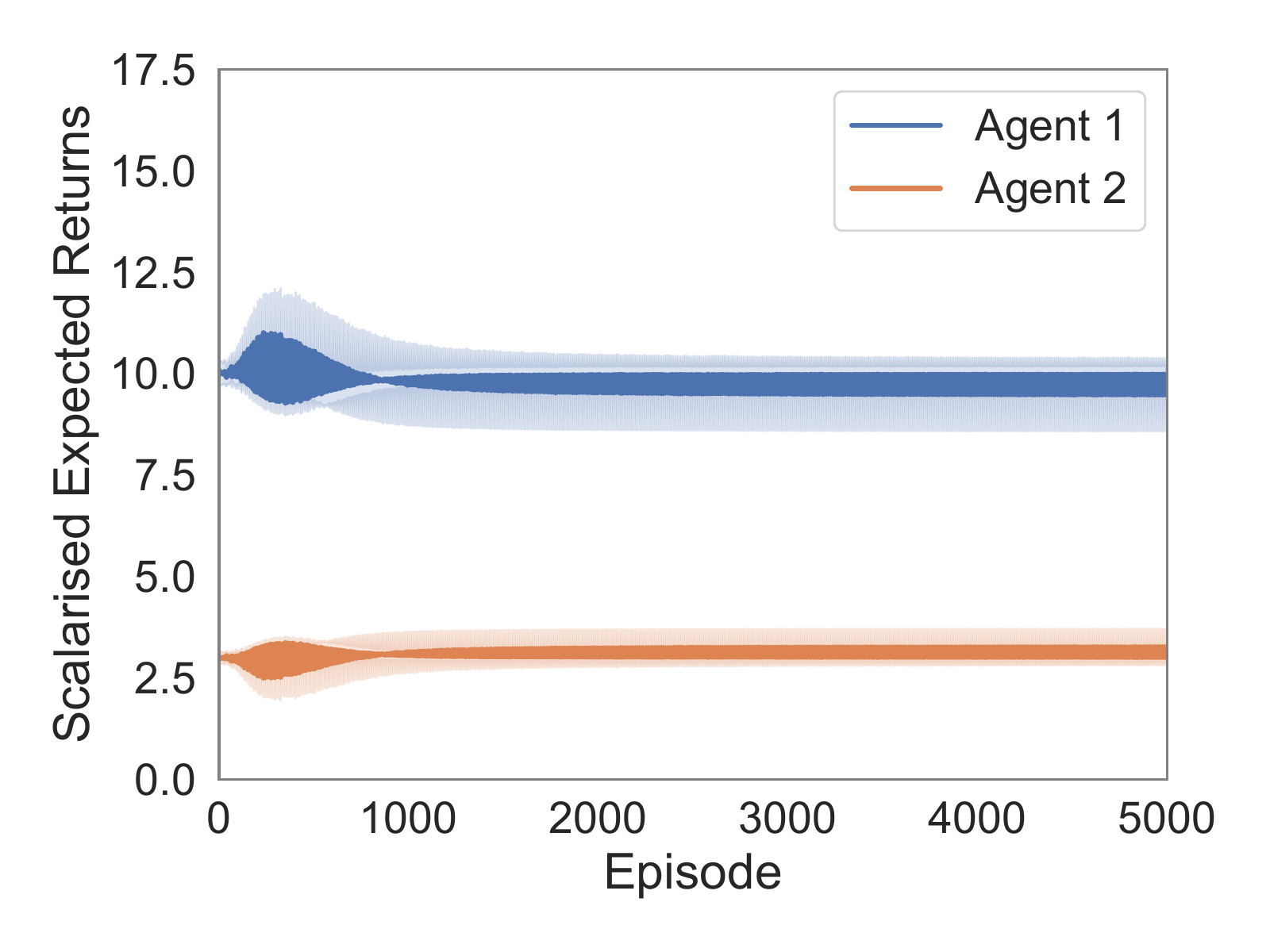}\label{fig:g3-ser-comp-action} }}%
    \quad
    \subfloat[Game 4]{{\includegraphics[width=.28\linewidth]{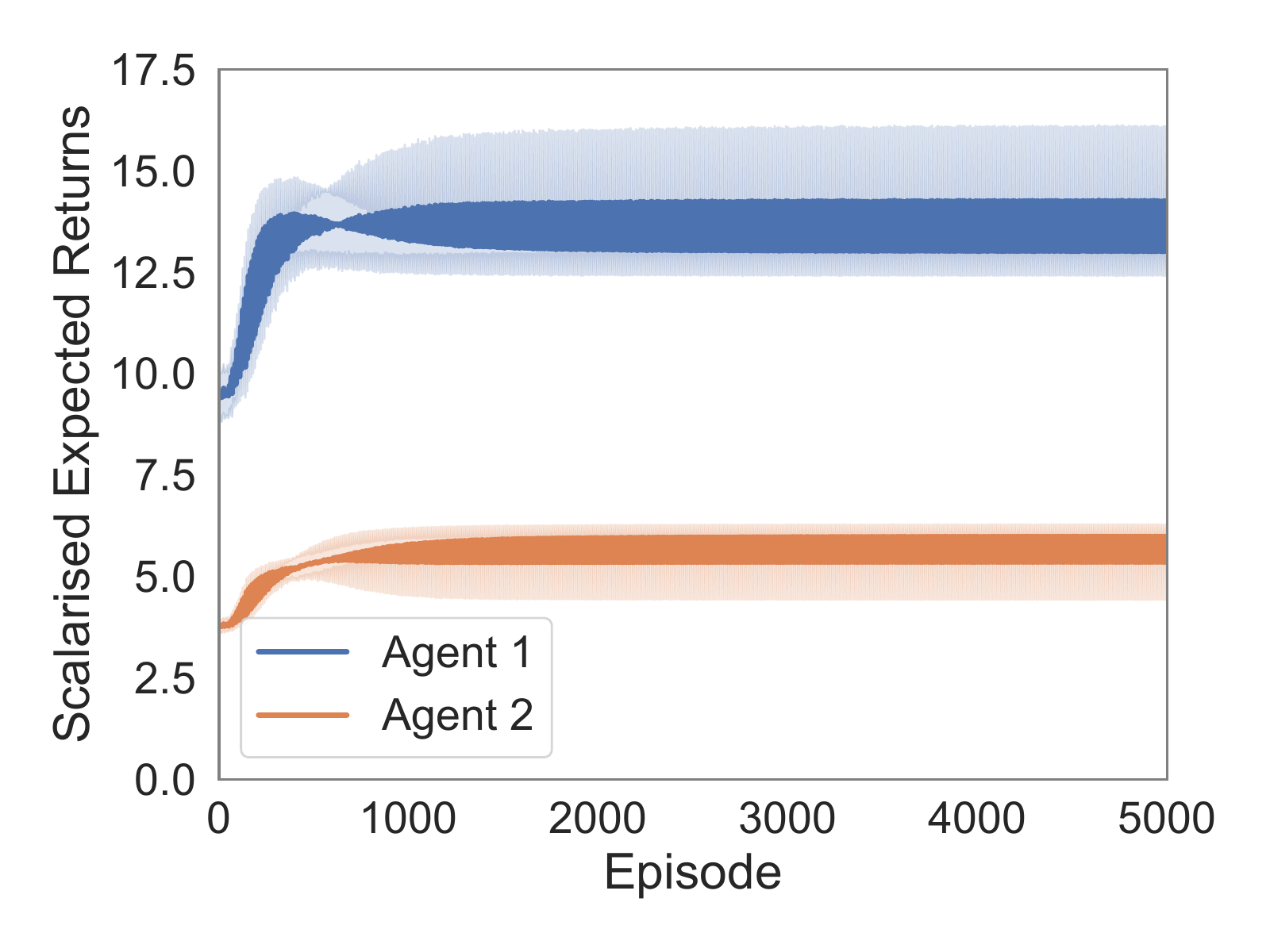}\label{fig:g4-ser-comp-action} }}%
    \quad
    \subfloat[Game 5]{{\includegraphics[width=.28\linewidth]{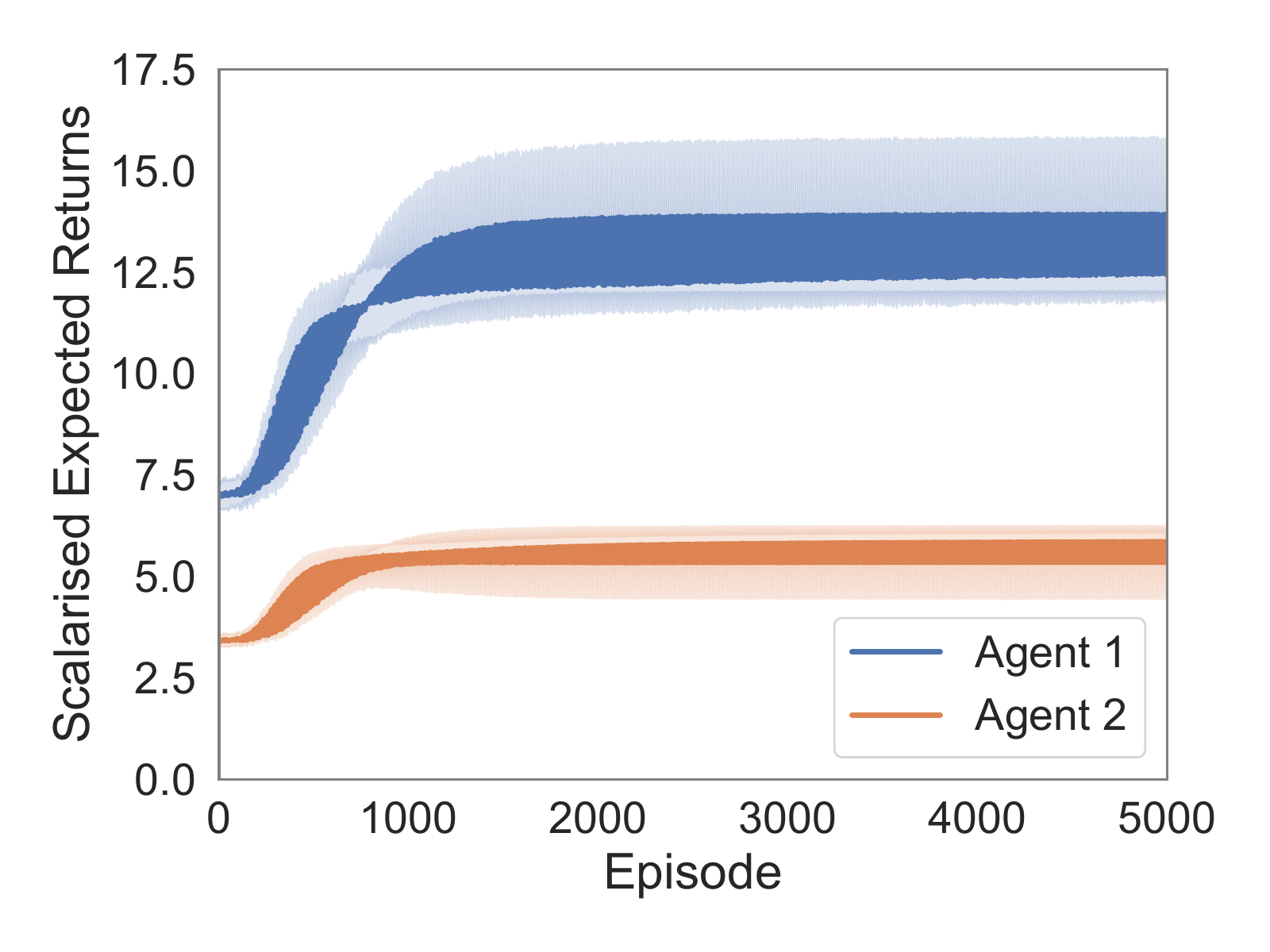}\label{fig:g5-ser-comp-action} }}%
    \caption{The SER for both agents when learning with self-interested action communication.}%
    \label{fig:comp-action-ser}%
\end{figure}

With self-interested action communication, agents learn a non-stationary policy that is conditioned on their current role and perceived communication. Note that we use distinct learning rates when leading ($\alpha_Q$ and $\alpha_\theta = 0.01$) or following ($\alpha_Q$ and $\alpha_\theta = 0.05$). The following learning rate is higher as multiple best-response policies need to be learned in the same time-span as one leading policy. We show the results for the SER in Figure \ref{fig:comp-action-ser}.

Observe that several plots exhibit oscillating SER. In games without NE this is less pronounced and average utility over different runs is relatively stable. This is because player 2 will always be able to force an expected return of (2, 2). The SER is not completely stable as agents continue updating their (best-response) strategies thus leading to slightly different results when leading and following. We note that while these games do not have an NE, agents using self-interested communication learn to cycle through leadership equilibria which is itself a stable solution (see Definition \ref{def:MOMA-LE-SER}). Concretely, both agents commit to their least exploitable action when leading and play a best-response when following.

In Game 3, there is a unique Nash equilibrium. However, we observe that the utility oscillates here as well. Upon further analysis, there are runs where player 1 does not learn their true best-response against a commitment of M by player 2. Similar to results in Sections \ref{sec:no-com-exp} and \ref{sec:coop-com-exp}, we attribute this to unbalanced payoffs being harder to learn than balanced payoffs which can steer players to suboptimal strategies. 

In Game 4 and 5, we see that agents learn to cycle through Nash equilibria. Concretely, whenever a player is the leader, they will commit to a strategy which is part of their preferred NE. Because players alternate between leader and follower, the utility also cycles between the two equilibria. This introduces the follow up question of whether this cycle is a cyclic Nash equilibrium. First, we note that cycling through leadership equilibria is not \emph{necessarily} a CNE as defined in Definition \ref{def:MOMA-CNE-SER}. However, it is straightforward to calculate that cycling through (L, L) and (M, M) is indeed a CNE. We believe this is the first time this solution concept has emerged in the context of MONFGs and suggests new possibilities for future work.

\begin{figure}[h!tb]%
    \centering
    \subfloat[Game 1]{{\includegraphics[width=.28\linewidth]{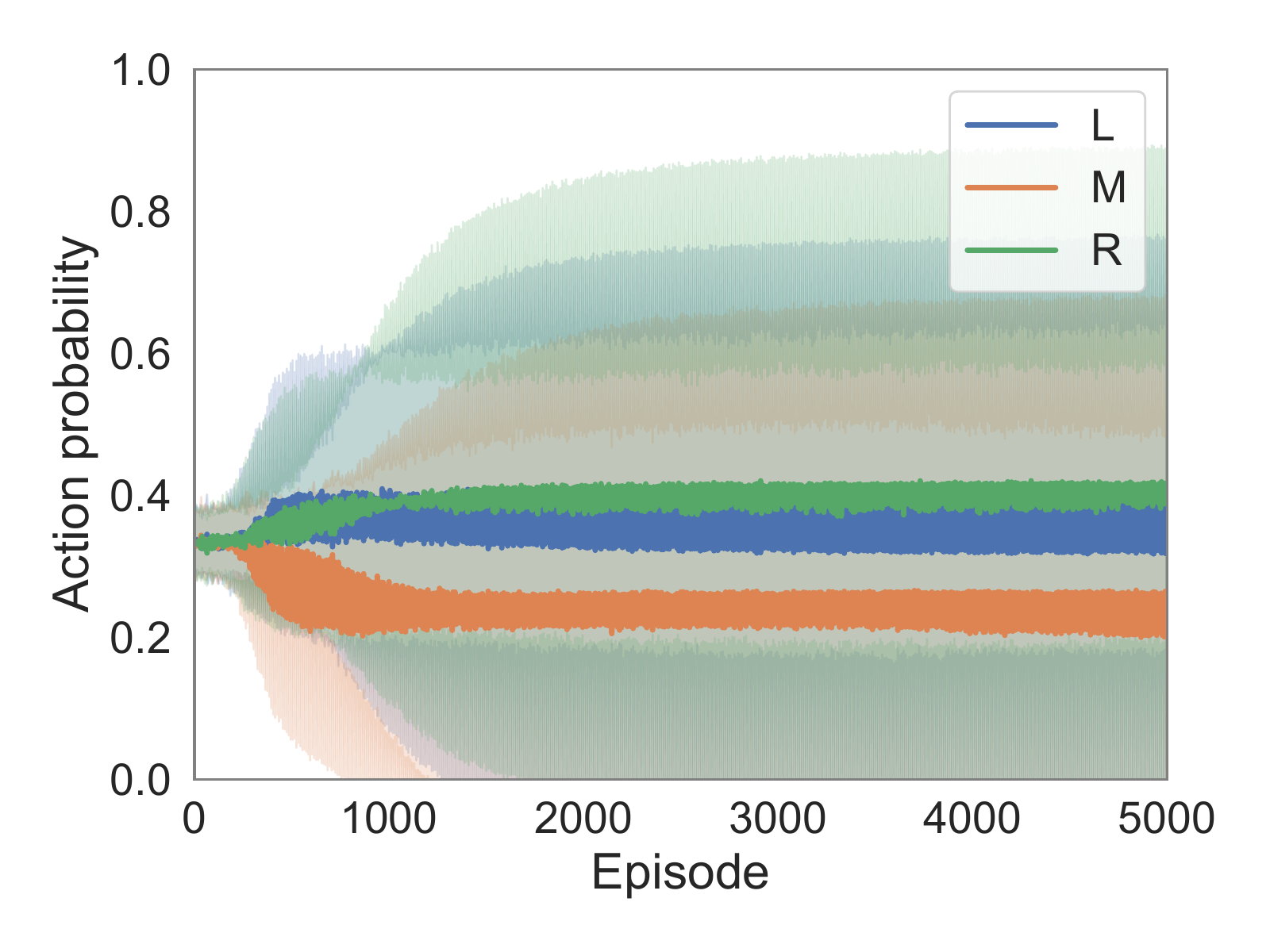}\label{fig:g1-A1-probs-comp-action} }}%
    \quad
    \subfloat[Game 2]{{\includegraphics[width=.28\linewidth]{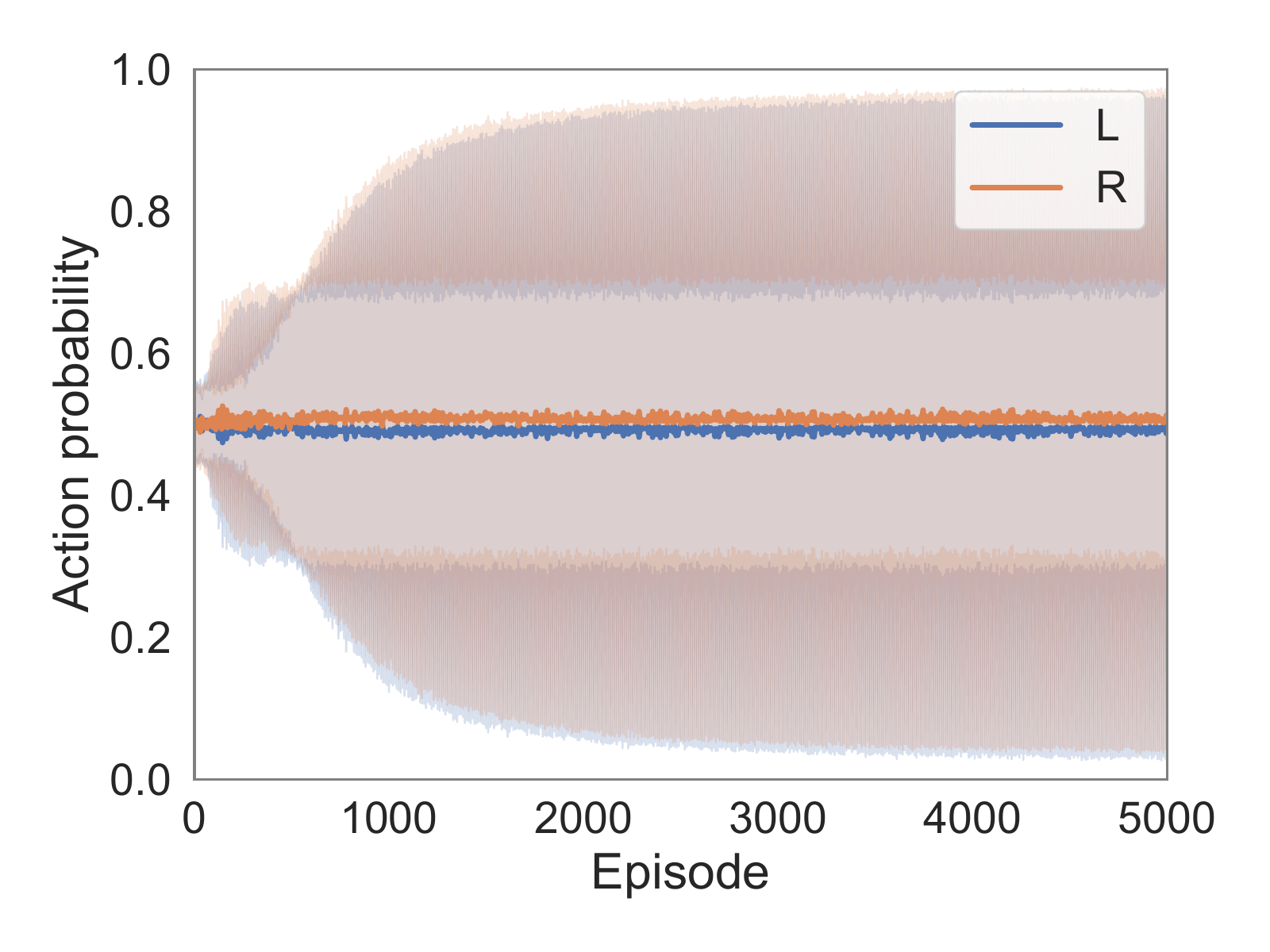}\label{fig:g2-A1-probs-comp-action} }}%
    \quad
    \subfloat[Game 3]{{\includegraphics[width=.28\linewidth]{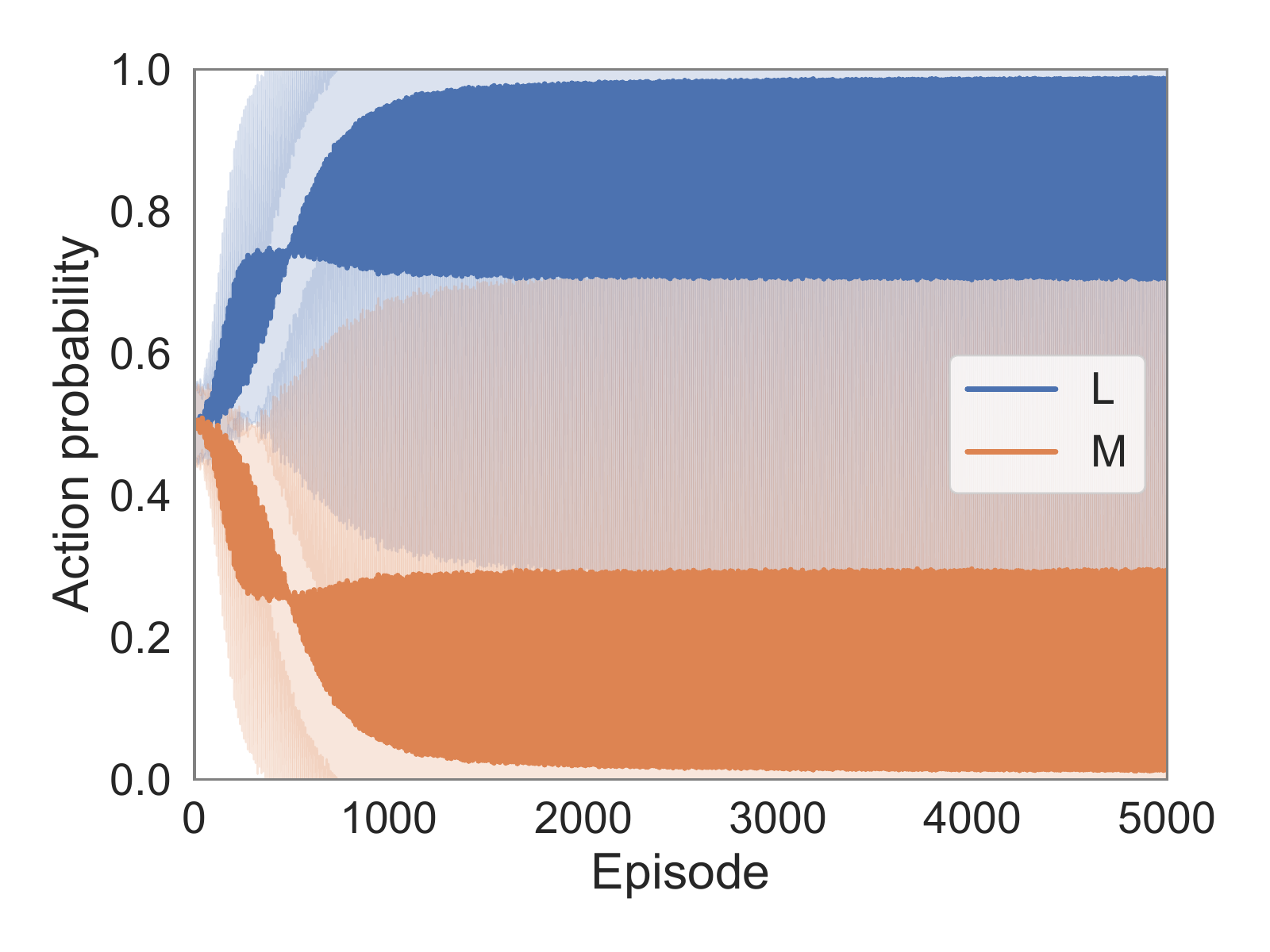}\label{fig:g3-A1-probs-comp-action} }}%
    \quad
    \subfloat[Game 4]{{\includegraphics[width=.28\linewidth]{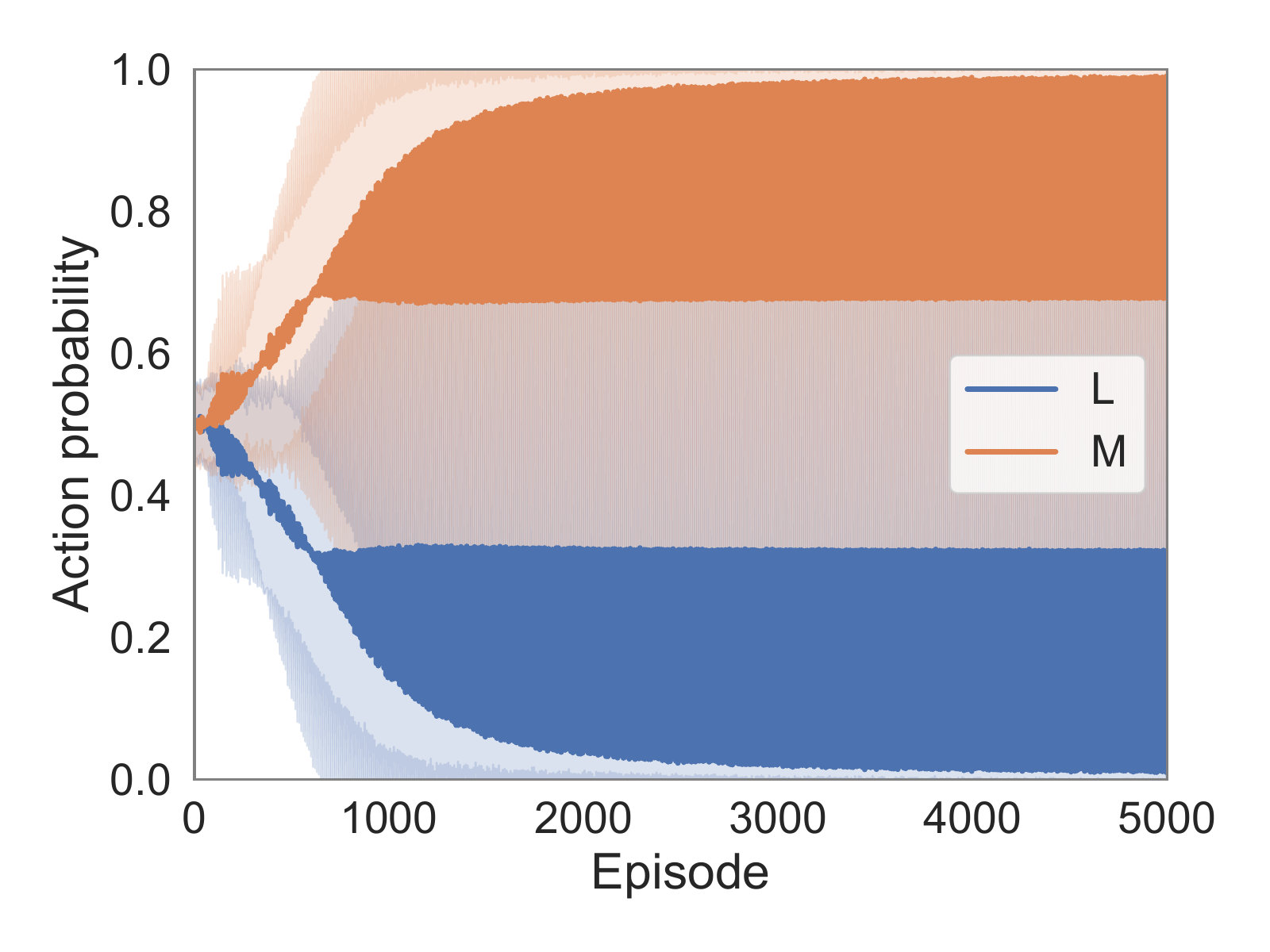}\label{fig:g4-A1-probs-comp-action} }}%
    \quad
    \subfloat[Game 5]{{\includegraphics[width=.28\linewidth]{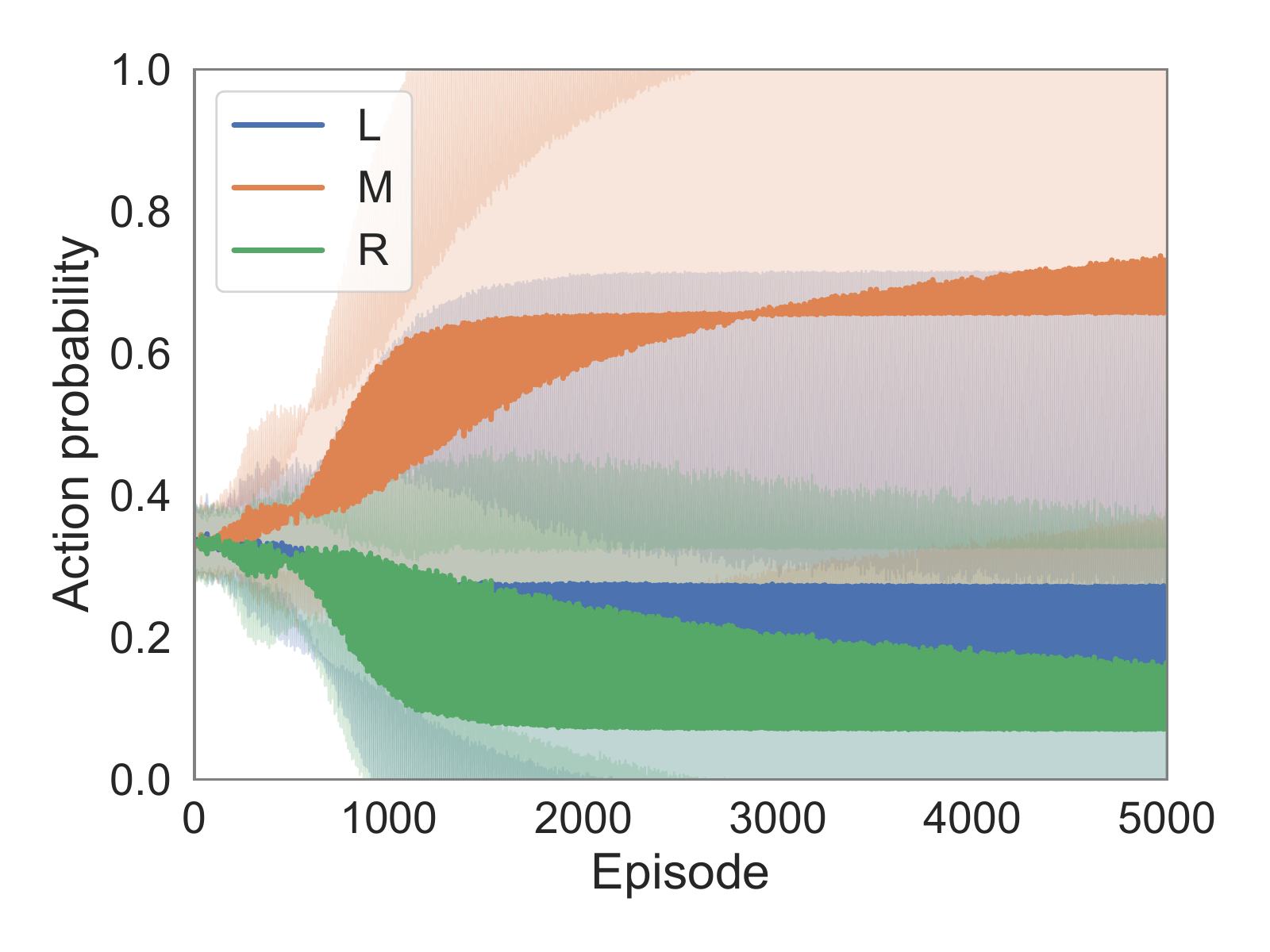}\label{fig:g5-A1-probs-comp-action} }}%
    \caption{The action probabilities for agent 1 when learning with self-interested action communication.}%
    \label{fig:comp-action-A1-probs}%
\end{figure}

The action probabilities show further evidence for the cycling through a set of stationary policies. Observe in Figure \ref{fig:comp-action-A1-probs} the ``thick'' lines. These are the action probabilities oscillating, as they are conditioned on the current role and observation. Note that we omit the action probabilities for player 2 as they show a similar pattern.

\begin{figure}[h!tb]%
    \centering
    \subfloat[Game 1]{{\includegraphics[width=.17\linewidth]{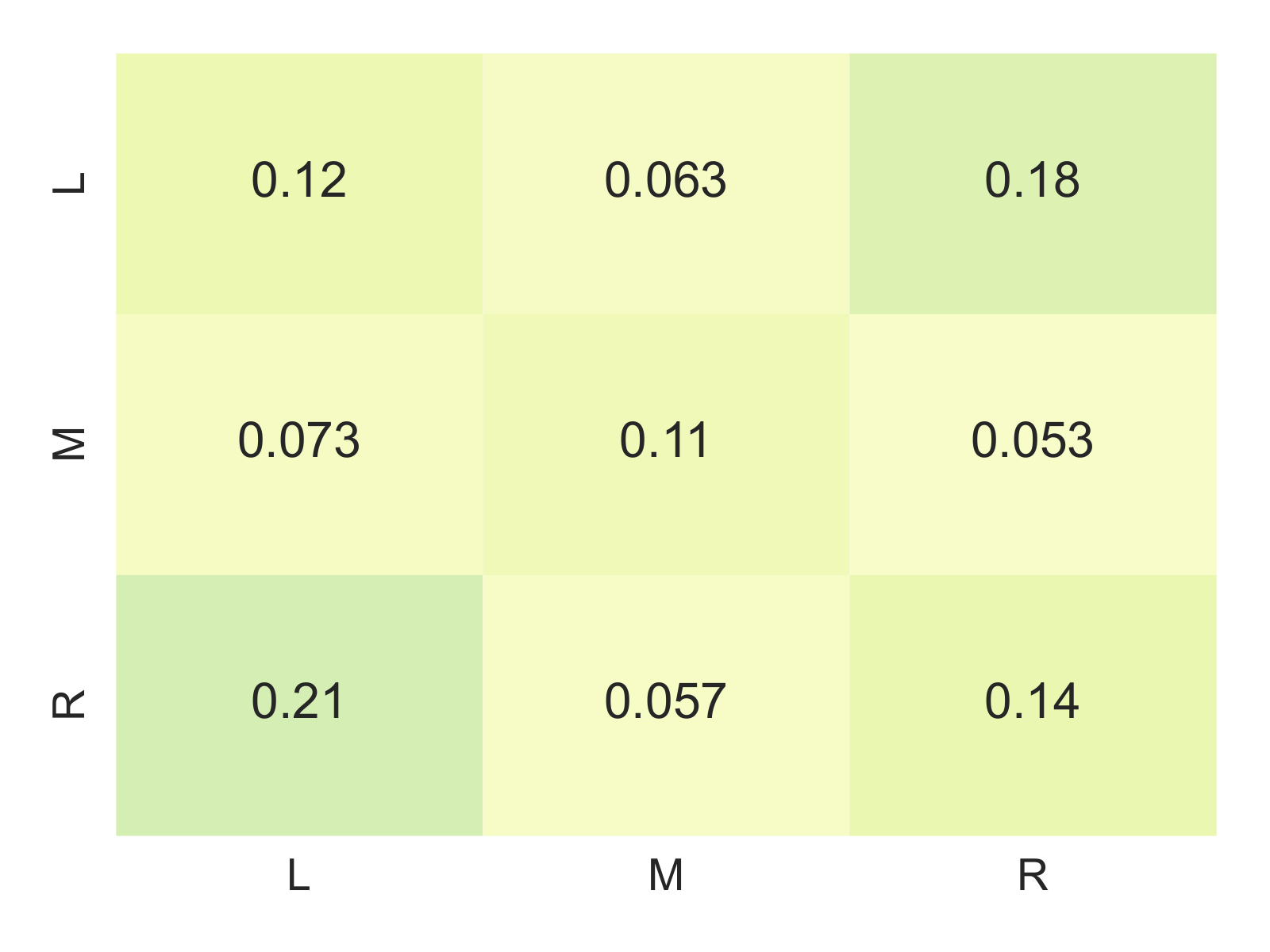}\label{fig:g1-states-comp-action} }}%
    \quad
    \subfloat[Game 2]{{\includegraphics[width=.17\linewidth]{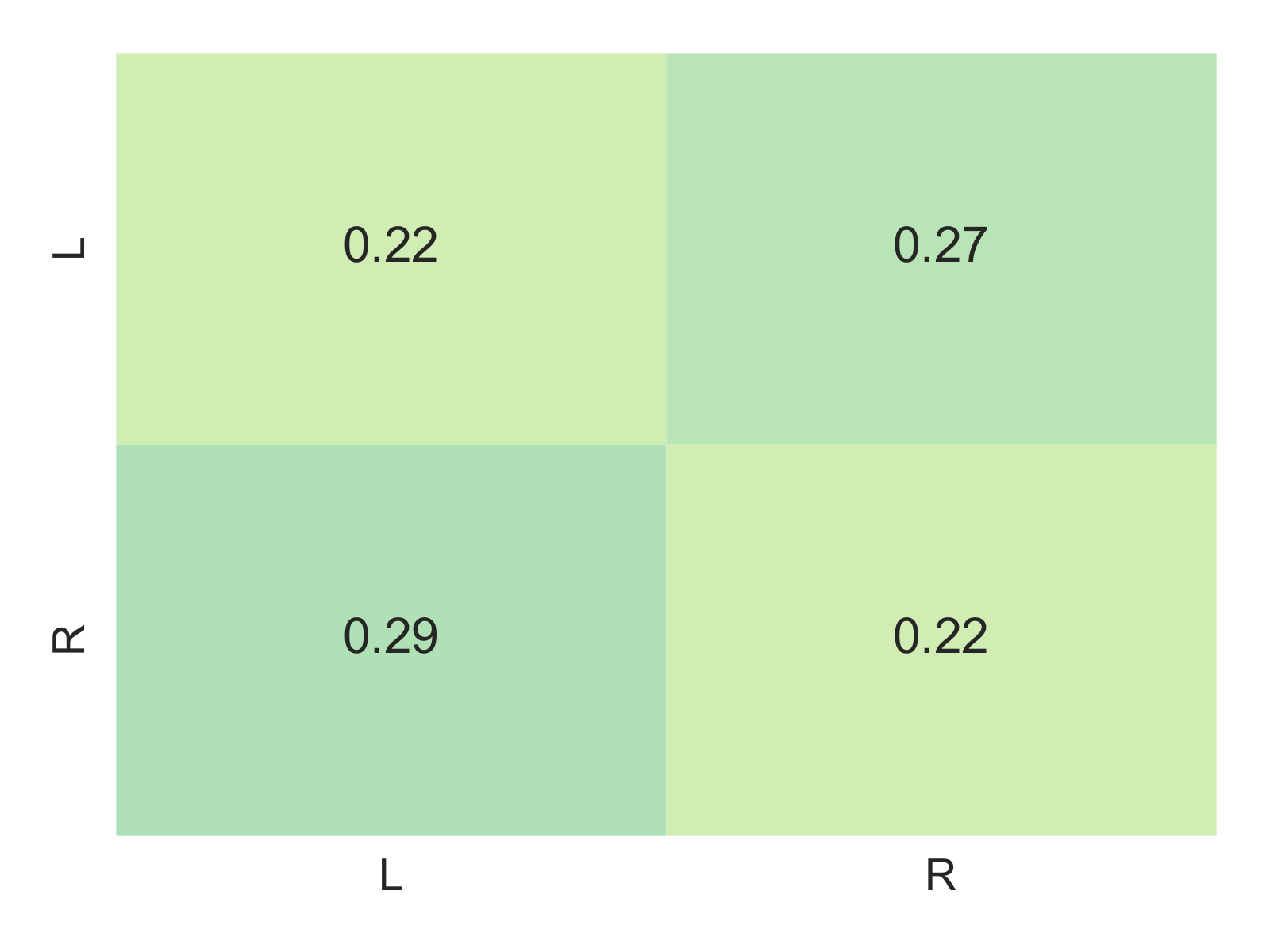}\label{fig:g2-states-comp-action} }}%
    \quad
    \subfloat[Game 3]{{\includegraphics[width=.17\linewidth]{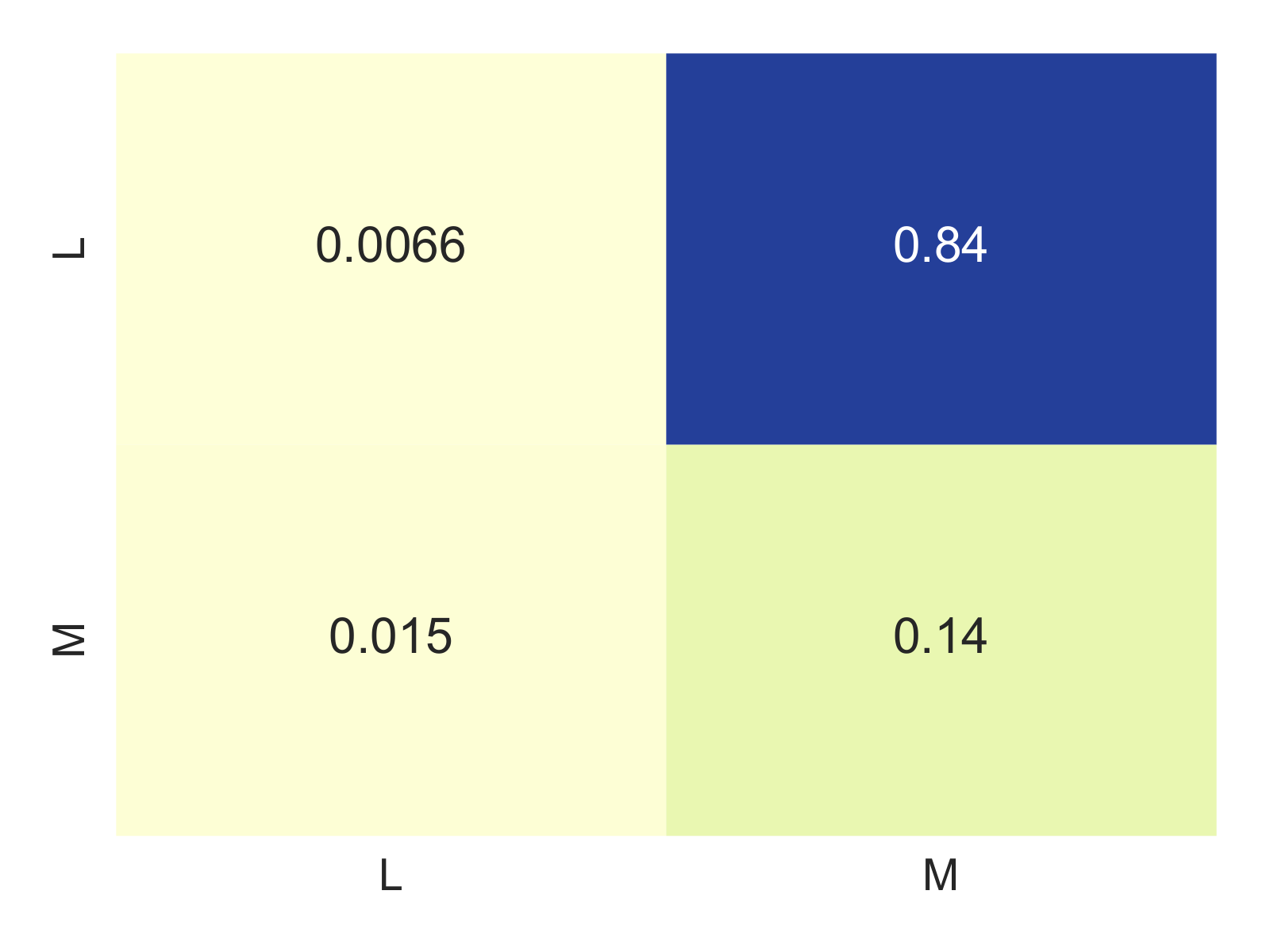}\label{fig:g3-states-comp-action} }}%
    \quad
    \subfloat[Game 4]{{\includegraphics[width=.17\linewidth]{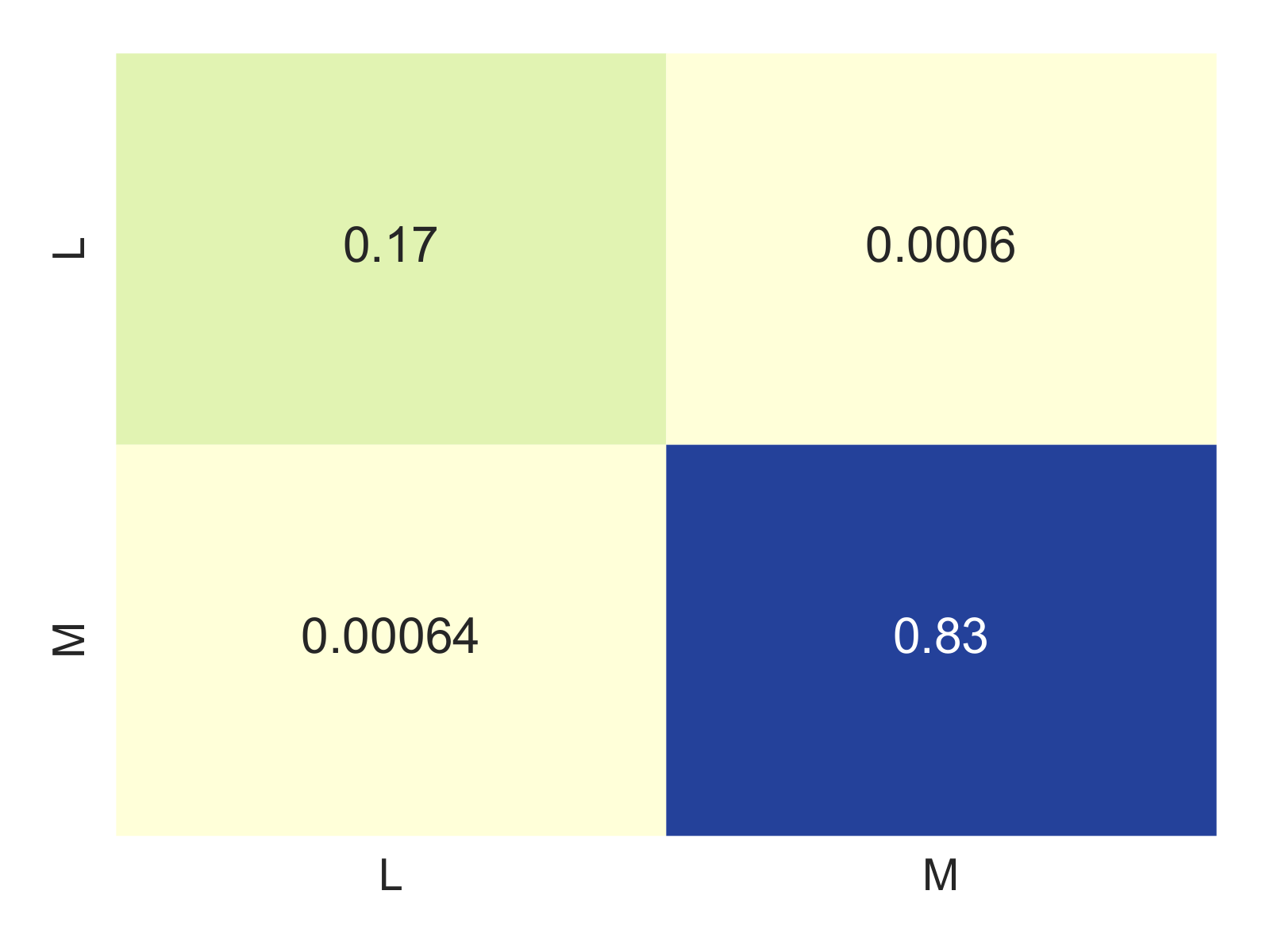}\label{fig:g4-states-comp-action} }}%
    \quad
    \subfloat[Game 5]{{\includegraphics[width=.17\linewidth]{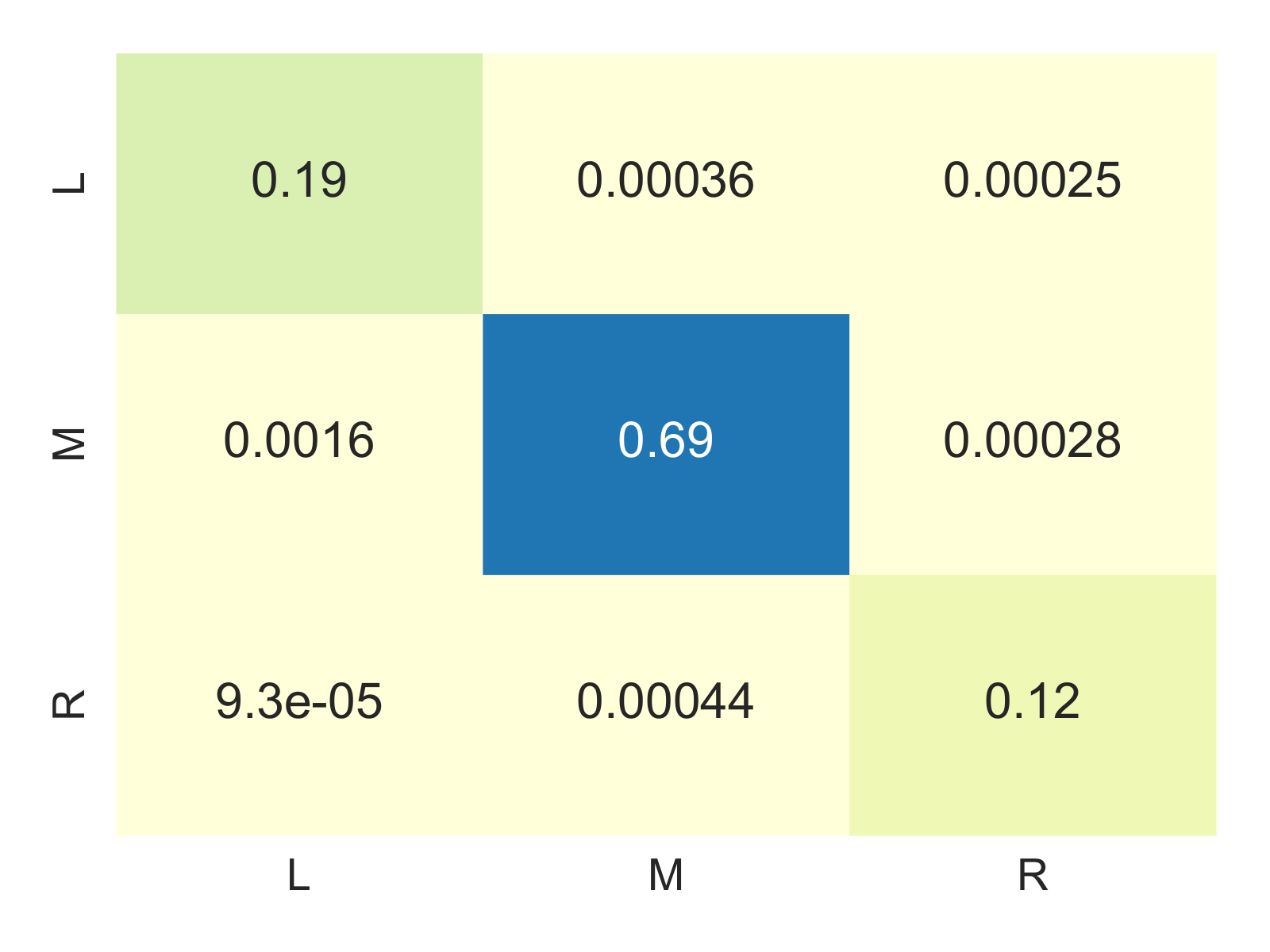}\label{fig:g5-states-comp-action} }}%
    \caption{The empirical joint-action distributions in the last 10\% of episodes when learning with self-interested action communication.}%
    \label{fig:comp-action-states}%
\end{figure}

Lastly, Figure \ref{fig:comp-action-states} shows the converged joint strategies. We highlight here the effects of self-interested communication on games without NE. Contrary to earlier experiments, agents do not primarily converge on (R, L) or (L, R). This is because every strategy may be exploited, leaving players to learn different strategies in each trial. Note however that the joint-actions which are most often played return the \emph{most balanced} or \emph{least balanced} payoffs. Such payoffs present the highest possible gain for an agent from a suboptimal best-response by their opponent. 

In games with NE, specifically Game 4, we see an increase for the NE (L, L). Again, this is because players are free to optimise a non-stationary policy. We stress that here too the proportion between (L, L) and (M, M) is still not balanced, pointing to the inherent difficulty of learning an NE with imbalanced payoffs. Lastly, in Game 5 we see an uptick in convergence to the dominated NE. A possible explanation is that because players can learn to cycle through two leadership equilibria, the probability of learning the dominated equilibrium as one of them is approximately doubled.

\subsection{Cooperative Policy Communication}
\label{sec:policy-com-exp}
The last experiments where communication is compulsory requires agents to optimise a single joint policy by communicating their current preferences over the actions. This approach is similar to the cooperative action protocol discussed in Section \ref{sec:coop-com-exp}, with the distinction that now entire policies are communicated. Because of this resemblance, we expect to see similar results as well. 

\begin{figure}[h!tb]%
    \centering
    \subfloat[Game 1]{{\includegraphics[width=.28\linewidth]{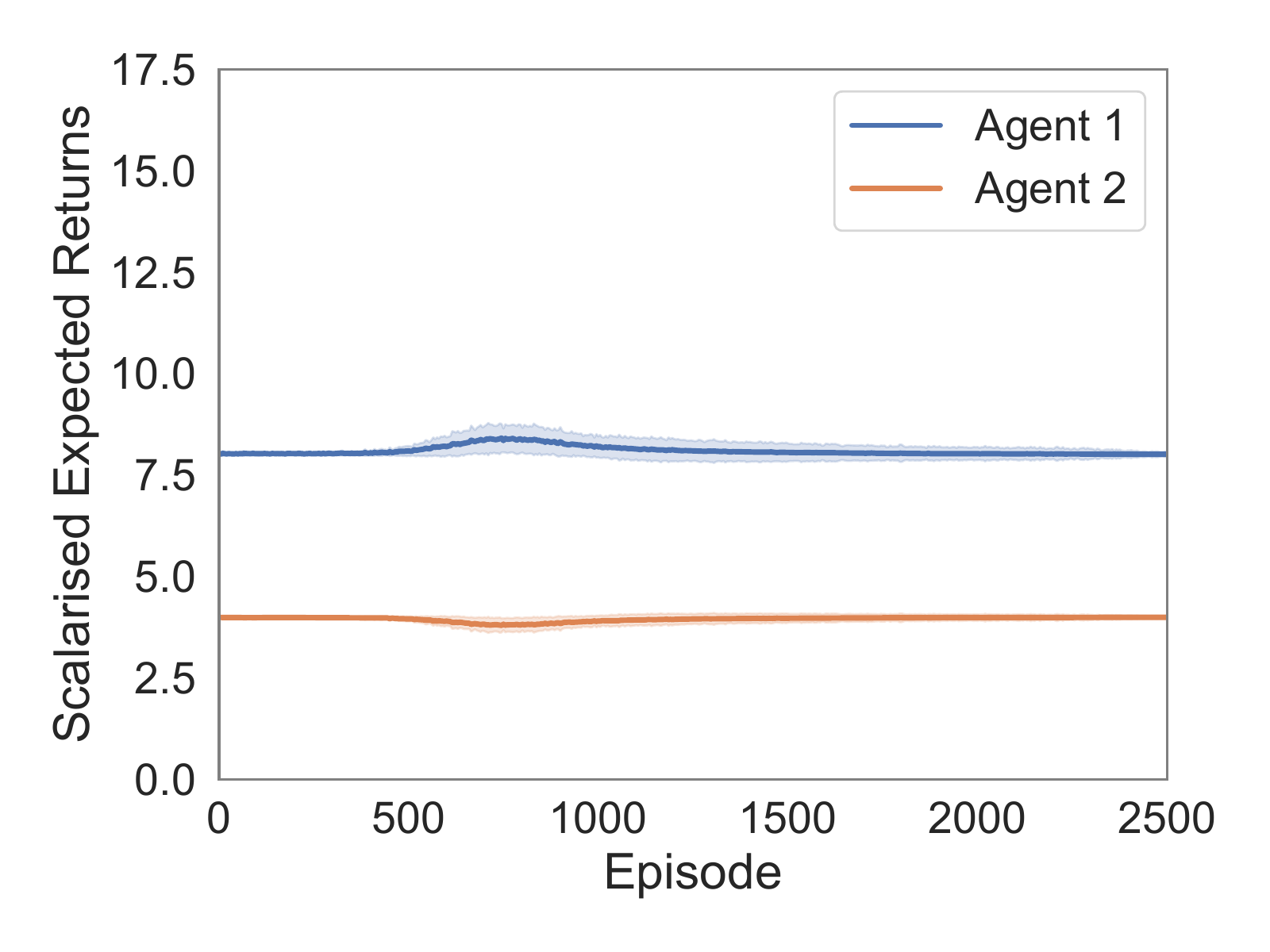}\label{fig:g1-ser-coop-policy} }}%
    \quad
    \subfloat[Game 2]{{\includegraphics[width=.28\linewidth]{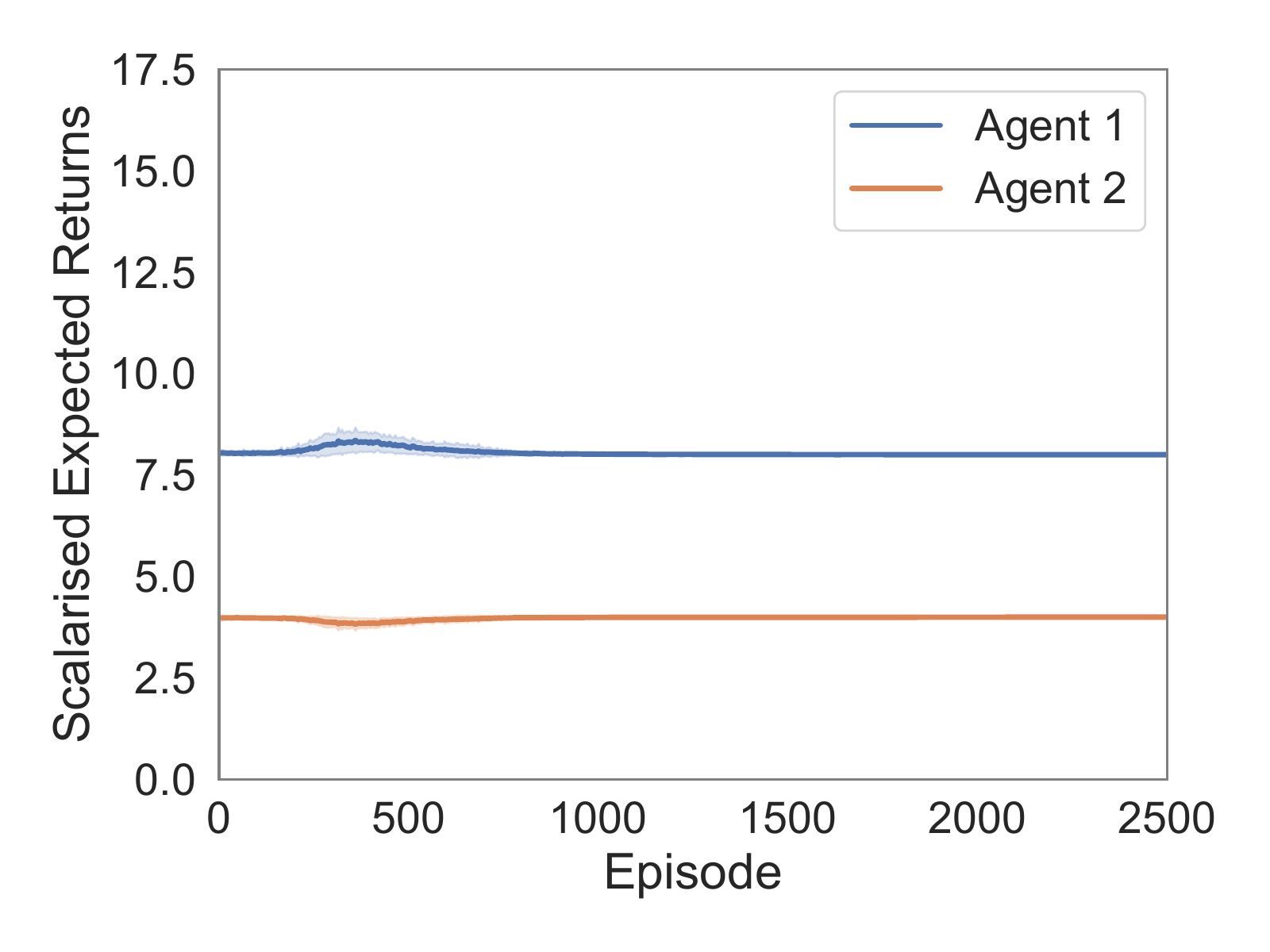}\label{fig:g2-ser-coop-policy} }}%
    \quad
    \subfloat[Game 3]{{\includegraphics[width=.28\linewidth]{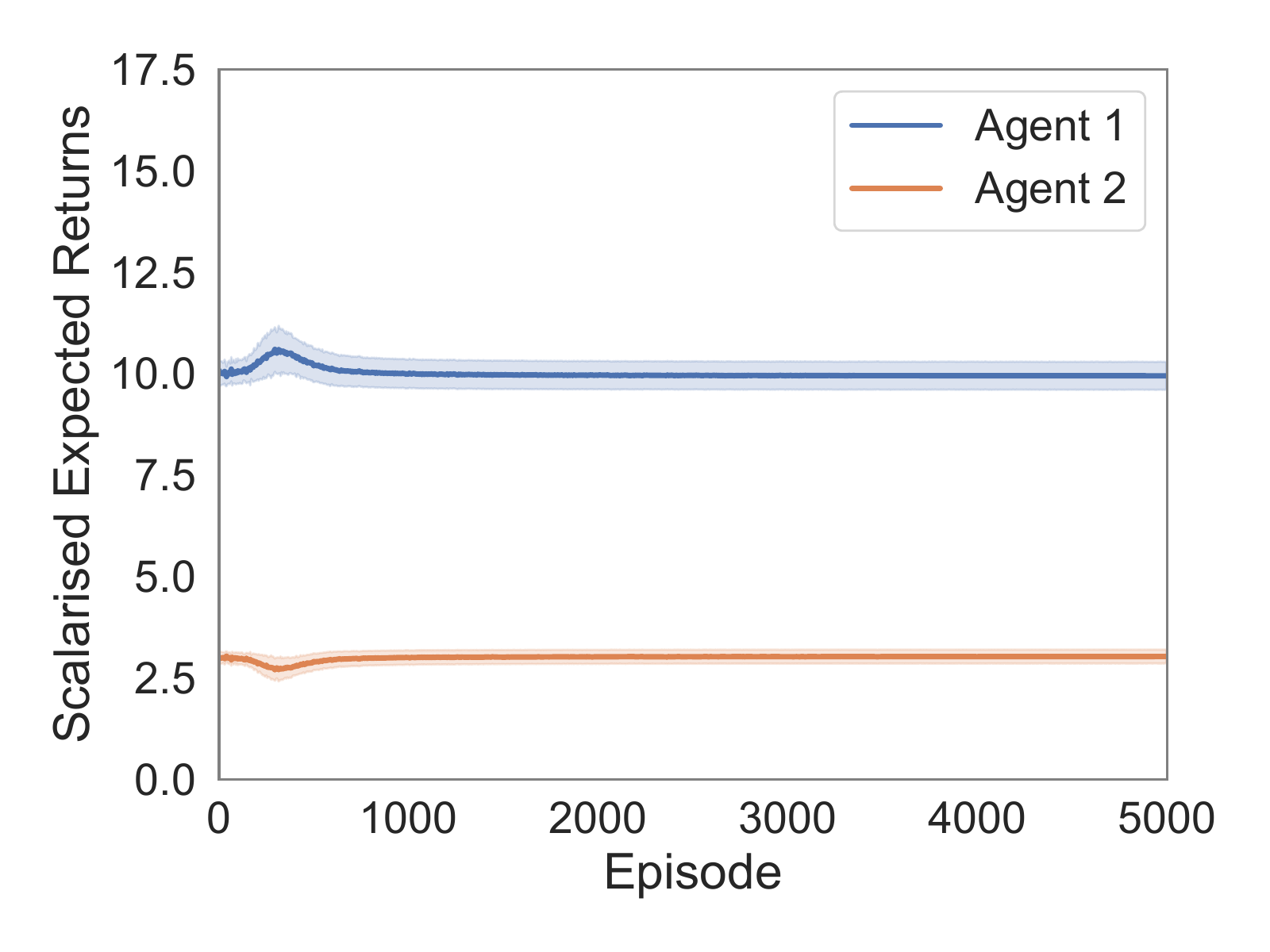}\label{fig:g3-ser-coop-policy} }}%
    \quad
    \subfloat[Game 4]{{\includegraphics[width=.28\linewidth]{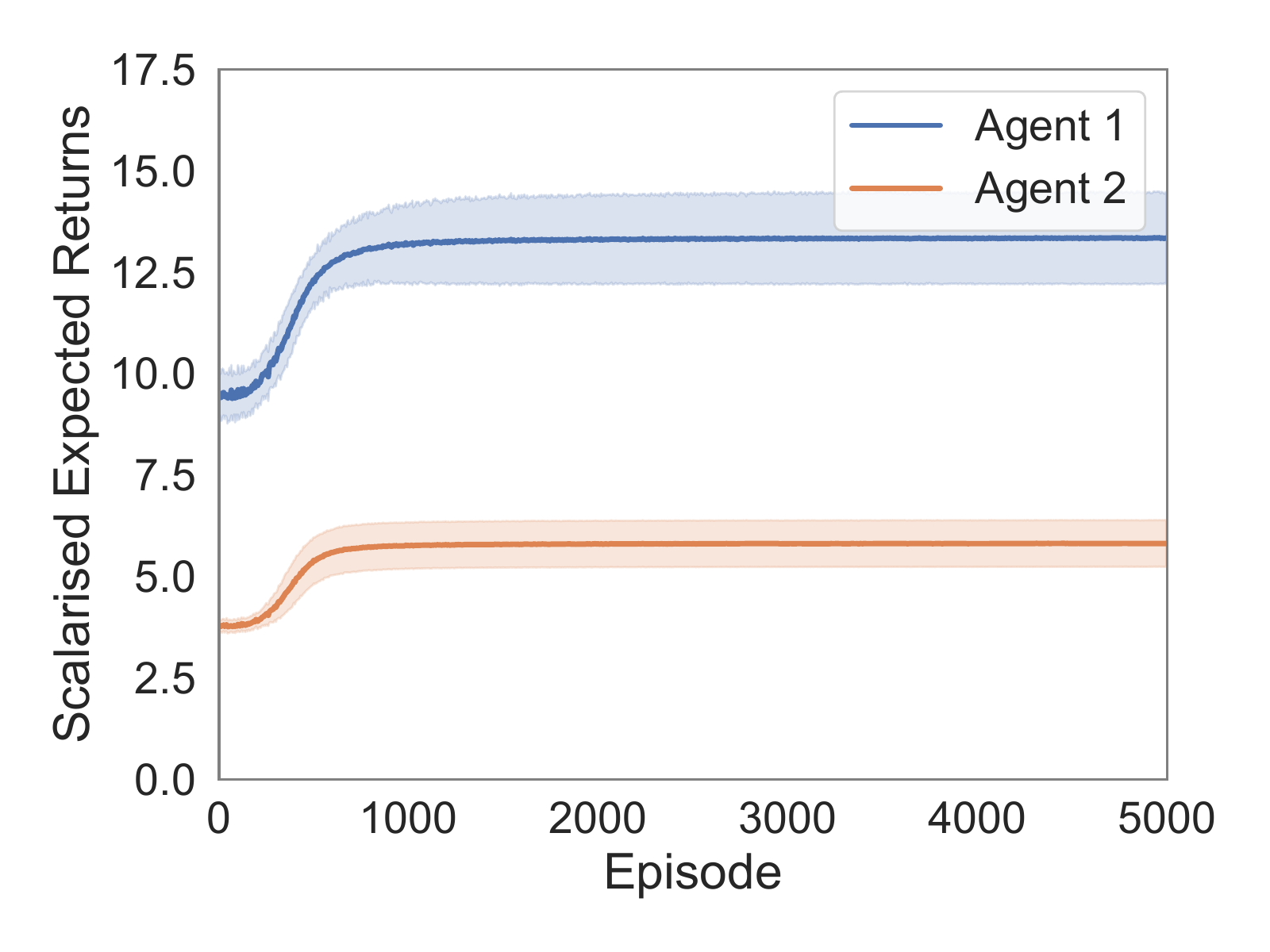}\label{fig:g4-ser-coop-policy} }}%
    \quad
    \subfloat[Game 5]{{\includegraphics[width=.28\linewidth]{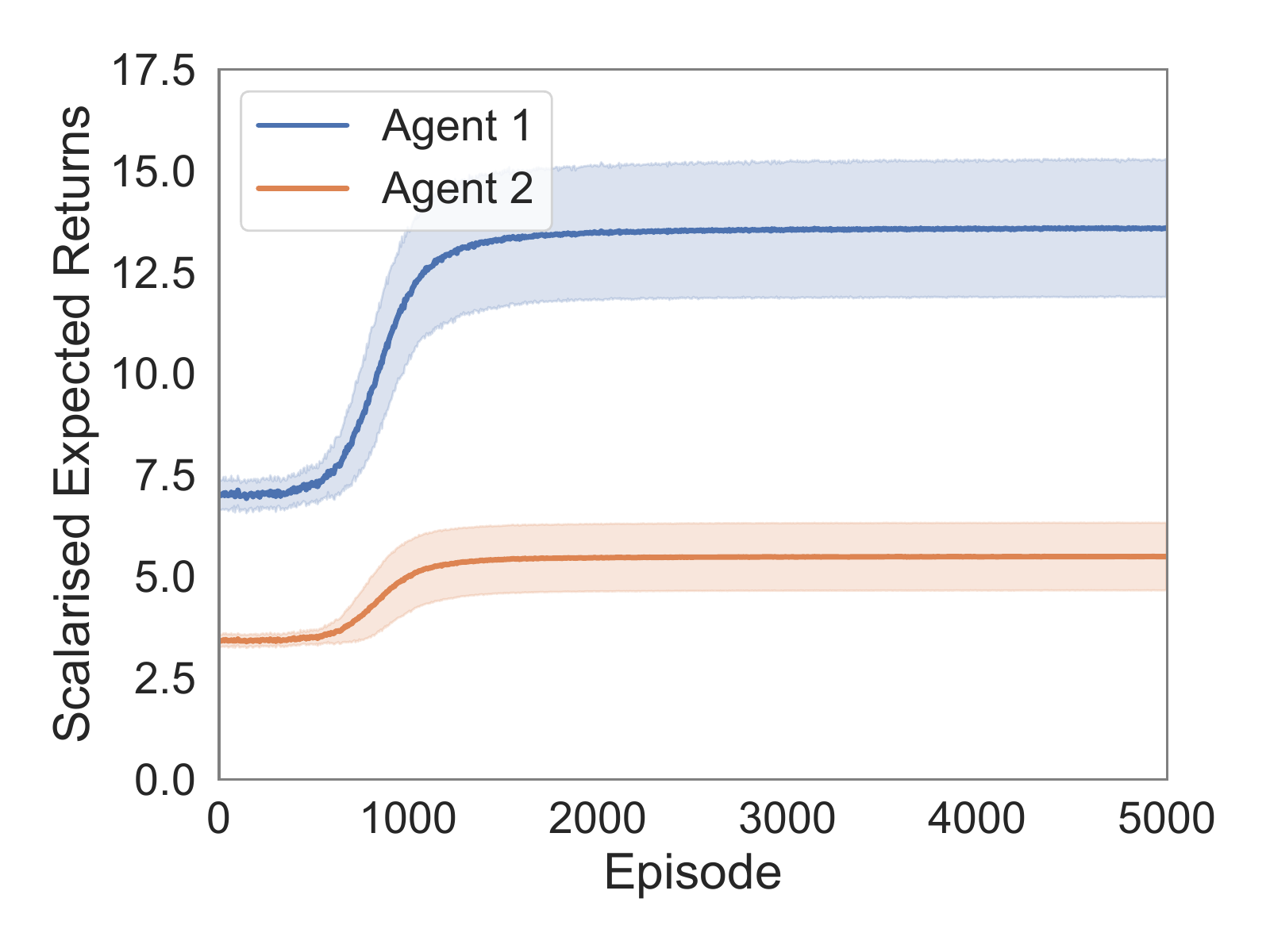}\label{fig:g5-ser-coop-policy} }}%
    \caption{The SER for both agents when learning with cooperative policy communication.}%
    \label{fig:coop-policy-ser}%
\end{figure}

We show the SER in Figure \ref{fig:coop-policy-ser}. The results are analogous to those from cooperative action communication (see Sec.~\ref{sec:coop-com-exp}). To summarise, in games without NE, cooperative communication decreases the level of divergence from the stable utility. In games with NE, it results in a steeper learning curve. We omit the action probabilities as they are again analogous to figures from cooperative action communication and no communication (see Sec.~\ref{sec:coop-com-exp} and Sec.~\ref{sec:no-com-exp}).

\begin{figure}[h!tb]%
    \centering
    \subfloat[Game 1]{{\includegraphics[width=.17\linewidth]{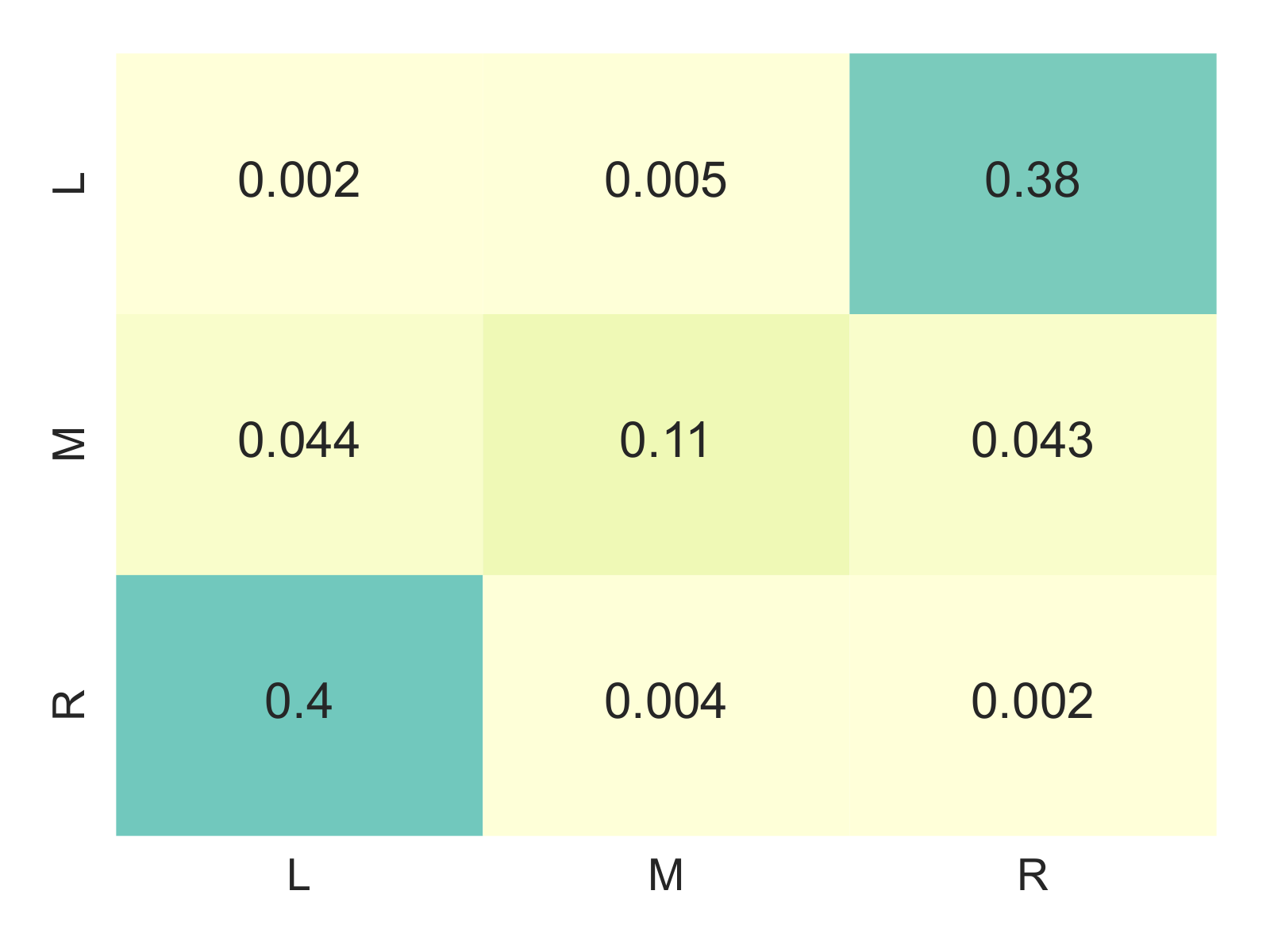}\label{fig:g1-states-coop-policy} }}%
    \quad
    \subfloat[Game 2]{{\includegraphics[width=.17\linewidth]{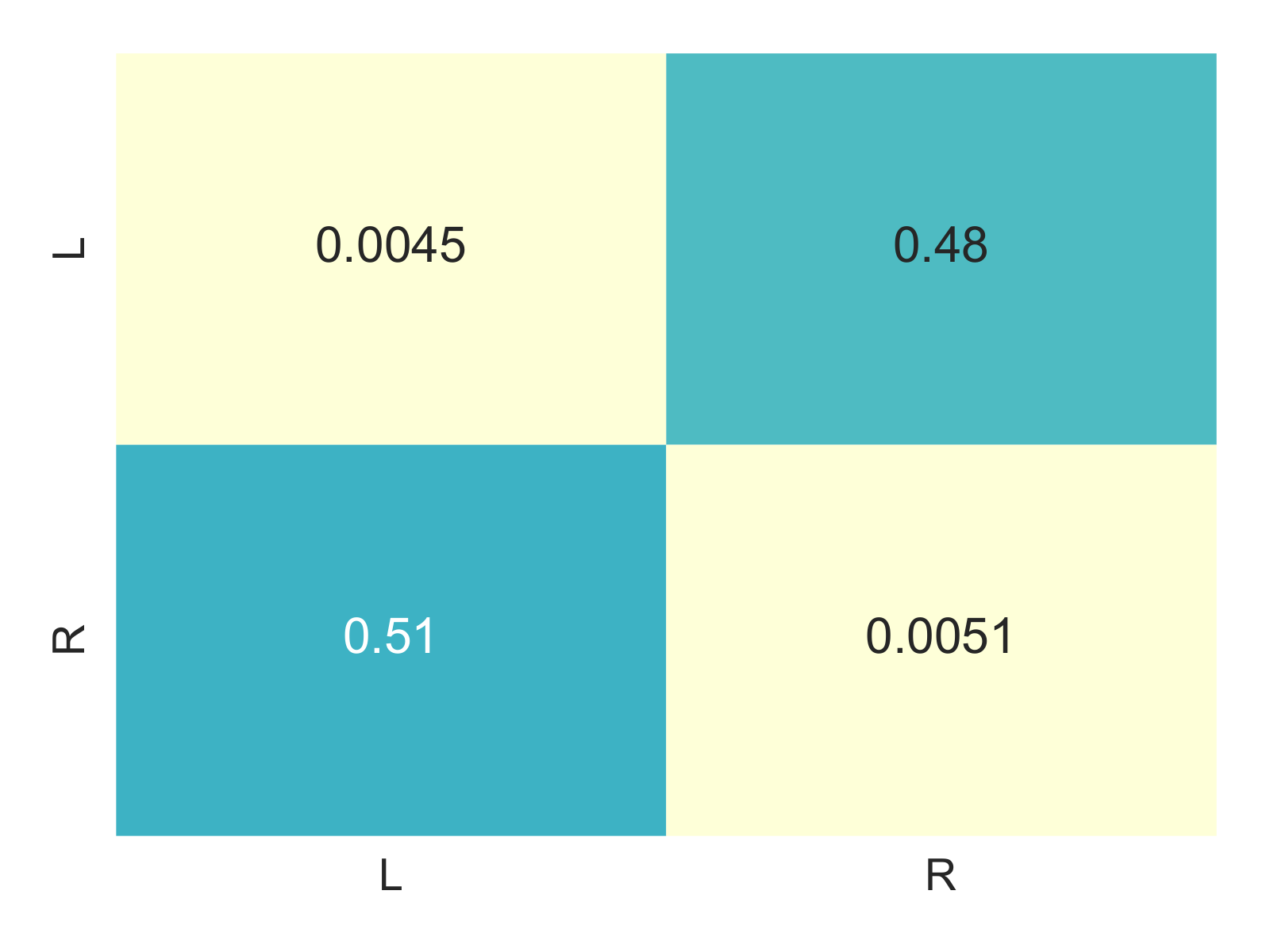}\label{fig:g2-states-coop-policy} }}%
    \quad
    \subfloat[Game 3]{{\includegraphics[width=.17\linewidth]{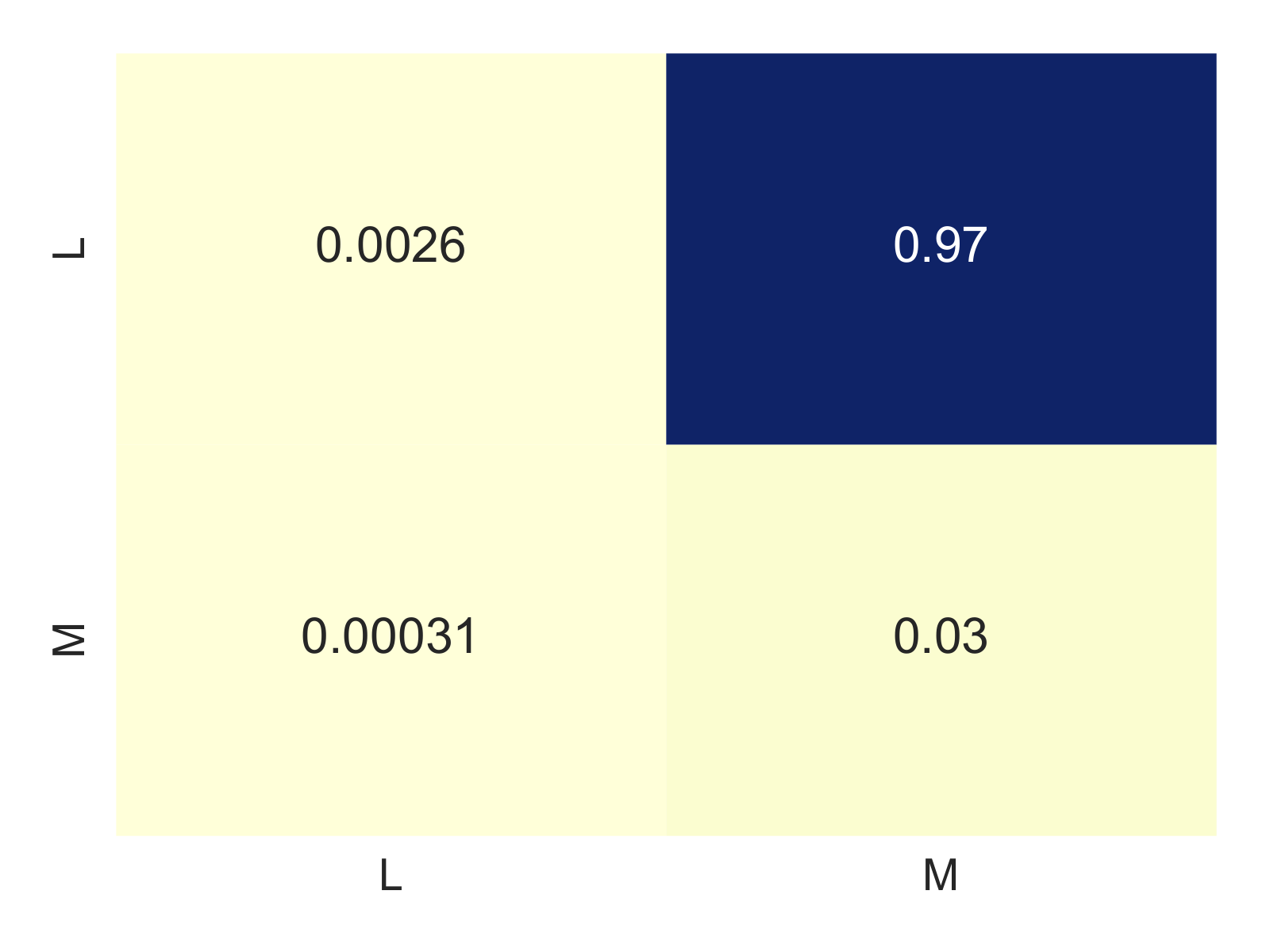}\label{fig:g3-states-coop-policy} }}%
    \quad
    \subfloat[Game 4]{{\includegraphics[width=.17\linewidth]{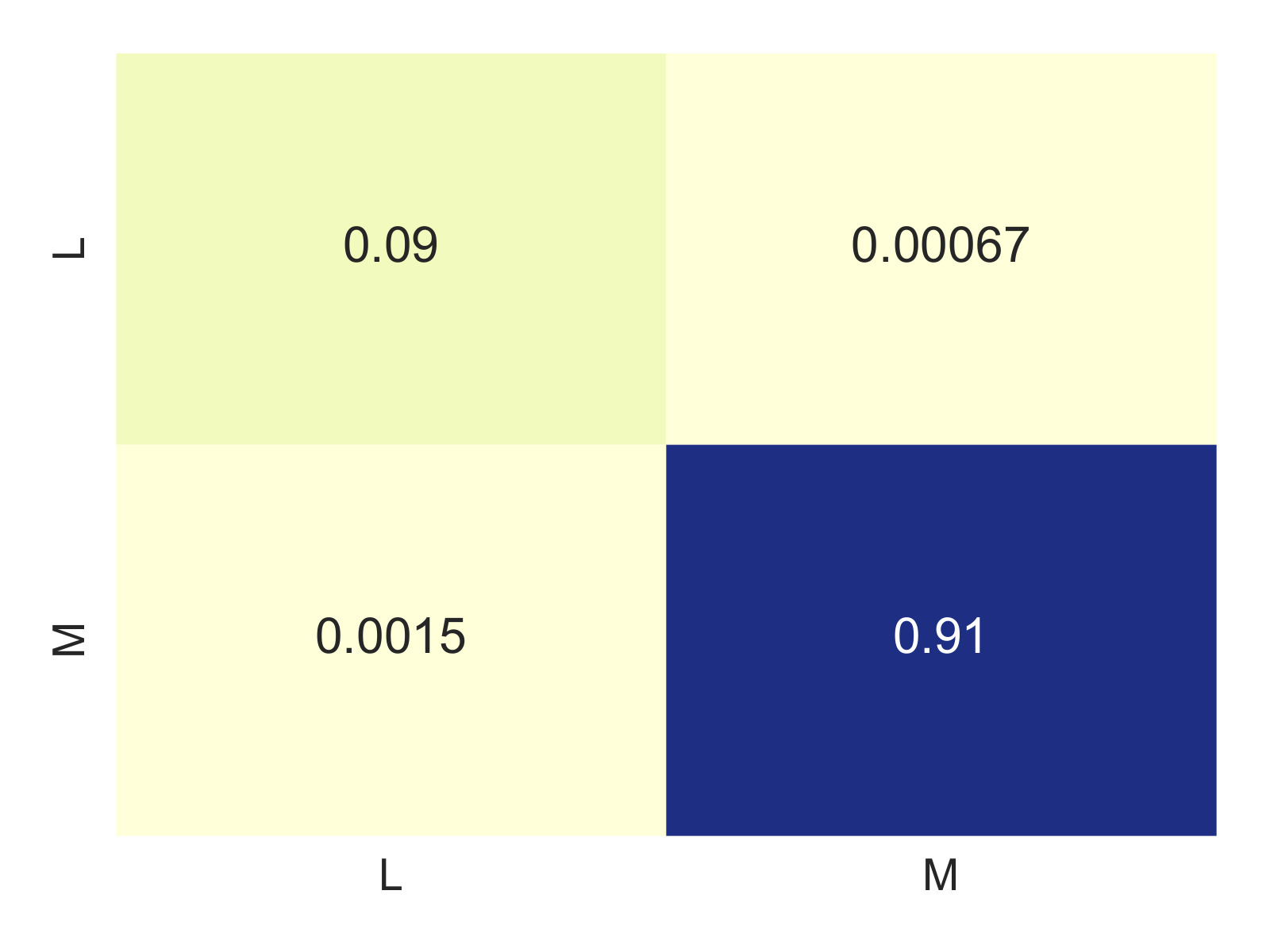}\label{fig:g4-states-coop-policy} }}%
    \quad
    \subfloat[Game 5]{{\includegraphics[width=.17\linewidth]{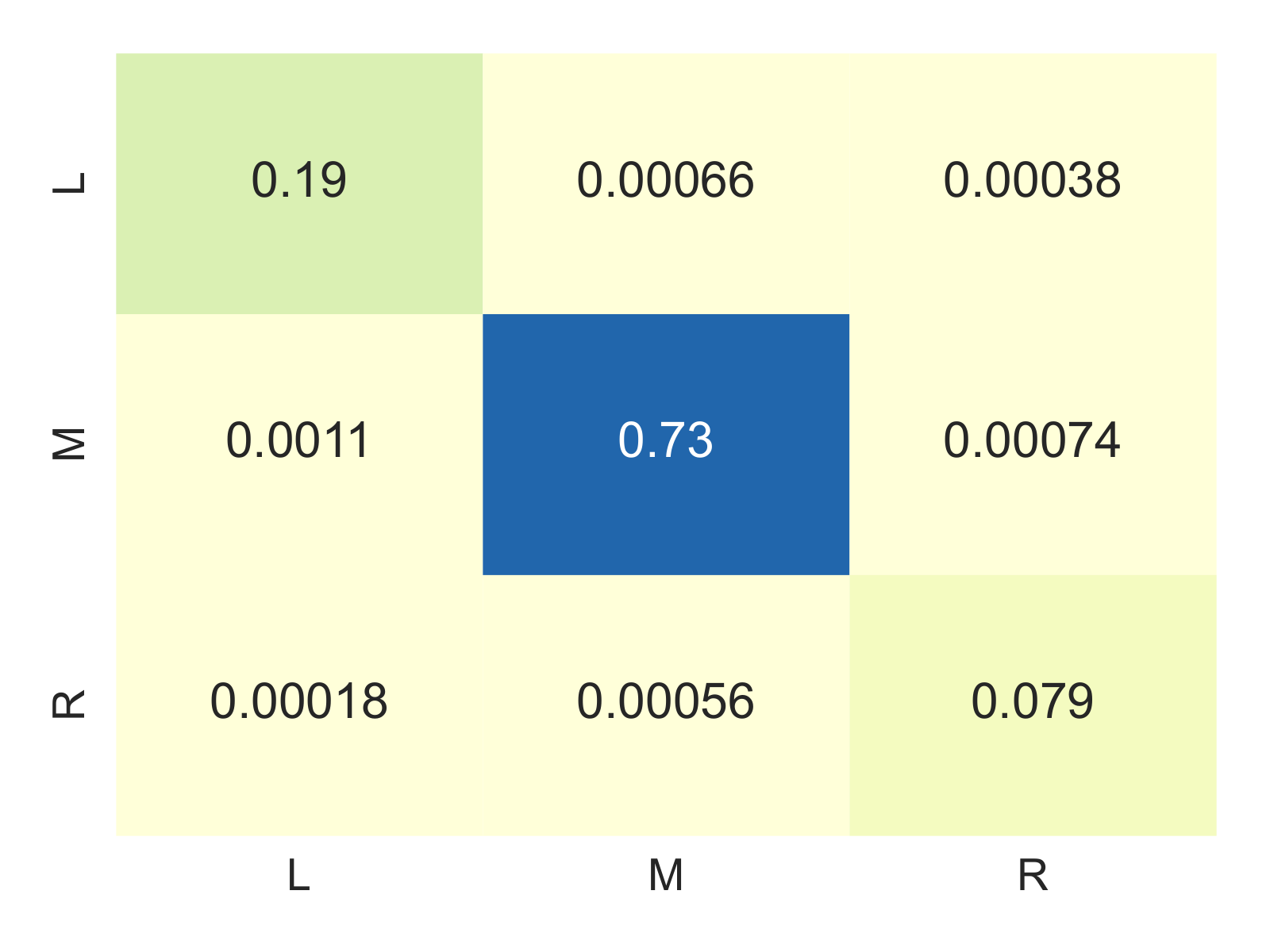}\label{fig:g5-states-coop-policy} }}%
    \caption{The empirical joint-action distributions in the last 10\% of episodes when learning with cooperative policy communication.}%
    \label{fig:coop-policy-states}%
\end{figure}

Lastly, we show the converged joint strategy distribution in Figure \ref{fig:coop-policy-states}. Overall, these results are analogous to the results from cooperative action communication. Game 4 however shows one difference with the NE of (L, L) being played at the same rate as without communication, while cooperative action communication had previously raised it. We attribute this to the reduced possibility for coordination to specific joint-actions. When committing to an action, this is much easier than when committing to policies. Depending on the exact scenario that is played, this could be a drawback or benefit and thus suggests the need for careful consideration. We expect additional differences between action and policy communication to surface when considering different types of games. For example, when considering a zero-sum game such as Rock Paper Scissors, it might be better to communicate policies rather than the next action as every individual action can be exploited by the opponent. In other cases, communicating the next action is better to avoid actions which are unacceptable for both players.

\subsection{Hierarchical Communication}
\label{sec:hierarchical-com-exp}
As described in Section \ref{sec:hierarchical-com}, we also explore a protocol in which agents concurrently learn when to communicate as well as what to communicate. This setting provides additional insight into the benefit of communication compared to independent learning and the applicability of different communication protocols. We design a two-layer protocol that contains both an independent learning protocol and a communication protocol in the bottom layer. The upper layer decides on when to communicate and thus which protocol to use.

We conduct three additional experiments so that each of the previous communication protocols is used once as the lower level communication protocol. We include in each experiment a figure of the communication probabilities for both agents over time. Important to note is that while we keep the learning rate for the top-level policy at 0.01 as in other experiments, we use a higher learning rate of 0.05 for all lower level protocols. We keep the top-level learning rate low to ensure that enough consideration is given to both communicating and not communicating. The low-level learning rate is relatively higher so that the advantages of using one protocol over the other is maximally exploited. For the figures in games without Nash equilibria we only show the results for the first 1000 episodes. In games with Nash equilibria, this cutoff point is at 1500 episodes. We do this as at this point agents had already converged and it allows for a clearer view of the learning dynamics.

\subsubsection{Hierarchical Cooperative Action Communication}
\label{sec:hierarchical-coop-com-exp}
\begin{figure}[h!tb]%
    \centering
    \subfloat[Game 1]{{\includegraphics[width=.28\linewidth]{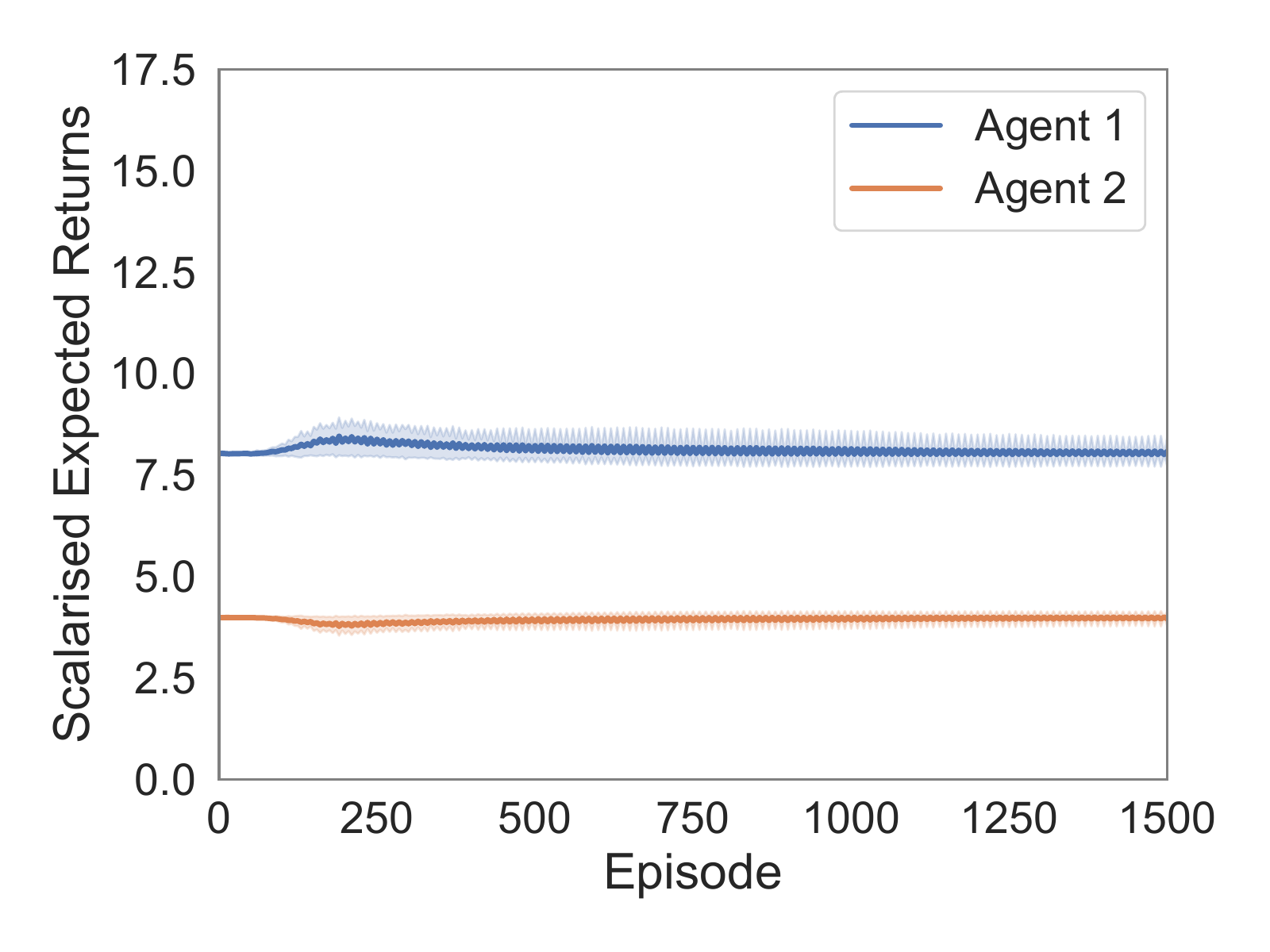}\label{fig:g1-ser-opt-coop-action} }}%
    \quad
    \subfloat[Game 2]{{\includegraphics[width=.28\linewidth]{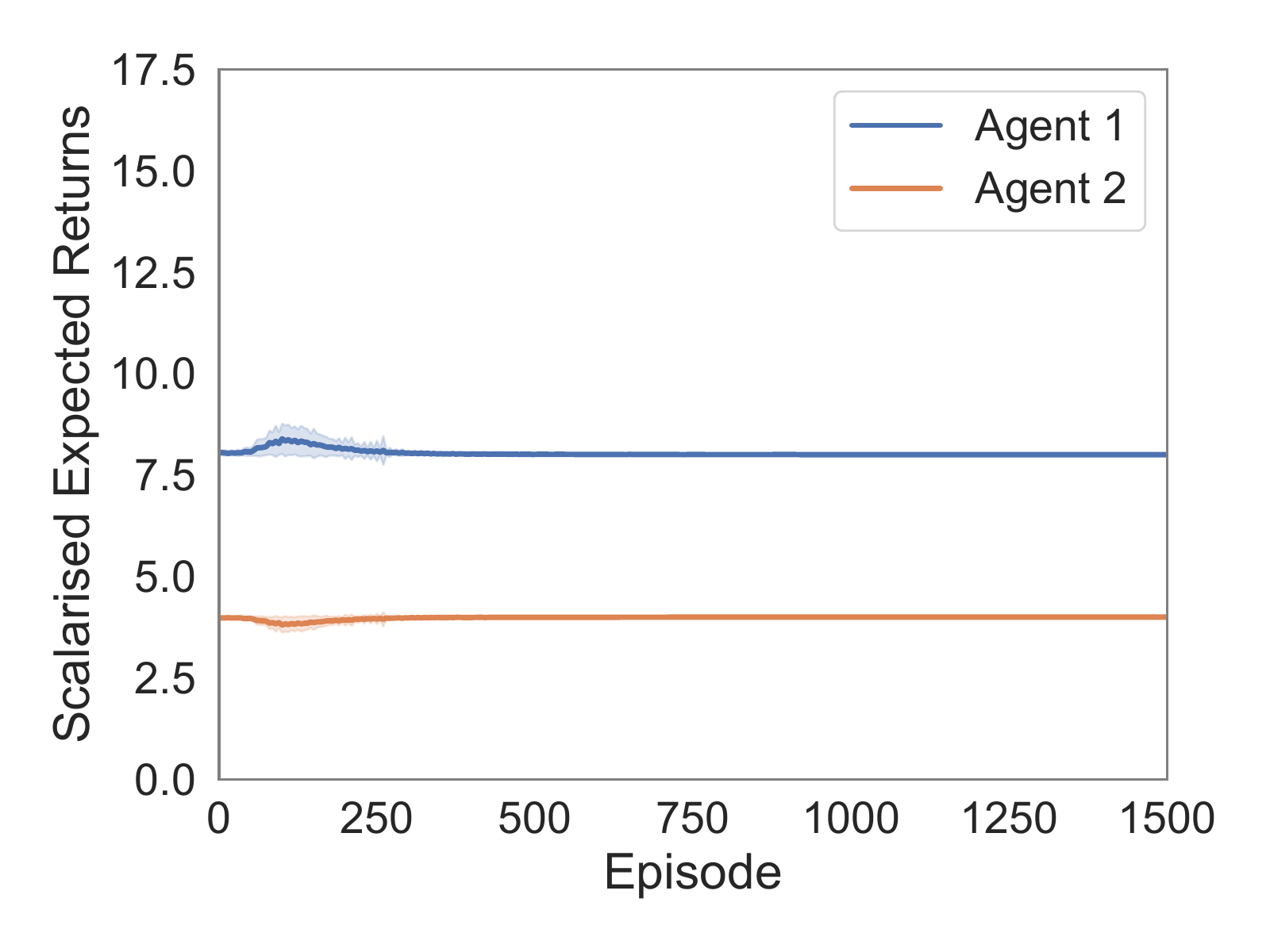}\label{fig:g2-ser-opt-coop-action} }}%
    \quad
    \subfloat[Game 3]{{\includegraphics[width=.28\linewidth]{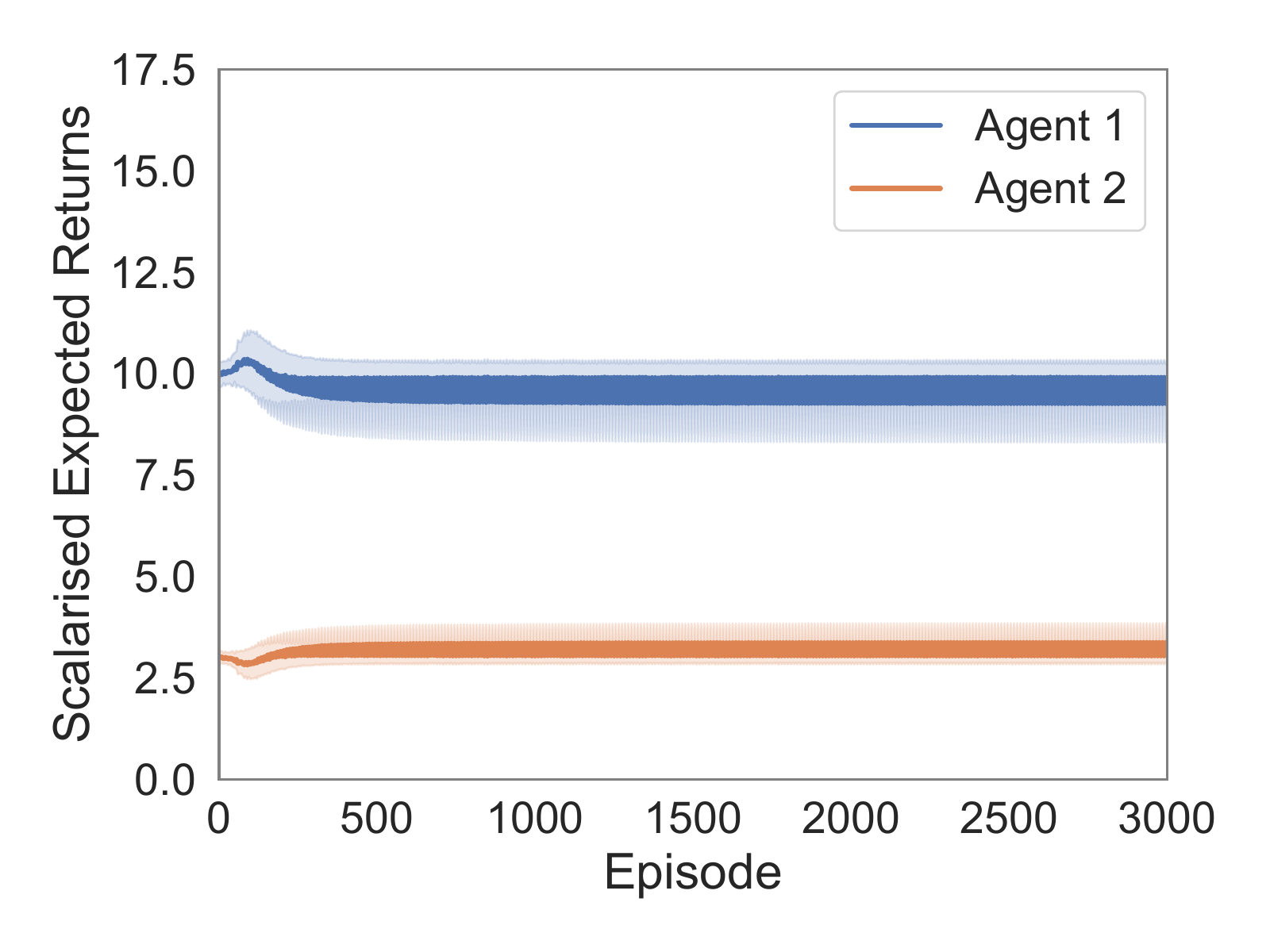}\label{fig:g3-ser-opt-coop-action} }}%
    \quad
    \subfloat[Game 4]{{\includegraphics[width=.28\linewidth]{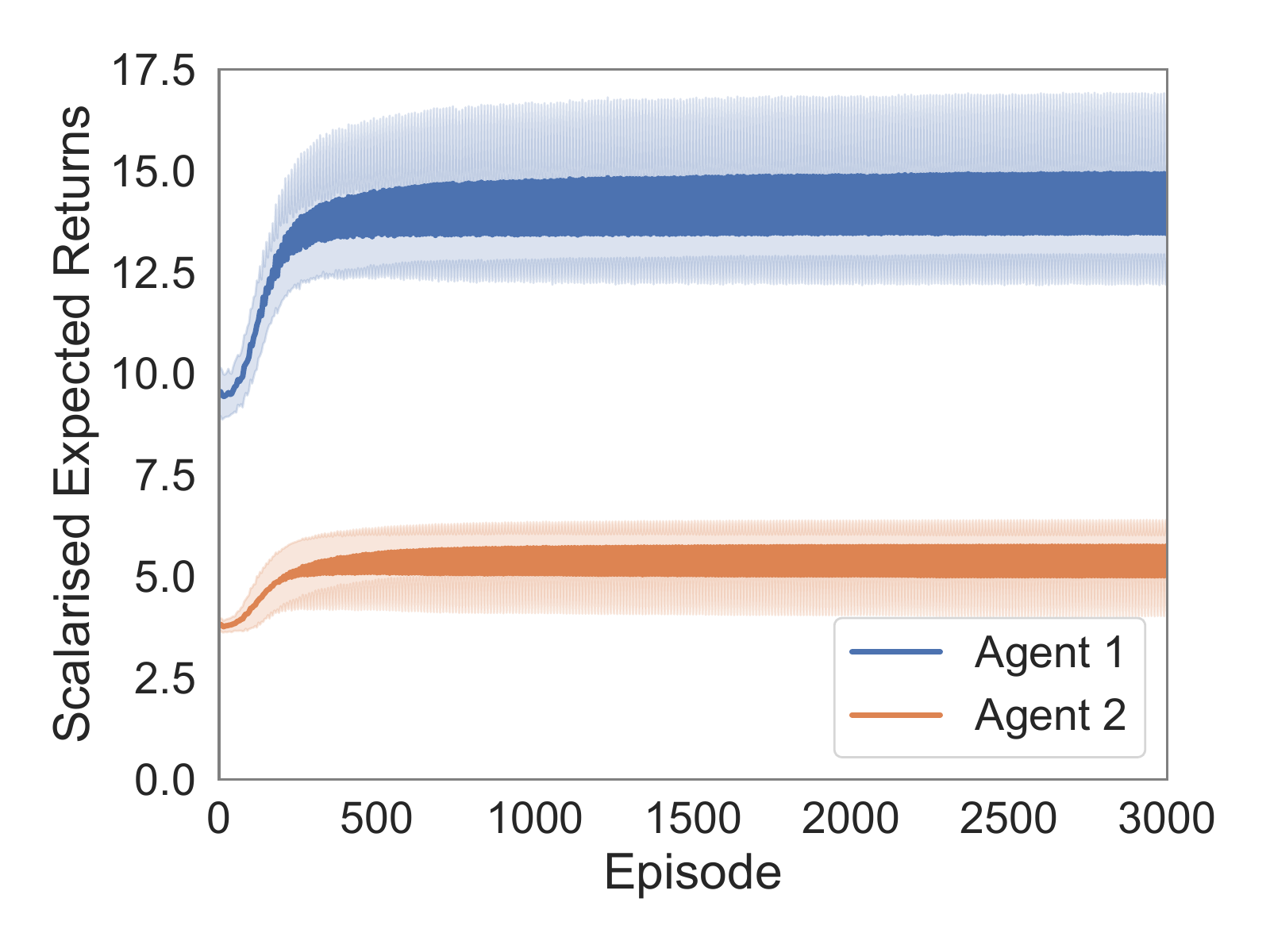}\label{fig:g4-ser-opt-coop-action} }}%
    \quad
    \subfloat[Game 5]{{\includegraphics[width=.28\linewidth]{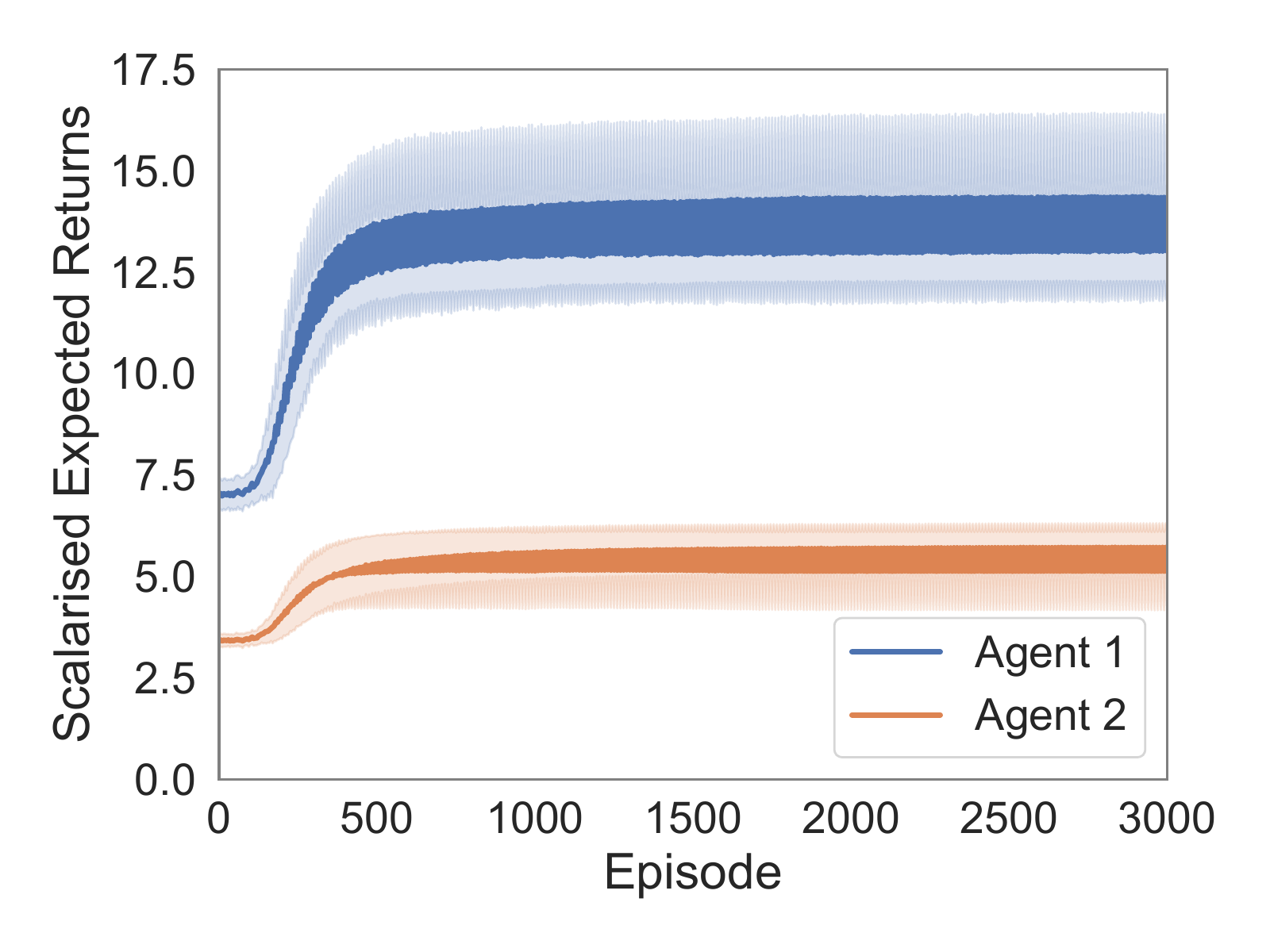}\label{fig:g5-ser-opt-coop-action} }}%
    \caption{The SER for both agents when learning with hierarchical cooperative action communication.}%
    \label{fig:opt-coop-action-ser}%
\end{figure}

We show the SER over time in Figure \ref{fig:opt-coop-action-ser}. The first pattern we identify is that outcomes from the same protocol with obligated communication (Sec.~\ref{sec:coop-com-exp}) do not necessarily translate to the hierarchical communication setting. Specifically, we find that hierarchical communication allows players to learn cyclic policies. We observe this pattern most clearly in games with multiple NE. Here, agents cycle through playing their preferred equilibria similar to results from self-interested action communication (Sec. \ref{sec:comp-com-exp}). Upon further analysis, we find that players use the additional layer to coordinate this cycle. Specifically, one agent learns to always communicate and play one equilibrium. The other agent learns to never communicate and play the second equilibrium. We thus conclude that leaving agents to learn non-stationary policies, conditioned on being the leader or not, can give rise to CNE in multi-objective games as defined in Section \ref{sec:solution-concepts}.

We stress that although non-stationary policies are a rational outcome for Game 4 and 5, this is not what is consistently learned. From the data we often see that agents often only employ one protocol. This occurs when one Nash equilibrium is found with a particular lower level protocol, thus increasing the likelihood of this protocol being used again in the future.

\begin{figure}[h!tb]%
    \centering
    \subfloat[Game 1]{{\includegraphics[width=.28\linewidth]{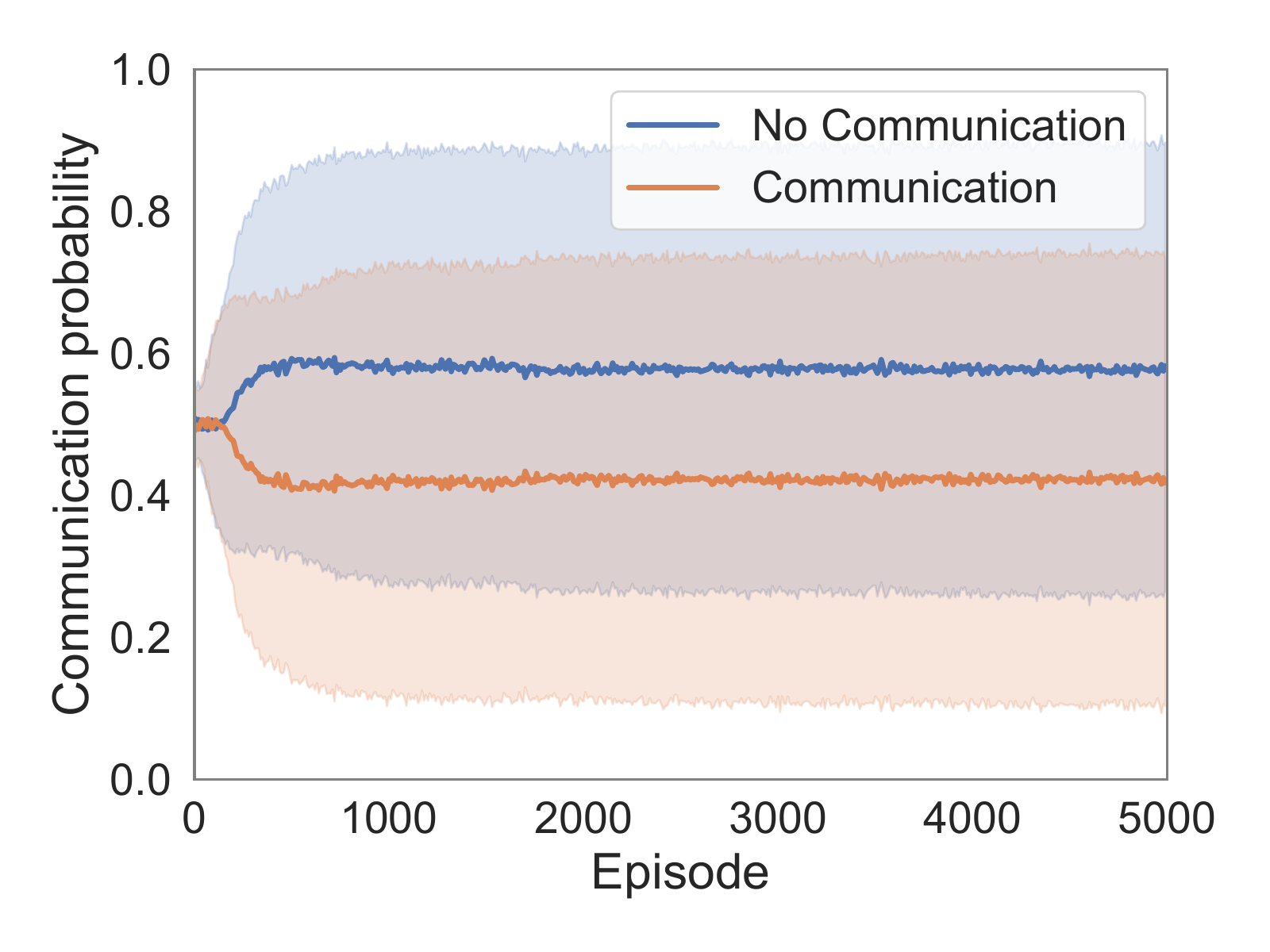}\label{fig:g1-A1-com-opt-coop-action} }}%
    \quad
    \subfloat[Game 2]{{\includegraphics[width=.28\linewidth]{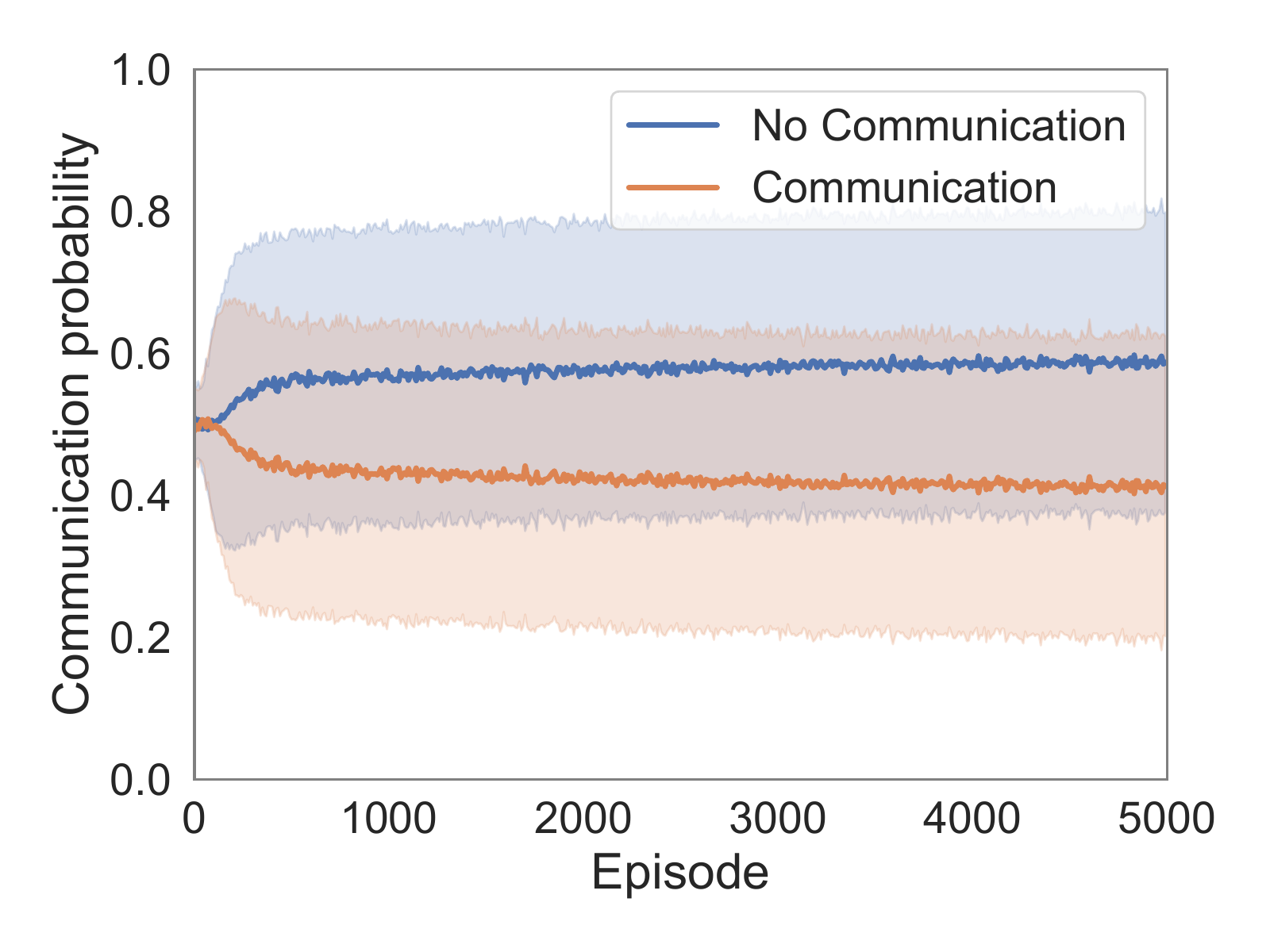}\label{fig:g2-A1-com-opt-coop-action} }}%
    \quad
    \subfloat[Game 3]{{\includegraphics[width=.28\linewidth]{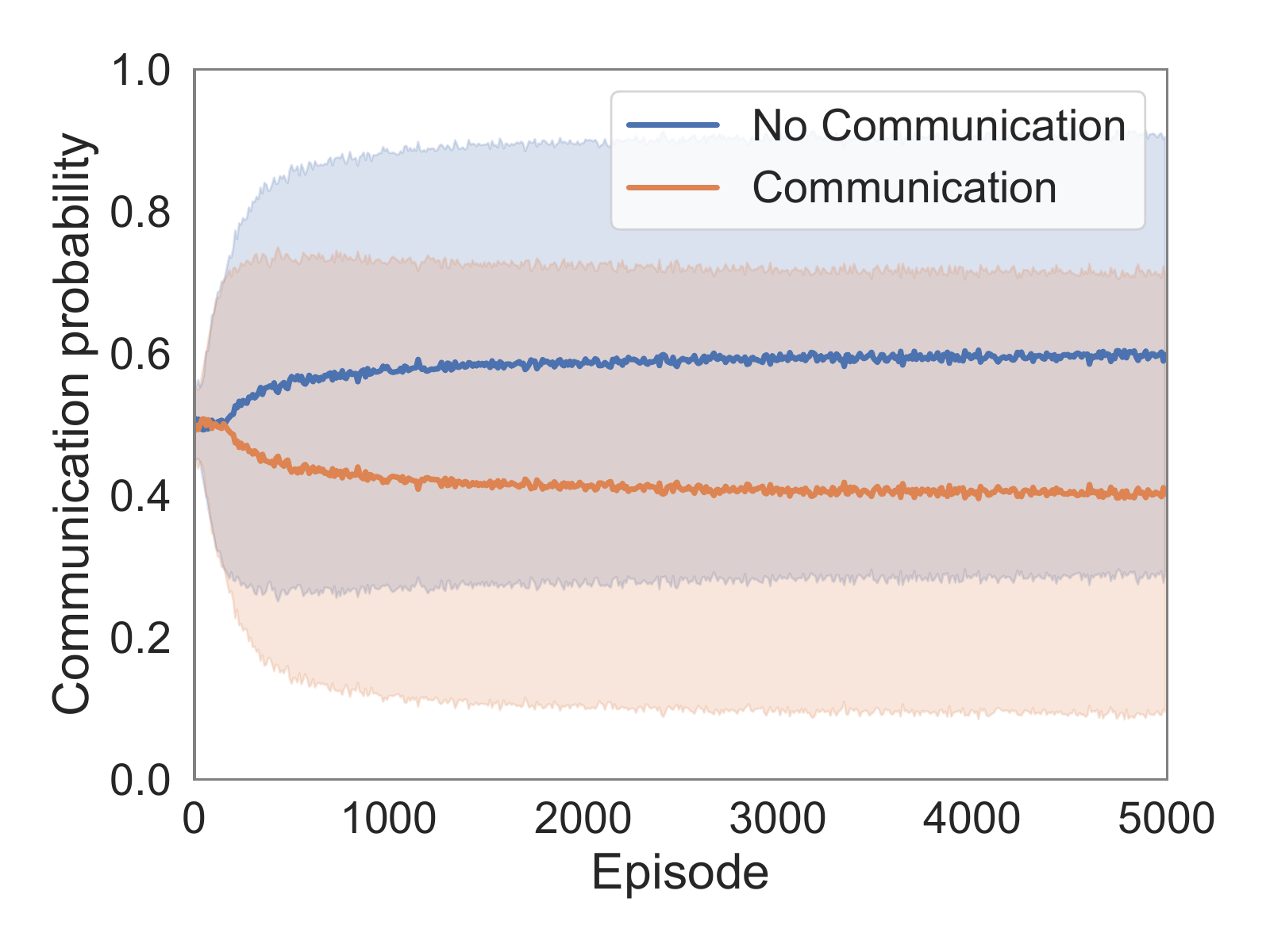}\label{fig:g3-A1-com-opt-coop-action} }}%
    \quad
    \subfloat[Game 4]{{\includegraphics[width=.28\linewidth]{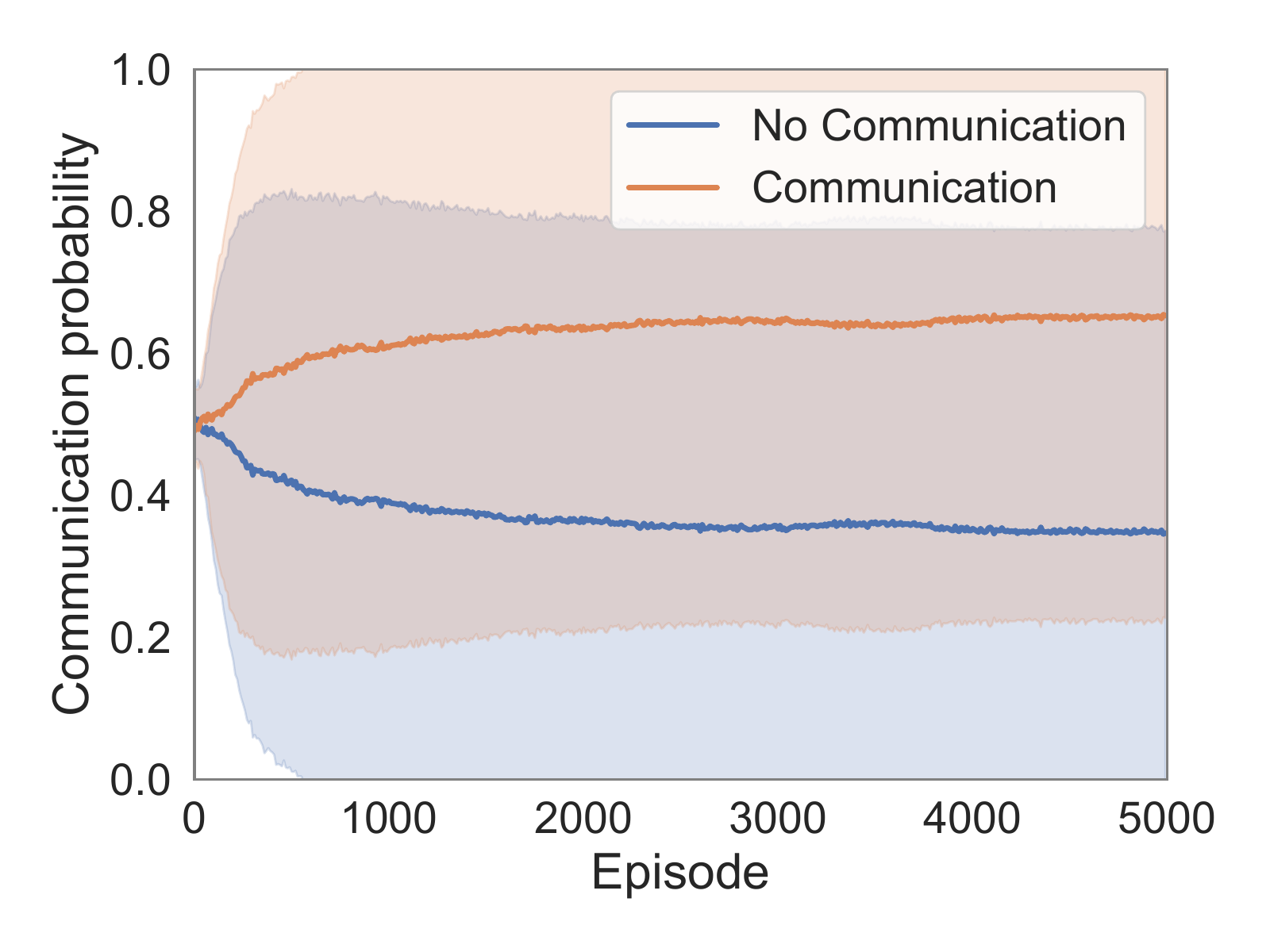}\label{fig:g4-A1-com-opt-coop-action} }}%
    \quad
    \subfloat[Game 5]{{\includegraphics[width=.28\linewidth]{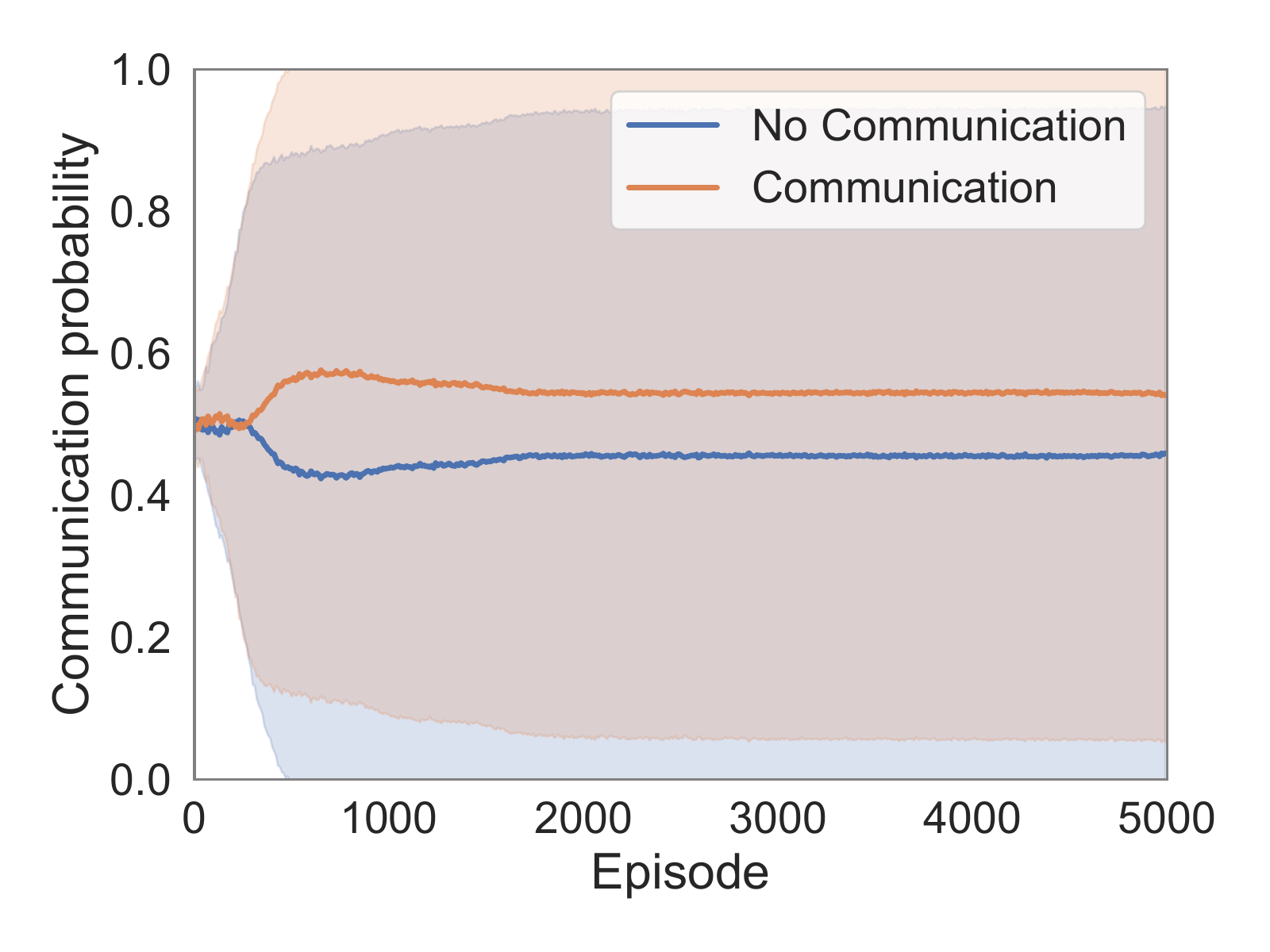}\label{fig:g5-A1-com-opt-coop-action} }}%
    \caption{The communication probabilities for agent 1 when learning with hierarchical cooperative action communication.}%
    \label{fig:opt-coop-action-A1-com}%
\end{figure}

\begin{figure}[h!tb]%
    \centering
    \subfloat[Game 1]{{\includegraphics[width=.28\linewidth]{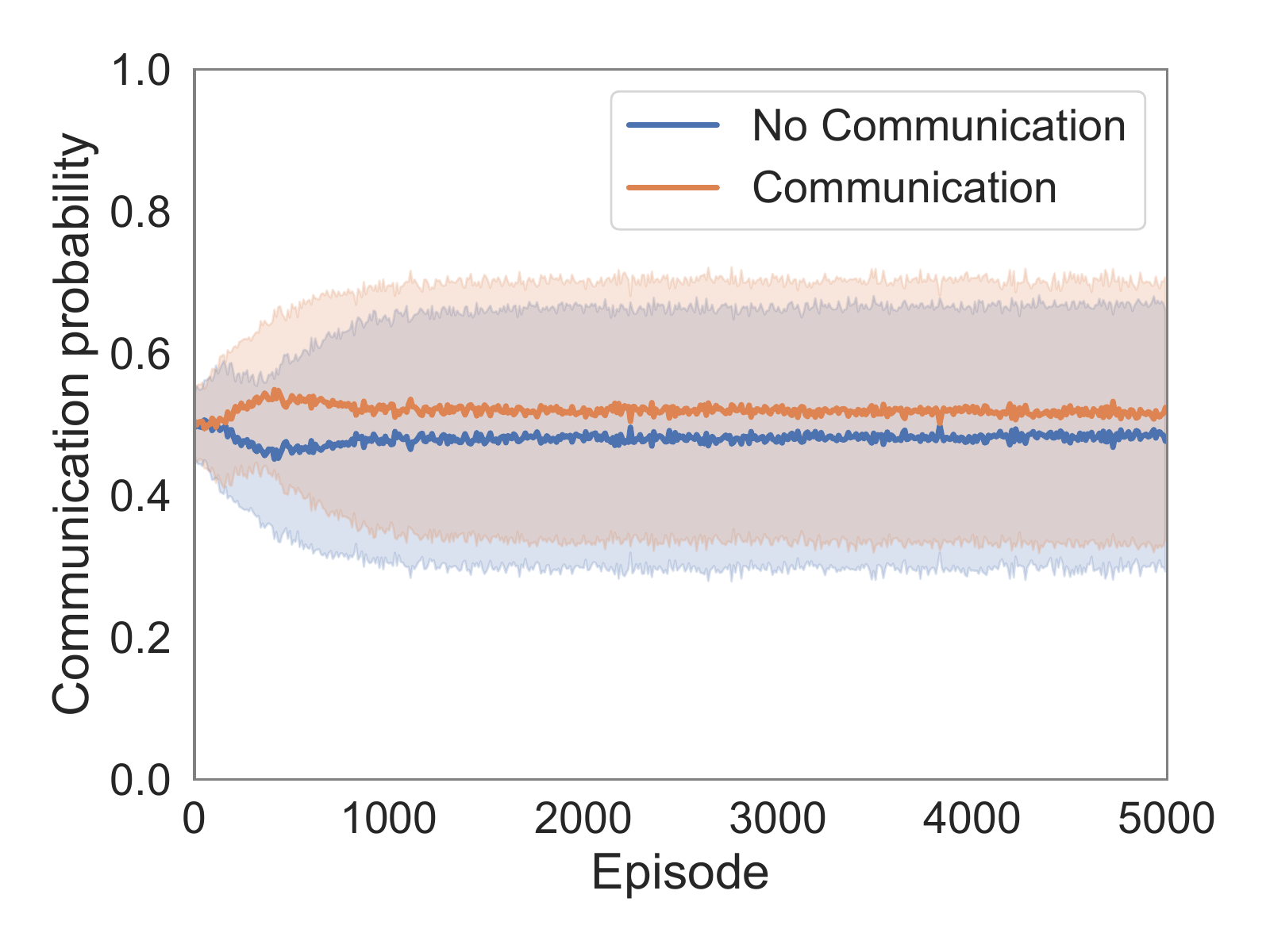}\label{fig:g1-A2-com-opt-coop-action} }}%
    \quad
    \subfloat[Game 2]{{\includegraphics[width=.28\linewidth]{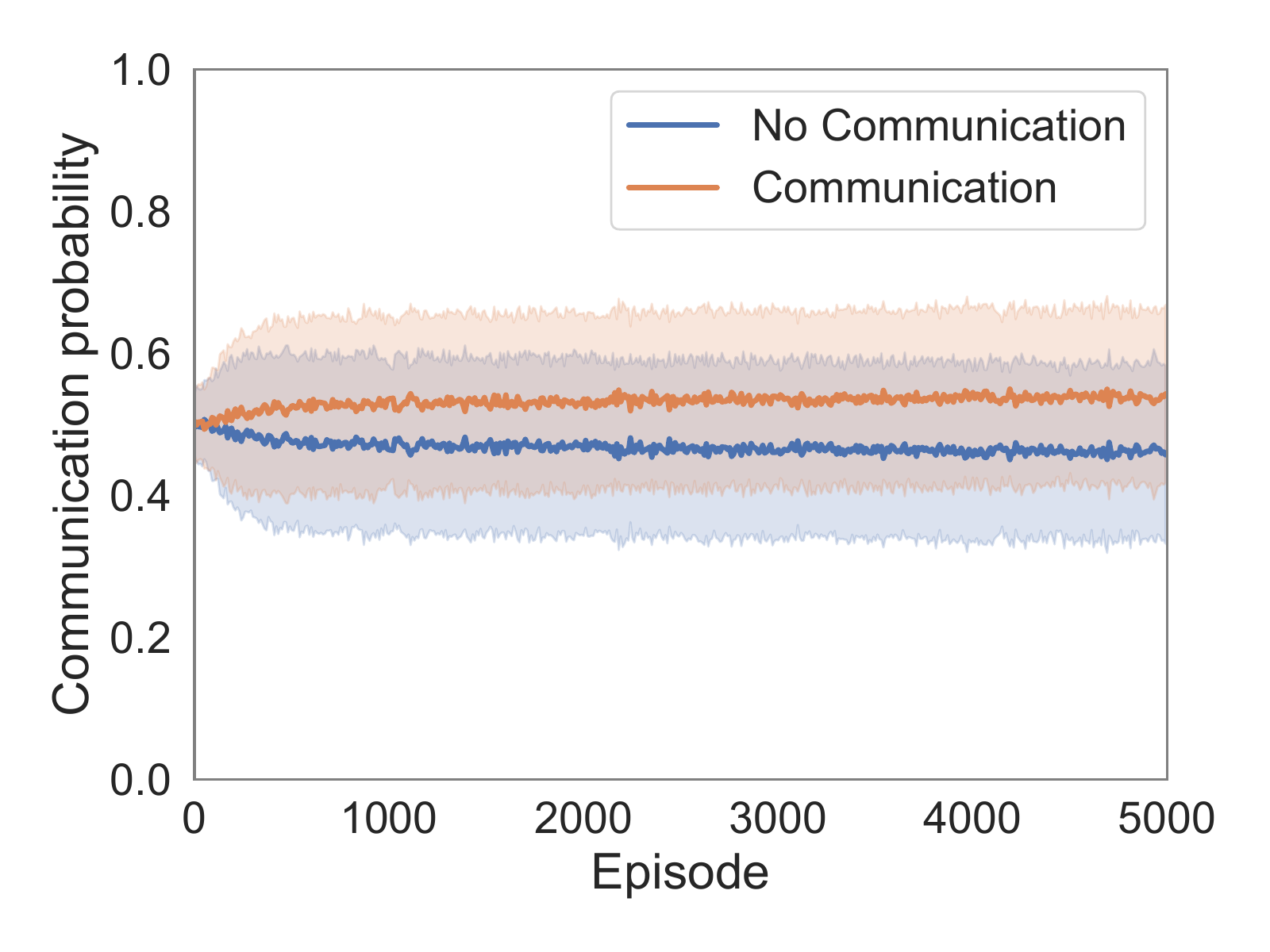}\label{fig:g2-A2-com-opt-coop-action} }}%
    \quad
    \subfloat[Game 3]{{\includegraphics[width=.28\linewidth]{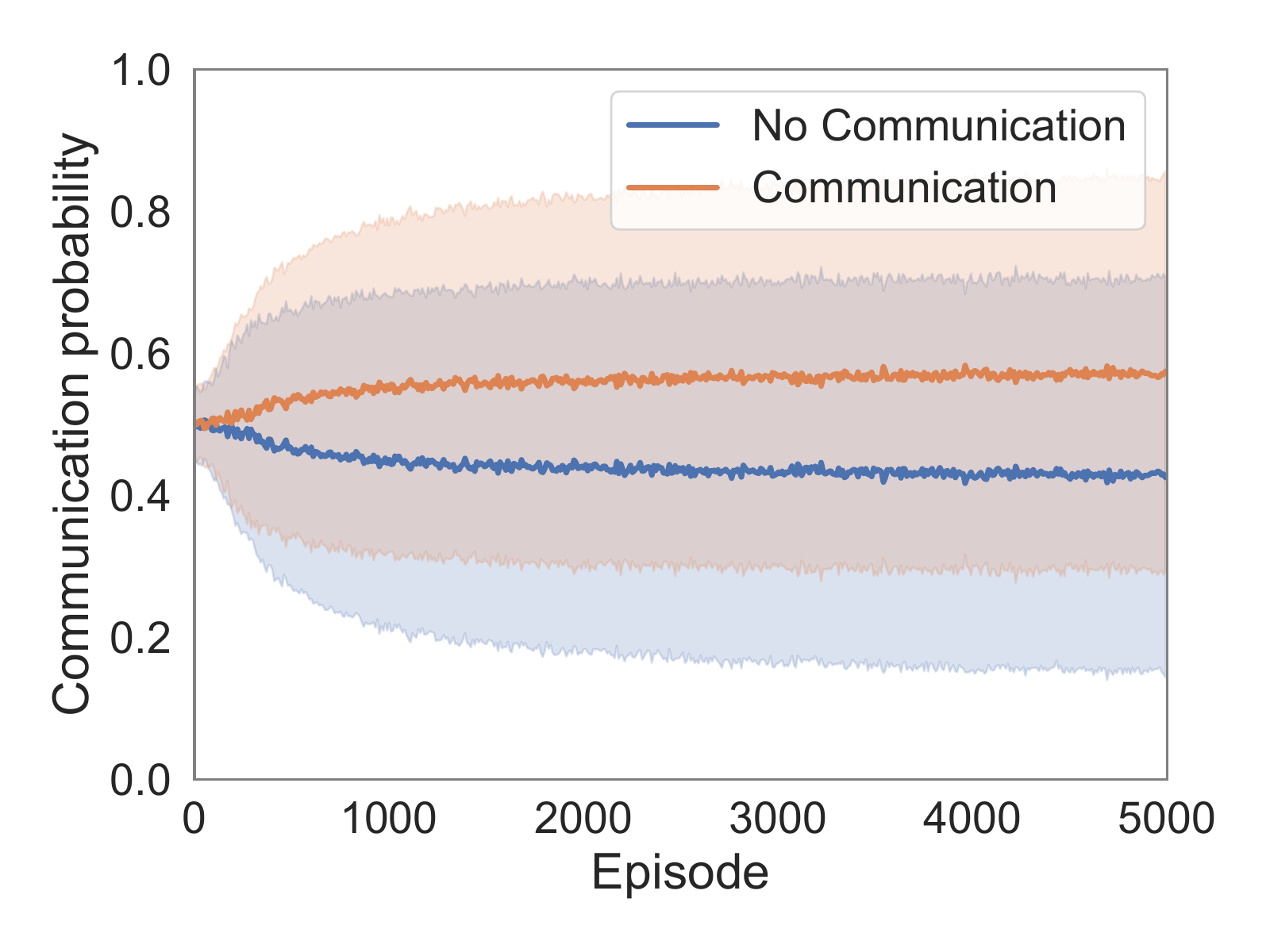}\label{fig:g3-A2-com-opt-coop-action} }}%
    \quad
    \subfloat[Game 4]{{\includegraphics[width=.28\linewidth]{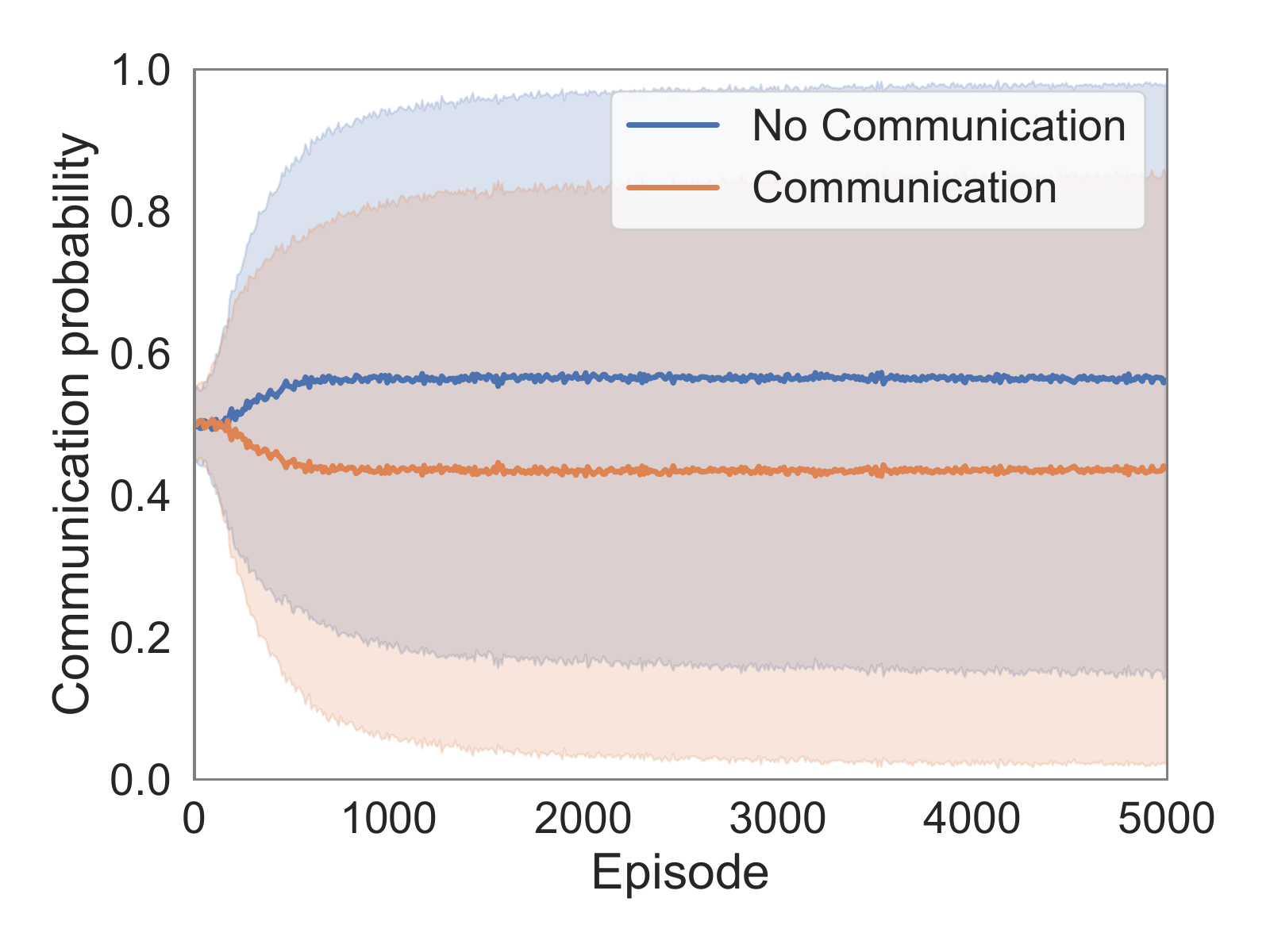}\label{fig:g4-A2-com-opt-coop-action} }}%
    \quad
    \subfloat[Game 5]{{\includegraphics[width=.28\linewidth]{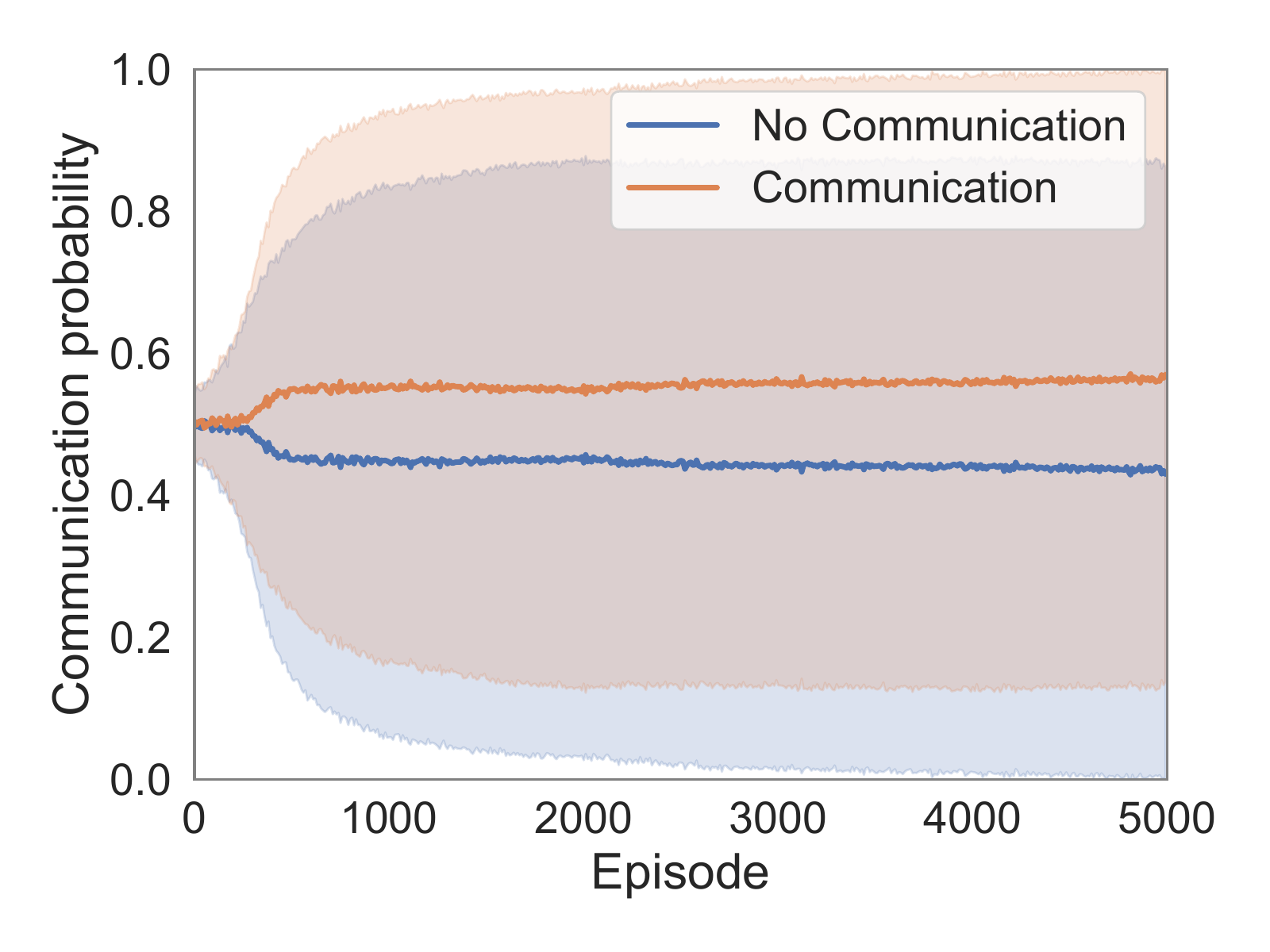}\label{fig:g5-A2-com-opt-coop-action} }}%
    \caption{The communication probabilities for agent 2 when learning with hierarchical cooperative action communication.}%
    \label{fig:opt-coop-action-A2-com}%
\end{figure}

These patterns can also be found in the communication strategies shown in Figures \ref{fig:opt-coop-action-A1-com} and \ref{fig:opt-coop-action-A2-com}. In Games 4 and 5, there is an average communication rate of around 50\%, but with a high standard deviation. As stated earlier, sometimes communication is used to cycle between NE. In other cases, agents learn a Nash equilibrium while following one protocol and adapt their communication strategy to always follow the successful protocol and avoid the other. This results in different communication strategies being learned in different runs of the experiment, thus explaining the high standard deviation.

In games without Nash equilibria, i.e. Game 1 and 2, the standard deviation around the communication probabilities is lower than in games with NE. Both agents learn to communicate approximately 50\% of the time. We attribute this to the underlying communication protocol and the actor-critic implementation. Firstly, recall from Section \ref{sec:coop-com-exp} that players learned the same strategies when employing cooperative action communication as without communication. Therefore, the utility from both protocols will be similar as long as both protocols are learned to a sufficient degree. This ensures that the gradient for the communication strategy is (approximately) zero and thus results in the observed stable communication strategies.



We note that this gradient dynamic results in higher randomisation in the communication strategies when executing the experiments with a lower learning rate in the low-level. Concretely, because the communication strategy is learning with the same rate as the lower level protocols, the communication strategy simply follows whichever protocol is preferred in the earlier episodes. This highlights an important link between the speed with which to learn a communication strategy and the speed with which to learn to play the game itself. Agents require a solid understanding of how to play optimally using the lower level protocols, before learning a definitive communication strategy.

While the proposed algorithm is not always able to learn the optimal (cyclical) policies, this does not imply that communication presents no benefit to agents. In fact, a learning algorithm capable of discovering the optimal cyclical policy in games with multiple NE would lead at least one agent to favour explicit communication. Furthermore, in games with larger action and state spaces, independent learning without any additional method becomes slow and in some cases even infeasible. In such settings, communicating one's preferences might lead to substantial speed ups in convergence times. We discuss studying this as a direction for future work in Section \ref{sec:conclusion}.

\subsubsection{Hierarchical Self-Interested Action Communication}
\label{sec:hierarchical-comp-com-exp}
\begin{figure}[h]%
    \centering
    \subfloat[Game 1]{{\includegraphics[width=.28\linewidth]{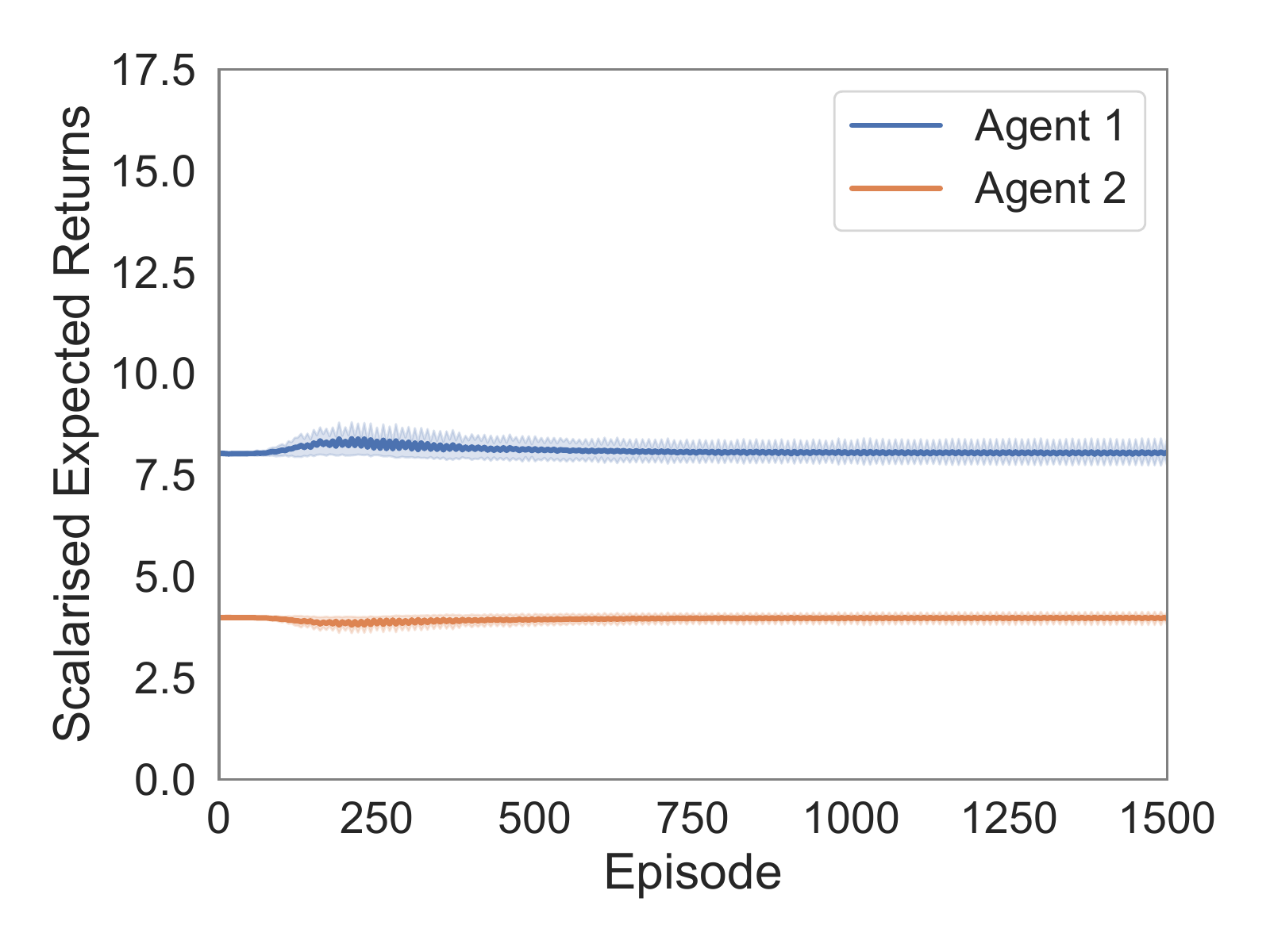}\label{fig:g1-ser-opt-comp-action} }}%
    \quad
    \subfloat[Game 2]{{\includegraphics[width=.28\linewidth]{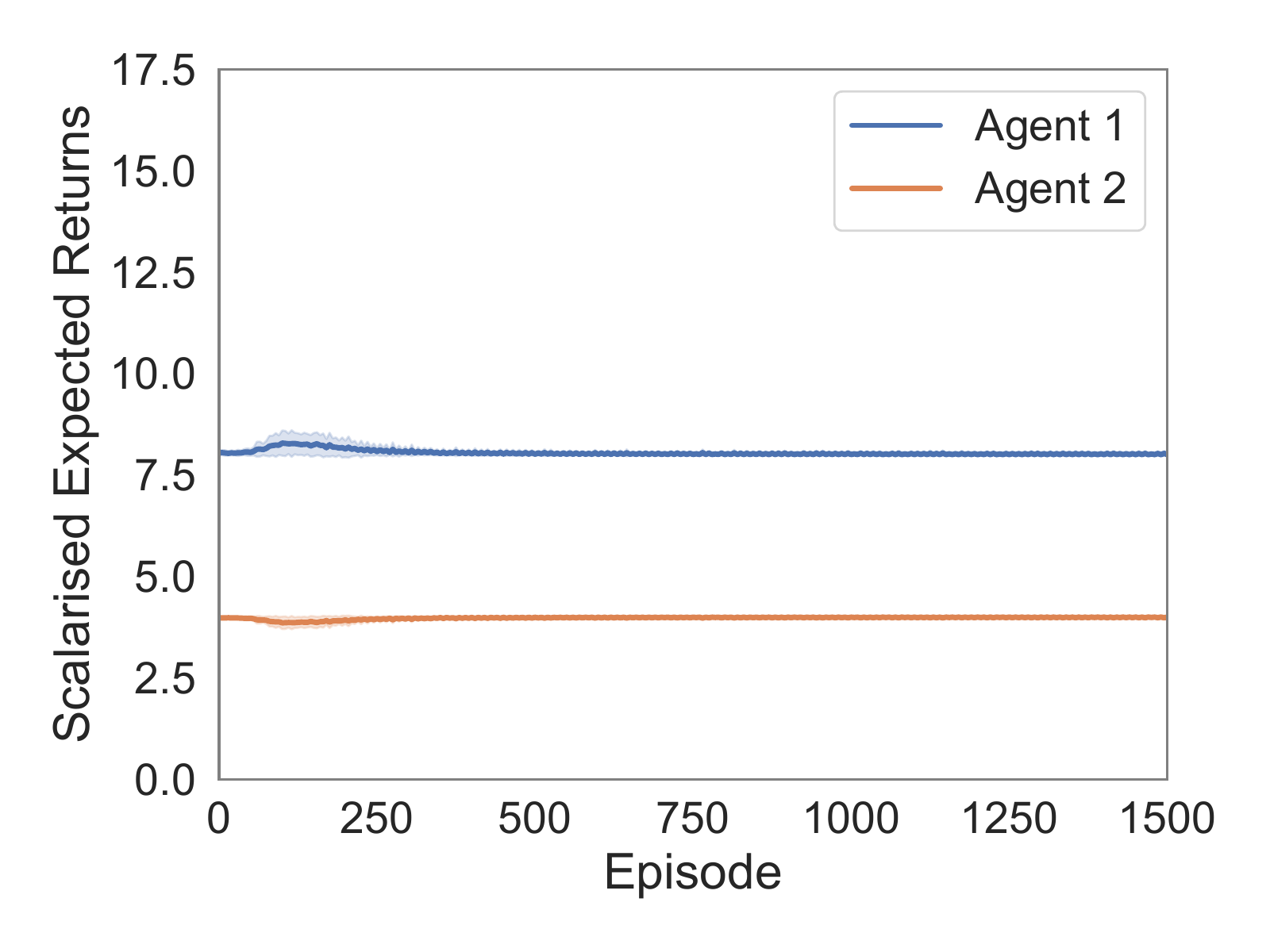}\label{fig:g2-ser-opt-comp-action} }}%
    \quad
    \subfloat[Game 3]{{\includegraphics[width=.28\linewidth]{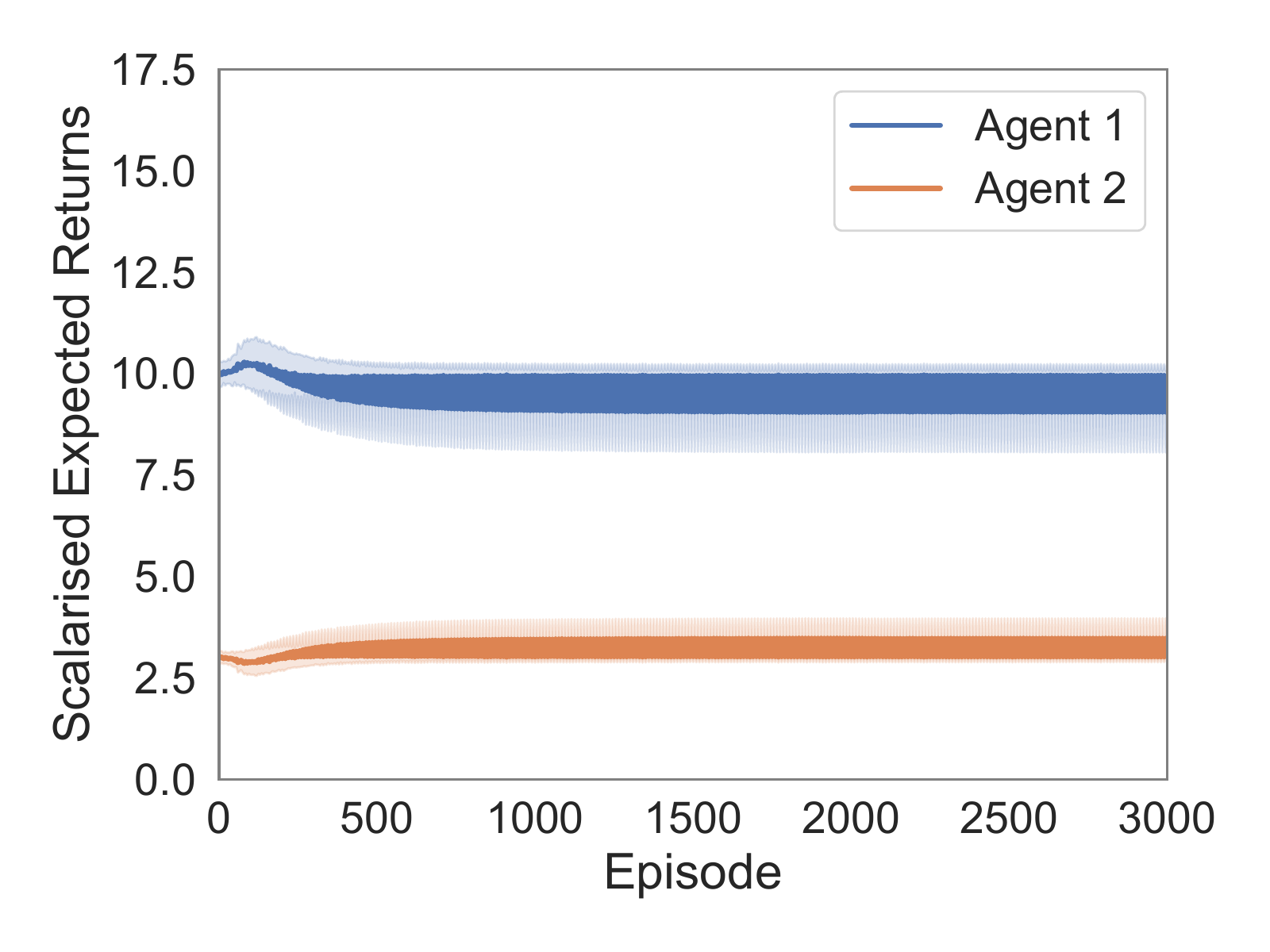}\label{fig:g3-ser-opt-comp-action} }}%
    \quad
    \subfloat[Game 4]{{\includegraphics[width=.28\linewidth]{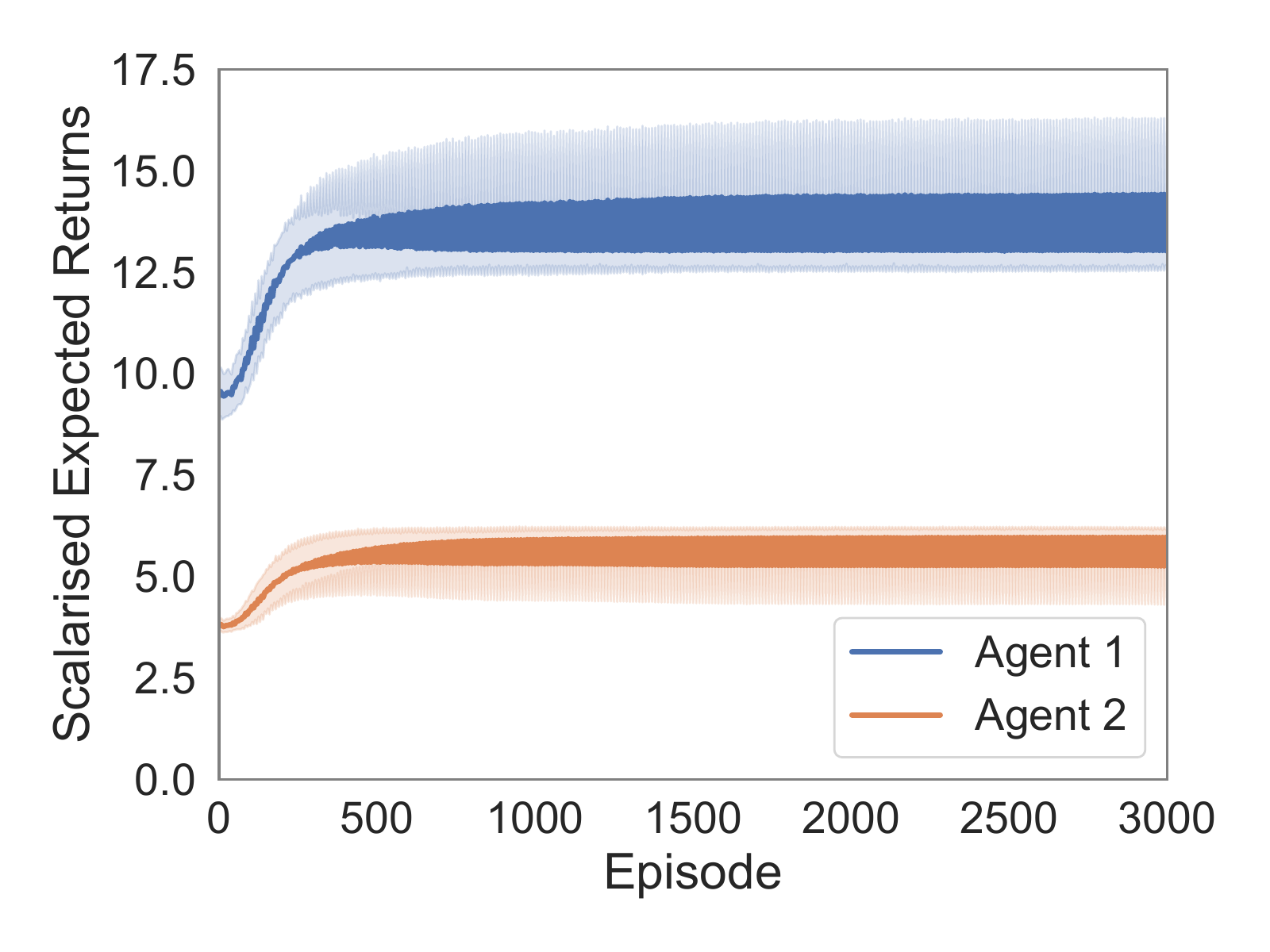}\label{fig:g4-ser-opt-comp-action} }}%
    \quad
    \subfloat[Game 5]{{\includegraphics[width=.28\linewidth]{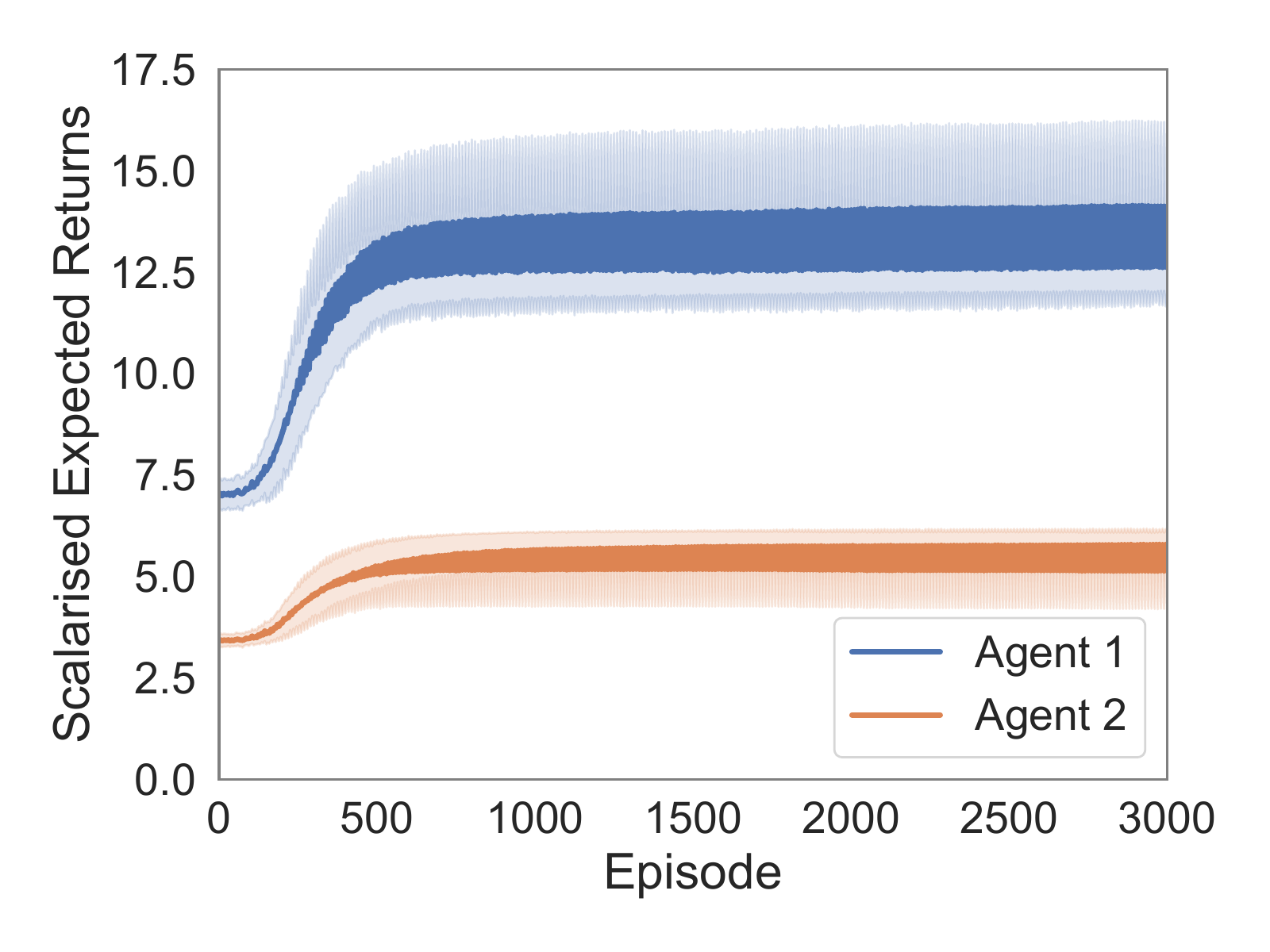}\label{fig:g5-ser-opt-comp-action} }}%
    \caption{The SER for both agents when learning with hierarchical self-interested action communication.}%
    \label{fig:opt-comp-action-ser}%
\end{figure}

In the hierarchical self-interested communication experiments, we employ the self-interested communication protocol shown in Algorithm \ref{alg:self-interested-alg} as the lower level communication policy. We generally observe the same results as in Section \ref{sec:comp-com-exp} with forced communication. Specifically, from Figure \ref{fig:opt-comp-action-ser} we see that cyclic policies and cyclic equilibria are again present. This is clear from the SER continuously going up and down, depending on the leader of the episode.

\begin{figure}[h!tb]%
    \centering
    \subfloat[Game 1]{{\includegraphics[width=.17\linewidth]{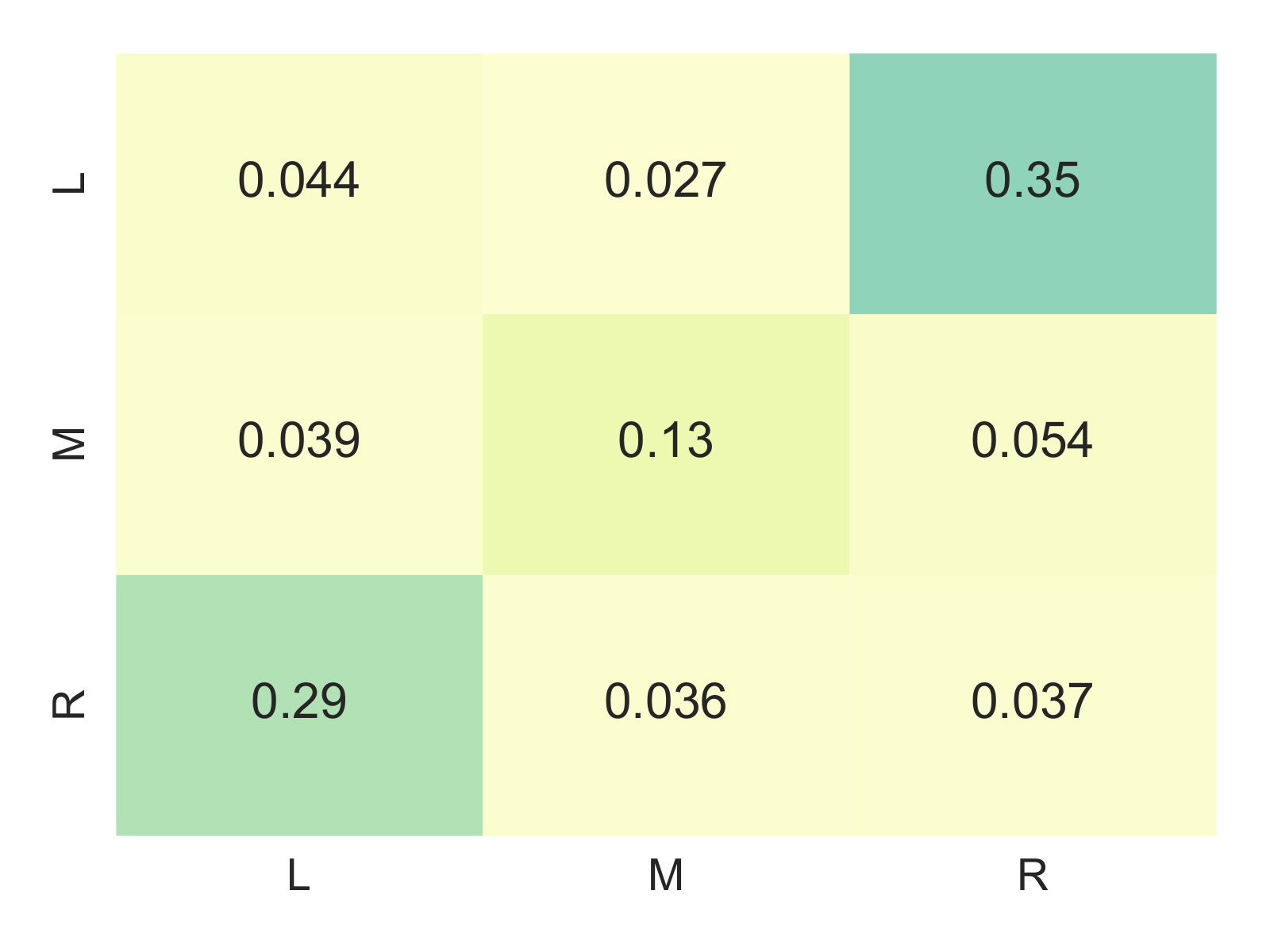}\label{fig:g1-states-opt-comp-action} }}%
    \quad
    \subfloat[Game 2]{{\includegraphics[width=.17\linewidth]{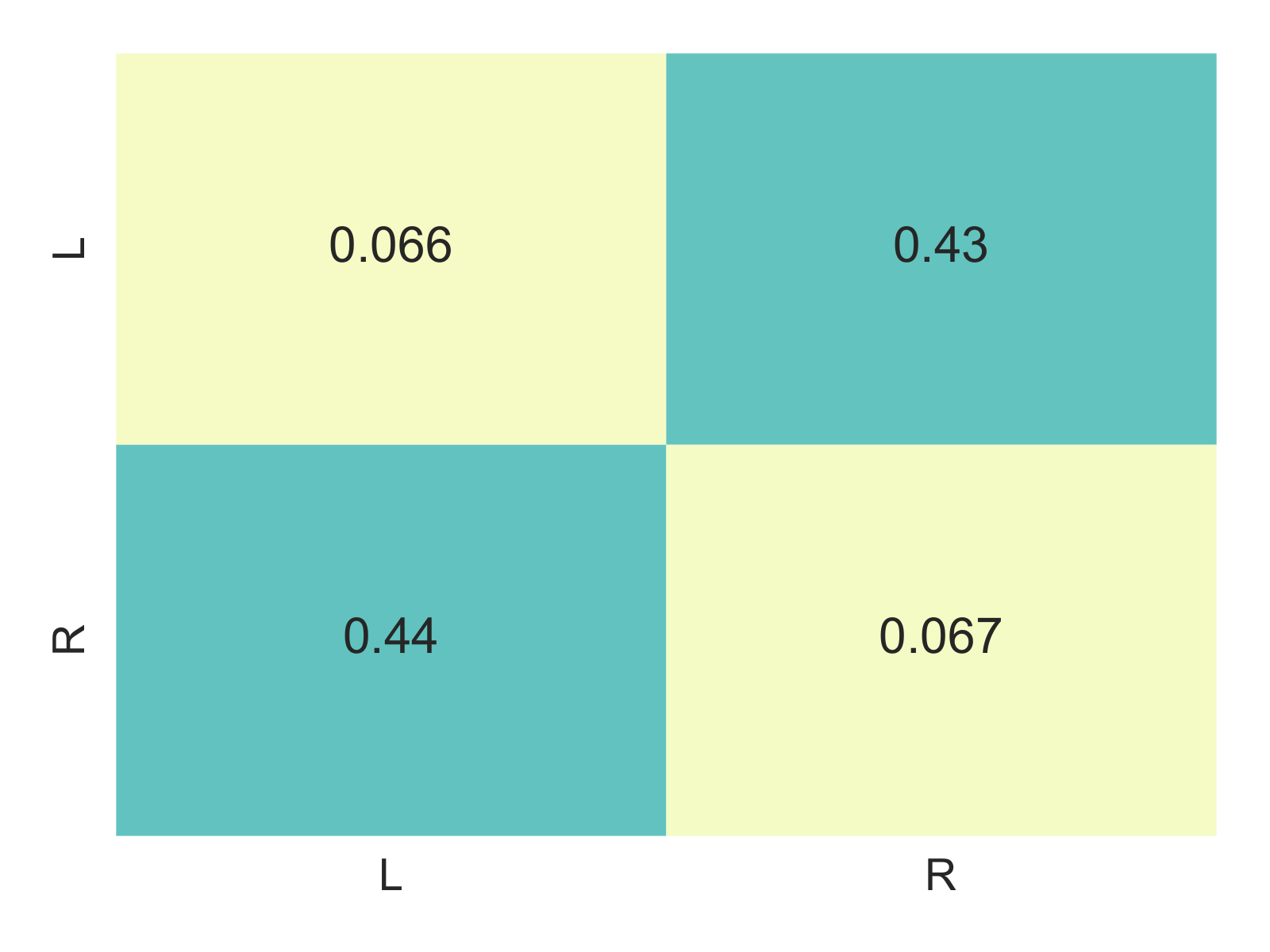}\label{fig:g2-states-opt-comp-action} }}%
    \quad
    \subfloat[Game 3]{{\includegraphics[width=.17\linewidth]{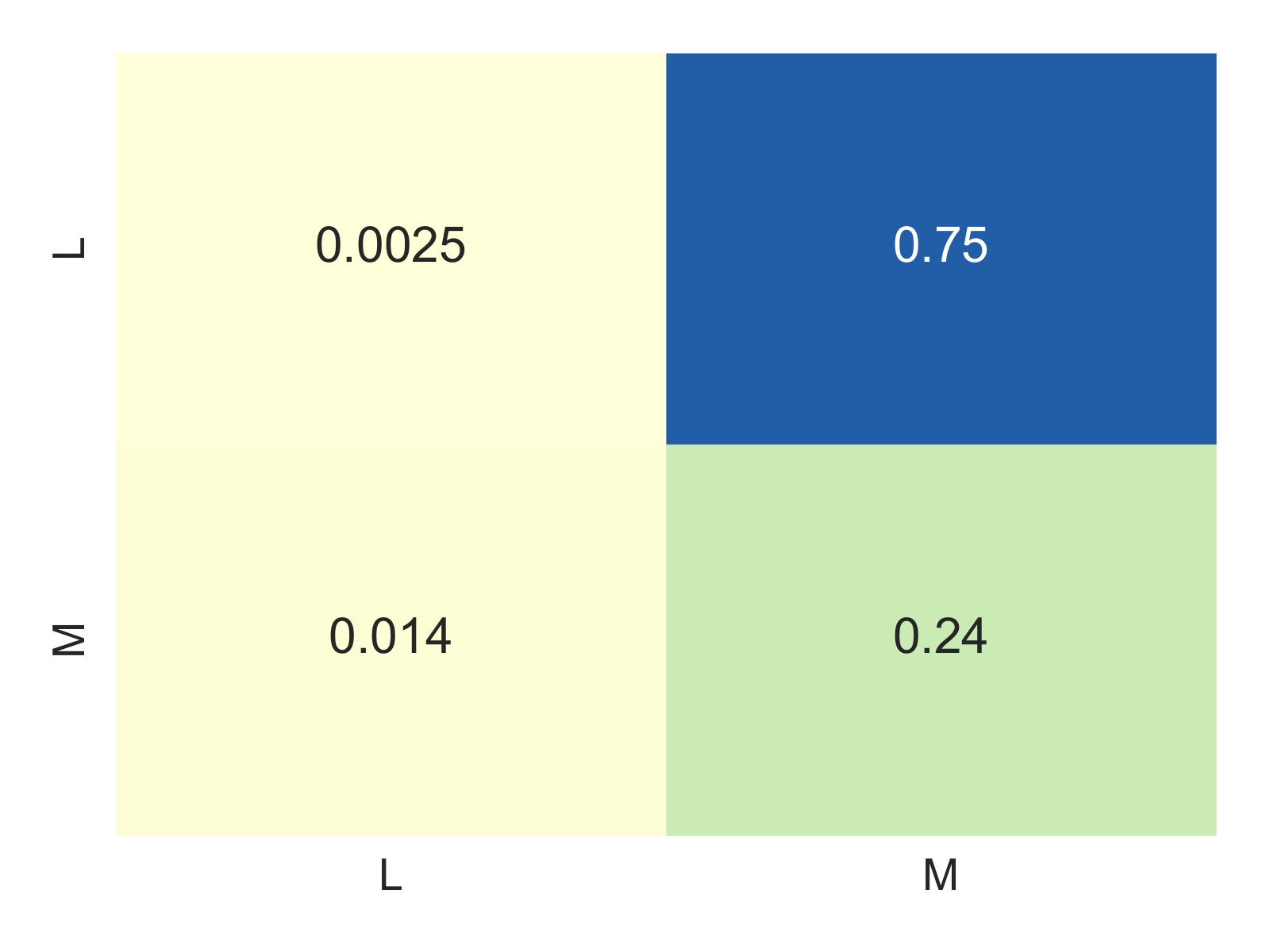}\label{fig:g3-states-opt-comp-action} }}%
    \quad
    \subfloat[Game 4]{{\includegraphics[width=.17\linewidth]{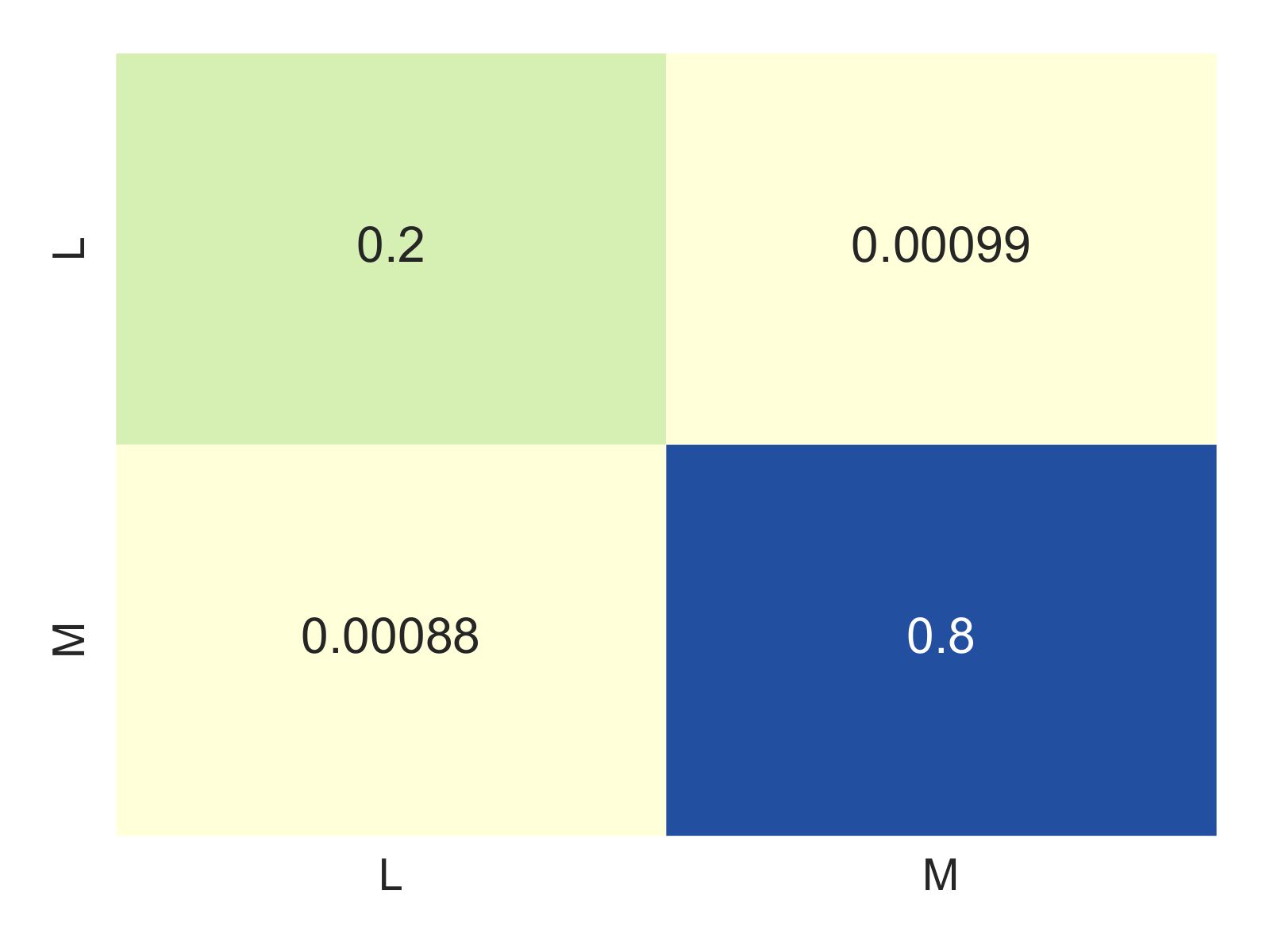}\label{fig:g4-states-opt-comp-action} }}%
    \quad
    \subfloat[Game 5]{{\includegraphics[width=.17\linewidth]{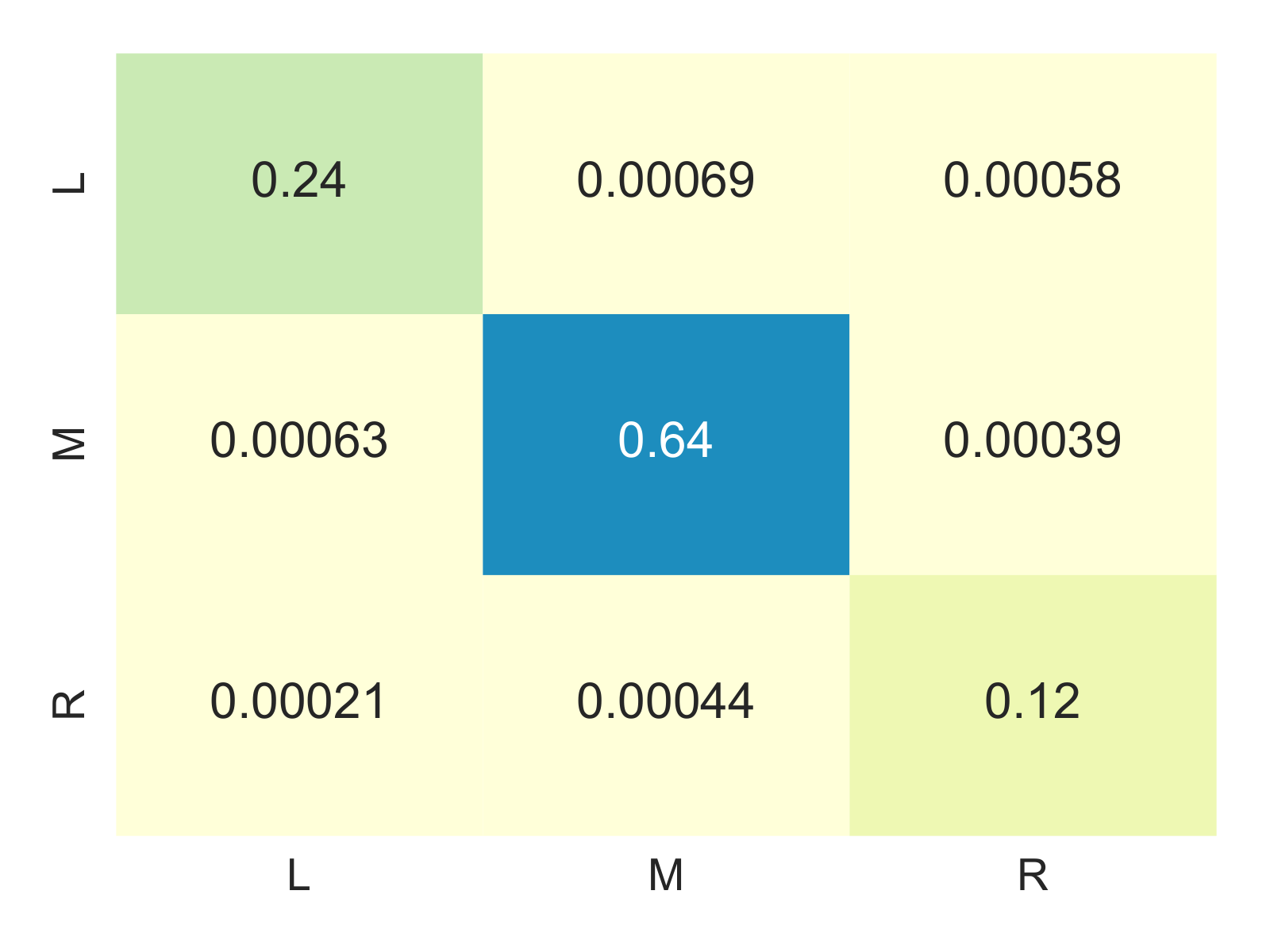}\label{fig:g5-states-opt-comp-action} }}%
    \caption{The empirical joint-action distributions in the last 10\% of episodes when learning with hierarchical self-interested action communication.}%
    \label{fig:opt-comp-action-states}%
\end{figure}

We also show the empirical state distribution for the last 10\% of episodes in Figure \ref{fig:opt-comp-action-states}. An important difference with results from Section \ref{sec:comp-com-exp} is that agents now converge on (L, R) and (R, L) again in the games without NE. This allows us to conclude that the introduction of hierarchical communication enables agents to more accurately coordinate their joint policies.

\begin{figure}[h!tb]%
    \centering
    \subfloat[Game 1]{{\includegraphics[width=.28\linewidth]{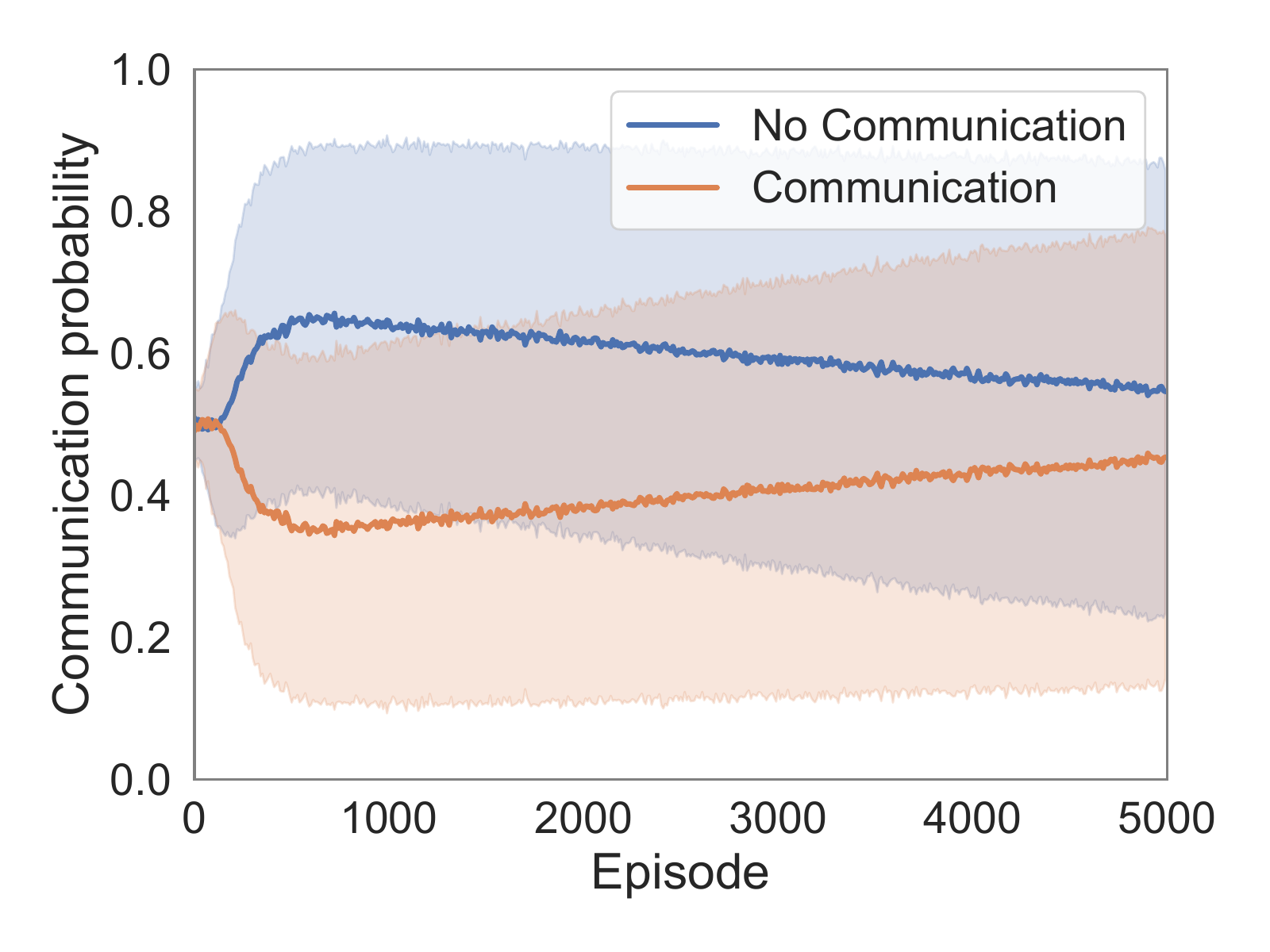}\label{fig:g1-A1-com-opt-comp-action} }}%
    \quad
    \subfloat[Game 2]{{\includegraphics[width=.28\linewidth]{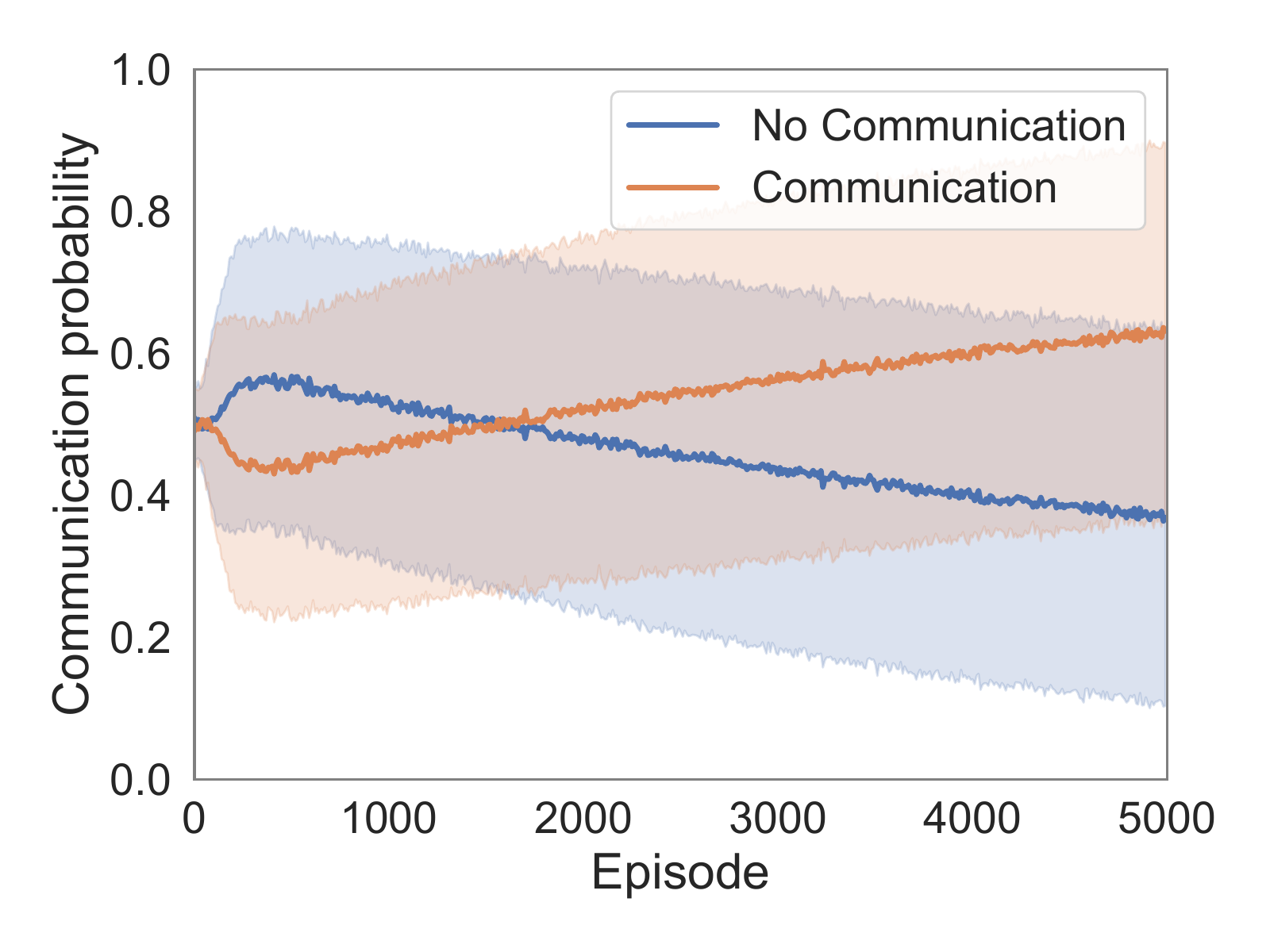}\label{fig:g2-A1-com-opt-comp-action} }}%
    \quad
    \subfloat[Game 3]{{\includegraphics[width=.28\linewidth]{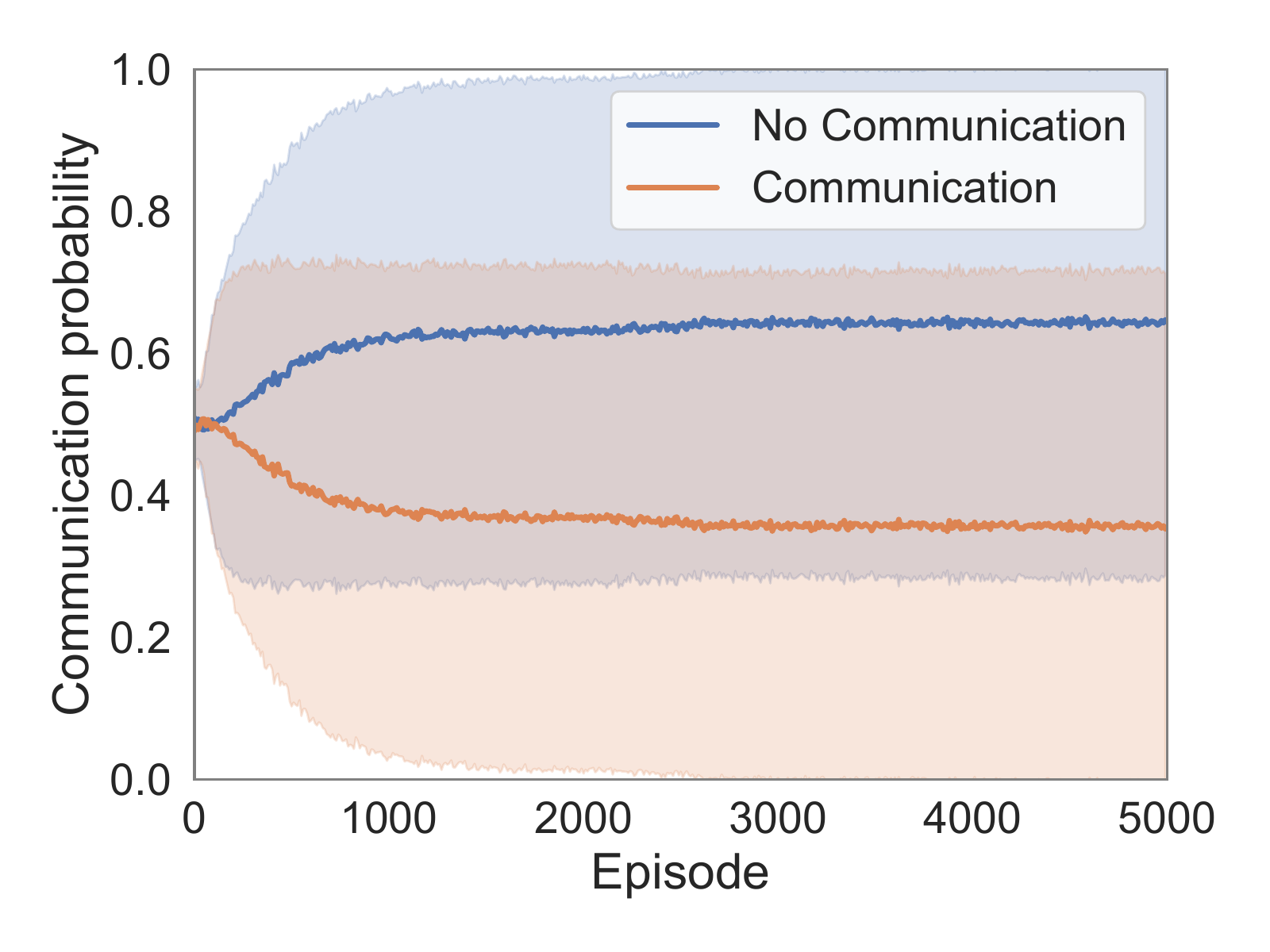}\label{fig:g3-A1-com-opt-comp-action} }}%
    \quad
    \subfloat[Game 4]{{\includegraphics[width=.28\linewidth]{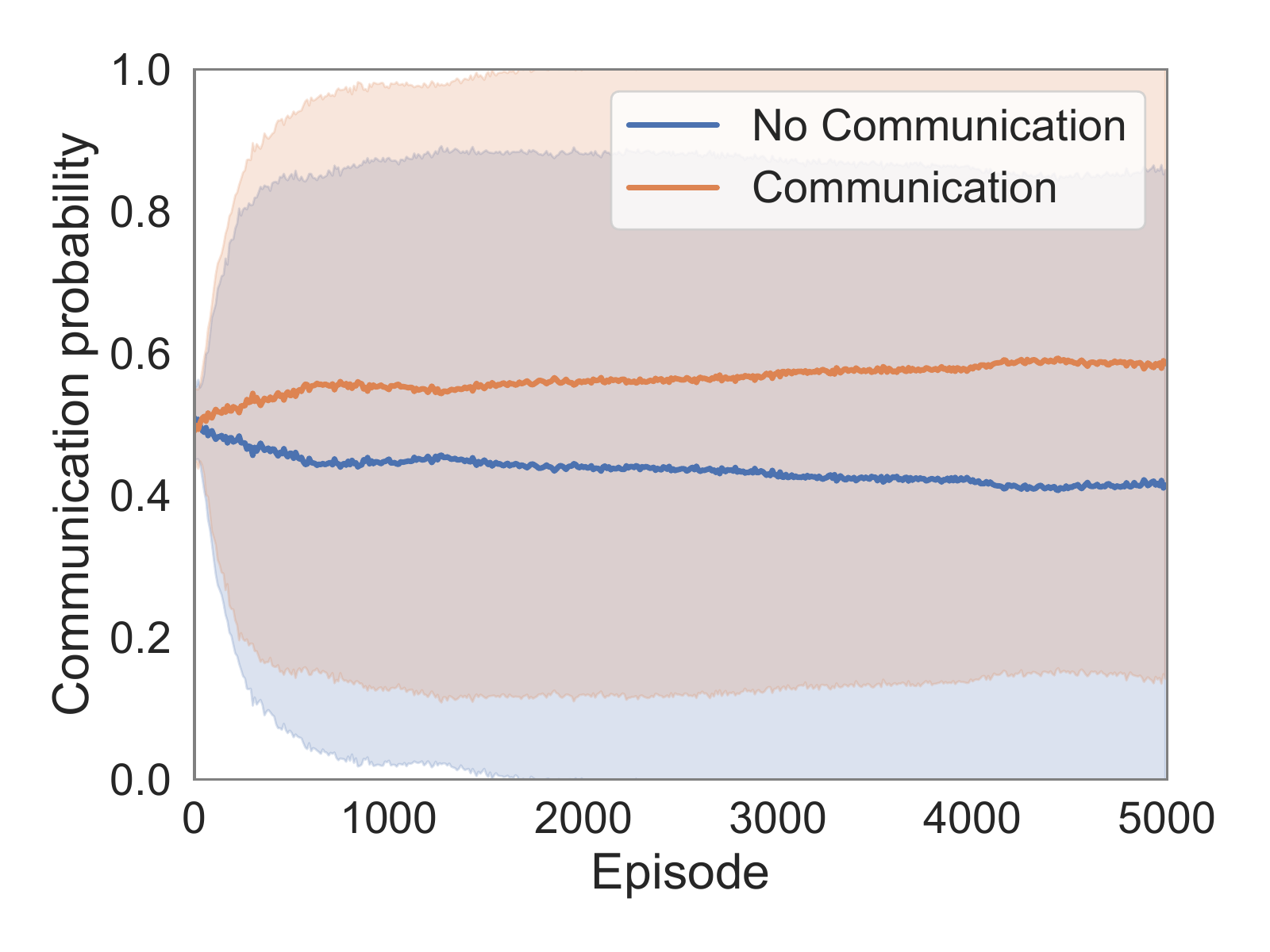}\label{fig:g4-A1-com-opt-comp-action} }}%
    \quad
    \subfloat[Game 5]{{\includegraphics[width=.28\linewidth]{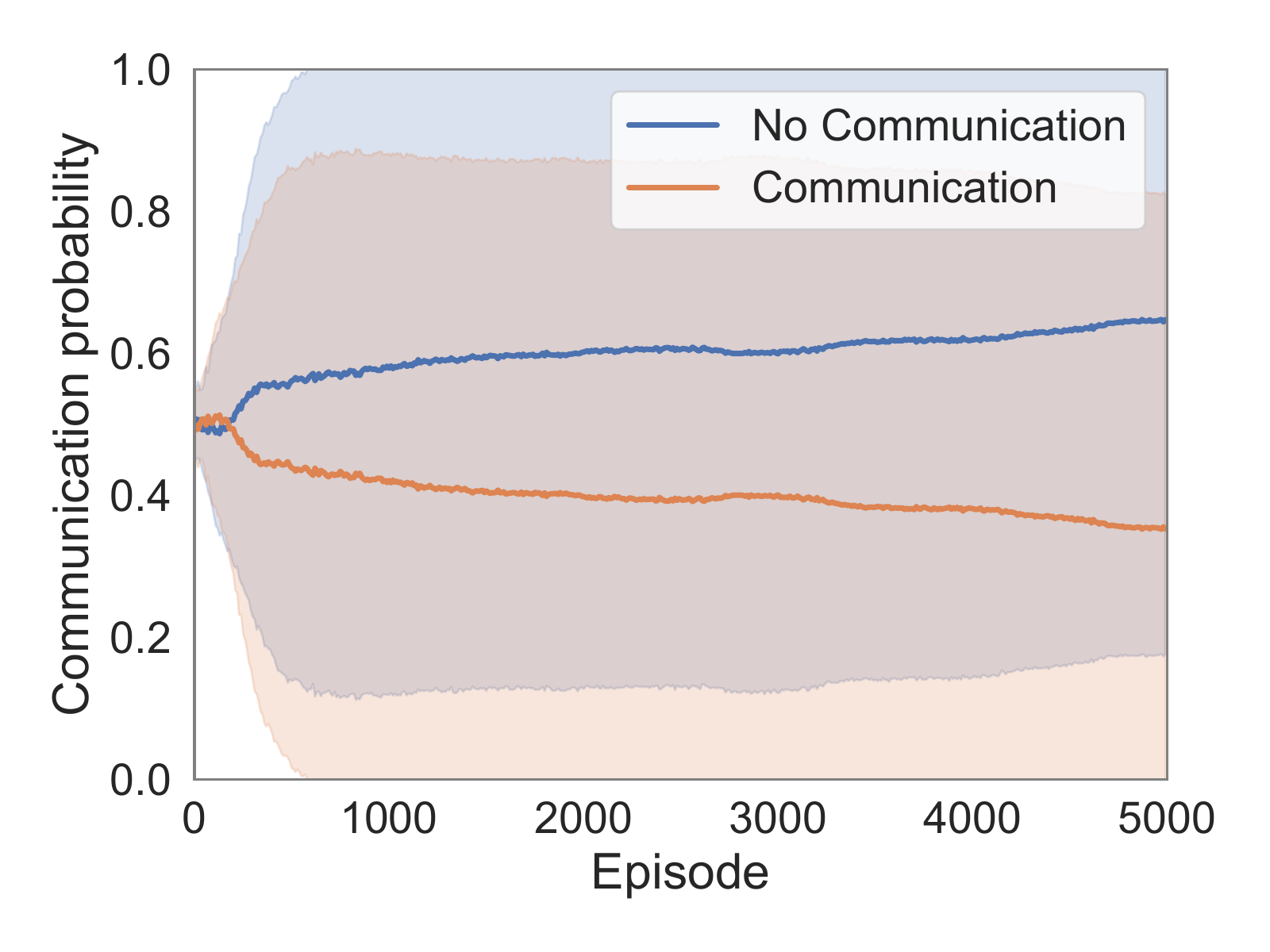}\label{fig:g5-A1-com-opt-comp-action} }}%
    \caption{The communication probabilities for agent 1 when learning with hierarchical self-interested action communication.}%
    \label{fig:opt-comp-action-A1-com}%
\end{figure}

\begin{figure}[h!tb]%
    \centering
    \subfloat[Game 1]{{\includegraphics[width=.28\linewidth]{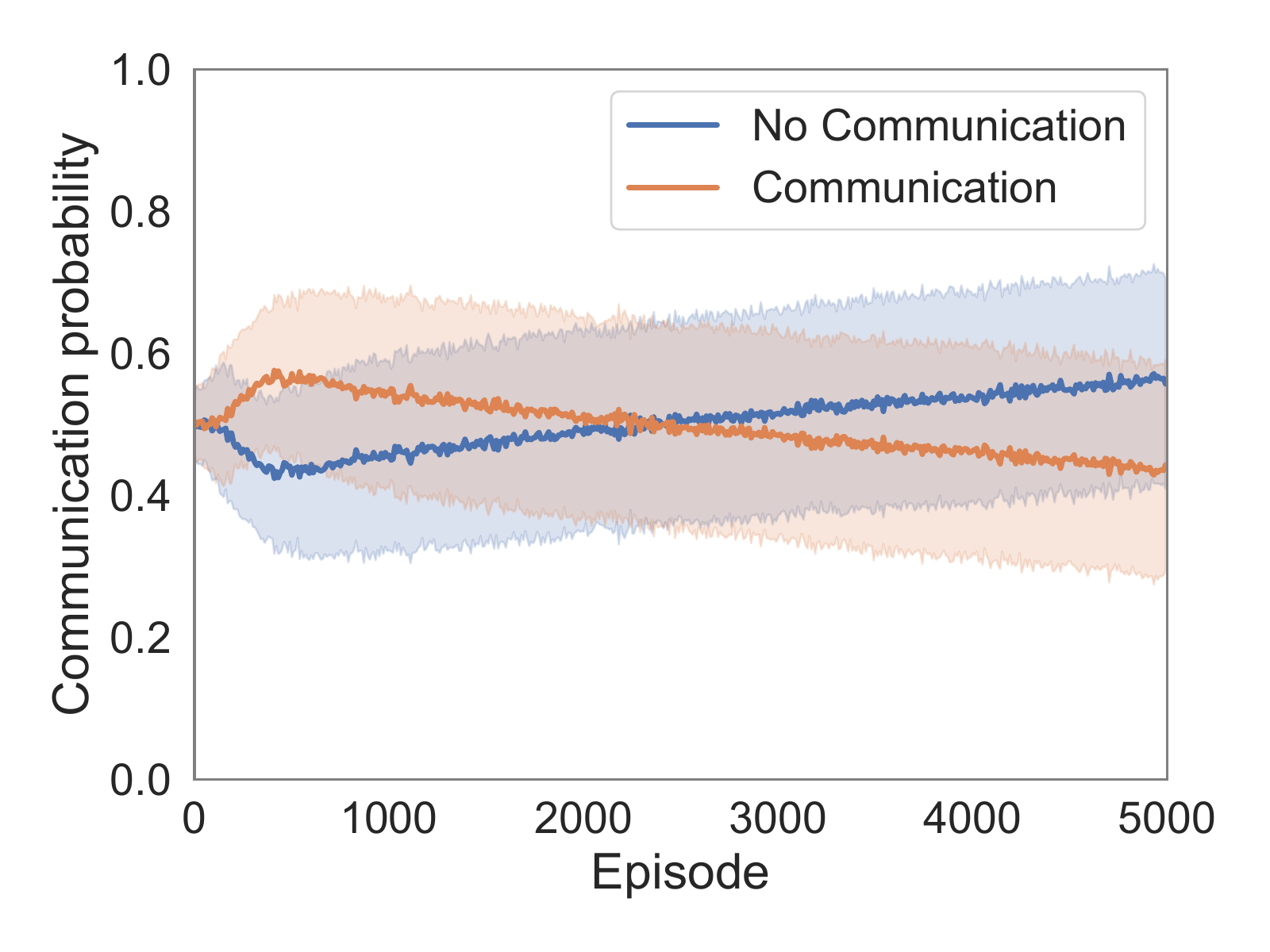}\label{fig:g1-A2-com-opt-comp-action} }}%
    \quad
    \subfloat[Game 2]{{\includegraphics[width=.28\linewidth]{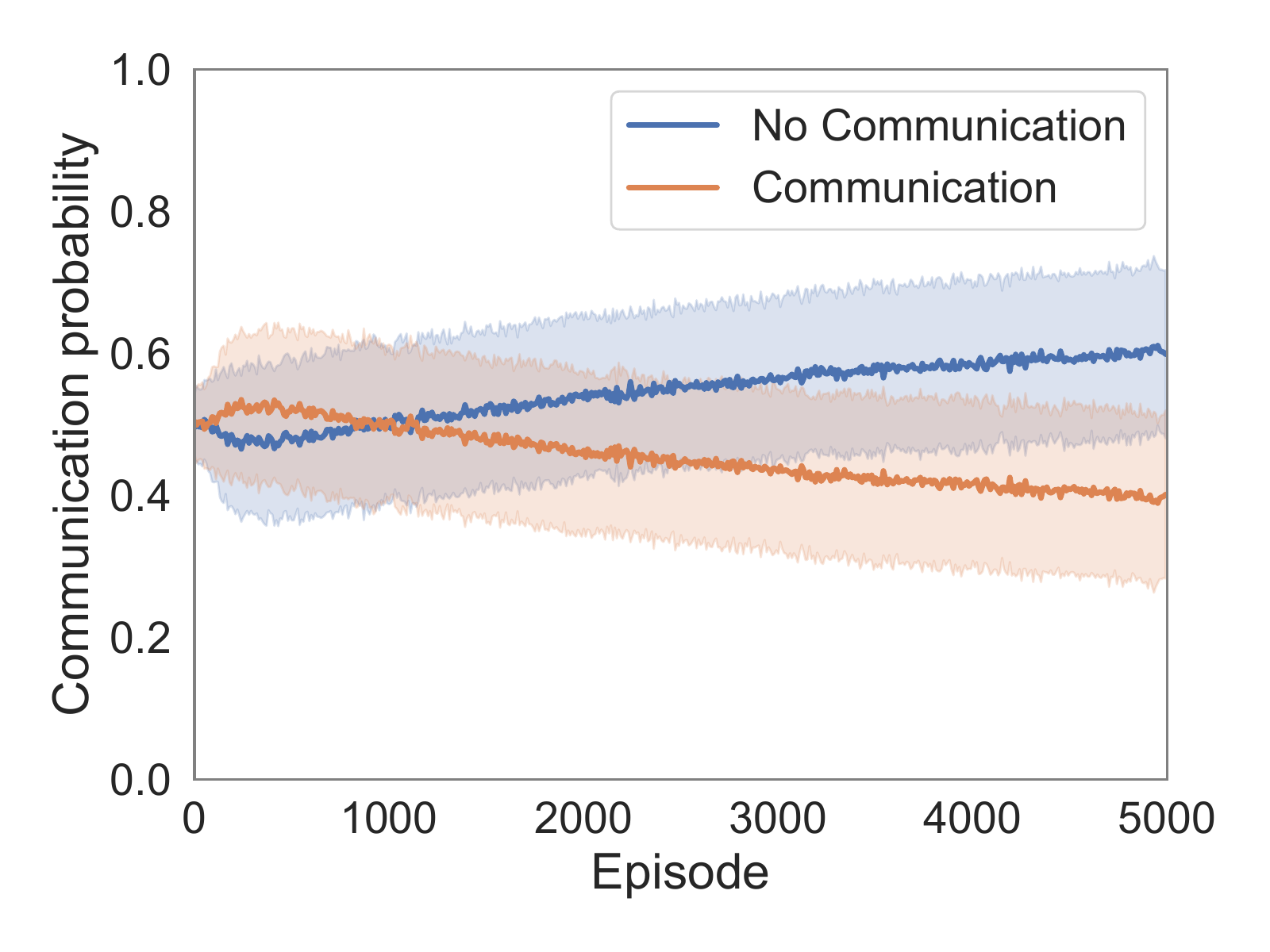}\label{fig:g2-A2-com-opt-comp-action} }}%
    \quad
    \subfloat[Game 3]{{\includegraphics[width=.28\linewidth]{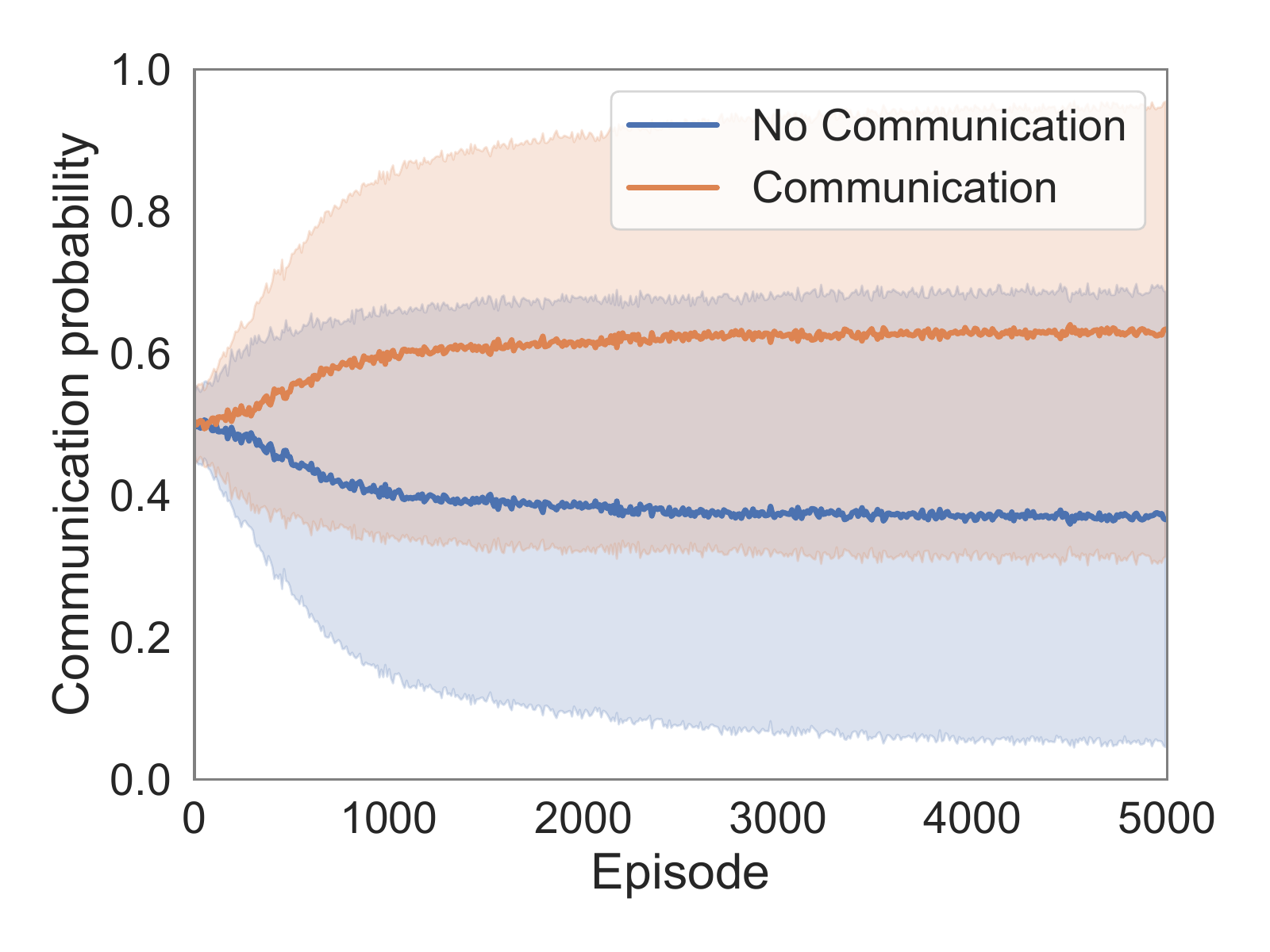}\label{fig:g3-A2-com-opt-comp-action} }}%
    \quad
    \subfloat[Game 4]{{\includegraphics[width=.28\linewidth]{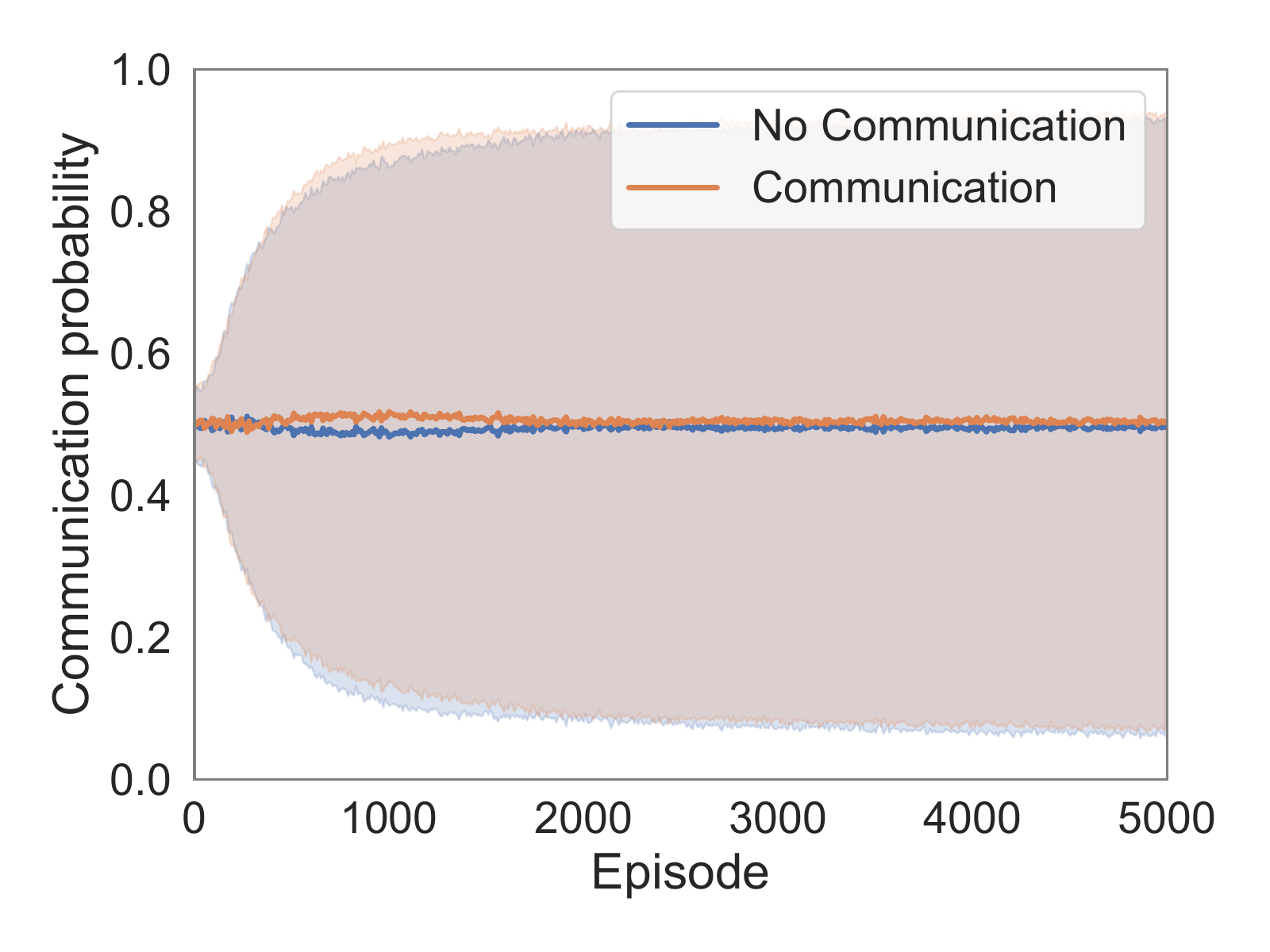}\label{fig:g4-A2-com-opt-comp-action} }}%
    \quad
    \subfloat[Game 5]{{\includegraphics[width=.28\linewidth]{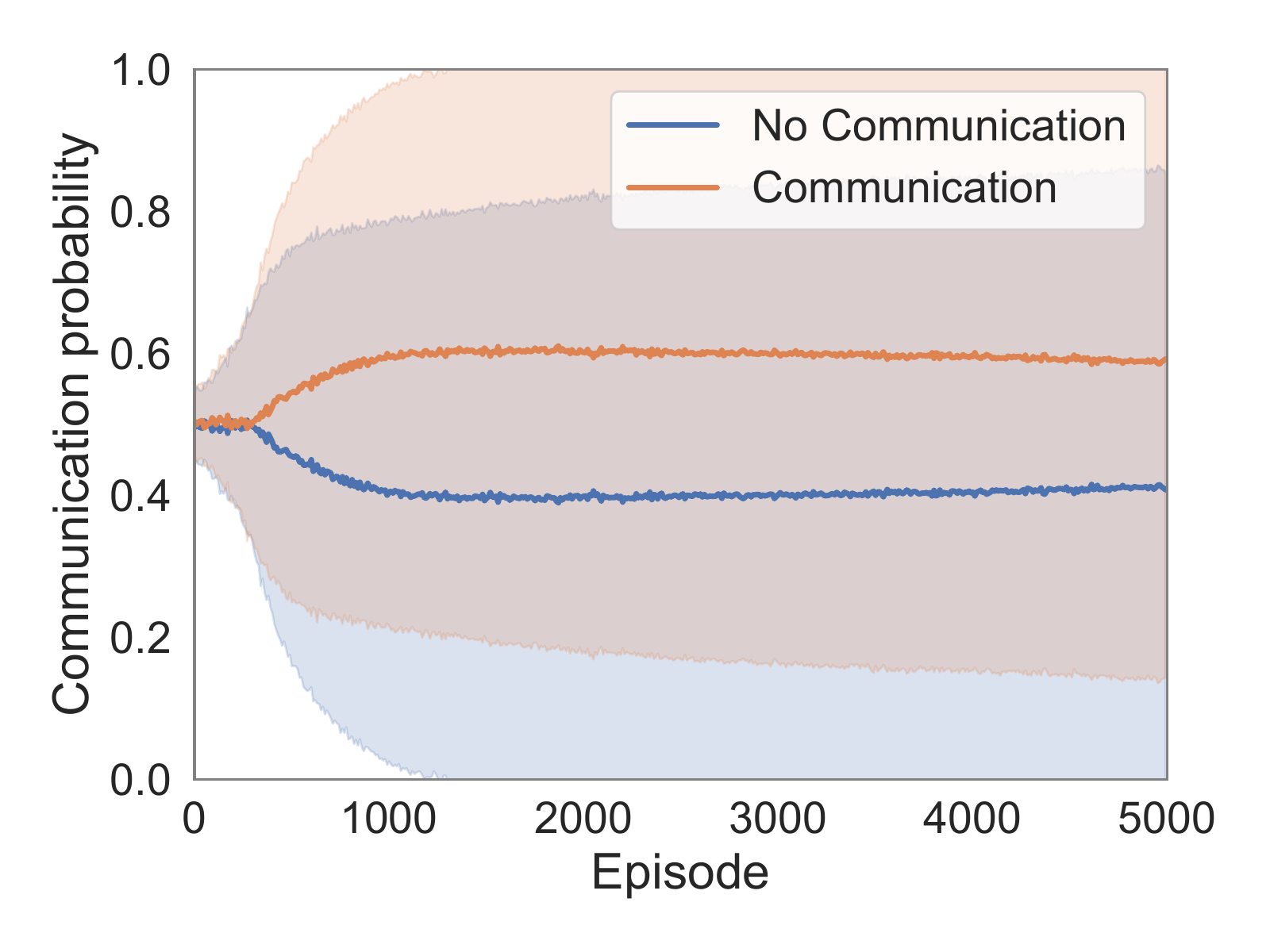}\label{fig:g5-A2-com-opt-comp-action} }}%
    \caption{The communication probabilities for agent 2 when learning with hierarchical self-interested action communication.}%
    \label{fig:opt-comp-action-A2-com}%
\end{figure}

This result is also reflected in Figures \ref{fig:opt-comp-action-A1-com} and \ref{fig:opt-comp-action-A2-com}. In the games without NE, agents consistently employ communication, shown by the smaller standard deviation around the communication probabilities. In addition, there appears an upwards trend for player 1 in their communication probabilities and a downwards trend for player 2. When executing for more episodes, we observe that this trend continues and converges at around 80\% communication for player 1 and 20\% communication for player 2. The most likely cause is that in earlier episodes player 1 benefits from a suboptimal best-response by player 2, while player 2 already reaches their globally optimal payoff of (2, 2) without communication. As such, communication only risks a suboptimal payoff for player 2. The top-level communication policy converges when all lower-level policies have converged to an expected payoff (2, 2) as gradients become (approximately) zero.

In games with NE, we see largely the same result as in the experiments with hierarchical cooperative action communication. Namely, communication can enable players to coordinate a CNE or players learn to either always communicate or never communicate in any given trial. These outcomes lead us to believe that communication can be useful in both cooperative, as well as self-interested settings. This phenomenon was only recently remarked in single-objective multi-agent settings as well \cite{noukhovitch2021emergent}. 

Lastly, we note that similar to previous hierarchical experiments, there is increased standard deviation in the communication strategies when lowering the low-level learning rates. We attribute this again to insufficient exploration of the low-level protocols, leading the communication strategy to follow the protocol that performs best in earlier episodes.

\subsubsection{Hierarchical Cooperative Policy Communication}
\label{sec:hierarchical-policy-com-exp}
\begin{figure}[h!tb]%
    \centering
    \subfloat[Game 1]{{\includegraphics[width=.28\linewidth]{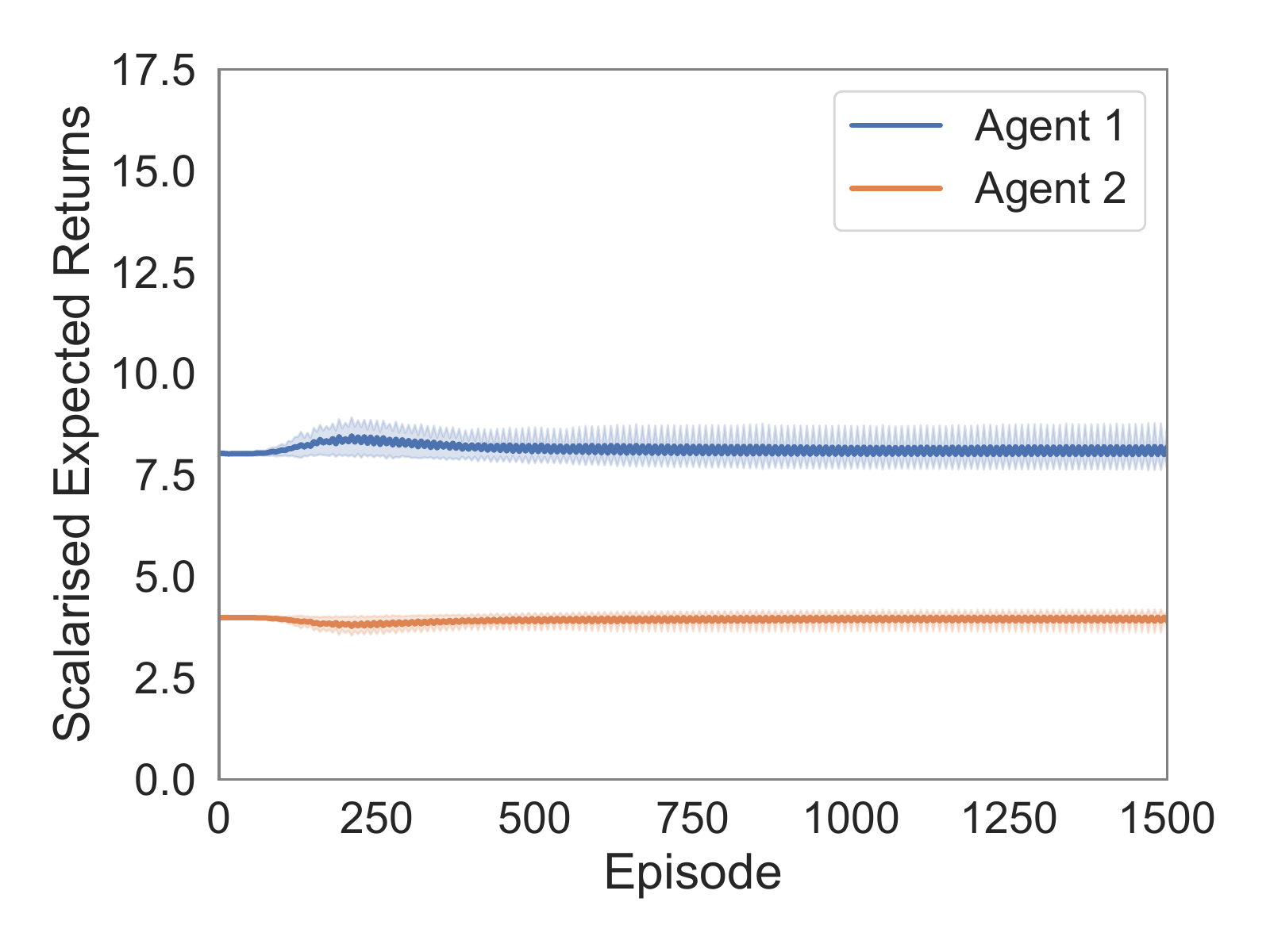}\label{fig:g1-ser-opt-coop-policy} }}%
    \quad
    \subfloat[Game 2]{{\includegraphics[width=.28\linewidth]{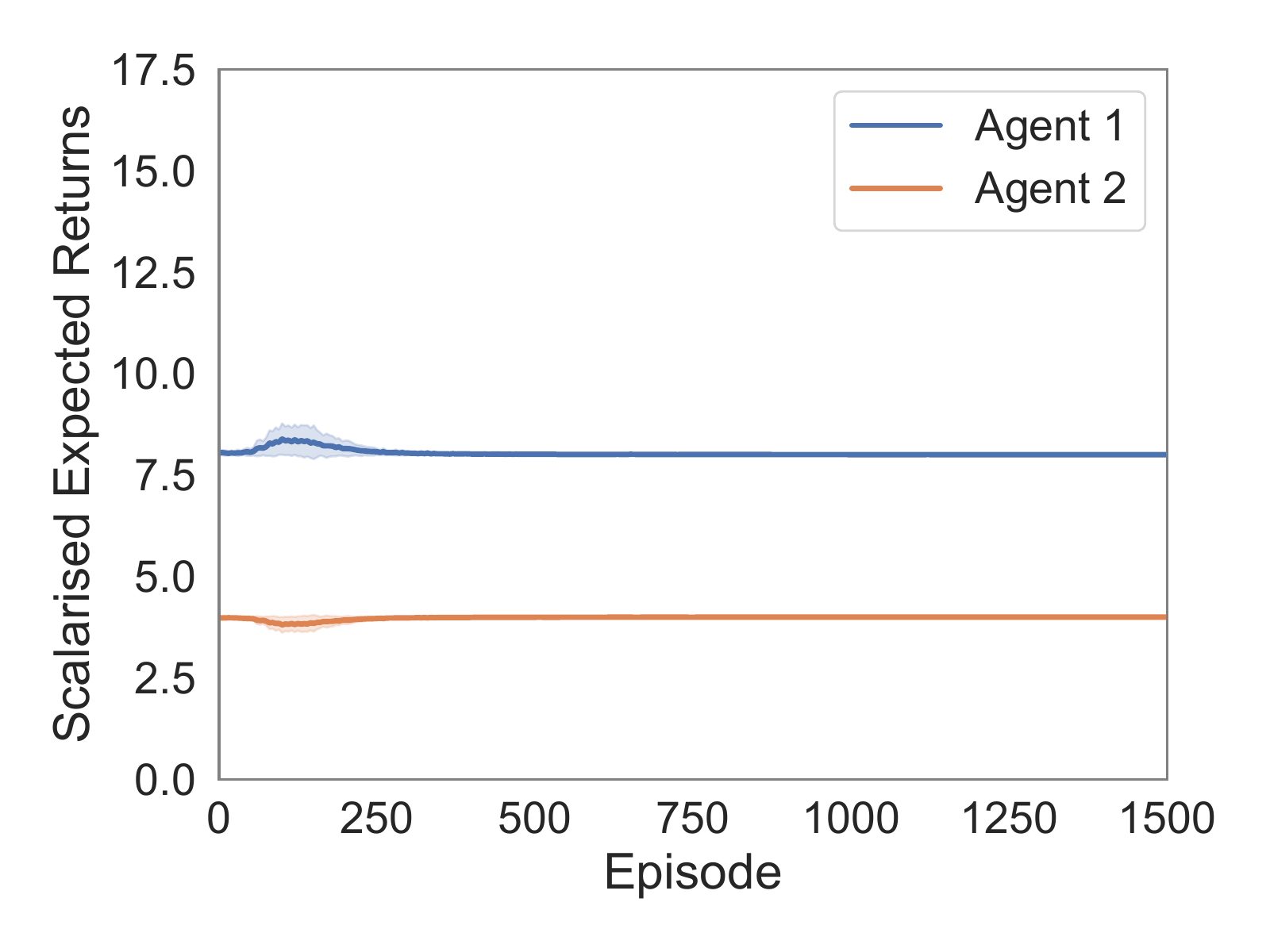}\label{fig:g2-ser-opt-coop-policy} }}%
    \quad
    \subfloat[Game 3]{{\includegraphics[width=.28\linewidth]{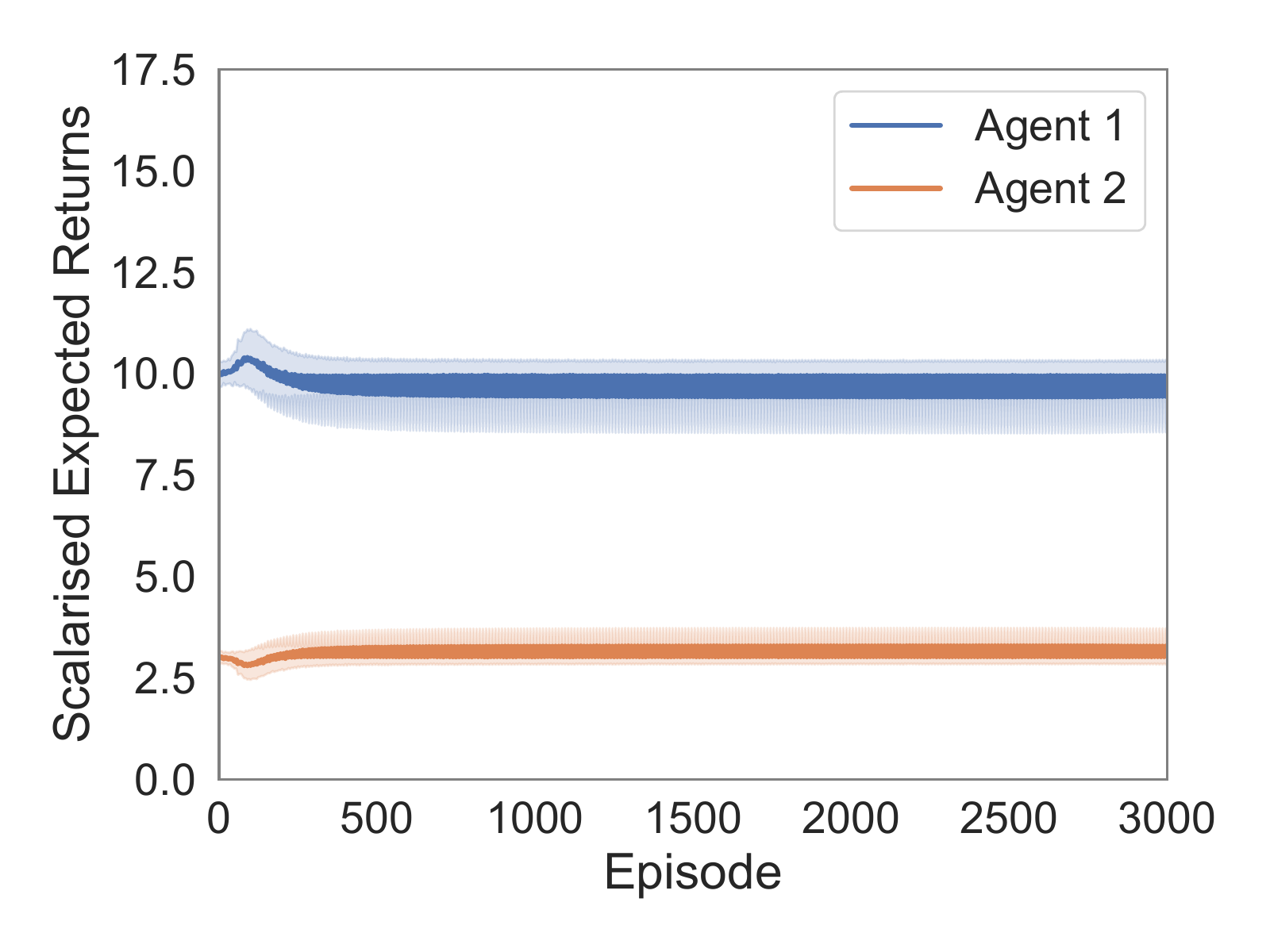}\label{fig:g3-ser-opt-coop-policy} }}%
    \quad
    \subfloat[Game 4]{{\includegraphics[width=.28\linewidth]{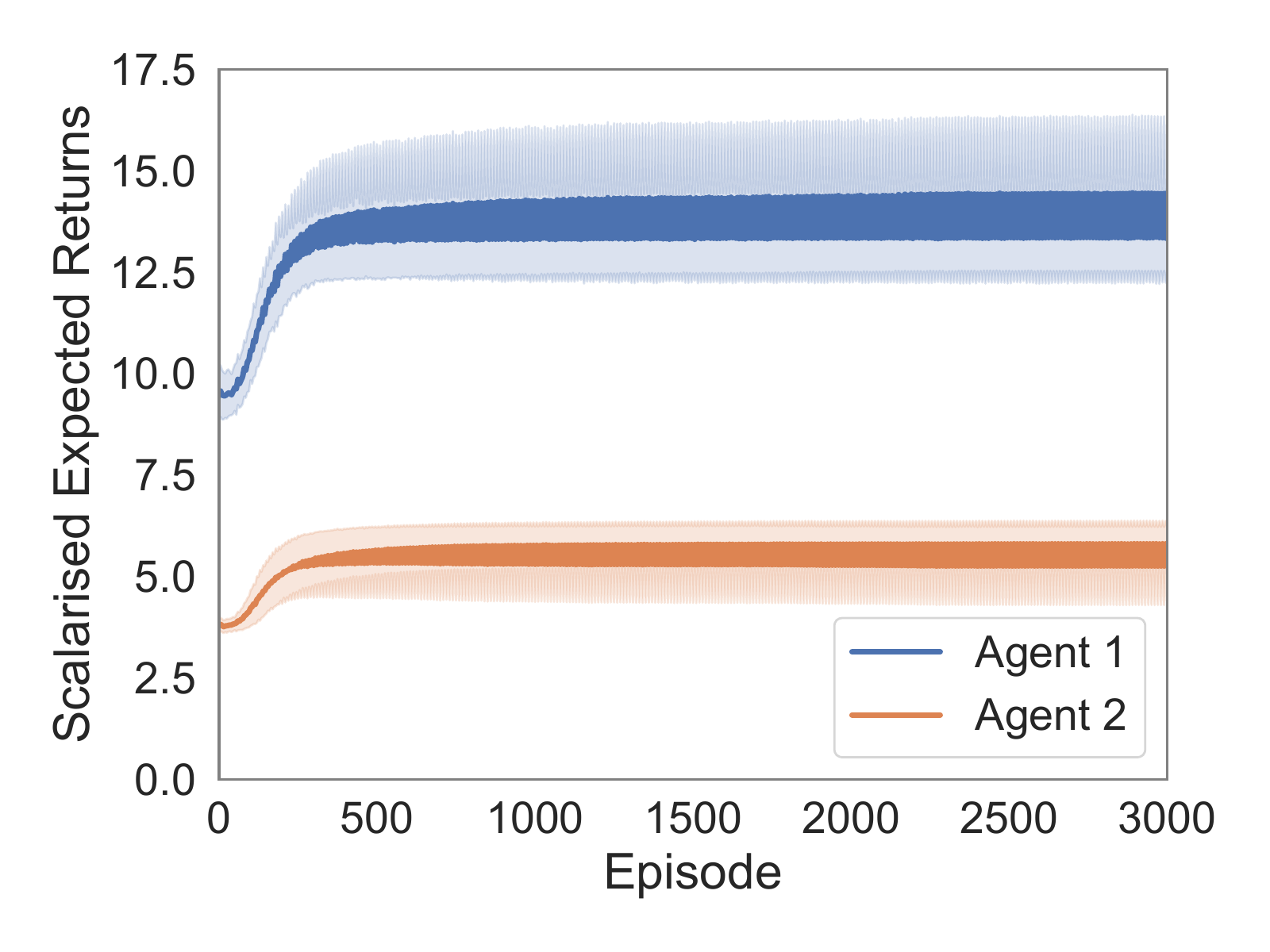}\label{fig:g4-ser-opt-coop-policy} }}%
    \quad
    \subfloat[Game 5]{{\includegraphics[width=.28\linewidth]{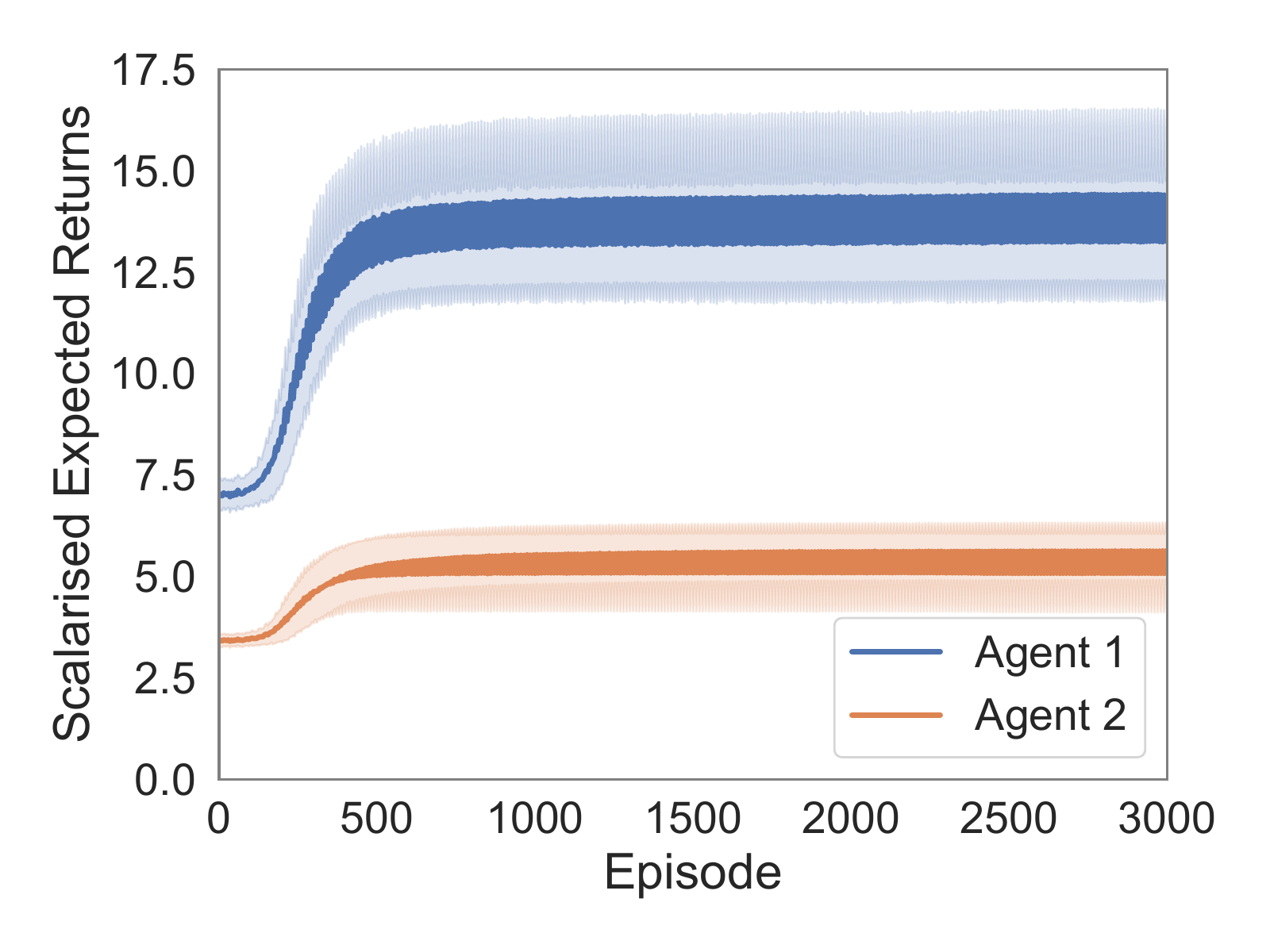}\label{fig:g5-ser-opt-coop-policy} }}%
    \caption{The SER for both agents when learning with hierarchical cooperative policy communication.}%
    \label{fig:opt-coop-policy-ser}%
\end{figure}

The last setting that we study places the agents in a cooperative setting with hierarchical policy communication. In Figure \ref{fig:opt-coop-policy-ser} we show the SER over time for both agents. The results are similar to the figures from hierarchical cooperative action communication setting (Sec.~\ref{sec:hierarchical-coop-com-exp}). We observe that enabling agents to learn non-stationary policies facilitates cyclic behaviour and cyclic equilibria. Recall further that using communication resulted in a more stable utility across the entire learning process in Section \ref{sec:policy-com-exp}, which is also observed here.

\begin{figure}[h!tb]%
    \centering
    \subfloat[Game 1]{{\includegraphics[width=.28\linewidth]{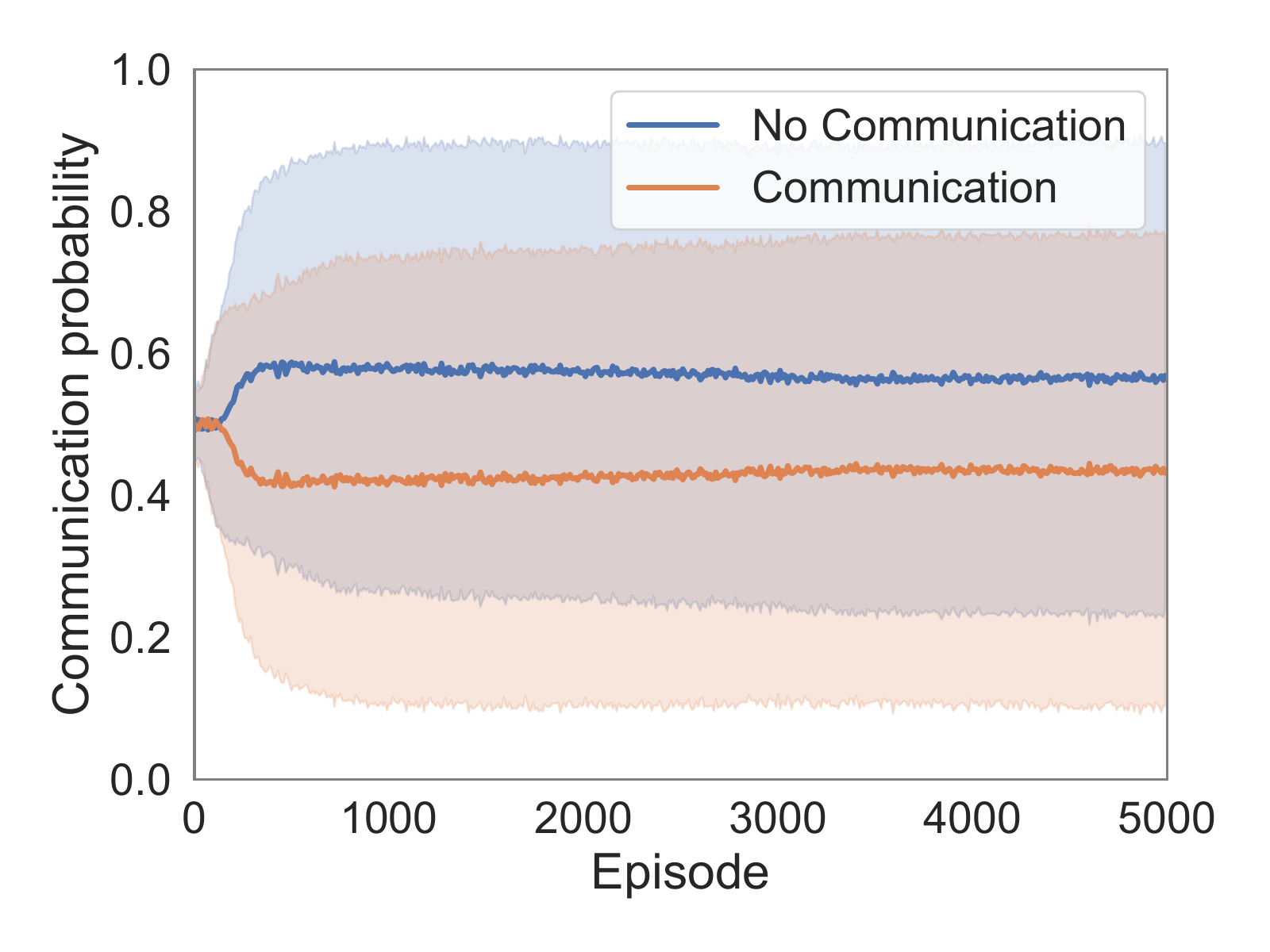}\label{fig:g1-A1-com-opt-coop-policy} }}%
    \quad
    \subfloat[Game 2]{{\includegraphics[width=.28\linewidth]{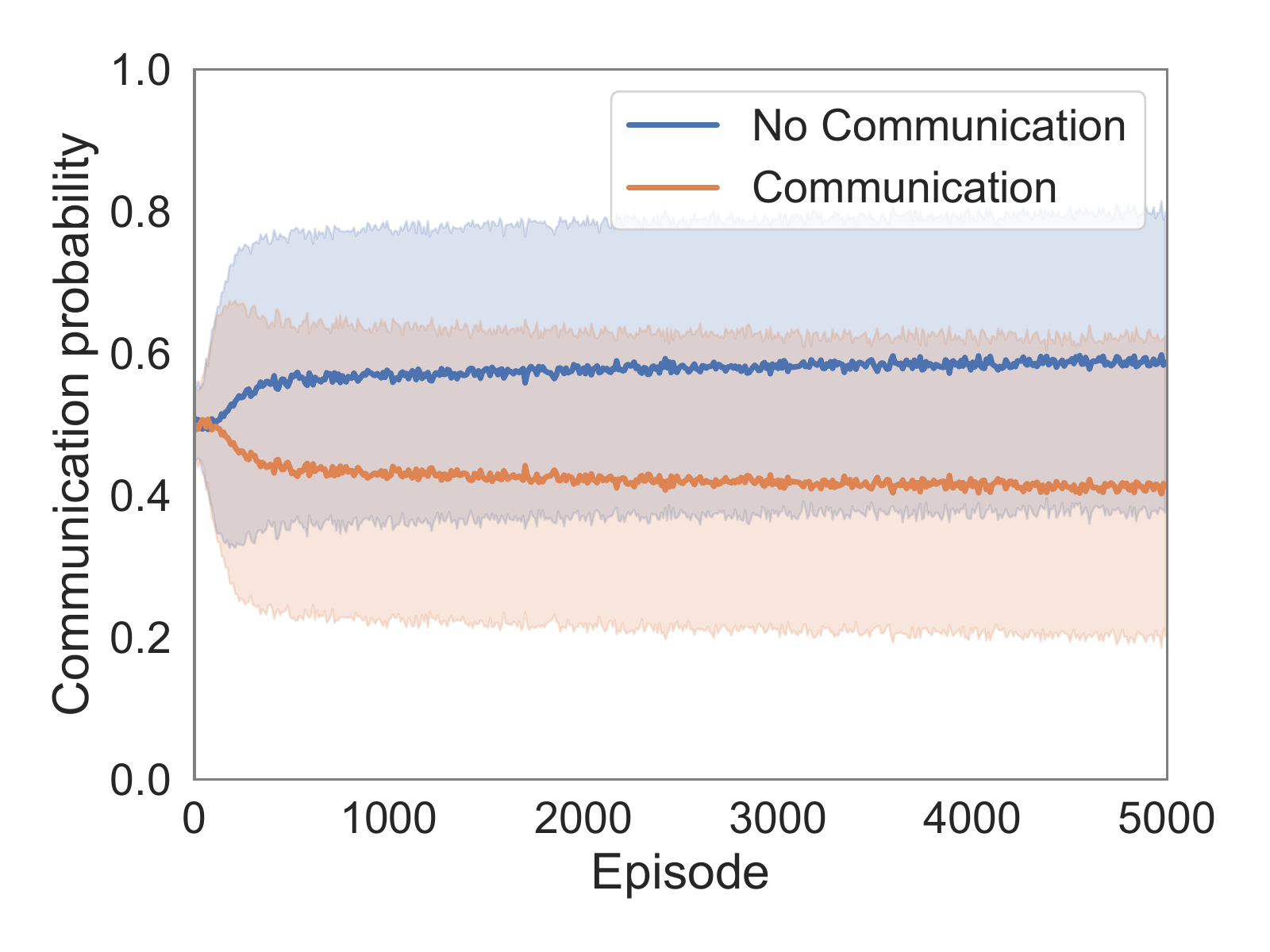}\label{fig:g2-A1-com-opt-coop-policy} }}%
    \quad
    \subfloat[Game 3]{{\includegraphics[width=.28\linewidth]{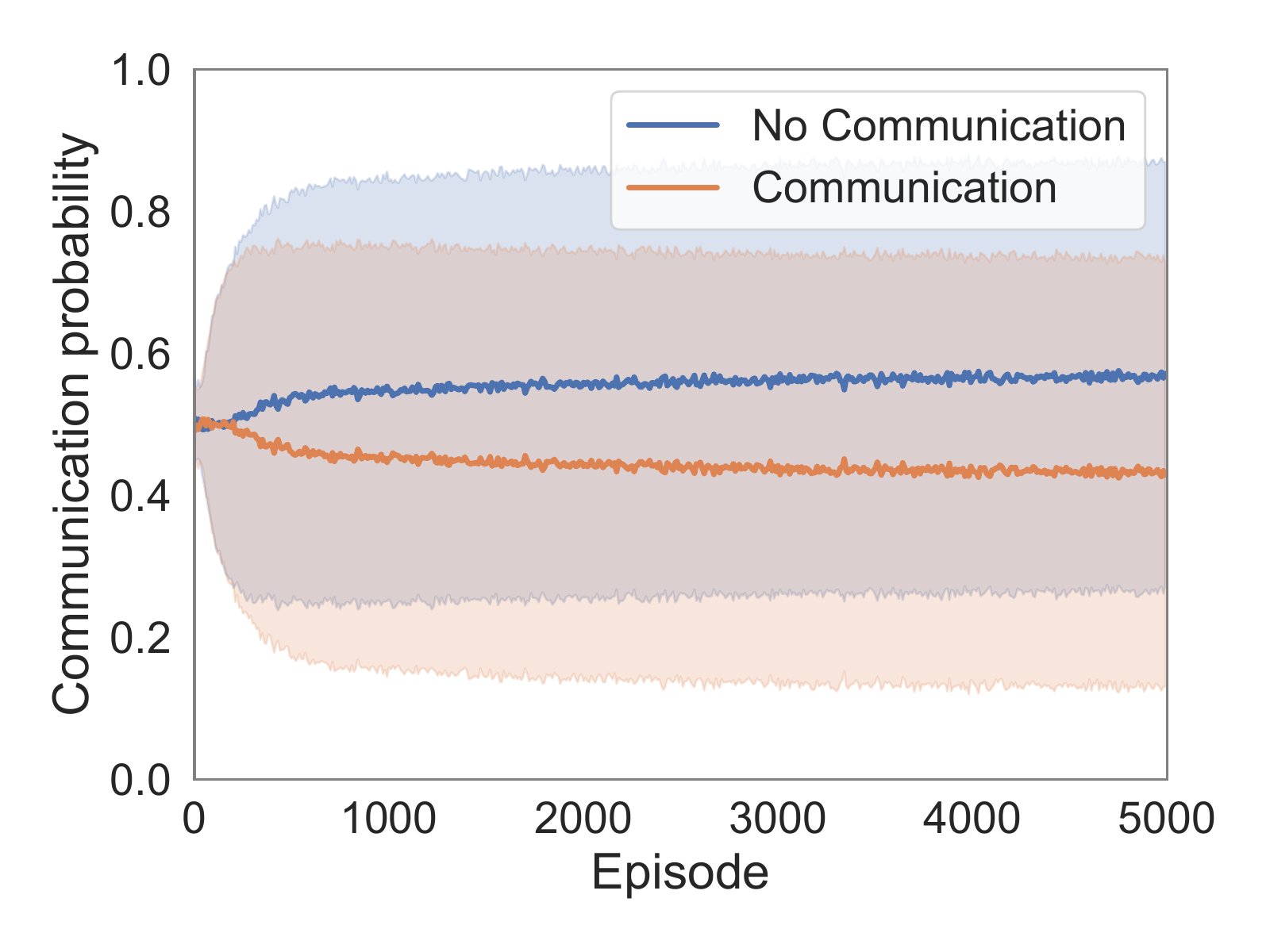}\label{fig:g3-A1-com-opt-coop-policy} }}%
    \quad
    \subfloat[Game 4]{{\includegraphics[width=.28\linewidth]{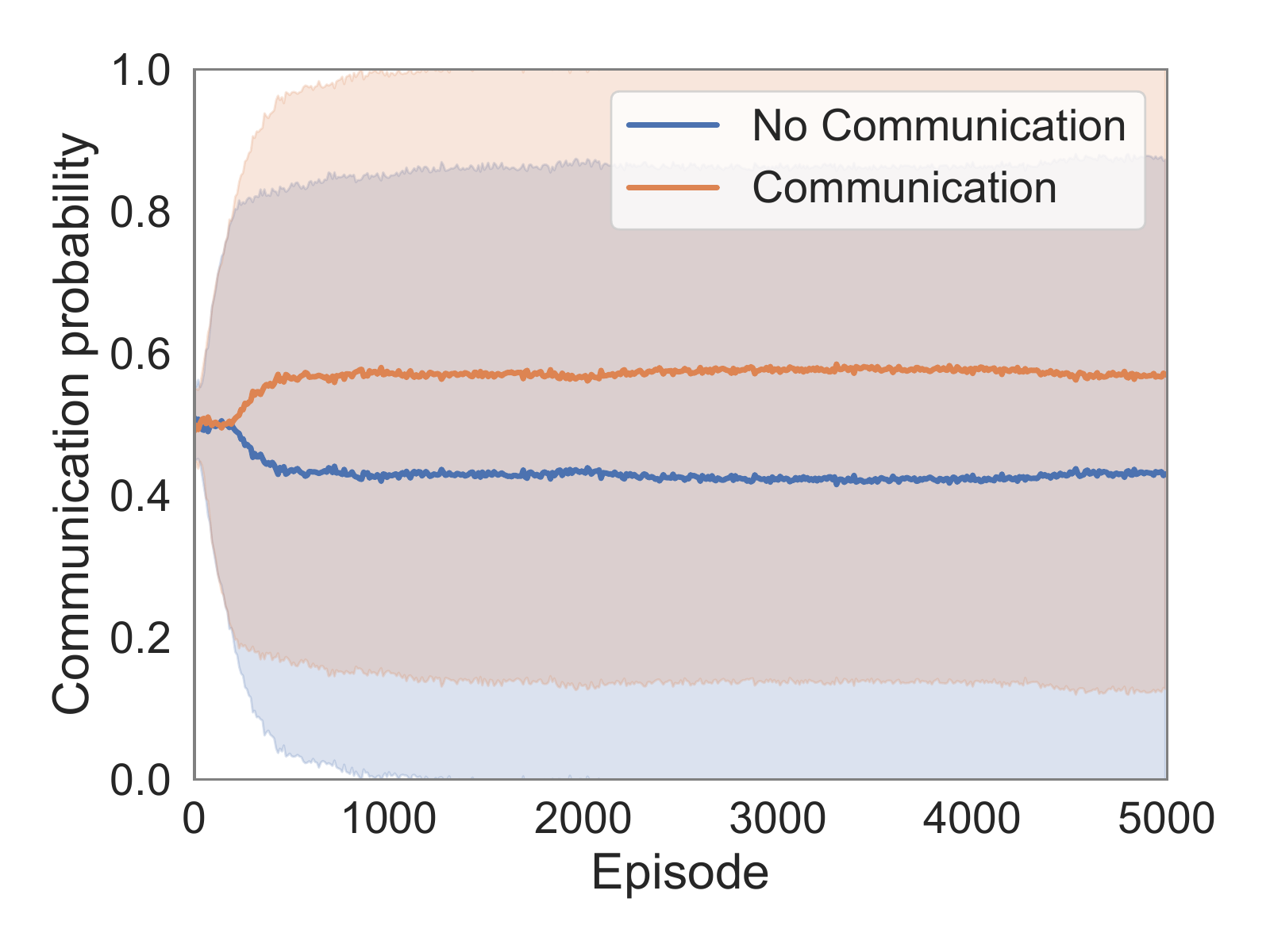}\label{fig:g4-A1-com-opt-coop-policy} }}%
    \quad
    \subfloat[Game 5]{{\includegraphics[width=.28\linewidth]{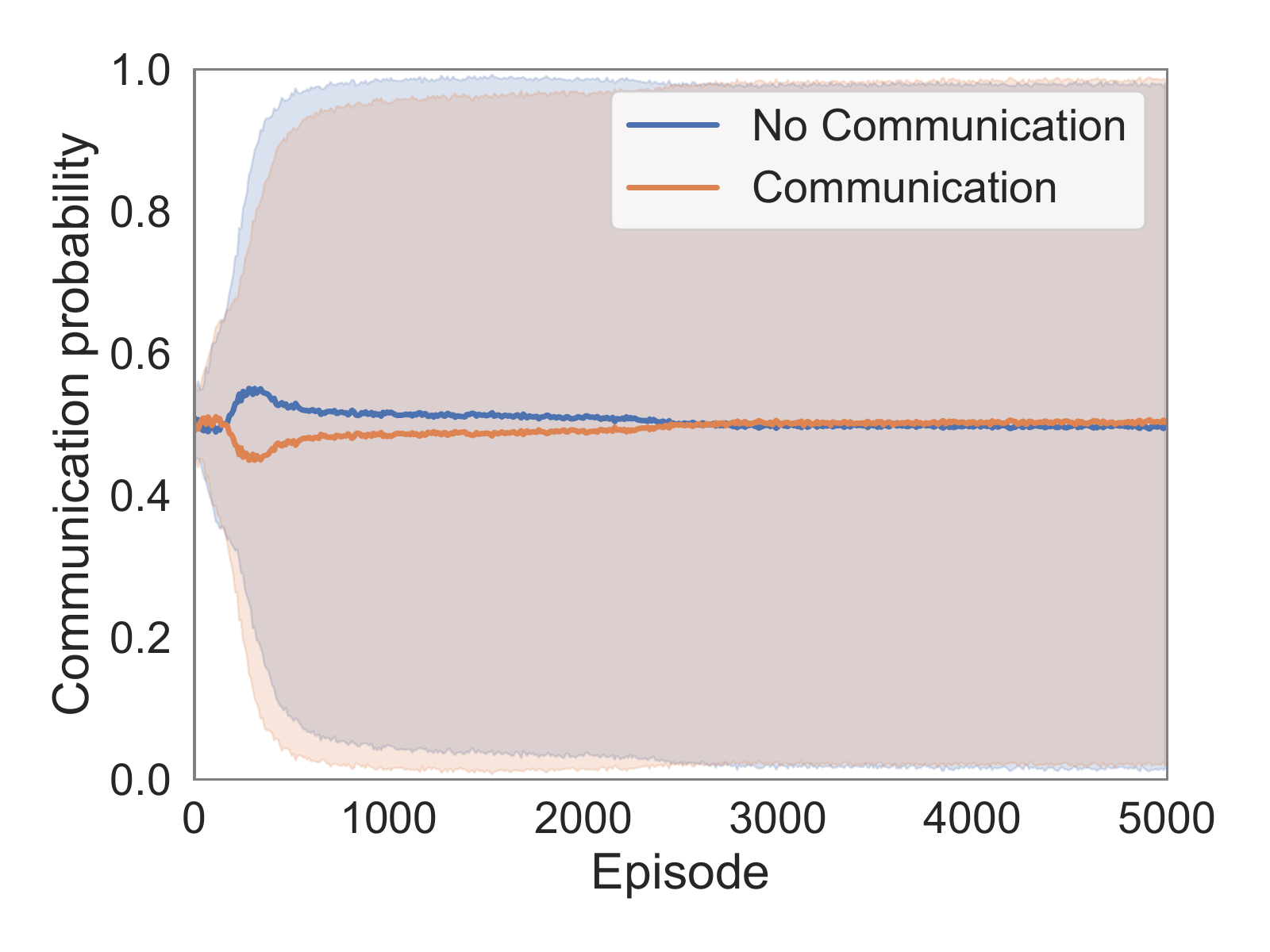}\label{fig:g5-A1-com-opt-coop-policy} }}%
    \caption{The communication probabilities for agent 1 when learning with hierarchical cooperative policy communication.}%
    \label{fig:opt-coop-policy-A1-com}%
\end{figure}

\begin{figure}[h!tb]%
    \centering
    \subfloat[Game 1]{{\includegraphics[width=.28\linewidth]{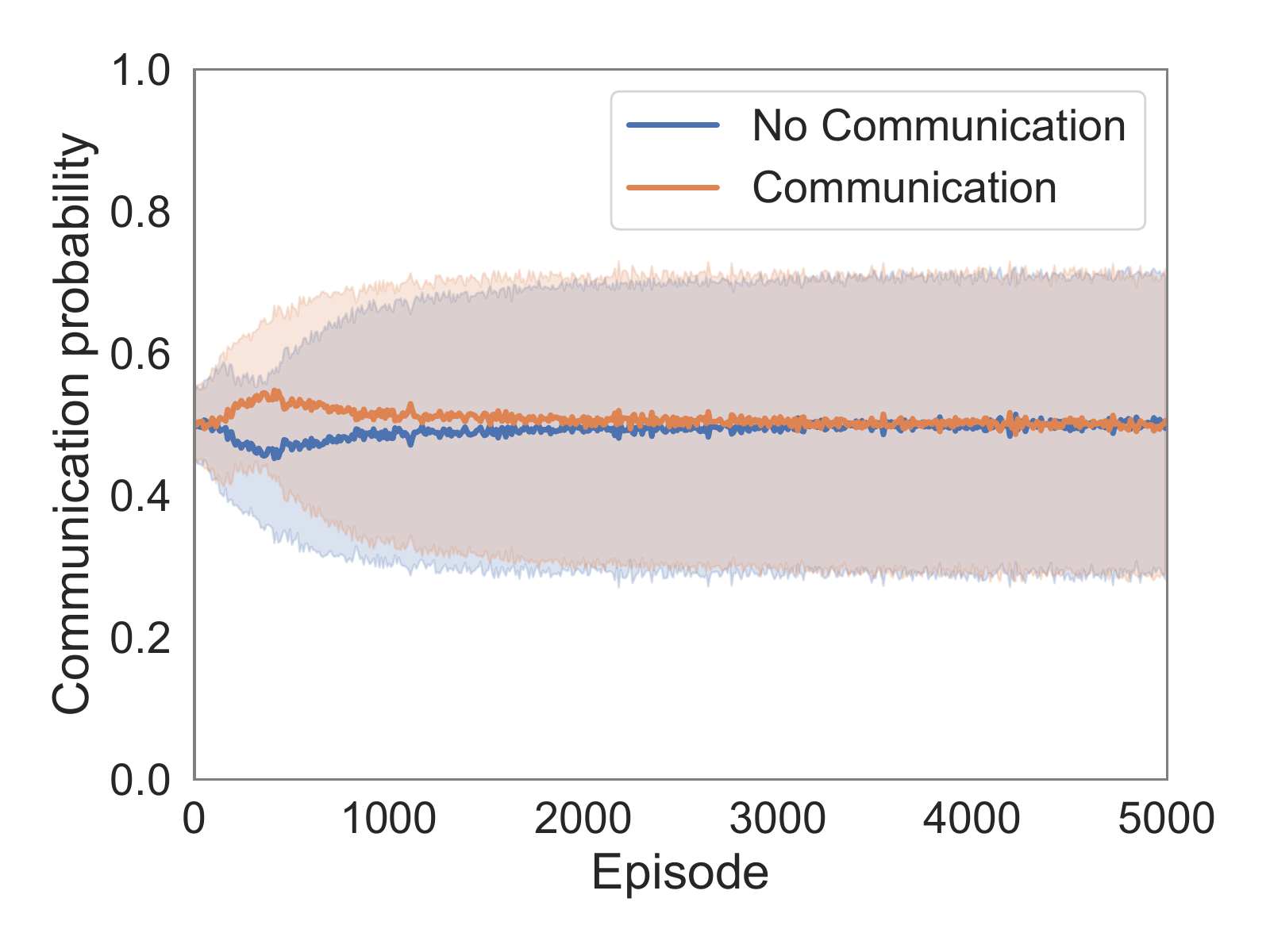}\label{fig:g1-A2-com-opt-coop-policy} }}%
    \quad
    \subfloat[Game 2]{{\includegraphics[width=.28\linewidth]{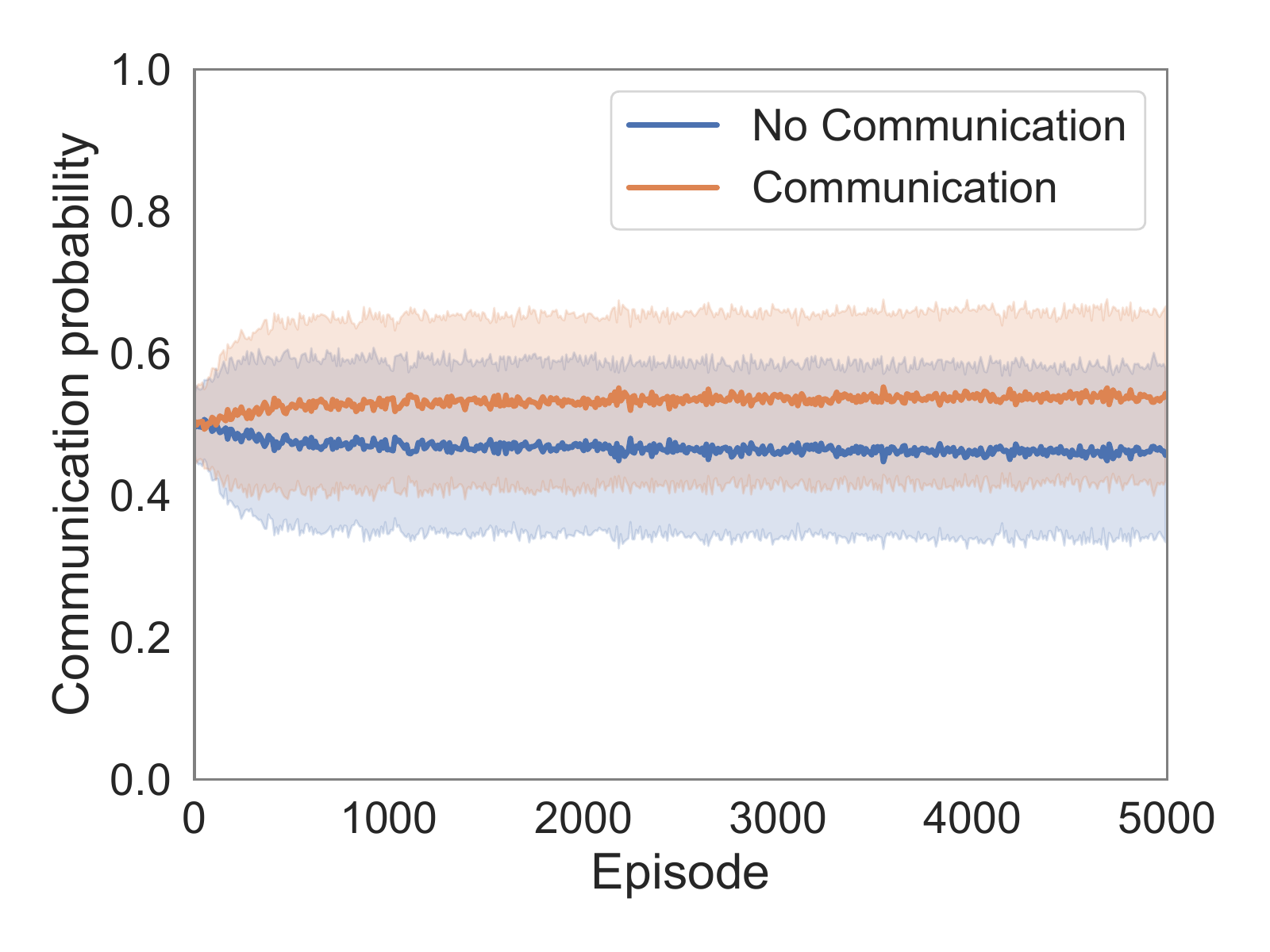}\label{fig:g2-A2-com-opt-coop-policy} }}%
    \quad
    \subfloat[Game 3]{{\includegraphics[width=.28\linewidth]{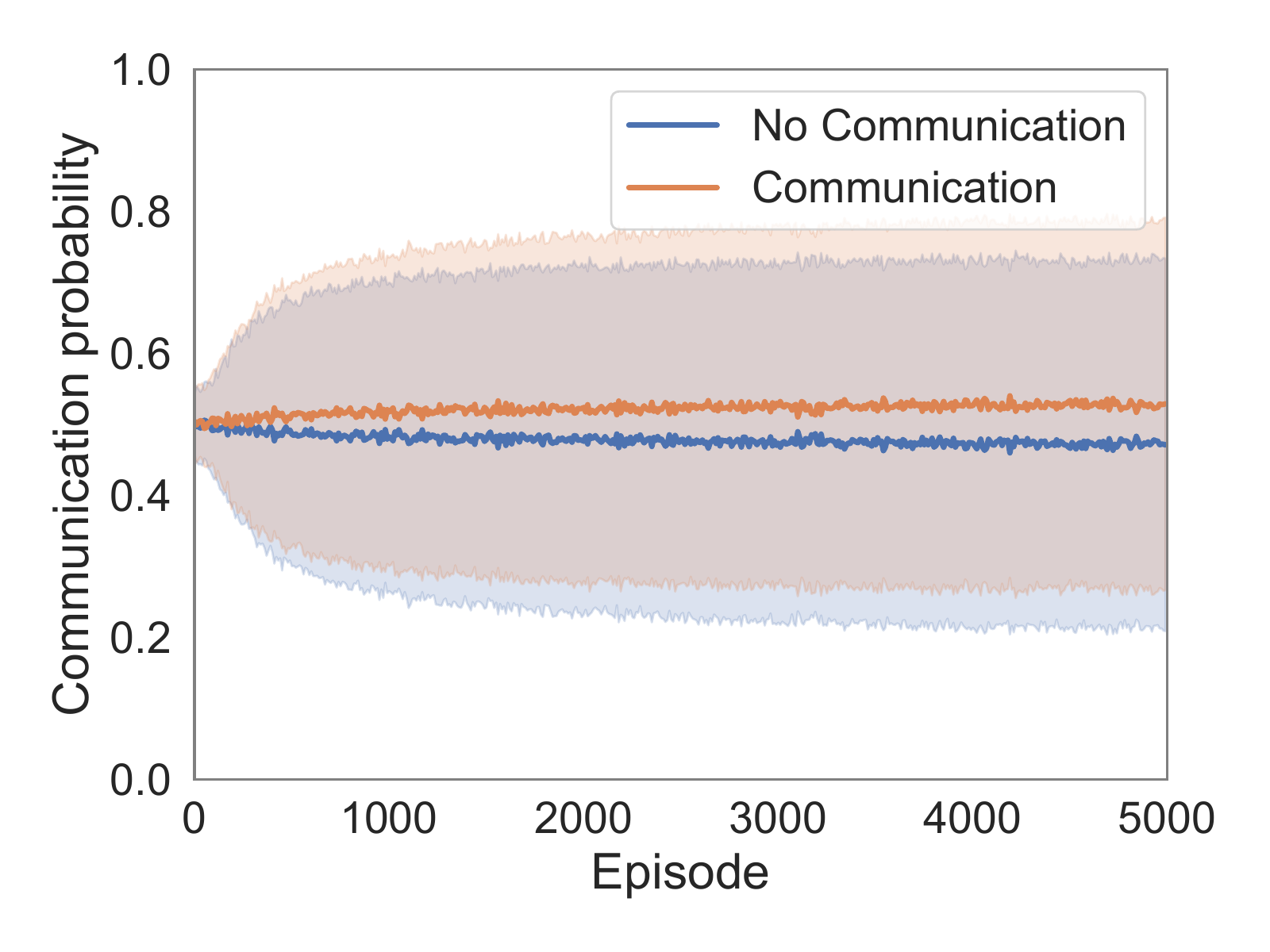}\label{fig:g3-A2-com-opt-coop-policy} }}%
    \quad
    \subfloat[Game 4]{{\includegraphics[width=.28\linewidth]{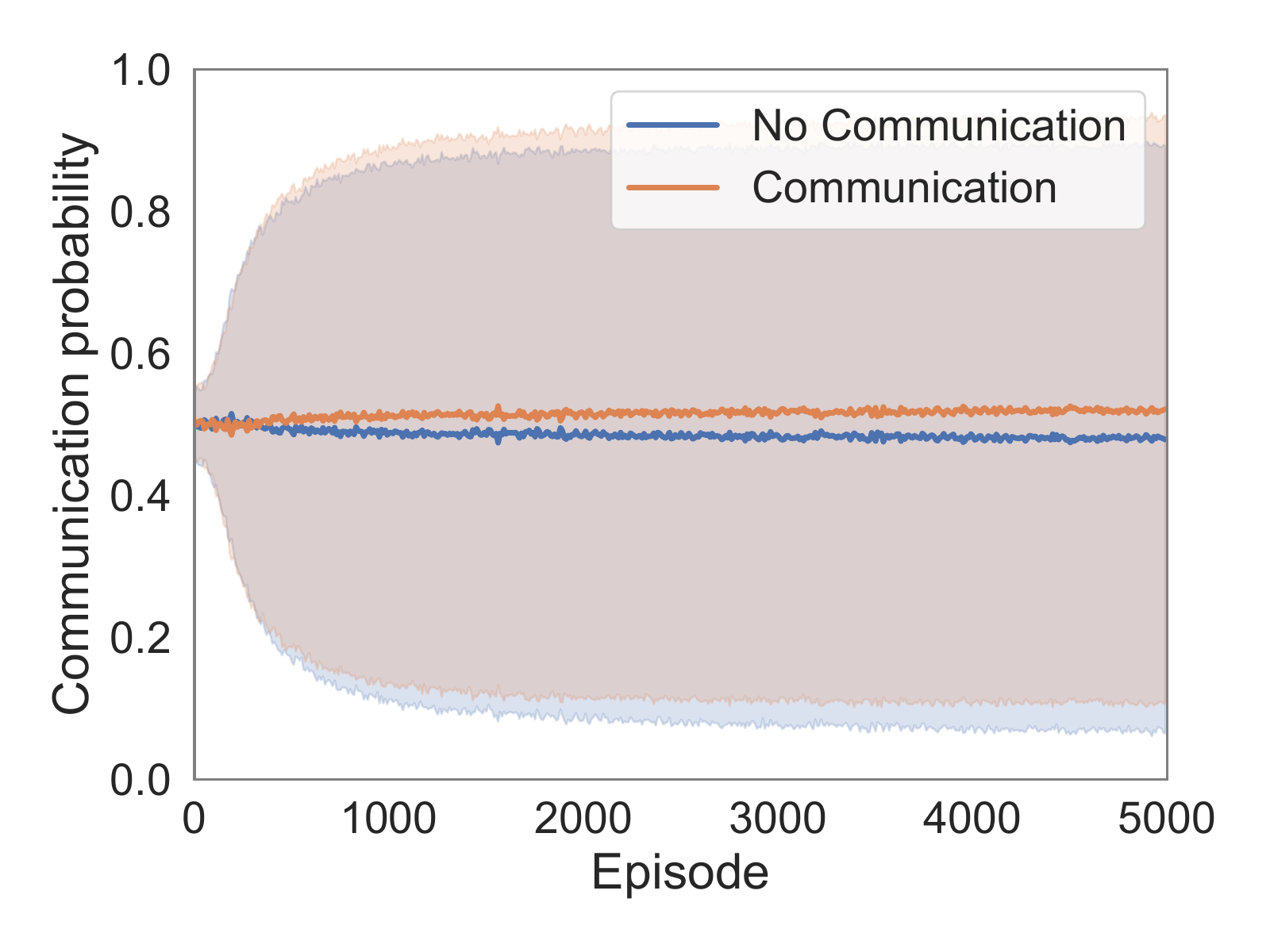}\label{fig:g4-A2-com-opt-coop-policy} }}%
    \quad
    \subfloat[Game 5]{{\includegraphics[width=.28\linewidth]{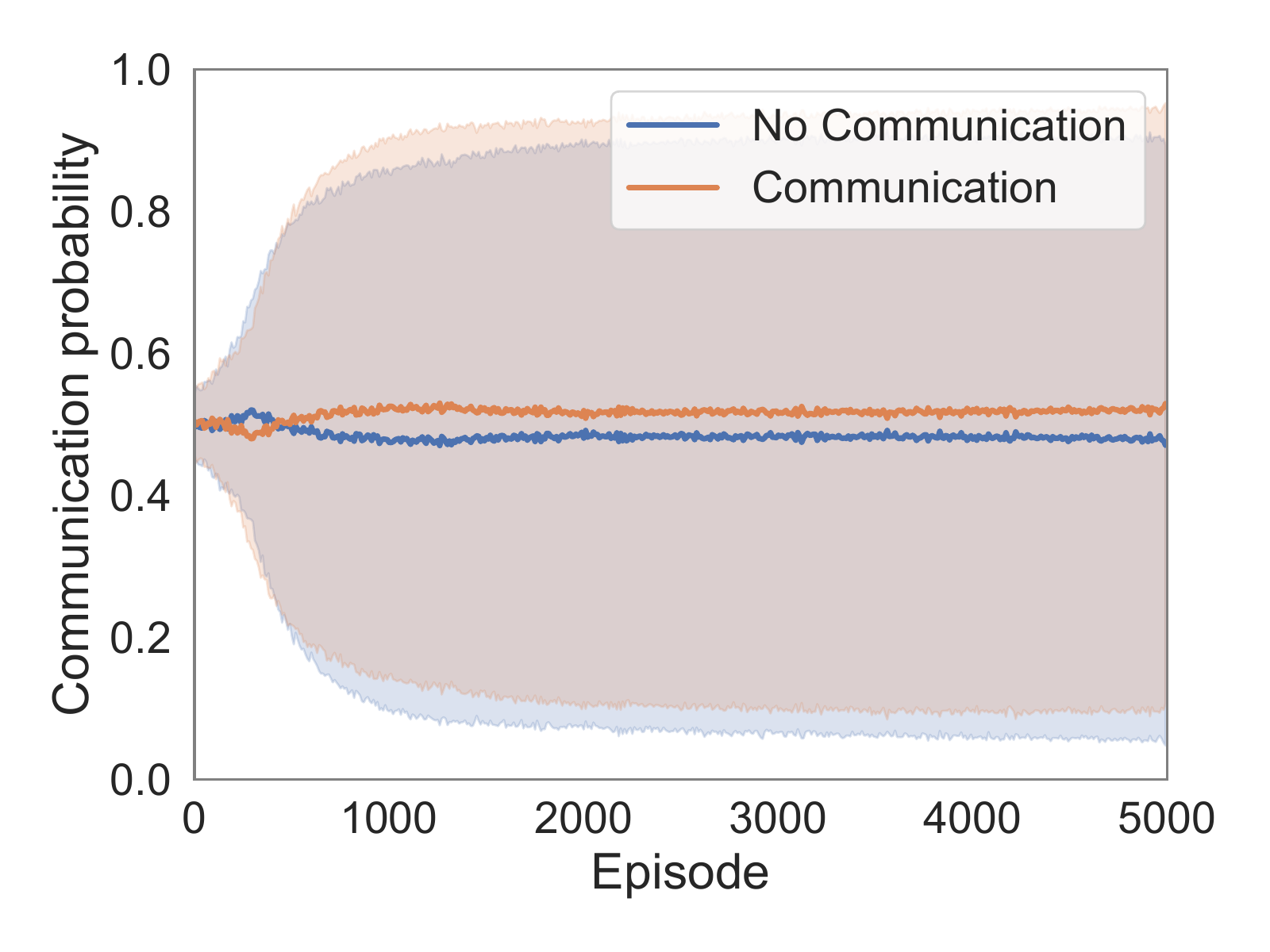}\label{fig:g5-A2-com-opt-coop-policy} }}%
    \caption{The communication probabilities for agent 2 when learning with hierarchical cooperative policy communication.}%
    \label{fig:opt-coop-policy-A2-com}%
\end{figure}

In Figures \ref{fig:opt-coop-policy-A1-com} and \ref{fig:opt-coop-policy-A2-com} we show the communication probabilities for both agents. These figures show similar patterns to the ones observed in other experiments with hierarchical communication. Specifically, agents in games without NE learn to communicate at least part of the time. Because both low-level protocols learn strategies that result in the same utility, the communication strategy does not strictly prefer either protocol. In games with NE, communication can be used to learn CNE where agents cycle through their preferred equilibria. However, as in the previous experiments with hierarchical communication, it is clear that agents have a harder time learning to use it to their advantage and often converge on exclusively utilising one protocol. Lastly, as also mentioned in the previous sections on hierarchical experiments, lower learning rates in the low-level protocols results in increased randomisation in the communication strategies.

\subsection{Summary}
In this section, we presented the empirical results for our communication protocols in a set of benchmark games. It is clear that different behaviour emerges depending on the exact communication protocol that is assumed. In addition, we find that the specific game that is played also introduces subtleties in the results. To help clarify this complexity, we present a concise summary of the key findings in Table \ref{tab:findings-summary}.

\begin{table}[h]
\centering
\begin{tabular}{L{.36\textwidth}|L{.57\textwidth}}
\textbf{Experiment}                                         & \textbf{Summary}                                                                                                                                                                      \\ \thickhline
No communication (Sec. \ref{sec:no-com-exp})                                  & Agents can reach a Nash equilibrium or stable utility after a learning period.                                                                                                        \\
Cooperative action (Sec. \ref{sec:coop-com-exp})                    & Speeds up learning and ensures less divergence from compromise strategies.                                                                                                       \\
Self-interested action (Sec. \ref{sec:comp-com-exp})                & Agents learn a non-stationary policy that cycles through leadership equilibria. In some cases, these cycles are cyclic Nash equilibria.                                               \\
Cooperative policy (Sec. \ref{sec:policy-com-exp})                    & Similar results to cooperative action communication. A benefit is that there is no need to commit to the exact next action. A drawback is less precise coordination to avoid joint-actions which are poor for both agents. \\
                                                            &                                                                                                                                                                                       \\
\textit{Hierarchical communication (Sec. \ref{sec:hierarchical-com-exp})}   & Communication emerges in games without NE when players sufficiently explore the low-level protocols. In games with NE, players learn non-stationary policies or exclusively follow the protocol that performs best in the earlier episodes.                                                    \\ \hline
Cooperative action (Sec. \ref{sec:hierarchical-coop-com-exp})     & Cyclic Nash equilibria can now also occur with cooperative communication.                                                                      \\
Self-interested action (Sec. \ref{sec:hierarchical-comp-com-exp}) & Communication naturally emerges in our self-interested settings. We observed similar results to forced communication, with additional coordination when NE do not exist.                                                                                                                                              \\
Cooperative policy (Sec. \ref{sec:hierarchical-policy-com-exp})     & Similar results to hierarchical cooperative action communication.                                                                                                                    
\end{tabular}
\caption{A summary of the empirical findings. Note that all rows under the italicised hierarchical communication assume the mentioned experiment as the low-level communication protocol.}
\label{tab:findings-summary}
\end{table}

\section{Related Work}
\label{sec:related-work}
\setlength{\parskip}{0.16cm}
Multi-objective multi-agent systems and MONFGs in particular have been considered in a broader context. Below, we provide a non-exhaustive overview of relevant related work. For an in-depth survey on the current state of the art in multi-objective multi-agent systems we refer the interested reader to \cite{radulescu2020multi}. 

Multi-objective normal-form games, also referred to as multi-criteria games, first surfaced in the prominent work by \cite{blackwell1954analog}. Early work on these games focused mostly on analysing relevant solution concepts such as the Pareto equilibrium \cite{shapley1959equilibrium,borm2003structure} and ideal equilibria \cite{voorneveld2000ideal}. Such equilibria remain relevant in more recent work \cite{ismaili2018existence} and are often defined on an expected payoff vector basis. This method closely relates to the SER criterion in the utility-based approach, where expected payoff vectors are considered as well. However, the introduction of utility functions allows for solution concepts that take into account the actual utility agents obtain from these vectors and thus are less limited than ones defined with a purely agnostic approach.

Another prevalent method was to assume only linear utility functions, which allows the vector-valued payoffs to be scalarised a priori \cite{shapley1959equilibrium,corley1985games}. Note that here too, this approach corresponds to a criterion from the utility-based approach. Specifically, it corresponds to ESR as the scalarisation occurs before taking the expectation. This implies that the approaches of scalarising a priori or expected payoff vectors are not always unifiable. Indeed, when the utility functions of the players are non-linear, solution concepts defined on expected payoff vectors such as with mixed strategy Pareto Nash equilibria are not equivalent to scalarising the game a priori and considering equilibria in the resulting game.

More recently, the utility-based approach proposed by \cite{roijers2013survey,roijers2017multi} is being applied with more success as the distinction between the SER and ESR criterion has been made explicit. Given this distinction, \cite{radulescu2020utility} show that these criteria are not generally equivalent and under SER no NE need necessarily exist. Later work managed to prove the existence of an NE when assuming only quasiconcave utility functions \cite{ropke2021nash}. Furthermore, this work has proven several novel properties on the relation between SER and ESR for MONFGs that can be applied in future algorithms.

The problem of learning in MONFGs has also surfaced as an important open question. Work by \cite{radulescu2020utility} explores the adaptations necessary to translate single-objective reinforcement learning algorithms to this setting. Specifically, a Q-learning approach is designed to learn vectorial Q-values that can be used to maximise the utility of expected returns. Later work by \cite{zhang2020opponent} studies this setting by using an actor-critic approach and further highlights the benefits opponent modelling can present in these settings. Lastly, \cite{radulescu2021opponent} studies opponent modelling with opponent awareness and shows that it can drastically alter the learning dynamics in these games. Important to note is that while several ideas from the traditional MARL literature are used in these works, dealing with the multi-objective nature of the proposed settings often requires fundamental changes. In addition, although most work focuses on the deterministic and stateless setting of MONFGs, attempts have been made to advance this to more complex games such as multi-objective stochastic games. Notably, \cite{mannion2017policy} proved that potential-based reward shaping does not alter the Pareto front in this setting and subsequently studies the effect of such reward shaping.

In this work, we assume that utility functions are known and private to the agents throughout all settings. In the real-world however, utility functions might not always be clearly defined. By taking inspiration from advances in utility modelling and preference elicitation, some recent works have attempted to capture the underlying utility function of human subjects to learn better policies \cite{zintgraf2018ordered}. This technique proved reliable and has been used in practice for aiding in decision making with regards to traffic regulation.

While our work more heavily relies on related work from the multi-objective multi-agent literature, there is also a history of Stackelberg games successfully being used in single-objective MARL. Such games in particular are one of the most well known examples of game theory being applied in the real-world, notably being used in security settings \cite{sinha2018stackelberg,pita2009using}. There has also been work using multi-objective Stackelberg games, specifically to model the interactions between a regulator and mining company \cite{sinha2013multi}. Stackelberg games provide a suitable model for these types of situation because regulators need to commit to their actions by default via public legislation. The multi-objective approach on the other hand proved necessary as regulators have two conflicting objectives namely to minimise pollution while still maximising tax-revenues. 

Lastly, communication has been explored in cooperative MARL settings, showing its applicability and benefits \cite{panait2008cooperative,foerster2016learning}. Learning to communicate between competitive (i.e. self-interested) agents on the other hand has long been regarded as ineffective. Recent work however showed that under the right circumstances, communication can arise in competitive two-player games when both agents benefit from it \cite{noukhovitch2021emergent}. 

\section{Conclusion and Future Work}
\label{sec:conclusion}
This work considered the problem of learning in multi-objective normal-form games with non-linear utility functions. This setting is notoriously difficult as Nash equilibria are not guaranteed to exist, which can lead to instability \cite{radulescu2020utility}. To increase coordination between agents, we introduced four novel preference communication protocols. This communication entails agents committing to their preferred actions or policies. Our approaches cover both cooperative as well as self-interested agents. We further contributed an approach where agents are required to learn when to communicate. 

In each setting, we assumed leader-follower dynamics inspired by Stackelberg games to model the communication. These settings, together with a baseline of agents without communication, were subsequently evaluated on five MONFGs. The results on our baseline show that agents in games without NE are able to obtain a stable utility. As no stable joint policy exists, agents continuously go through cycles of policies which results in a stable utility over time. We note that this cyclical behaviour is also known to occur when learning in single-objective games \cite{anthony2020learning}. On the other hand, agents in games with NE are able to consistently converge to these. 

Our first contribution considered agents following a cooperative protocol with the aim of optimising for a single joint policy. In every episode, the leader commits to playing an action sampled from their current policy. The follower anticipates this action by updating their policy in the direction of a best response. Our findings showed that agents in such settings get a boost in learning speed in games with NE and diverge less from the compromise strategy in games without NE.

Next, in the self-interested setting agents were free to learn a non-stationary policy conditioned on their current role and perceived communication. In practice, this means that agents learn a different policy while leading and a different best response policy for each message that can be observed while following. In this setting, we observe for the first time the occurence of cyclic policies and cyclic Nash equilibria in MONFGs.

Our third contribution again equipped agents with a cooperative protocol. This time, rather than simply committing to the next action, the leader communicates their entire current policy. The follower uses the opponent policy to marginalise over a joint-action Q-table in order to calculate expected Q-values. The expected Q-values are then used to update their policy before action selection. In our experiments, we generally observed a boost in learning speed as was the case with cooperative action communication. 

Our final contribution introduced a hierarchical approach to communication with the goal of studying what dynamics lead to the emergence of communication. In the top-level, each agent learns a policy that decides whether to communicate or not. In the lower level, a no communication and communication protocol are learned. Which lower level protocol is used in any given round gets decided by the top-level policy. We found that allowing agents to learn non-stationary policies can give rise to cyclic equilibria, whether with cooperative or self-interested dynamics. In games without NE, communication emerges when both protocols are sufficiently learned. In games with NE, agents can learn to cycle through their preferred NE by alternating between protocols. In other cases, agents adapted their communication strategy to exclusively follow the protocol that performed best in early episodes.

For future work, we propose several different directions. First of all, we aim to extend our set of benchmarks to MONFGs with larger action spaces and including noisy reward signals. Additionally, we aim to design a self-interested setting where agents share their entire policy. As stated earlier, this is not possible with our current approach, as we would need to learn a best-response policy to each possible policy. This problem could potentially be solved by using deep neural networks which have previously been applied with success in both multi-objective \cite{mossalam2016multiobjective} as well as multi-agent reinforcement learning \cite{tampuu2017multi}.

The final direction that we wish to explore are conceptually more complex games such as multi-objective stochastic games. By introducing stochasticity and statefulness, independent agents will struggle to learn optimal strategies \cite{bowling2002multi}. In such settings, communication can provide a benefit to increase coordination and arrive at better policies. Furthermore, this setting has also been understudied from a theoretical standpoint. It is unclear if or in what situations NE exist. Previous work on MONFGs has shown that theoretical results from single-objective game theory do not necessarily carry over to the multi-objective setting \cite{radulescu2020utility}. We thus also highlight the need for explicit work on studying solution concepts for multi-objective game theory. 

\section{Acknowledgements}
The first author is supported by the Research Foundation – Flanders (FWO), grant number 1197622N. This research was supported by funding from the Flemish Government under the ``Onderzoeksprogramma Artifici\"{e}le Intelligentie (AI) Vlaanderen'' program.

\section*{Conflict of interest}
The authors declare that they have no conflict of interest.

\bibliographystyle{plainnat}
\bibliography{bibliography}  






\end{document}